\documentclass[12pt]{book}
\usepackage[english]{babel}
\usepackage{epsf,fancyhdr}
\usepackage{times}
\usepackage{latexsym} 
\usepackage{amsmath}
\usepackage{amssymb}
\usepackage{version}

\makeatletter

\renewcommand{\bibname}{Bibliography}
\def\thebibliography#1{\vskip10mm{\Large\textbf\bibname}

\list
 {\arabic{enumi}.}{\settowidth\labelwidth{[#1]}
 \leftmargin\labelwidth
 \advance\leftmargin\labelsep
 \usecounter{enumi}}
 \def\newblock{\hskip .11em plus .33em minus .07em}
 \sloppy\clubpenalty4000\widowpenalty4000
 \sfcode`\.=1000\itemsep -1pt\relax}

\renewcommand{\sectionmark}[1]{\markright{\thesection\ #1}}

\def\section{\@startsection{section}{1}{\z@}{-3.5ex plus-1ex minus
    -.2ex}{7mm}{\reset@font\large\bf\raggedright}}

\def\thesection{\arabic{section}.}

\@addtoreset{equation}{section}
\renewcommand{\theequation}%
{\thesection\arabic{equation}}

\def\ps@logunov{\let\@mkboth\@gobbletwo
 \def\@oddhead{}%
 \def\@oddfoot{{}\hfil \rm\thepage}%
 \def\@evenhead{}%
 \def\@evenfoot{\rm\thepage \hfil {}}%
 \def\sectionmark##1{}\def\subsectionmark##1{}}

\makeatother

\newcommand{\grad}{\mathop{\rm grad}\nolimits}
\newcommand{\rot}{\mathop{\rm rot}\nolimits}
\newcommand{\arcsh}{\mathop{\rm arcsh}\nolimits}

\newcommand{\be}{\begin{equation}}
\newcommand{\ee}{\end{equation}}
\newcommand{\ba}{\begin{eqnarray}}
\newcommand{\ea}{\end{eqnarray}}
\newcommand{\pa}{\partial}
\let\f\frac

\newcommand{\ds}{\displaystyle}

\begin{document}
\newpage
\thispagestyle{empty}
\addtolength{\leftmargin}{0.6cm}

\begin{flushleft}
{\LARGE A.A.\,Logunov}
\end{flushleft}
\vspace*{45mm}

\begin{center}
{\LARGE HENRI POINCAR\'{E}}\\[2mm]
{\large AND}\\[2mm]
{\LARGE RELATIVITY THEORY}\\[25mm]
{\it Translated by G.\,Pontecorvo and V.O.\,Soloviev\\
edited by V.A.\,Petrov}
\end{center}
\newpage
\thispagestyle{empty}
\noindent
\thispagestyle{empty}

\vspace*{15mm}

\noindent
{\normalsize\bf Logunov A.\,A.}\\
\hspace*{5mm} Henri Poincar'{e} and relativity theory. -- M.: Nauka, 2005. -- 252 pp.\\[2mm]
{\footnotesize
\hspace*{5mm}The book presents ideas by H.\,Poincar\'{e} and H.\,Minkowski according to those the essence and the main
content of the relativity theory are the following: the space and time form а unique four-dimensional
continuum supplied by the pseudo-Euclidean geometry. All physical processes take place just in this four-
dimensional space. Comments to works and quotations related to this subject by L.\,de Broglie, P.\,A.\,M.\,Dirac, A.\,Einstein, V.\,L.\,Ginzburg, S.\,Goldberg, P.\,Langevin, H.\,A.\,Lorentz, L.\,I.\,Mandel'stam, 
H.\,Min\-kowski, A.\,Pais, W.\,Pauli, M.\,Planck, A.\,Sommerfeld and H.\,Weyl are given in the book. It is also
shown that the special theory of relativity has been created not 
by A.\,Einstein only but even to а greater
extent by H.\,Poincar\'{e}. \\
\hspace*{5mm}The book is designed for scientific workers, post-graduates and upper-year
students majoring in theoretical physics.
}

\vspace*{15mm}

\noindent
ISBN\hspace*{56mm} All rights reserved\\
\hspace*{60mm}\copyright\quad A.\,A.\,Logunov \\
 \hspace*{60mm}\copyright\quad Translated by G.\,Pontecorvo \\
\hspace*{68mm} and V.\,O.\,Soloviev,\\
\hspace*{68mm} edited by V.\,A.\,Petrov\\
\hspace*{60mm}\copyright\quad Designed by Nauka\\ 
\hspace*{68mm}Pubishers, 2005.\\

\newpage
\addtolength{\leftmargin}{0.6cm}
\setcounter{page}{3}
\begin{flushright}
{\bf \textit{Devoted to 150th birthday of Henri Poincar\'e --- }}\\
{\bf \textit{the greatest mathematician, mechanist,}}\\
{\bf \textit{theoretical physicist}}
\end{flushright}
\vspace*{2mm}
\section*{Preface}\addcontentsline{toc}{section}{Preface}
\textbf{The special theory of relativity} {\bf \textit{``resulted from the joint efforts of
a group of great researchers: 
Lorentz,\index{Lorentz} 
Poincar\'{e},
\index{Poincar\'e} 
Einstein,
\index{Einstein} 
Minkowski}''}
\index{Minkowski} 
 (\textbf{Max Born}).\index{Born} 

{\bf \textit{``Both Einstein
\index{Einstein} and Poincar\'{e},
\index{Poincar\'e} took their stand
on the preparatory work of 
H.\,A.\,Lorentz\index{Lorentz}, who 
had already come quite
close to  the result, 
without however quite reaching it.
In the agreement between the
results of the methods followed independently of each other by 
Einstein\index{Einstein} and
Poincar\'{e}\index{Poincar\'e} I 
discern a deeper significance
of a harmony between the mathematical
method and analysis by means of conceptual
experiments (Gedankenexperimente), which rests on general features
of physical experience''}} (\textbf{W.\,Pauli, 1955}).\index{Pauli}

H.\,Poincar\'e\index{Poincar\'e}, being based upon the relativity
principle\index{relativity principle} formulated by him for all physical phenomena and upon the 
Lorentz\index{Lorentz}  work, has discovered and formulated everything that composes  the essence of
the special theory of relativity. 
A.\,Einstein\index{Einstein} was coming to the theory
of relativity from the side of relativity principle\index{relativity principle} formulated earlier by
H.\,Poincar\'e\index{Poincar\'e}.  At that he relied upon ideas by
H.\,Poincar\'e\index{Poincar\'e} on definition of the simultaneity\index{simultaneity} of
events occurring in different spatial points by means of the light signal. Just for this reason he introduced an
additional postulate --- the constancy of the velocity of 
light.\index{principle of constancy of velocity of light}
This book presents a comparison of the article by 
A.\,Einstein\index{Einstein} of 1905
with the articles by H.\,Poincar\'e\index{Poincar\'e} and clarifies what is the
\textbf{new} content contributed by each of them.  Somewhat later H.\,Min\-kow\-ski
\index{Minkowski} further developed Poincar\'{e}'s\index{Poincar\'e} approach. Since
Poincar\'{e}'s\index{Poincar\'e} approach was more general and profound, our presentation
will precisely follow Poincar\'{e}.\index{Poincar\'e} 

\markboth{thesection\hspace{1em}}{Preface}
According to Poincar\'{e}\index{Poincar\'e}  and Minkowski,
\index{Minkowski} the essence of relativity theory consists in the following: {\bf the special theory of relativity is
the pseudo-Euclidean geometry of space-time.\index{pseudo-Euclidean geometry of space-time} All physical processes
take place just in such a space-time}. The consequences of this postulate are energy-momentum and angular momentum
conservation laws, the existence of inertial reference 
systems,\index{inertial reference systems} the relativity
principle\index{relativity principle} for all physical phenomena, 
Lorentz\index{Lorentz} transformations,\index{Lorentz transformations} the constancy of  the velocity of light\index{principle of
constancy of velocity of light} in Galilean coordinates
\index{Galilean (Cartesian) coordinates} of the inertial
frame, the retardation of time, the Lorentz\index{Lorentz} 
contraction\index{Lorentz contraction}, the possibility to exploit non-inertial reference
systems,\index{non-inertial reference systems} the clock 
paradox,\index{``clock paradox''} the Tho\-mas\index{Thomas} precession, the Sagnac\index{Sagnac} effect, and so on. Series of fundamental consequences have been obtained on the base of this
postulate and the quantum notions, and the quantum field theory has been constructed. The preservation
(form-invariance) of physical equations in all inertial reference 
systems\index{inertial reference systems} means
that all \textbf{physical processes} taking place in these systems under the same conditions are
\textbf{identical}. Just for this reason all \textbf{natural standards} are {\bf the same} in all inertial reference
systems\index{inertial reference systems}.

The author expresses profound gratitude to Academician of the Russian Academy of Sciences Prof.
S.\,S.\,Ger\-shtein,\index{Gershtein}  Prof. V.\,A.\,Pet\-rov,\index{Petrov}
Prof. N.\,E.\,Tyurin,\index{Tyurin} Prof.
Y.\,M.\,Ado,\index{Ado} senior research associate 
A.\,P.\,Samokhin\index{Samokhin} who read the manuscript and made a num\-ber of va\-lu\-able com\-ments,
and, al\-so, to G.\,M.\,Aleksandrov for sig\-ni\-fi\-cant work in
preparing the manuscript for publication and completing Author and Subject Indexes.

\vspace*{5mm} \noindent
\hspace*{8.6cm} {\it A.A.\,Logunov} \\
\hspace*{8.5cm} {\it January 2004}

\newpage
\markboth{thesection\hspace{1em}}{}
\section{Euclidean geometry}

In the third century BC {\bf Euclid\index{Euclid} published a treatise on mathematics, the {\sf
``Elements''}}, in which he summed up the preceding de\-ve\-lop\-ment of {\bf mathematics in antique Greece}. It
was precisely in this work that the geometry of our three-dimensional space --- Euclidean\index{Euclid} geometry --- was formulated.

This happened to be a most important step in the development of both mathematics and physics. The point is that
geometry ori\-gi\-na\-ted from observational data and practical experience, i.\,e. it arose via the study of Nature.
But, since all natural phenomena take place in space and time, the importance of geometry for physics cannot be
overestimated, and, moreover, geometry is actually a part of physics.

{\bf In the modern language of mathematics the essence of 
Eu\-cli\-de\-an\index{Euclid}
geometry\index{Euclidean geometry} is determined by the 
Py\-tha\-go\-re\-an\index{Pythag\'oras}
theorem}.\index{Pythagorean theorem} In accordance with the 
Pythagorean\index{Pythag\'oras}
theorem, the distance of a point with Cartesian coordinates $x, y, z$ from the origin of the re\-fe\-rence system
is determined by the formula \be
\ell^2=x^2+y^2+z^2,\;
\ee or in differential form, the distance between two infinitesimally close points is \be
(d\ell)^2=(dx)^2+(dy)^2+(dz)^2.
\ee Here $dx, dy, dz$ are differentials of the Cartesian coordinates. 
Usu\-al\-ly,  the proof of the
Pythagorean\index{Pythag\'oras} theorem is based on 
Euclid's\index{Euclid}
axioms, but it turns out to be that it can actually be considered a definition of Euclidean\index{Euclid}
geometry. Three-dimensional space, determined by Euclidean
\index{Euclid} geometry,
possesses the properties of homogeneity and isotropy. This means that there exist no singular points or singular
directions in Euclidean\index{Euclid} geometry. By performing transformations of coordinates
from one Cartesian reference system, $x,y,z$, to another, $x^\prime,y^\prime,z^\prime$, we obtain \be
\ell^2=x^2+y^2+z^2=x^\prime{^2}+y^\prime{^2}+z^\prime{^2}.
\ee \markboth{thesection\hspace{1em}}{1. Euclidean geometry} This means that the square distance $\ell^2$ is an
invariant, while its projections onto the coordinate axes are not. We especially note this obvious circumstance,
since it will further be seen that such a situation also takes place in 
four-dimensional\index{four-dimensional space-time} space-time, so, consequently, depending on the choice of reference system in space-time the
projections  onto spatial and time axes will be relative. Hence arises the relativity of time and length. But this
issue will be dealt with later.

Euclidean\index{Euclid} geometry\index{Euclidean geometry} became a composite part of
Newtonian\index{Newton} mechanics. For about two thousand years 
Euclidean\index{Euclid} geometry was thought to be the unique  and unchangeable geometry, in spite of the rapid
development of mathematics, mechanics, and physics.

{\bf It was only at the beginning of the 19-th century that the Russian
mathematician Nikolai Ivanovich
Lo\-ba\-chev\-sky\index{Lobachevsky} 
made the revolutionary step ---
a new geometry was constructed --- the
Lobachevsky\index{Lobachevsky} 
geometry.\index{Lobachevsky geometry}
Somewhat
later it was discovered by
the Hungarian ma\-the\-ma\-ti\-ci\-an
Bolyai}.\index{Bolyai} 

About 25 years later Riemannian\index{Riemann} 
geometries\index{Riemannian geometry} were developed by the German ma\-the\-ma\-ti\-ci\-an Riemann.\index{Riemann}  Numerous geometrical constructions arose. As new geometries came
into being the issue of the geometry of our space was raised. What kind was it? Euclidean\index{Euclid} or non-Euclidean?

\newpage
\markboth{thesection\hspace{1em}}{}
\section{Classical Newtonian mechanics}

All natural phenomena proceed in space and time. Precisely for this
reason, in formulating the
laws of mechanics\index{Newton's laws}
in the 17-th cen\-tury, Isaac
Newton\index{Newton} 
first of all defined these concepts:
\begin{quote}
{\it\hspace*{5mm}``Absolute Space,\index{absolute space}
 in its own nature, without regard
to any thing external, remains always similar and im\-mo\-vable''.}
\end{quote}\index{absolute space}
\begin{quote}
{\it\hspace*{5mm}``Absolute, True, and Mathematical Time, of 
itself, and from its own nature flows equably without regard to
any thing external, and by another name is called Duration''.}
\end{quote}\index{absolute time}

As the geometry of three-dimensional space Newton
\index{Newton} actually applied
Euclidean\index{Euclid} geometry, and he chose a Cartesian reference system with its origin at
the center of the Sun, while its three axes were directed toward distant stars. Newton\index{Newton} considered precisely such a reference system to be ``motionless''. The introduction of absolute
motionless space and of absolute time\index{absolute time} turned out to be extremely fruitful at the time.

The first law of mechanics, or the law of inertia, was formulated 
by Newton\index{Newton} 
as follows:
\begin{quote}
{\it\hspace*{5mm}``Every body perseveres in its state of rest, or of
uniform motion in a right line, unless it is compelled to change that
state by forces impressed the\-re\-on''.}
\end{quote}
\markboth{thesection\hspace{1em}}{2. Classical Newtonian mechanics} The law of inertia was first discovered by
Galileo.\index{Galilei} If, in motionless space, one defines a Cartesian reference
system, then, in accordance with the law of inertia, a solitary body will move along a trajectory determined by
the following equations: \be
x=v_xt,\qquad y=v_yt,\qquad z=v_zt.
\ee
Here, $v_x, v_y, v_z$ are the constant velocity projections, their
values may, also, be equal to zero.

In the book {\sf ``Science and Hypothesis''}
H.\,Poincar\'{e}\index{Poincar\'e} 
formulated
the following general principle:
\begin{quote}
{\it\hspace*{5mm}``The acceleration of a body depends only
on its position and that of neighbouring bodies, and on their 
velocities. Mathematicians would say that
the movements of all the material molecules of the universe\index{Universe} depend on 
differential equations of the
seconal order. To make it clear that this is really а generalisation of 
the law of inertia we may again have
recourse to our imagination. The law of inertia, as I have said above, 
is not imposed on us \'{а} priori; other
laws would be just as compatible with the principle of sufficient 
reason. If а body is not acted upon by a force, instead of supposing
that its velocity is unchanged we may suppose that its position
or its acceleration is unchanged. Let us for moment suppose that one
of these two laws is a law of nature, and substitute it for the law of 
inertia: what will be the natural generalisation? А moment's reflection 
will show us. In the first case, we
may suppose that the velocity of а body depends only on its position 
and that of neighbouring bodies; in
the second case, that the variation of the acceleration of а body 
depends only on the position of the body
and of neighbouring bodies, on their velocities and accelerations; or, 
in mathematical terms, the
differential equations of the motion would be of the first order in the 
first case and of the third order in
the second''.}
\end{quote}

Newton\index{Newton} 
formulated the second
law\index{Newton's laws}
of mechanics as follows:
\begin{quote}
{\it\hspace*{5mm}``The alteration of motion is ever proportional to
the motive force impressed; and is made in the di\-rection of the
right line in which that force is impressed''.}
\end{quote}

And, finally, the
Newton's\index{Newton} 
third law\index{Newton's laws}
of mechanics:
\begin{quote}
{\it\hspace*{5mm}``To every Action there is always opposed an equal
Reaction: or the mutual actions of two bodies upon each other are
always equal, and directed to contrary parts''.}
\end{quote}

On the basis of these laws of mechanics, in the case of central forces,
the equations for a system of two particles in a reference system ``at
rest'' are:
\vspace*{-2mm}
\ba
&&M_1\ds\f{d^2\vec{r}_1}{dt^2}=F(\vert\vec{r}_2-\vec{r}_1\vert)
\ds\f{\vec{r}_2-\vec{r}_1}{\vert\vec{r}_2-\vec{r}_1\vert}\,,\nonumber\\[-2mm]
\label{2.2}\\[-1mm]
&&M_2\ds\f{d^2\vec{r}_2}{dt^2}=-F(\vert\vec{r}_2-\vec{r}_1\vert)
\ds\f{\vec{r}_2-\vec{r}_1}{\vert\vec{r}_2-\vec{r}_1\vert}\,.\nonumber
\ea
Here $M_1$ and $M_2$ are the respective masses of the first and second
particles, $\vec{r}_1$ is the vector radius of the first particle,
$\vec{r}_2$ is the vector radius of the second particle. The function
$F$ reflects the character of the forces acting between bodies.

In Newtonian\index{Newton} 
mechanics, mostly forces of two types are con\-si\-de\-red: of
gravity and of elasticity.

For the forces\index{Newton's force of gravity}
of
Newtonian\index{Newton} 
gravity
\be
F(\vert\vec{r}_2-\vec{r}_1\vert)=
G\f{M_1M_2}{|\vec{r}_2-\vec{r}_1|^2},
\ee
$G$ is the gravitational constant.

For elasticity forces
Hooke's\index{Hooke} 
law\index{Hooke's law}
is
\be
F(\vert\vec{r}_2-\vec{r}_1\vert)=k\vert\vec{r}_2-\vec{r}_1\vert,
\ee
$k$ is the elasticity coefficient.

Newton's\index{Newton} 
equations are written in vector form, and, con\-seq\-u\-ent\-ly, they
are independent of the choice of three-dimensional reference system.
From equations (2.2) it is seen that the momentum of a closed system is
conserved.

As it was earlier noted,
Newton\index{Newton} 
considered equations (2.2) to hold valid
only in a reference system at rest. But, if one takes a reference system
moving with respect to the one at rest with a constant velocity $\vec{v}$
\be
\vec{r}\;^\prime=\vec r-\vec{v}\,t,
\ee it turns out that equations (2.2) are not altered, i.\,e. {\bf they remain form-invariant, and this means that
no mechanical phenomena could permit to ascertain
whether we are in a state of rest or of uniform and rectilinear motion. This is the essence of the relativity
principle\index{relativity principle} first discovered by Galileo.
\index{Galilei} The
trans\-for\-ma\-tions {\rm (2.5)} have been termed Galilean}.
\index{Galilei} 

Since the velocity $\vec{v}$ in (2.5) is arbitrary, there exists an infinite number of reference systems, in which
the equations retain their form. This means, that in each reference system the law of inertia holds valid. If in
any one of these reference systems a body is in a state of rest or in a state of uniform and rectilinear motion,
then in any other reference system, related to the first by transformation (2.5), it will also be either in a
state of uniform rectilinear motion or in a state of rest.

{\bf All such reference systems have been termed inertial.\index{inertial reference systems} The
principle\index{relativity principle} of relativity consists in conservation of the form of the equations of
mechanics in any inertial reference system}.\index{inertial reference systems} We are to emphasize that {\bf in
the base of definition of an inertial reference system lies the law of inertia by Galileo.} According to it in the
absence of forces a body motion is described by linear functions of time.

But how has an inertial reference system\index{inertial reference systems} to be defined?
New\-to\-ni\-an\index{Newton} mechanics gave no answer to this question. Nevertheless, the
reference system chosen as such an inertial 
system\index{inertial reference systems} had its origin at the center
of the Sun, while the three axes were directed toward distant stars.

In classical
Newtonian\index{Newton} 
mechanics time is independent of the choice of
reference system, in other words, three-dimensional space and time are
separated, they do not form a unique four-dimensional continuum.

Isaac
Newton's\index{Newton} 
ideas concerning
absolute space\index{absolute space}
and
absolute motion\index{absolute motion}
were criticized in the 19-th century by Ernst
Mach.\index{Mach} 
Mach wrote:
\begin{quote}
{\it\hspace*{5mm}``No one is competent to predicate things
about absolute 
space\index{absolute space} and absolute
motion;\index{absolute motion} 
they are pure things of thought, pure mental constructs, that
cannot be produced in experience''.}
\end{quote}
And further:
\begin{quote}
{\it\hspace*{5mm} ``Instead, now, of referring a moving body K to space 
(that is to say to a system of coordinates) let us view directly its
relation to the  
bodies of the universe, by which alone a 
system of coordinates can be determined.

\ldots even in the simplest case, in which apparently we deal with the
mutual action of only 
{\bf\textit{two}}  masses, the neglecting of the rest of the world is
{\bf\textit{impossible}}. 
\ldots If a body rotates with respect to the sky of motionless stars, then there arise
centrifugal forces, while if it rotates around {\bf\textit{another}} $\>$ body, instead of the sky of motionless
stars, no centrifugal forces will arise. I have nothing against calling the first revolution {\bf\textit{abslute,}} $\>$ if only one does not forget that this signifies nothing but revolution {\bf\textit{relative}} $\>$
to the sky of motionless stars''.}
\end{quote}
Therefore
Mach\index{Mach} 
wrote:
\begin{quote}
{\it``\ldots there is no necessity for relating the Law of inertia to some
special absolute space''.}\index{absolute space}
\end{quote}

All this is correct, since
Newton\index{Newton} 
did not define the relation of an
inertial reference system\index{inertial reference systems}
to the distribution of matter, and, actually,
it was quite impossible, given the level of physics development
at the time. By
the way,
Mach\index{Mach} 
also did not meet with success. But his criticism was useful,
it drew the attention of scientists to the analysis of the main concepts
of physics.

Since we shall further deal with field concepts, it will be useful to consider the methods of analytical mechanics
developed during the 18-th and 19-th centuries. Their main goal, set at the time, consisted in finding the most
general formulation for classical me\-cha\-nics. Such research turned out to be extremely important, since it gave
rise to methods that were later quite readily generalized to systems with an infinite number of degrees of
free\-dom. Precisely in this way was a serious theoretical start created, that was successfully  used of in the
19-th and 20-th centuries.

In his {\sf ``Analytic Mechanics''}, published in 1788, Joseph
La\-grange\index{Lagrange} 
obtained his famous equations. Below we shall present their derivation.
In an
inertial reference system,\index{inertial reference systems}
Newton's\index{Newton} 
equations for a set of $N$
material points moving in a
potential\index{potential}
field $U$ have the form
\be
m_\sigma\f{d\vec v_\sigma}{dt}=
-\f{\pa U}{\pa \vec r_\sigma}, \quad \sigma=1, 2,\ldots, N.
\ee
In our case the force $\vec f_\sigma$ is
\be
\vec f_\sigma=-\f{\pa U}{\pa \vec r_\sigma}.
\ee To determine the state of a mechanical system at any moment of time it is necessary to give the coordinates
and velocities of all the material points at a certain moment of time. Thus, the state of a mechanical system is
fully determined by the coordinates and velocities of the material points. In a Cartesian reference system
Eqs.~(2.6) assume the form \be m_\sigma\f{dv_\sigma^1}{dt}=f_\sigma^1,\quad
m_\sigma\f{dv_\sigma^2}{dt}=f_\sigma^2,\quad
m_\sigma\f{dv_\sigma^3}{dt}=f_\sigma^3.
\ee

If one passes to another
inertial reference system\index{inertial reference systems}
and makes use of
coordinates other than Cartesian, then it is readily seen that the
equations written in the new coordinates differ essentially in form from
equations (2.8).
Lagrange\index{Lagrange} 
found for
Newton's\index{Newton} 
mechanics such a covariant formulation for the equations
 of motion that they retain their form, when
transition is made to new variables.

Let us introduce, instead of coordinates $\vec r_\sigma$, new {\bf generalized coordinates}\index{generalized
coordinates} $q^\lambda,\,\lambda=1, 2,\ldots ,n$, here $n=3N$. Let us assume relations \be
\vec r_\sigma=\vec r_\sigma(q_1,\ldots ,q_n,t).
\ee After scalar multiplication of each equation (2.6) by vector \be
\f{\pa \vec r_\sigma}{\pa q_\lambda}
\ee and performing addition we obtain \be m_\sigma\f{d\vec v_\sigma}{dt}\cdot\f{\pa \vec r_\sigma}{\pa q_\lambda}
=-\f{\pa U}{\pa \vec r_\sigma}\cdot\f{\pa \vec r_\sigma}{\pa q_\lambda}, \quad
\lambda=1, 2,\ldots, n.
\ee
Here summation is performed over identical indices $\sigma$.\\
We write the left-hand part of equation (2.11) as
\be
\f{d}{dt}\left[m_\sigma \vec v_\sigma
\f{\pa \vec r_\sigma}{\pa q_\lambda}\right]
-m_\sigma \vec v_\sigma \f{d}{dt}
\left(\f{\pa \vec r_\sigma}{\pa q_\lambda}\right).
\ee
Since
\be
\vec v_\sigma=\f{d\vec r_\sigma}{dt}
=\f{\pa \vec r_\sigma}{\pa q_\lambda}\dot q_\lambda
+\f{\pa \vec r_\sigma}{\pa t},
\ee
hence, differentiating (2.13) with respect to $\dot q_\lambda$ we obtain
the equality
\be
\f{\pa \vec r_\sigma}{\pa q_\lambda}
=\f{\pa \vec v_\sigma}{\pa \dot q_\lambda}.
\ee
Differentiating (2.13) with respect to $q_\nu$ we obtain
\be
\f{\pa \vec v_\sigma}{\pa q_\nu}
=\f{\pa^2 \vec r_\sigma}{\pa q_\nu \pa q_\lambda}\dot q_\lambda
+\f{\pa^2 \vec r_\sigma}{\pa t\pa q_\nu}.
\ee But, on the other hand, we have \be \f{d}{dt}\left(\f{\pa \vec r_\sigma}{\pa q_\nu}\right) =\f{\pa^2 \vec
r_\sigma}{\pa q_\nu \pa q_\lambda}\dot q_\lambda
+\f{\pa^2 \vec r_\sigma}{\pa t\pa q_\nu}.
\ee
Comparing (2.15) and (2.16) we find
\be
\f{d}{dt}\left(\f{\pa \vec r_\sigma}{\pa q_\nu}\right)
=\f{\pa \vec v_\sigma}{\pa q_\nu}.
\ee In formulae (2.13), (2.15) and (2.16) summation is performed over identical indices $\lambda$.

Making use of equalities (2.14) and (2.17) we represent ex\-pression (2.12)
in the form
\be
\f{d}{dt}\left[\f{\pa}{\pa \dot q_\lambda}
\left(\f{m_\sigma v_\sigma^2}{2}\right)\right]
-\f{\pa}{\pa q_\lambda}\left(\f{m_\sigma v_\sigma^2}{2}\right).
\ee Since (2.18) is the left-hand part of equations (2.11) we obtain 
Lagrangian\index{Lagrange} 
equations \be \f{d}{dt}\left(\f{\pa T}{\pa \dot q_\lambda}\right) -\f{\pa T}{\pa q_\lambda}
=-\f{\pa U}{\pa q_\lambda},\;\lambda=1, 2,\ldots, n.
\ee Here $T$ is the kinetic energy of the system of material points \be
T=\f{m_\sigma v_\sigma^2}{2},
\ee summation is performed over identical indices $\sigma$. If one introduces the Lagrangian\index{Lagrange}
function $L$\index{Lagrangian density} as follows \be
L=T-U,
\ee
then the
Lagrangian\index{Lagrange} 
equations\index{Lagrangian equations}
assume the form
\be
\f{d}{dt}\left(\f{\pa L}{\pa \dot q_\lambda}\right)
-\f{\pa L}{\pa q_\lambda}=0,\;\lambda=1, 2,\ldots, n.
\ee

The state of a mechanical system is fully determined by the generalized 
coordinates\index{generalized coordinates}
and velocities. The form of  Lagrangian\index{Lagrange} equations (2.22) is
independent of the choice of {\bf generalized co\-or\-di\-na\-tes}. Although these equations are totally
equivalent to the set of equations (2.6), this form of the equations of classical mechanics, however, turns out to
be extremely  fruitful,
 since it opens up the possibility of
its generalization to phenomena which lie far beyond the limits of classical mechanics.

The most general formulation of the law of motion of a me\-cha\-ni\-cal system
is given by the {\bf principle of least action} (or the principle of
stationary action). The action is composed as follows
\be
S=\int\limits_{t_1}^{t_2} L(q,\dot q)dt.
\ee The integral (functional) (2.23) depends on the behaviour  of func\-ti\-ons $q$ and $\dot q$ within the given
limits. Thus, these functions are functional arguments of the integral (2.23). The least action prin\-ciple is
written in the form \be
\delta S=\delta \int\limits_{t_1}^{t_2}L(q,\dot q)dt=0.
\ee The equations of motion of mechanics are obtained from (2.24) by varying  the integrand expression \be
\int\limits_{t_1}^{t_2}\left( \f{\pa L}{\pa q}\delta q
+\f{\pa L}{\pa \dot q}\delta \dot q\right)dt=0.
\ee Here $\delta q$ and $\delta \dot q$ represent infinitesimal variations in the form of the functions. The
variation commutes with differentiation, so \be
\delta\dot q=\f{d}{dt}(\delta q).
\ee
Integrating by parts in the second term of (2.25) we obtain
\be
\delta S=\f{\pa L}{\pa\dot q}\delta q
\Bigg|_{t_1}^{t_2}+\int\limits_{t_1}^{t_2}
\left(\f{\pa L}{\pa q}-
\f{d}{dt}\cdot\f{\pa L}{\pa\dot q}\right)
\delta qdt=0.
\ee
Since the variations $\delta q$ at points $t_1$ and $t_2$ are zero,
expression (2.27) assumes the form
\be
\delta S=\int\limits_{t_1}^{t_2}\left(\f{\pa L}{\pa q}
-\f{d}{dt}\cdot\f{\pa L}{\pa\dot q}\right)
\delta qdt=0.
\ee The variation $\delta q$ is arbitrary within the interval of integration, so, by virtue of the main lemma of
variational calculus, from here the {\bf necessary condition for an extremum} follows in the form of the equality
to zero of the variational derivative\index{variational derivative} 
\be \f{\delta L}{\delta q}=\f{\pa L}{\pa q}
-\f{d}{dt}\left(\f{\pa L}{\pa\dot q}\right)=0.
\ee Such equations were obtained by Leonard Euler\index{Euler} in the course of
development of variational calculus. For our choice of function $L$, these equations in accordance with (2.21)
coincide with the Lagrangian\index{Lagrange} equations.
\index{Lagrangian equations}

From the above consideration it is evident that mechanical mo\-ti\-on satisfying the 
Lagrangian\index{Lagrange} equations provides for extremum of the integral (2.23), and, consequently, the
action has a stationary value.

The application of the Lagrangian\index{Lagrange} 
function\index{Lagrangian density} for describing a mechanical system with a finite number of degrees of freedom turned out to be fruitful,
also, in describing a physical field po\-sse\-ssing an infinite number of degrees of freedom. In the case of a
field, the function $\psi$ describing it depends not only on time, but also on the space coordinates. This means
that, instead of the variables $q_\sigma, \dot q_\sigma$ of a mechanical system, it is necessary to introduce the
variables $\ds\psi (x^\nu),\, \f{\pa \psi}{\pa x^\lambda}$. Thus, the field is considered as a me\-cha\-ni\-cal
system with an infinite number of degrees of freedom. We shall see further (Sections 10 and 15) how the principle
of stationary action is applied in electrodynamics and classical field theory.

The formulation of classical mechanics within the framework of 
Hamiltonian\index{Hamilton} 
approach has become very important. Consider a certain quantity determined as follows \be
H=p_\sigma\dot q_\sigma - L,
\ee
and termed the
Hamiltonian.\index{Hamilton} \index{Hamiltonian}
In (2.30) summation is performed over identical indices $\sigma$. We define
the {\bf generalized momentum}\index{generalized momentum}
as follows:
\be
p_\sigma=\f{\pa L}{\pa \dot q_\sigma}\,.
\ee
Find the differential of expression (2.30)
\be
dH=p_\sigma d\dot q_\sigma+\dot q_\sigma dp_\sigma
-\f{\pa L}{\pa q_\sigma}dq_\sigma
-\f{\pa L}{\pa\dot q_\sigma}d\dot q_\sigma
-\f{\pa L}{\pa t}dt.
\ee
Making use of (2.31) we obtain
\be
dH=\dot q_\sigma dp_\sigma
-\f{\pa L}{\pa q_\sigma}dq_\sigma
-\f{\pa L}{\pa t}dt.
\ee
On the other hand, $H$ is a function of the independent variables
$q_\sigma,\,p_\sigma$ and $t$, and therefore
\be
dH=\f{\pa H}{\pa q_\sigma}dq_\sigma
+\f{\pa H}{\pa p_\sigma}dp_\sigma
+\f{\pa H}{\pa t}dt.
\ee
Comparing (2.33) and (2.34) we obtain
\be
\dot q_\sigma=\f{\pa H}{\pa p_\sigma},\quad
\f{\pa L}{\pa q_\sigma}=-\f{\pa H}{\pa q_\sigma},\quad
\f{\pa L}{\pa t}=-\f{\pa H}{\pa t}.
\ee These relations were obtained by transition from  independent variables $q_\sigma,\,\dot q_\sigma$ and $t$ to
 independent variables $q_\sigma,\,p_\sigma$ and $t$.

Now, we take into account the
Lagrangian\index{Lagrange} 
equations (2.22)\index{Lagrangian equations}
in relations
(2.35) and obtain the
Hamiltonian\index{Hamilton} 
equations
\be
\dot q_\sigma=\f{\pa H}{\pa p_\sigma},\quad
\dot p_\sigma=-\f{\pa H}{\pa q_\sigma}.
\ee
When the
Hamiltonian\index{Hamilton}\index{Hamiltonian}
$H$ does not depend explicitly on time,
\be
\f{\pa H}{\pa t}=0,
\ee
we have
\be
\f{dH}{dt}=\f{\pa H}{\pa q_\sigma}\dot q_\sigma
+\f{\pa H}{\pa p_\sigma}\dot p_\sigma.
\ee Taking into account equations (2.36) in the above expression, we obtain \be
\f{dH}{dt}=0;
\ee this means that the Hamiltonian\index{Hamilton} remains constant during the
mo\-ti\-on.

We have obtained the
Hamiltonian\index{Hamilton} 
equations\index{Hamiltonian equations}
(2.36) making use of the
Lagrangian\index{Lagrange} 
equations.\index{Lagrangian equations}
But they can be found also directly with the aid
of the least action principle (2.24), if, as $L$, we take, in accordance
with (2.30), the expression
\[
L=p_\sigma\dot q_\sigma-H, \nonumber
\]
\ba
&&\delta S=\ds\int\limits_{t_1}^{t_2}\delta p_\sigma
\left(dq_\sigma-\ds\f{\pa H}{\pa p_\sigma}dt\right)- \nonumber \\
&&-\ds\int\limits_{t_1}^{t_2}\delta q_\sigma \left(dp_\sigma+\ds\f{\pa H}{\pa q_\sigma}dt\right) +\ds
p_\sigma\delta q_\sigma\Bigg|_{t_1}^{t_2}=0.\nonumber \ea Since  variations $\delta q_\sigma$ at the points $t_1$
and $t_2$ are zero, while inside the interval of integration  variations $\delta q_\sigma,\,\delta p_\sigma$ are
arbitrary, then, by virtue of the main lemma of variational calculus, we obtain the Hamiltonian\index{Hamilton} equations\index{Hamiltonian equations}
\[
\dot q_\sigma=\f{\pa H}{\pa p_\sigma},\quad
\dot p_\sigma=-\f{\pa H}{\pa q_\sigma}.
\]
If during the motion the value of a certain function remains constant \be
f(q, p, t)={\textrm{const},}
\ee  then it is called as integral of motion. Let us find the equation of motion for function $f$.

Now we take the total derivative with respect to time of ex\-pression
(2.40):
\be
\f{df}{dt}=\f{\pa f}{\pa t}
+\f{\pa f}{\pa q_\sigma}\dot q_\sigma
+\f{\pa f}{\pa p_\sigma}\dot p_\sigma=0.
\ee
Substituting the
Hamiltonian\index{Hamilton} 
equations (2.36) into (2.41), we obtain
\be
\f{\pa f}{\pa t}
+\f{\pa f}{\pa q_\sigma}\cdot\f{\pa H}{\pa p_\sigma}
-\f{\pa f}{\pa p_\sigma}\cdot\f{\pa H}{\pa q_\sigma} =0.
\ee
The expression
\begin{equation}
(f, g)= \begin{vmatrix}
 \ds \f{\pa f}{\pa q_\sigma}&\ds\f{\pa f}{\pa p_\sigma}\\[5mm]
  \ds\f{\pa g}{\pa
q_\sigma}&\ds\f{\pa g}{\pa p_\sigma}
\end{vmatrix}
=\f{\pa f}{\pa q_\sigma}\cdot\f{\pa g}{\pa p_\sigma} -\f{\pa f}{\pa p_\sigma}\cdot\f{\pa g}{\pa q_\sigma
}
\end{equation}
has been termed the Poisson\index{Poisson} bracket. In (2.43) summation is
per\-for\-med over the index $\sigma$.

On the basis of (2.43), Eq.~(2.42) for function $f$ can be written in the form \be
\f{\pa f}{\pa t}+(f, H)=0.
\ee Poisson\index{Poisson} brackets have  the following properties 
\ba
&&(f, g)=-(g, f),\nonumber\\
&&(f_1+f_2, g)=(f_1, g)+(f_2, g),\\
&&(f_1 f_2, g)=f_1(f_2, g)+f_2(f_1, g),\nonumber
\ea
\be
\bigl(f,(g,h)\bigr)+\bigl(g,(h,f)\bigr)+\bigl(h,(f,g)\bigr)\equiv 0.
\ee
Relation (2.46) is called the
{\bf Jacobi\index{Jacobi} 
identity}.
On the basis of (2.43)
\be
(f,q_\sigma)=-\f{\pa f}{\pa p_\sigma},\quad
(f,p_\sigma)=\f{\pa f}{\pa q_\sigma}.
\ee
Hence we find
\be
(q_\lambda, q_\sigma)=0,\quad (p_\lambda, p_\sigma)=0,\quad (q_\lambda, p_\sigma)
=\delta_{\lambda\sigma}.
\ee

In the course of  development of the quantum mechanics, by analogy with the classical Poisson\index{Poisson}
brackets (2.43), there originated quantum Poisson\index{Poisson} 
brackets, which also satisfy all the conditions (2.45), (2.46). The application of relations (2.48)
for quantum Poisson\index{Poisson} brackets\index{Poisson bracket} has permitted to
establish the commutation re\-lations between coordinates and momenta.

The discovery of the Lagrangian\index{Lagrange} and 
Hamiltonian\index{Hamilton} methods\index{Lagrangian method} in classical mechanics permitted, at the
time, to generalize and extend them to other physical phenomena. The search for various re\-pre\-sen\-ta\-tions of
the physical theory is always extremely important, since on their basis the possibility may arise of their
generalization for describing new physical phenomena. Within the depths of the theory created there may be found
formal sprouts of the future theory. The experience of classical and quantum mechanics bears witness to this
assertion.

\newpage
\markboth{thesection\hspace{1em}}{}
\section{Electrodynamics. Space-time geometry}

 Following the discoveries made by Faraday\index{Faraday} in
electromagnetism, Maxwell\index{Maxwell} combined magnetic, electric and optical
phenomena and, thus, completed the construction of electrodynamics by writing out his famous equations.

H.\,Poincar\'{e}\index{Poincar\'e} 
in the book {\sf ``The Value of Science``} wrote the
following about
Maxwell's\index{Maxwell} 
studies:
\begin{quote}
{\it\hspace*{5mm}``At the time, when
Maxwell\index{Maxwell} 
initiated his studies, the
laws of electrodynamics adopted before him ex\-plai\-ned all known phenomena.
He started his work not because some new experiment limited the importance
of these laws. But, considering them from a new stand\-point,
Maxwell\index{Maxwell} 
noticed
that the equations became more symmetric, when a certain term was
introduced into them, although, on the other hand, this term was too
small to give rise to phenomena, that could be es\-ti\-ma\-ted by the previous
methods. \\
\hspace*{5mm} A priori ideas of Maxwell\index{Maxwell} are known to have waited for
their experimental confirmation for twenty years; if you prefer another expression, --- Maxwell\index{Maxwell}
an\-ti\-ci\-pa\-ted the experiment by twenty years.
How did he achieve such triumph?\\
\hspace*{5mm}This happened because
Maxwell\index{Maxwell} 
was always full of a sense
of mathematical symmetry \ldots''}
\end{quote}

According to
Maxwell\index{Maxwell} 
{\bf there exist no currents, except closed
currents.} He achieved this by introducing a small term ---
{\bf a dis\-pla\-ce\-ment current},\index{displacement current}
which resulted in the law of electric
charge conservation\index{equations of charge conservation}
following from the new equations.
\markboth{thesection\hspace{1em}}{3. Electrodynamics \ldots}

In formulating the equations of electrodynamics, Maxwell\index{Maxwell} ap\-pli\-ed
the Euclidean\index{Euclid} geometry\index{Euclidean geometry} of three-di\-men\-si\-o\-nal
space and ab\-so\-lute\index{absolute time} time, which is identical for all points of this space. Guided by a
profound sense of symmetry, he supplemented the equations of electrodynamics in such a way that, in the same time
 explaining available experimental facts, they were the equations  of elec\-tro\-mag\-ne\-tic
waves.\index{electromagnetic waves} He, naturally, did not suspect that the information on
 the geometry of space-time was
concealed in the equations. But his supplement of the equations of
electrodynamics turned out to be so indispensable and precise, that it
clearly led
H.\,Poincar\'{e},\index{Poincar\'e} 
who relied on the work of
H.\,Lorentz,\index{Lorentz} 
to
the discovery of the
pseudo-Euclidean\index{pseudo-Euclidean geometry of space-time}
geometry of space-time. Below, we
shall briefly describe, how this came about.

In the same time we will show that the striking desire of some authors to prove that 
H.\,Poincar\'e\index{Poincar\'e} ``\textbf{has not made the decisive step}'' to create the theory of relativity is
based upon both  misunderstanding of the essence of the theory of relativity and the shallow knowledge of
Poincar\'e works.\index{Poincar\'e}  We will show this below in our comments to such
statements. Just for this reason in this book I present results, first discovered and elucidated by the light of
consciousness by H.\,Poincar\'e,\index{Poincar\'e}  minutely enough. Here the need to
compare the content of A.\,Einstein's\index{Einstein} work of 1905 both with results of
publications [2, 3] by H.\,Poincar\'e,\index{Poincar\'e} and with his earlier works
naturally arises. After such a comparison it becomes clear what \textbf{new} each of them has produced.

{\bf How could it be happened that the outstanding research of Twentieth Century ---  works [2,3] by
H.\,Poincar\'e\index{Poincar\'e} --- were used in many ways at  in the same time were industriously consigned to
oblivion?} It is high time at least now, a hundred years later, to return everyone his  property. It is also our
duty.

Studies of the properties of the equations of electrodynamics revealed them not to retain their form under the
Galilean\index{Galilei} trans\-for\-ma\-tions (2.5),\index{Galilean transformations}
i.\,e. not to be form-invariant with respect to Galilean\index{Galilei} 
trans\-for\-ma\-tions.\index{Galilean transformations} Hence the conclusion follows that the Galilean
\index{Galilei} relativity principle\index{relativity principle} is violated, and, consequently,
 the ex\-pe\-ri\-men\-tal possibility arises to distinguish between one inertial reference
system\index{inertial reference systems} and another with the aid of electromagnetic or optical phenomena.
However, various experiments performed, es\-pe\-ci\-al\-ly Mi\-chel\-son's
\index{Michelson} 
experiments, showed that it is impossible to find out even by electro\-magnetic (optical)
experiments, with a precision up to $(v/c)^2$, whether one is in a state of rest or of uniform and rectilinear
motion. H.\,Lorentz\index{Lorentz} found an explanation for the results of these
ex\-pe\-ri\-ments, as H.\,Poincar\'{e}\index{Poincar\'e} noted, ``{\bf only by piling up
hypotheses}''.

In his book {\sf ``Science and Hypothesis''} (1902)
H.\,Poincar\'{e}\index{Poincar\'e} 
noted:
\begin{quote}
{\it\hspace*{5mm}``And now allow me to make a digression; I
must explain why I do not believe, in spite of Lorentz,\index{Lorentz}
that more exact observations will ever
make evident anything else but the relative displacements of material 
bodies. Experiments have been
made that should have disclosed the terms of the first order; the 
results were nugatory. Could that have
been by chance? No one has admitted this; а general explanation was 
sought, and Lorentz\index{Lorentz}  found it. He
showed that the terms of the first order should cancel each other, but 
not the terms of the second order.
Then more exact experiments were made, which were also negative; 
neither could this be the result of
chance. An explanation was necessary, and was forthcoming; they always 
are; hypotheses are what we
lack the least. But this is not enough. Who is there who does not think 
that this leaves to chance far too 
important а role? Would it not also be а chance that this singular 
concurrence should cause а certain
circumstance to destroy the terms of the first order, and that а 
totally different but very opportune
circumstance should cause those of the second order to vanish? No; the 
same explanation must be found
for the two cases, and everything tends to show that this explanation 
would serve equity well for the terms of the higher order, and that the 
mutual destruction of these terms will be rigorous and absolute''.}
\end{quote}

{\baselineskip=13.5pt
In 1904, on the basis of experimental facts, Henri
Poincar\'{e}\index{Poincar\'e} 
generalized
the Galilean\index{Galilei} 
relativity principle\index{relativity principle}
to all natural phe\-no\-me\-na.
He wrote~\cite{1}:\label{wash}
\begin{quote}
{\it\hspace*{5mm}``The principle of relativity, according to which the laws of physical phenomena
should be the same, whe\-ther to an observer fixed, or
for an observer carried along in a uniform motion of translation, so that we have not and could not have any means of discovering whether or not we are carried along in such a motion''.}
\end{quote}
Just this  \textbf{principle has become the key one} for the subsequent development of both electrodynamics and
the theory of relativity.  It can be formulated as follows. {\bf The principle of relativity\index{relativity
principle} is the preservation of form by all physical equations in any inertial reference system}.\index{inertial
reference systems}

But if this formulation uses the notion of 
the \textit{inertial\index{inertial reference systems} reference
system} then it means that the physical law of inertia by Galilei\index{Galilei} is already incorporated into this
formulation of the relativity principle\index{relativity principle}. This is just the difference between this
formulation and formulations given by Poincar\'e\index{Poincar\'e} and Einstein.\index{Einstein}
}

Declaring this principle Poincar\'e\index{Poincar\'e} precisely knew that one of its consequences was the
impossibility of {\bf absolute motion},\index{absolute motion} because {\bf all inertial reference systems were
equitable}. It follows from here that the principle of 
relativity\index{relativity principle} by
Poincar\'e\index{Poincar\'e} does not require a denial of  \textbf{ether}\index{ether} in general, it only
deprives ether\index{ether} of relation to any system of reference. In other words, it removes the \textbf{ether}\index{ether}
in Lorentz sense.\index{Lorentz} Poincar\'e\index{Poincar\'e} does not exclude the concept of
\textbf{ether}\index{ether} because it is difficult to imagine more absurd thing than  empty space. Therefore the
word \textbf{ether},\index{ether} which  can be found in the Poincar\'e\index{Poincar\'e} articles even after his
formulation of the relativity principle, has another meaning, different of the \textbf{Lorentz}\index{Lorentz}
 \textbf{ether}\index{ether}. Just this  \textbf{Poincar\'{e}'s}\index{Poincar\'e} 
 \textbf{ether}\index{ether} has to satisfy the relativity
 principle\index{relativity principle}. Also Einstein\index{Einstein} has come to the idea of ether\index{ether}
 in 1920.

In our time such a role is played by physical vacuum.\index{physical vacuum} Namely this point
is up to now not understood by some physicists (we keep silence about philosophers and historians of science). So
they erroneously attribute to Poincar\'{e}\index{Poincar\'e} the interpretation of
relativity principle as impossibility to register  the translational uniform motion relative to ether.\index{ether} Though, as
the reader can see, there is no the  word ``ether''\index{ether} in the formulation of the relativity principle.

One must distinguish between the {\bf Galilean\index{Galilei} relativity
principle}\index{relativity principle} and {\bf Galilean\index{Galilei} 
transformations}.\index{Galilean transformations} 
While Poincar\'{e}\index{Poincar\'e} 
extended the {\bf Ga\-li\-le\-an\index{Galilei} relativity principle}\index{relativity
principle} to all physical phenomena {\bf without al\-te\-ring its physical essence}, the {\bf
Ga\-li\-le\-an\index{Galilei} transformations} turned out to hold valid only when the
velocities of bodies are small as compared to the velocity of light.

Applying this
relativity principle\index{relativity principle}
to electrodynamical phe\-no\-me\-na in
ref.[3],
H.\,Poincar\'{e}\index{Poincar\'e} 
wrote:
\begin{quote}
{\it\hspace*{5mm}``This impossibility of revealing experimentally the
Earth's motion seems to represent a general law of Nature; we naturally
come to accept this law, which we shall term the {\bf\textit{relativity postulate}},
and to accept it without reservations. It is irrelevant, whether this
postulate, that till now is consistent with experiments, will or will not
later be confirmed by more precise measurements, at present, at any rate,
it is interesting to see, what consequences can be deduced from it''.}
\end{quote}

In 1904, after the critical remarks made by Poincar\'{e},\index{Poincar\'e} H.\,Lo\-rentz\index{Lorentz} made a most important step by attempting again to
write electrodynamics equations in a moving reference system and showing that the {\bf wave equation of
electrodynamics}\index{wave equation}  remained {\bf unaltered} (form-invariant) under the following
trans\-for\-ma\-tions of the coordinates and time: \be X^\prime=\gamma(X-vT),\; T^\prime=\gamma\Biggl(\ds
T-\f{v}{c^2}X\Biggr),
Y^\prime=Y,\; Z^\prime=Z\,,
\ee Lorentz named $T'$ as the {\bf modified local time} in contrast to {\bf local time} $\tau=T'/\gamma$
introduced earlier in 1895; 
\be
\gamma=\f{1}{\sqrt{1-\ds\f{v^2}{c^2}}}\,,
\ee where $c$ is the electrodynamic constant.

H.\,Poincar\'{e}\index{Poincar\'e} termed
these transformations the Lorentz\index{Lorentz} trans\-for\-ma\-tions. The
Lorentz\index{Lorentz} transformations, as it is evident from (3.1), are related
to two inertial reference systems.\index{inertial reference systems} 
H.\,Lorentz\index{Lorentz} 
did not establish the relativity principle\index{relativity principle} for electromagnetic
phenomena, since he did not succeed in demonstrating the form-invariance of all the 
Maxwell-Lo\-rentz\index{Lorentz}\index{Maxwell} equations under these
trans\-for\-ma\-tions.

From formulae (3.1) it follows that the wave equation\index{wave equation} being independent of translational
uniform motion of the reference system is achieved only by changing the time. Hence, the conclusion arises,
naturally, that for each inertial reference 
system\index{inertial reference systems} it is necessary to introduce
its own physical time.\index{physical time}

In 1907,
A.\,Einstein\index{Einstein} 
wrote on this:
\begin{quote}
{\it\hspace*{5mm}$\ll$Surprisingly, however, it turned out that a
sufficiently sharpened conception of time was all that was
needed to overcome the difficulty discussed.
One had only to realize that an auxiliary
quantity introduced by H.\,A.\,Lorentz,\index{Lorentz} 
and named by him 
``local time'',\index{local time} 
could be defined as ``time'' in general. 
If one adheres to this definition of time, the basic 
equations of Lorentz's,\index{Lorentz} theory correspond
to the principle of relativity
\ldots$\gg$}\index{relativity principle}
\end{quote}
Or, speaking more precisely, instead of the {\bf true time} there arose the {\bf modified local 
time}\index{local time} by Lorentz\index{Lorentz} different for each inertial reference
system.\index{inertial reference systems}

But H.\,Lorentz\index{Lorentz} did not notice this, and in 1914 he wrote on that
in detailed article ``The two papers by Henri Poincar\'e\index{Poincar\'e} on mathematical physics'':
\begin{quote}
\hspace*{5mm}``\textit{These considerations published by myself in 1904, have stimulated
Poincar\'e\index{Poincar\'e} to write his article on the dynamics of electron where he has given my name to the
 just mentioned transformation\index{Lorentz transformations}. I have to note as regards this  that a similar
 transformation have been already given in an article by Voigt\index{Voigt} published in 1887 and I have not taken
 all possible benefit from it.
 Indeed I have not given the most appropriate transformation for some physical quantities encountered in the formulae.
This was done by Poincar\'e\index{Poincar\'e} and later by Einstein\index{Einstein} and
Minkowski.\index{Minkowski} \ldots I had not thought of the straight path leading to them, since I considered
there was an essential dif\-fe\-rence between the reference systems $x, y, z, t$ and $x^\prime, y^\prime,
z^\prime, t^\prime$. In one of them were used --- such was my reasoning --- coordinate axes with a definite position
in ether\index{ether} and what could be termed {\bf \textit{true time}}; in the other, on the contrary, one simply dealt with
subsidiary quantities introduced with the aid of a mathematical trick. Thus, for instance, the variable $t^\prime$
could not be called {\bf \textit{time}} in the same sense as the variable $t$. Given such reasoning, I did not
think of describing phenomena in the reference system $x^\prime, y^\prime, z^\prime, t^\prime$ {\bf \textit{in
precisely the same way}}, as in the reference system $x, y, z, t$ \ldots I later saw from the article by
Poincar\'{e}\index{Poincar\'e} that, if I had acted in a more systematic manner, I could
have achieved an even more significant simplification. Having not noticed this, I was not able to achieve total
invariance of the equations; my formulae remained cluttered up with excess terms, that should have vanished. These
terms were too small to influence phenomena noticeably, and by this fact I could explain their independence of the\break
Earth's motion, revealed by observations, but I did not establish the relativity principle\index{relativity principle} as a rigorous and universal truth. On the contrary, 
Poincar\'{e}\index{Poincar\'e} achieved total invariance of the equations of electrodynamics and formulated the {\bf
\textit{relativity postulate}} --- a term first introduced by him \ldots I may add that, while thus cor\-rec\-ting
the defects of my work, he never reproached me for them.\\ 
\hspace*{5mm} I am unable to present here all the
beautiful results obtained by  Poincar\'{e}\index{Poincar\'e}. Nevertheless  let me stress
some of them. First, he did not restrict himself by demonstration that the relativistic transformations left
  the form of electromagnetic equations unchangeable. He explained this success of transformations by the
  opportunity to present these equations as a consequence of the least action  principle and by the fact that
   the fundamental
  equation expressing this principle and  the operations used in derivation of the field equations are identical in systems
  \( x, y, z, t \) and 
  \(x^\prime, y^\prime, z^\prime, t^\prime \)\ldots\break There are some new notions in this part
  of the article, I should especially mark them. Poincar\'{e}\index{Poincar\'e} notes, for
  example, that in consideration  of quantities
   \( x, y, z, t\sqrt{-1} \) as coordinates of a point in 
   four-dimensional\index{four-dimensional space-time}
space the relativistic transformations reduces to rotations in this space. He also comes to idea to add to the
three  components \( X, Y, Z \)  of the force a quantity
\[
T=X\xi+Y\eta+Z\zeta,
\]
which is nothing more than the work of the force at a unit of time, and which may be treated as a fourth component
of the force in some sense.  When dealing with the force acting at a unit of volume of a body the relativistic
transformations change  quantities   \( X, Y, Z,\break T\sqrt{-1} \) in a similar way to quantities \( x, y,
z,t\sqrt{-1}. \)\break I remind on these ideas by Poincar\'{e}\index{Poincar\'e} because they
are closed to methods later used by Minkowski and other scientists to easing  mathematical actions in the theory
of relativity}.''
\end{quote}

As one can see, in the course of studying the article 
by Poincar\'{e},\index{Poincar\'e} 
H.\,Lorentz\index{Lorentz} sees and accepts the possibility of {\bf describing
phe\-no\-me\-na in the reference system} {\mathversion{bold}\( x^\prime, y^\prime, z^\prime, t^\prime \)}
{\bf in exactly the same way as in the reference system} {\bf \textit{x, y, z, t}},
and that all this fully complies with the relativity 
principle,\index{relativity principle} formulated by
Poincar\'{e}.\index{Poincar\'e}\linebreak Hence it follows that {\bf physical phenomena
are identical}, if they take place in identical conditions in inertial reference sys\-tems\index{inertial reference systems} ($x, y,\break z, t $) and ($x^\prime, y^\prime, z^\prime, t^\prime $), moving with respect to each
other with a velocity $v$. All this was a direct consequence of the {\bf physical equations not altering} under
the Lorentz\index{Lorentz} trans\-for\-mations,\index{Lorentz transformations}
that together with space rotations form a group.\index{group} Precisely all this is contained, also, in articles
by Poincar\'{e} [2, 3].\index{Poincar\'e} 

H.\,Lorentz\index{Lorentz} writes in 1915 in a new edition of his book \textsf{``Theory of electrons''}, in comment
\( 72^\ast \):
\begin{quote}
\hspace*{5mm}``\textit{The main reason of my failure was I always thought  that only quantity  \( t \) could be
treated  as a true time  and that my local time\index{local time} \( t^\prime \) was considered only as an
auxiliary mathematical value. In the Einstein\index{Einstein} theory, just opposite,  \( t^\prime \) is playing
the same role as \( t \). If we want to describe phenomena as dependent on \( x^\prime, y^\prime, z^\prime,
t^\prime \), then we should operate with these variables  in just the same way as with \( x, y, z, t \) ''.}
\end{quote}
{\baselineskip=14pt 
Compare this quotation with the detailed analysis of the Poincar\'{e}\index{Poincar\'e}
article given by Lorentz\index{Lorentz} in 1914.

Further he demonstrates in this comment the derivation of velocity composition formulae, just in the same form as
it is done in article  [3] by Poincar\'{e}.\index{Poincar\'e} In comment \( 75^\ast \) he
discusses the transformation of forces, exploits invariant (3.22) in the same way as it is done by
Poincar\'{e}.\index{Poincar\'e}  The Poincar\'{e}\index{Poincar\'e} 
work is cited only in connection with a particular point. It is surprising but Lorentz\index{Lorentz} 
in his dealing with the theory of relativity even does not cite Poincar\'e articles [2; 3].
What may happen with Lorentz\index{Lorentz} in the period after 1914? How we can
explain this?  To say the truth, we are to mention that because of the war the Lorentz article written in 1914 has
appeared in print only in  1921. But it was printed  in the same form as Lorentz wrote it in 1914. In fact he
seems to confirm by this that his opinion has not been changed. But \textbf{all this} in the long run \textbf{does
not mean nothing substantial, because now we can ourselves examine deeper and in more detail who has done the
work,  what has been done and what is the level of this work, being informed on the modern state of the theory}
and comparing article of 1905 by Einstein\index{Einstein} to articles by Poincar\'{e}.\index{Poincar\'e}

 \textbf{The scale of works can be better estimated from the time distance}. Recollections of contemporaries
 are valuable for us as a testimony on how new ideas have been admitted by the physical community of that time.
But moreover one may obtain some knowledge on the ethic of science for some scientists, on group interests, and
maybe even something more, which is absolutely unknown to us.
\par}

It is necessary to mention that Lorentz\index{Lorentz} in his article of 1904 in
calculating his transformations has made an error  and as a result Maxwell-Lorentz\index{Lorentz} \index{Maxwell} equations in a moving reference frame have become
different than electrodynamics  equations in the rest frame. These equations were overloaded by
\textbf{superfluous} terms. But Lo\-rentz\index{Lorentz} has not been troubled by
this. He would easily see the error if he were not \textbf{keep away of the relativity principle}. After all, just
the relativity principle\index{relativity principle} requires that equations have to be the same in both two
reference frames.
 But he singled out \textbf{one} reference frame directly connected
with the ether.\index{ether}

Now, following the early works of
H.\,Poincar\'{e}\index{Poincar\'e} 
we shall deal with the
definition of
simultaneity,\index{simultaneity}
on the
synchronization\index{synchronization of clocks}
of clocks occupying
different points of space, and we shall clarify the physical sense of
{\bf local time},\index{local time}
introduced by
Lorentz. In the article {\sf ``Me\-a\-su\-re\-ment
of time''}, published in 1898,  
Poincar\'{e}\index{Poincar\'e} 
discusses
the issue of time measurement in detail. This article was especially
noted in the book {\sf ``Science and hypothesis''} by
Poincar\'{e},\index{Poincar\'e} 
and,
therefore, it is quite com\-pre\-hen\-sible to an inquisitive reader.

In this article, for instance the following was said:
\begin{quote}
{\it\hspace*{5mm}``
But let us pass to examples less artificial; to 
understand the definition implicitly supposed by the savants,
let us watch them at work and look for the rules by which
they investigate simultaneity.\index{simultaneity} \ldots\\
\hspace*{5mm}When an astronomer tells me that some stellar
phenomenon, which his telescope reveals to him
at this moment, happened, nevertheless,
fifty years ago, I seek his meaning, and to that end I shall ask him 
first how he knows it, that is, how he has
measured the velocity of light.\\
\hspace*{5mm}He has begun by {\bf \textit{supposing}} that light
has a constant velocity, and in
particular that its velocity is the same in all directions. That is 
a postulate without which no measurement of this
velocity could be attempted. This postulate could never be verified
directly by experiment; it might be contradicted by it if the results
of different measurements were not concordant.
We should think ourselves fortunate
that this contradiction has not happened
and that the slight discordances which may happen
can be re\-a\-di\-ly explained.\\
\hspace*{5mm}The postulate, et all events, resembling the principle
of sufficient reason,  has been
accepted by everybody; what I wish to emphasize is that {\bf \textit{it 
furnishes us with a new rule for the investigation of
simultaneity}},\index{simultaneity} ({\rm singled out by me.} --- A.L.) 
entirely different from that which we have enunciated above''.}
\end{quote}

\textbf{It follows from this postulate} that \textbf{the value of light velocity does not depend on velocity of
the source of this light}. This statement is also a straightforward consequence of Maxwell\index{Maxwell}
electrodynamics.  \textbf{The above postulate together with the relativity principle\index{relativity principle}}
formulated by H.\,Poincar\'{e}\index{Poincar\'e} in
 1904 for all physical phenomena precisely  \textbf{become the initial statements} in Einstein\index{Einstein} work of 1905.

Lorentz\index{Lorentz} 
dealt with the
Maxwell-Lorentz\index{Lorentz} 
\index{Maxwell} 
equations\index{Maxwell-Lorentz equations}
in a ``mo\-ti\-on\-less''
reference system related to the ether.\index{ether} He considered the coordinates
$X, Y, Z$ to be {\bf absolute}, and the time $T$ to be the
{\bf true time}.

In a reference system moving along the $X$ axis with a velocity $v$
relative to a reference system ``at rest'', the co\-or\-di\-nates with respect
to the axes moving together with the reference system have the values
\be
x=X-vT,\quad y=Y,\quad z=Z,
\ee
while the time in the moving reference system was termed by
Lo\-rentz\index{Lorentz} 
{\bf local time}\index{local time}
(1895) and defined as follows:
\be
\tau = T-\f{v}{c^2}X.
\ee He introduced this time so as to be able, in agreement with ex\-pe\-ri\-men\-tal data, to exclude from the
theory the influence of the Earth's motion on optical phenomena in the first order over $v/c$.

This time, as he noted, ``{\bf was introduced with the aid of a mathematical
trick}''. The physical meaning of
{\bf local time}\index{local time}
was uncovered by
H.\,Poincar\'{e}.\index{Poincar\'e} 

In the article {\sf ``The theory of Lorentz\index{Lorentz} and the principle of
equal action and reaction``}, published in 1900, he wrote  about the \textbf{local time}\index{local time}
{\mathversion{bold} \( \tau \)},
defined as follows (Translation from French by 
V.\,A.\,Petrov\index{Petrov}):
\begin{quote}
{\it\hspace*{5mm}``I assume observers, situated at different points, to  compare their clocks with the aid of
light signals; they correct these signals for the transmission time, but, without knowing the relative motion they
are un\-der\-go\-ing and, consequently, considering the signals to propagate with the same velocity in both
directions, they limit themselves to performing observations by sending sig\-nals from $A$ to $B$ and, then, from
$B$ to $A$. The {\bf \textit{local time}}\index{local time} $\tau$ is the time read from the clocks thus
controlled. Then, if {\bf \textit{c}}
is the velocity of light,
and $v$ is the velocity of the Earth's motion, which I assume to be
parallel to the positive $X$ axis, we will have:
\be\tau=T-\f{v}{c^2}X\;\mbox{''}.
\ee}
\end{quote}
Taking into account (3.3) in (3.5) we obtain
\be
\tau= T\left(1-\f{v^2}{c^2}\right)-\f{v}{c^2}x.
\ee
The velocity of light in a reference system ``at rest`` is
{\bf \textit{c}}.
In a moving reference system, in the variables $x, T$, it will be equal,
in the direction parallel to the $X$ axis, to
\be
c-v
\ee
in the positive, and
\be
c+v
\ee
--- in the negative direction.

This is readily verified, if one recalls that the velocity of light in a
reference system ``at rest`` is, in all directions, equal to
{\bf \textit{c}},
i.\,e.
$$
c^2=\left(\f{dX}{dT}\right)^2
+\left(\f{dY}{dT}\right)^2
+\left(\f{dZ}{dT}\right)^2.\eqno(\lambda)
$$
In a moving reference system $x=X-vT$ the upper expression assumes,
in the variables $x, T$, the form
\[
c^2=\left(\f{dx}{dT}+v\right)^2
+\left(\f{dY}{dT}\right)^2
+\left(\f{dZ}{dT}\right)^2.
\]
Hence it is evident that in a moving reference system the coordinate velocity\index{coordinate velocity of light}
of a light signal parallel to the $X$ axis $\ds\f{dx}{dT}$ is given as follows
\[
\f{dx}{dT}=c-v
\]
in the positive direction,
\[
\f{dx}{dT}=c+v
\]
--- in the negative direction.

The coordinate velocity of light\index{coordinate velocity of light}
in a moving reference system
along the $Y$ or $Z$ axis equals the quantity
\[
\sqrt{c^2-v^2}.
\]
We note that if we had made use of the
Lorentz\index{Lorentz} 
transformations\index{Lorentz transformations}
inverse to (3.1), then taking into account the equality
\[
c^2\gamma^2\left(dT^\prime+\f{v}{c^2}dX^\prime\right)^2
-\gamma^2\left(dX^\prime+vdT^\prime\right)^2
=c^2(dT^\prime)^2-(dX^\prime)^2,
\]
we would have obtained from Eq.~($\lambda$)the expression
$$
c^2=\left(\f{dX^\prime}{dT^\prime}\right)^2
+\left(\f{dY^\prime}{dT^\prime}\right)^2
+\left(\f{dZ^\prime}{dT^\prime}\right)^2,\eqno(\rho)
$$
which would signify that the velocity of light equals
{\bf \textit{c}}
in all directions in a moving reference system, too. Let us mention also that the light cone equation remains the
same after multiplying r.h.s. of Eqs. (3.1) (Lorentz transformations\index{Lorentz} ) by arbitrary function $\phi(x)$. The light cone equation preserves its form under conformal
transformations.

Following
Poincar\'{e},\index{Poincar\'e} 
we shall perform
synchronization\index{synchronization of clocks}
of the clocks in
a moving reference system with the aid of
Lorentz's\index{Lorentz} 
{\bf local time}.\index{local time}
Consider a light signal leaving point $A$ with coordinates ($0, 0, 0$) at
the moment of time $\tau_a$:
\be
\tau_a=T\left(1-\f{v^2}{c^2}\right).
\ee
This signal will arrive at point $B$ with coordinates ($x, 0, 0$) at
the moment of time
$\tau_b$
\be
\tau_b=\left(T+\f{x}{c-v}\right)\left(1-\f{v^2}{c^2}\right)
-\f{v}{c^2}x=\tau_a+\f{x}{c}.
\ee Here, we have taken into account the transmission time of the signal from $A$ to $B$. The signal was reflected
at point $B$ and arrived at point $A$ at the moment of time $\tau_a^\prime$ \be
\tau_a^\prime=\left(T+\f{x}{c-v}+\f{x}{c+v}\right)
\left(1-\f{v^2}{c^2}\right)=\tau_b+\f{x}{c}.
\ee
On the basis of (3.9), (3.11) and (3.10) we have
\be
\f{\tau_a+\tau_a^\prime}{2}=\tau_b.
\ee Thus  the definition of simultaneity\index{simultaneity} has been introduced, which was later applied by
A.\,Einstein\index{Einstein} for deriving the Lorentz\index{Lorentz} 
transformations.\index{Lorentz transformations} We have verified that the Lorentz\index{Lorentz}
``{\bf local time}''\index{local time} (3.6) satisfies con\-di\-tion (3.12). Making use
of (3.12) as the initial equation for defining time in a moving reference system, Einstein\index{Einstein}
arrived at the same Lorentz\index{Lorentz} ``{\bf local
time}''\index{local time} (3.6) multiplied by an arbitrary function de\-pen\-ding only on the velocity $v$. From
(3.10), (3.11) we see that in a reference system moving along the $X$ axis with the \textbf{local
time}\index{local time} {\mathversion{bold} \( \tau \)}
the light sig\-nal has velocity  {\bf \textit{c}}
along any direction parallel to the $X$ axis. The transformations, inverse to (3.3) and (3.4), will be as follows
\be T=\f{\tau+\ds\f{v}{c^2}x}{1-\ds\f{v^2}{c^2}},\; X=\f{x+v\tau}{1-\ds\f{v^2}{c^2}},\;
Y=y,\;Z=z.
\ee
Since the velocity of light in a reference system ``at rest'' is
{\bf \textit{c}},
in the new variables $\tau, x, y, z$ we find from Eqs.  $(\lambda)$ and (3.13) \be
\gamma^2\left(\f{dx}{d\tau}\right)^2+\left(\f{dy}{d\tau}\right)^2+
\left(\f{dz}{d\tau}\right)^2=\gamma^2c^2.
\ee We can see from the above that to have the velocity of light equal to {\bf \textit{c}}
in any direction in the moving reference system, also, it is necessary to multiply the right-hand sides of
transformations (3.3) and (3.4) for $x$ and $\tau$ by $\gamma$ and to divide the right-hand sides in
transformations (3.13) for $T$ and $X$
 by $\gamma$. Thus, this requirement leads to appearance of 
 the Lorentz\index{Lorentz} transformations\index{Lorentz transformations}  here.

H.\,Lorentz\index{Lorentz} in 1899 used transformation of the following form
\[
X^\prime=\gamma(X-vT),\quad Y^\prime=Y,\quad Z^\prime=Z,\quad T^\prime=\gamma^2\left(T-\ds\f{v}{c^2}X\right),
\]
to explain the Michelson\index{Michelson} experiment. The inverse transformations are
\[
X=\gamma X^\prime+vT^\prime,\quad Y=Y^\prime,\quad Z=Z^\prime,\quad T=T^\prime+\ds\f{v}{c^2}\gamma X^\prime.
\]
If H.\,Lorentz\index{Lorentz} would proposed the relativity
principle\index{relativity principle} for all physical phenomena and required in this connection that a spherical
wave should have the same form in unprimed and primed  systems of reference, then he would come to
Lorentz\index{Lorentz} transformations\index{Lorentz transformations}. Let we have in unprimed system of reference
\[
c^2T^2-X^2-Y^2-Z^2=0,
\]
then according to his formulae this expression in new variables is as follows
\[
c^2\left(T^\prime +\ds\f{v}{c^2}\gamma X^\prime\right)^2 -(\gamma X^\prime+vT^\prime)^2-Y^{\prime\, 2}-Z^{\prime\,
2}=0,
\]
and after some simplifications we obtain
\[
c^2T^{\prime\, 2}\left(1-\ds\f{v^2}{c^2}\right)-X^{\prime\, 2} -Y^{\prime\, 2}-Z^{\prime\, 2}=0.
\]
We see that to guarantee the same form of a spherical wave in new variables as in the old ones it is necessary to
change variable  \( T^\prime \) replacing it by new variable  \( \tau \)
\[
\f{1}{\gamma} T^\prime =\tau.
\]
After transition to the new variable we obtain Lorentz\index{Lorentz} 
transformations\index{Lorentz transformations}
\[
X^\prime=\gamma(X-vT),\quad Y^\prime=Y,\quad Z^\prime=Z,\quad \tau =\gamma\left(T-\ds\f{v}{c^2}X\right),
\]
and the inverse transformations
\[
X=\gamma(X^\prime+v\tau),\quad Y=Y^\prime,\quad Z=Z^\prime,\quad T=\gamma\left(\tau+\ds\f{v}{c^2}X^\prime\right).
\]
But H.\,Lorentz\index{Lorentz} has not seen this in 1899.  He obtained these
transformations in 1904 only, then he also came closely to the theory of relativity, but did not make the decisive
step. Lorentz\index{Lorentz} transformations\index{Lorentz transformations} (3.1)  were obtained in 1900 by
Larmor\index{Larmor}. But he also did not propose the principle of relativity for all physical phenomena and did
not require form-invariance of Maxwell\index{Maxwell} equations under these transformations. Therefore Larmor\index{Larmor} also
has not made a decisive step to construct the theory of relativity.

Precisely the constancy\index{principle of constancy of velocity of light} of the velocity of light in any
inertial reference system\index{inertial reference systems} is what 
A.\,Einstein\index{Einstein} 
chose to underlie his approach to the electrodynamics of moving bodies. But it is provided for not by
transformations (3.3) and (3.4), but by the Lorentz\index{Lorentz} 
transformations\index{Lorentz transformations}.

A.\,Einstein\index{Einstein}  started from the relativity principle and from the principle of constancy of the light velocity. Both
principles were formulated as follows:
\begin{quote}
$\ll$ 1. The laws governing the changes of the 
 state of any physical system do not depend on which
one of two coordinate systems in uniform translational
motion relative to each other these changes of the state are 
referred to.\\
\hspace*{5mm}2. Each ray of light moves in the coordinate system 
``at rest'' with the definite velocity $V$ independent of whether
this ray of  light is emitted by a body at rest or a body in motion$\gg$.
\end{quote}
\label{cyte:postulates} Let us note that \textbf{Galilean principle of relativity} is not included into these
principles.

It is necessary to specially emphasize that the \textbf{principle of constancy of velocity of
light}\index{principle of constancy of velocity of light}, suggested by 
A.\,Einstein\index{Einstein} 
as the \textbf{second independent postulate}, is really a special consequence of requirements of the
relativity principle\index{relativity principle} by H.\,Poincar\'{e}.\index{Poincar\'e} This principle was extended by him on all physical phenomena. To be convinced in this it is sufficient to consider
requirements of the relativity principle\index{relativity principle} for an elementary process --- propagation of
the electromagnetic spherical wave. We will discuss this later.

In 1904, in the article {\sf ``The present and future of ma\-the\-ma\-ti\-cal
physics''},
H.\,Poincar\'{e}\index{Poincar\'e} 
formulates the
relativity principle\index{relativity principle}
for all
natural phenomena, and in the same article he again returns to
Lorentz's\index{Lorentz} 
idea of {\bf local time}.\index{local time}
He writes:
\begin{quote}
{\it $\ll$Let us imagine two observers, who
wish to regulate their watches by means of optical signals; they 
exchange signals, but as they know that the transmission of light is not instantaneous, they are careful
to cross them. When station $B$ sees the signal from station $A$, its timepiece should not mark the same hour as that of station $A$ at the moment the signal was sent, but this hour increased by а constant representing the time of transmission. Let us suppose, for example, that station $A$ sends it signal at the moment when its time-piece marks the hour zero, and that station 
$B$ receives it when its time-piece
marks the hour $t$. The watches will be set, if the time $t$ is the time of transmission,
and in order to verify it, station $В$ in turn sends а signal at the instant when its time-piece is at zero; station $А$ must then see it when its time-piece is at $t$. Then the watches are regulated.\\
\hspace*{5mm}And, indeed, they mark the same hour at the same physical instant, but under one condition, namely, that the two stations are stationary. Otherwise, the time of transmission will not be the вате in the two directions, since the station $А$, for example, goes to meet the disturbance emanating from $В$, whereas station $В$ flees before the disturbance emanating from $A$. Watches regulated in this way, therefore,
will not mark the true time; 
({\rm the time in the reference system ``at rest``} --- A.L.)
they will mark what might be called the local time,\index{local time}
so that one will gain on the other. It matters little, since we have по means of perceiving it. All the phenomena which take place at $А$, for example, will be behind time, but all just the вате amount, and the observer will not notice it since his watch is also behind time; thus, in accordance with the principle of relativity\index{relativity principle} he will have по means of ascertaining whether he is at rest or in absolute 
motion.\index{absolute motion} Unfortunately this is not sufficient; additional hypotheses are necessary. We must
admit that the moving bodies undergo a uniform contraction in the 
direction of motion$\gg$.}
\end{quote}

Such was the situation before the work of Lorentz,\index{Lorentz} which also
appeared in 1904. Here Lorentz presents again the transformations connecting a reference system ``at
rest'' with a re\-fe\-rence system moving with a velocity $v$ relative to the one ``at rest'', which were termed
by Poincar\'{e}\index{Poincar\'e} the Lorentz\index{Lorentz} 
transformations.\index{Lorentz transformations} In this work, Lorentz,\index{Lorentz} 
instead of the \textbf{local time}\index{local time} (3.4) introduced the time $T^\prime$,
equal to \be
T^\prime=\gamma\tau.
\ee Lorentz called time $T^\prime$ as the \textbf{modified local time}. Precisely this time will be present in any
inertial reference system\index{inertial reference systems} in Galilean
\index{Galilei} 
coordinates.\index{Galilean (Cartesian) coordinates} It does not violate the condition of syn\-chro\-ni\-za\-tion
(3.12)

Below we shall see following Lorentz that the wave equation\index{wave equation} does indeed not alter its form
under the Lorentz\index{Lorentz} 
transformations (3.1).\index{Lorentz transformations} Let us check this. The wave equation\index{wave equation} of electrodynamics has the form: \be
\Box\phi=\Biggl(\f{1}{c^2}\cdot \f{\pa\,^2}{\pa\,t^2}- \f{\pa\,^2}{\pa\,x^2}- \f{\pa\,^2}{\pa\,y^2}-
\f{\pa\,^2}{\pa\,z^2}\Biggr)\,\phi=0.
\ee
Here $\phi$ is a scalar function in
four-dimensional\index{four-dimensional space-time}
space, which changes
under coordinate-time transformations according to the rule
${\phi}^\prime ({x}^\prime)=\phi (x)$,
{\bf \textit{c}}
is the {\bf electrodynamic constant},\index{electrodynamic constant} that has the dimension of velocity.

Let us establish the form-invariance of the operator $\Box$ with res\-pect to transformations (3.1). We represent
part of the operator $\Box$ in the form \be \f{1}{c^2}\cdot \f{\pa\,^2}{\pa\,t^2}- \f{\pa\,^2}{\pa\,x^2}=
\Biggl(\f{1}{c}\cdot \f{\pa}{\pa\,t}- \f{\pa}{\pa\,x}\Biggr) \Biggl(\f{\,1\,}{\,c\,}\cdot \f{\pa}{\pa\,t}+
\f{\pa}{\pa\,x}\Biggr)\,.
\ee
We calculate the derivatives in the new coordinates, applying
for\-mu\-lae (3.1)
\[
\f{\,1\,}{\,c\,}\cdot
\f{\pa}{\pa\,t}=
\f{\,1\,}{\,c\,}\cdot
\f{\pa\,{t^\prime}}{\pa\,t}\cdot
\f{\pa}{\pa\,{t^\prime}}+
\f{\,1\,}{\,c\,}\cdot
\f{\pa\,{x^\prime}}{\pa\,t}\cdot
\f{\pa}{\pa\,{x^\prime}}=
\gamma\,\Biggl(\f{\,1\,}{\,c\,}\cdot\f{\pa}{\pa\,{t^\prime}}-
\f{\,v\,}{\,c\,}\cdot\f{\pa}{\pa {x^\prime}}\Biggr)\,,
\]
\[
\f{\pa}{\pa\,x}=
\f{\pa\,{t^\prime}}{\pa x}\cdot
\f{\pa}{\pa\,{t^\prime}}+
\f{\pa\,{x^\prime}}{\pa\,x}\cdot
\f{\pa}{\pa\,{x^\prime}}=-
\gamma\,\Biggl(\f{v}{c^2}\cdot\f{\pa}{\pa\,{t^\prime}}-
\f{\pa}{\pa\,{x^\prime}}\Biggr)\,.
\]
Hence we find
\be
\f{\,1\,}{\,c\,}\cdot
\f{\pa}{\pa\,t}-
\f{\pa}{\pa\,x}=
\gamma\,\Biggl(1+\f{\,v\,}{\,c\,}\Biggr)
\Biggl(\f{\,1\,}{\,c\,}\cdot
\f{\pa}{\pa\,{t^\prime}}-
\f{\pa}{\pa\,{x^\prime}}\Biggr)\,,
\ee
\be
\f{\,1\,}{\,c\,}\cdot
\f{\pa}{\pa\,t}+
\f{\pa}{\pa\,x}=
\gamma\,\Biggl(1-\f{\,v\,}{\,c\,}\Biggr)
\Biggl(\f{\,1\,}{\,c\,}\cdot
\f{\pa}{\pa\,{t^\prime}}+
\f{\pa}{\pa\,{x^\prime}}\Biggr)\,.
\ee
Substituting these expressions into (3.17) we obtain
\be
\f{1}{c^2}\cdot
\f{\pa\,^2}{\pa\,t^2}-
\f{\pa\,^2}{\pa\,x^2}=
\f{1}{c^2}\cdot
\f{\pa\,^2}{\pa\,{t^\prime}^2}-
\f{\pa\,^2}{\pa\,{x^\prime}^2}\,.
\ee
Taking into account that the variables $y$ and $z$ in accordance with
(3.1) do not change, on the basis of (3.20) we have
\ba
\ds\f{1}{c^2}\cdot
\f{\pa\,^2}{\pa\,t^2}-
\f{\pa\,^2}{\pa\,x^2}-
\f{\pa\,^2}{\pa\,y^2}-
\f{\pa\,^2}{\pa\,z^2}=\nonumber\\
\label{3.21}\\
=\ds\f{1}{c^2}\cdot
\f{\pa\,^2}{\pa\,{t^\prime}^2}-
\f{\pa\,^2}{\pa\,{x^\prime}^2}-
\f{\pa\,^2}{\pa\,{y^\prime}^2}-
\f{\pa\,^2}{\pa\,{z^\prime}^2}\,.\nonumber
\ea
This means that the {\bf wave equation} (3.16)\index{wave equation} {\bf remains form-\-in\-va\-ri\-ant with
respect to the Lorentz\index{Lorentz} transformations} 
(3.1).\index{Lorentz transformations} In other words, it is the same in both inertial reference systems.\index{inertial reference
systems} Hence, for instance, it follows that the velocity of a light wave equals {\bf \textit{c}},
both in
a re\-fe\-rence system ``at rest'' and in any other reference system moving
relative to the one ``at rest'' with a velocity $v$.

We have shown that the Lorentz\index{Lorentz} 
transformations\index{Lorentz transformations} leave the ope\-ra\-tor $\Box$ unaltered, i.\,e. they conserve the form-invariance of the wave
equation.\index{wave equation} On the other hand, this computation can be considered as an exact derivation of the
Lorentz\index{Lorentz} trans\-for\-mations\index{Lorentz transformations} based on
the form-invariance of the operator $\Box$.

In electrodynamics, the wave equation\index{wave equation} holds valid outside the source both for the scalar and
vector potentials,\index{potential} $\varphi$ and $\vec A$, res\-pec\-tive\-ly. In this case, $\varphi$ is defined
as a scalar with respect to three-dimensional coordinate transformations, and $\vec A$ is defined as a vec\-tor
with respect to the same trans\-for\-mations. For the wave 
equation\index{wave equation} to be
form-\-in\-va\-ri\-ant under the Lorentz\index{Lorentz} 
transformations\index{Lorentz transformations} it is ne\-ce\-ssa\-ry to consider the quantities $\varphi$ and
$\vec A$ as components of the four-di\-men\-si\-onal vector $A^\nu=(\varphi,\,\vec A$\,)
$$
\Box A^\nu=0,\qquad \nu=0,1,2,3.
$$

In 1905 Henri Poincar\'{e}\index{Poincar\'e} first es\-tab\-li\-shed~\cite{2,3} the
invariance of the Maxwell-Lorentz\index{Lorentz}\index{Maxwell} 
equations\index{Maxwell-Lorentz equations} and of the equations of motion of charged particles under
the action of the Lorentz\index{Lorentz} force with respect to the
Lorentz\index{Lorentz} transformations (3.1)\index{Lorentz transformations}  on
the basis of the 1904 work by Lorentz,\index{Lorentz} in which the
Lorentz\index{Lorentz} transformations\index{Lorentz transformations} were
discovered, and on the relativity principle,\index{relativity principle} formulated 
by Poincar\'{e}\index{Poincar\'e}  in the same year for all natural phenomena. All the above will be demonstrated in
detail in Sections 8 and 9.

{\bf H.\,Poincar\'{e}\index{Poincar\'e} discovered that these transformations, together
with  spatial rotations form a group.\index{group}\index{group, Poincar\'e} He was the first to in\-tro\-duce
the notion of four-dimensionality of a number of physical quantities}. The discovery of this group\index{group}
together with quantum ideas created the foundation of modern theoretical physics.

Poincar\'{e}\index{Poincar\'e} established that the scalar and vector
potential\index{potential} $(\varphi,\,\vec A)$, the charge density and current $(c\rho,\,\rho\vec v),$  the
four-ve\-lo\-ci\-ty\index{four-velocity}
 $\bigl(\gamma, \gamma\,\vec v/c\bigr),$ 
  the work per unit time and
force normalized to unit volume,
$\bigl(\vec f,\vec v/c,\break \vec f\,\bigr),$ 
 as well as the
four-force\index{four-force} transform like the quantities $(ct, \vec x)$. The existence of 
the Lorentz\index{Lorentz} group\index{group}\index{group, Lorentz} signifies that in all inertial
reference systems\index{inertial reference systems} the Maxwell-Lorentz\index{Lorentz} \index{Maxwell} equations\index{Maxwell-Lorentz equations} in
Galilean\index{Galilei} coordinates\index{Galilean (Cartesian) coordinates} remain
form-\-in\-va\-ri\-ant, i.\,e. the relativity principle\index{relativity principle} is satisfied. Hence it directly
follows that the descriptions of phenomena are the same both in the reference system $x,y,z,t$ and in the
reference system $x^\prime, y^\prime, z^\prime, t^\prime$, so, consequently, time $t$, like the other variables
$x,y,z,$ is relative. Thus, time being relative is a direct con\-se\-quence of the existence of the
group,\index{group} which itself arises as a consequence of the requirement to fulfil the relativity
principle\index{relativity principle}  for elec\-tro\-mag\-ne\-tic phenomena. The existence of this
group\index{group} led to the discovery of the geometry of space-time.

{\bf H.\,Poincar\'{e}\index{Poincar\'e} 
discovered a number of invariants of the
group\index{group}
and among these --- the fundamental invariant}
\be
J=c^2T^2-X^2-Y^2-Z^2\,,
\ee which arose in exploiting the Lorentz\index{Lorentz} trans\-for\-ma\-tion.
{\bf It testifies that space and time form a unique four-di\-men\-si\-onal continuum of events with metric
pro\-per\-ti\-es determined by the invariant} (3.22). The four-dimensional space-time discovered by
\textbf{H.\,Poincar\'e},\index{Poincar\'e}  and defined by invariant (3.22), was later
called the \textbf{Minkowski space}.\index{Minkowski space}\index{Minkowski} 
Precisely this is the essence of special relativity theory. This is why it is related to all physical phenomena.
It is space-time determined by the invariant (3.22) that provides for the existence of physically equal inertial
reference sys\-tems\index{inertial reference systems} in Nature. However, as earlier in classical mechanics, it
remains unclear, how the inertial reference 
systems\index{inertial reference systems} are related to the
distribution of matter in the Universe.\index{Universe} From expression (3.22) it follows that in any inertial
reference system\index{inertial reference systems} a given quantity $J$ in Galilean\index{Galilei} (Cartesian) coordinates\index{Galilean (Cartesian)
coordinates} remains unaltered
(form-\-in\-va\-ri\-ant), while its projections onto the axes change. Thus, depending on the choice of inertial
reference system\index{inertial reference systems} the pro\-jec\-tions $X,Y,Z,T$ \textbf{are relative quantities},
while the {\bf quantity \textit{J}}
for any given $X,Y,Z,T$ \textbf{has an absolute value}. A positive interval $J$ can be measured by a clock whereas
a negative one --- by a rod.  According to (3.22), in differential form we have \be
(d\sigma)^2=c^2(dT)^2-(dX)^2-(dY)^2-(dZ)^2.
\ee
The quantity $d\sigma$ is called an
{\bf interval}.\index{interval}

{\bf The geometry of space-time, i.\,e. the space of events} (the Minkowski\index{Minkowski} space)\index{Minkowski space} {\bf with the measure} (3.23) {\bf has been termed pseu\-do-Euclidean
geometry}.

As it could be seen  from the structure of invariant $J$, written in orthogonal (Ga\-li\-le\-an)\index{Galilei} coordinates, it is always possible to introduce a unique time $T$ for all  points of
the three-dimensional space. This means that the three-dimensional space of a given iner\-ti\-al reference
system\index{inertial reference systems} is orthogonal to the lines of time. Since, as we shall see below, the
invariant $J$ in another inertial re\-fe\-rence 
system\index{inertial reference systems} assumes the form (3.27),
it hence follows that in this reference system the unique time will already be different, it is determined by the
variable $T^\prime$. But length will change simultaneously. Thus, the possibility to introduce
si\-mul\-ta\-neity\index{simultaneity} for all the points of three-di\-men\-si\-onal space is a direct consequence
of the pseudo-Euclidean\index{pseudo-Euclidean geometry of space-time} geometry of the four-di\-men\-si\-onal
space of events.

Drawing a conclusion to all the above, we see that 
H.\,Lo\-rentz\index{Lorentz} 
found the transformations (3.1),\index{Lorentz
transformations} which conserve the form of the wave equation 
(3.16).\index{wave equation} On the basis of the
relativity principle\index{relativity principle} for all physical phenomena formulated by him in 1904 and of the
Lorentz\index{Lorentz} transformations,\index{Lorentz transformations} Henri
Poincar\'{e}\index{Poincar\'e} es\-tab\-li\-shed form-invariance of the
Maxwell-Lorentz\index{Lorentz}\index{Maxwell} 
equations\index{Maxwell-Lorentz equations} and discovered the pseudo-Eu\-cli\-de\-an
\index{pseudo-Euclidean geometry of space-time} geometry of space-time, determined by the invariant (3.22) or (3.23).

A short exposition of the detailed article [3] was given by\linebreak 
H.\,Poincar\'{e}\index{Poincar\'e}
in the reports to the French academy of sciences [2] and pub\-li\-shed even before the work by
Einstein\index{Einstein} was sub\-mit\-ted for publication. This paper contained a
precise and rigorous de\-scrip\-tion of the solution to the problem of the elec\-tro\-dy\-na\-mics of moving
bodies and, at the same time, an extension of the Lorentz\index{Lorentz} transformations\index{Lorentz transformations} to all natural forces, independently of their origin. In this
publication H.\,Poincar\'{e}\index{Poincar\'e}  discovered Lorentz
group\index{group}\index{group, Lorentz} with accordance to that a whole set of four-dimensional physical values
transforming similar to $t, x, y, z$ arose. The presence of Lorentz 
group\index{group}\index{group, Lorentz}
automatically provides the synchronization of 
clocks\index{synchronization of clocks} in any inertial reference
system\index{inertial reference systems}. So the proper physical  time arises in any inertial system of
reference\index{inertial reference systems} --- the \textbf{modified local  time} by Lorentz. In paper [2]
relativistic formulae for adding velocities and the transformation law for forces arose for the first
time\index{Poincar\'{e}'s equations of mechanics}. There existence of \textbf{gravitational
waves}\index{gravitational waves} propagating with light velocity was \textbf{predicted}.

It should be emphasized that just the discovery of Lorentz 
group\index{group}\index{group, Lorentz} provided the
uniformity of description of all physical effects in all the inertial reference systems\index{inertial reference systems}
in full accordance with the relativity principle\index{relativity principle}. Just all this automatically provided the
relativity of time and length.

H.\,Poincar\'{e}\index{Poincar\'e} discovered the invariant (3.22) on the basis of the
Lorentz\index{Lorentz} 
transformations (3.1).\index{Lorentz transformations}
On the other hand, applying the invariant
(3.22) it is easy to derive the actual
Lorentz\index{Lorentz} 
trans\-for\-ma\-tions (3.1).
Let the invariant $J$ in an
inertial reference system\index{inertial reference systems}
have the form (3.22)
in Galilean\index{Galilei} 
coordinates.\index{Galilean (Cartesian) coordinates}
Now, we pass to another inertial reference
system\index{inertial reference systems}
\be
x=X-vT,\quad Y^\prime =Y,\quad Z^\prime=Z,
\ee
then the invariant $J$ assumes the form
\be
J=c^2\Bigl(1-\f{v^2}{c^2}\Bigr)T^2-
2xvT-x^2-Y^{\prime 2}-Z^{\prime 2}.
\ee
Hence we have
\ba
&&J=c^2\left[\sqrt{1-\ds\f{v^2}{c^2}}\,T-\ds
\f{xv}{c^2 \sqrt{1-\ds\f{v^2}{c^2}}}\right]^2-\nonumber\\
\label{3.26}\\
&&-x^2\Biggl[1+\ds\f{v^2}{c^2-v^2}\Biggr]
-Y^{\prime 2}-Z^{\prime 2}.\nonumber
\ea
Expression (3.26) can be written in the form
\be
J=c^2T^{\prime 2}-X^{\prime 2}-Y^{\prime 2}-Z^{\prime 2},
\ee
where
\be
T^\prime =\sqrt{1-\f{v^2}{c^2}}\,T-
\f{xv}{c^2 \sqrt{1-\ds\f{v^2}{c^2}}}=
\f{\ds T-\f{v}{c^2}\,X}{\sqrt{1-\ds\f{v^2}{c^2}}},
\ee
\be
X^\prime =\f{x}{\sqrt{1-\ds\f{v^2}{c^2}}}=
\f{X-vT}{\sqrt{1-\ds\f{v^2}{c^2}}}.
\ee

We see from expression (3.27) that the form-invariance of the invariant $J$ is provided for by the
Lorentz\index{Lorentz} transformations (3.28) and 
(3.29).\index{Lorentz transformations} In deriving the Lorentz\index{Lorentz} 
transformations\index{Lorentz transformations} from the ex\-pres\-sion for the invariant (3.22) we took advantage
of the fact that the invariant $J$ may assume an arbitrary real value. Precisely this circumstance has permitted
us to con\-si\-der quantities $T$ and $X$ as independent variables, that can assume any real values. If we,
following Einstein,\index{Einstein} knew only one value of $J$, equal to zero, we could
not, in principle, obtain Lorentz\index{Lorentz} 
transformations\index{Lorentz transformations} of the ge\-ne\-ral form, since the space variables would be related to the time variable.

In this case the following \textbf{heuristic} approach can be realized. The equation of spherical electromagnetic
wave having its center in the origin of the coordinate system has the following form
\[ c^2T^2-X^2-Y^2-Z^2=0, \]
where  {\bf \textit{c}} is the electrodynamic constant, if we use Galilean coordinates of the ``rest'' system of
reference  \textit{K}. This fact follows from the Maxwell\index{Maxwell} -Lorentz\index{Lorentz} equations\index{Maxwell-Lorentz equations}.

Let us consider two inertial reference systems \textit{K} and \( K^\prime \) with Galilean coordinates\index{Galilean
(Cartesian) coordinates} moving relative to each other with velocity  \( v \) along axis  \textit{X}. Let their origins
coincide at the moment  \( T=0 \) and let a spherical electromagnetic wave is emitted just at this moment from their
common origin. In reference system \textit{K} it  is given by equation
\[ c^2T^2-X^2-Y^2-Z^2=0. \]
As system of reference   \( K^\prime \) is moving with velocity  \( v \), we can use Galilean
transformations\index{Galilean transformations} \index{Galilei} 
\[ x=X-vT,\quad Y=Y^\prime,\quad Z=Z^\prime \]
and rewrite the preceding equation of spherical wave in the following form
\[ c^2T^2\left(1-\f{v^2}{c^2}\right)-2xvT-x^2-Y^{\prime 2}-Z^{\prime 2}=0. \]
The requirement of relativity principle\index{relativity principle} here is reduced to necessity that the electromagnetic
wave in a new inertial reference system\index{inertial reference systems} \( K^\prime \)  has to be also spherical having
its center at the origin of this reference system.

Having this in mind we transform the above equation (as done before) to the following form
\[ c^2T^{\prime 2}-X^{\prime 2}-Y^{\prime 2}-Z^{\prime 2}=0. \]
So, we derive the Lorentz\index{Lorentz} transformations\index{Lorentz transformations}.
\[ T^{\prime}=\f{T-\ds\f{v}{c^2}X}{\sqrt{1-\ds\f{v^2}{c^2}}},\quad
X^{\prime}=\f{X-vT}{\sqrt{1-\ds\f{v^2}{c^2}}},\quad
Y^{\prime}=Y,\quad  Z^{\prime}=Z, \]
but at the light cone only.

Now, we go to the most important stuff. Let us treat variables \textit{T, X, Y, Z,} appearing in the derived
transformations  as \textbf{independent}. Then after inserting these expressions into r.h.s. of equation (3.27) we
can see that they leave  quantity \[ c^2T^2-X^2-Y^2-Z^2\] unchanged due to the linear character of
transformations. Therefore we come to the fundamental invariant \( J \), and so to
pseudo-Euclidean\index{pseudo-Euclidean geometry of space-time} geometry of space-time. It follows from the above,
in particular, that velocity of light both in system  \textit{K}, and in system  \( K^\prime \) is the same and
therefore the principle of constancy of velocity of light\index{principle of constancy of velocity of light} is a
particular consequence of the relativity principle\index{relativity principle}. Precisely this circumstance
remained un\-no\-ti\-ced by A.\,Einstein\index{Einstein} in his 1905 work, in which the
Lorentz\index{Lorentz} trans\-for\-ma\-tions\index{Lorentz transformations} were derived.


Earlier we have shown, following Poincar\'{e},\index{Poincar\'e} that 
Lorentz's\index{Lorentz} ``{\bf local time}''\index{local time} permits  to perform
synchronization\index{synchronization of clocks} of clocks  in a moving reference system at different spatial
points with the aid of a light signal. Precisely expression (3.12) is the condition for the
synchronization\index{synchronization of clocks} of clocks in a moving re\-fe\-rence system. It introduces the
definition of simultaneity\index{simultaneity} of events at different points of space. Poincar\'{e}\index{Poincar\'e} established that Lorentz's\index{Lorentz} ``{\bf
local time}''\index{local time} satisfies this condition.

\textbf{So, the definition of simultaneity\index{simultaneity}
of events in different spatial points by means of a light signal as well as the definition of time in a moving reference
system by means of light signal both were considered 
by Poincar\'e\index{Poincar\'e} 
in his papers of 1898, 1900 and 1904. Therefore nobody has any ground to believe that these ideas have been first treated
by A.\,Einstein\index{Einstein} 
in  1905.}

But let us see, for example, what is written by Academician 
L.\,I.\,Mandel'stam\index{Mandel'stam} 
in his lectures [8]:
\begin{quote}
{\it\hspace*{5mm}``So, the great achievement of Einstein\index{Einstein} consists in
discovering that the  concept of simultaneity\index{simultaneity} is a concept \dots  that we have to define.
People had the knowledge of space, the knowledge of time, had this knowledge many centuries, but nobody guessed
that idea.''.}
\end{quote}
And the following was written by H.\,Weyl:\index{Weyl} 
\begin{quote}
{\it\hspace*{5mm}``\textbf{\ldots we are to discard our belief in the objective
meaning of simultaneity; it was the great achievement 
of Einstein\index{Einstein} in the field of the
theory of knowledge that he banished
this dogma from our minds}, and this is what leads us to rank
his name with that of Copernicus''.}\index{Copernicus}
\end{quote}
Is it possible that L.\,I.\,Mandel'stam\index{Mandel'stam} and 
H.\,Weyl\index{Weyl} have not read  articles and  books by 
Poincar\'e?\index{Poincar\'e}

Academician V.\,L.\,Ginzburg\index{Ginzburg} in his book
 \textsf{``On physics and astrophysics''} (Moscow: Nauka, 1985)
in the article ``{\sf How and who created Special Relativity Theory?}''\footnote{All the citations of Academician
V.\,L.\,Ginzburg\index{Ginzburg} presented here and below are taken from this article. --- {\sl A.\,L.}}\label{wash15} wrote:
\begin{quote}
{\it\hspace*{5mm}``From the other side, in earlier works, in articles and reports by Poincar\'e\index{Poincar\'e}
there are a set of comments which sound almost prophetical. I mean both the necessity to define a concept of
simultaneity,\index{simultaneity} and an opportunity to use light signals for this purpose, and on the principle
of relativity\index{relativity principle}. But Poincar\'e\index{Poincar\'e} have not developed these ideas and
followed Lorentz\index{Lorentz} in his works of 1905-1906''.}
\end{quote}
Let us give some comments to this citation.

To be precise it should be said that \textbf{Poincar\'e\index{Poincar\'e} was the first who formulated the
relativity principle\index{relativity principle} for all physical processes}. He also \textbf{defined the concept
of simultaneity\index{simultaneity} at different spatial points by means of the light signal} in his papers 1898,
1900 and 1904.  In Poincar\'e\index{Poincar\'e} works  [2; 3]\textbf{these concepts} have been adequately realized
in the language of Lorentz group\index{group}\index{group, Lorentz}\index{Lorentz} which provides fulfilment  both
the requirement of relativity principle\index{relativity principle}, and the introduction of his own
\textbf{modified local} Lorentz\index{Lorentz} time\index{local time} in every inertial system of
reference\index{inertial reference systems}. All that automatically provided a unique synchronization of
clocks\index{synchronization of clocks} by means of the light signal in every inertial reference
system\index{inertial reference systems}. Just due to this not any
 \textbf{further development of these concepts} were required after H.\,A.\,Lorentz\index{Lorentz}  work of 1904.

It was necessary only to introduce these concepts into the bosom of the theory. It was precisely realized in works [2; 3]
by means of the Lorentz\index{Lorentz} group\index{group}
\index{group, Lorentz}, discovered by H.\,Poincar\'e. Poincar\'e
does not follow Lorentz, he develops \textbf{his own ideas} by using Lo\-rentz achievements and he completes the
creation of the theory of relativity in this way. Exactly in papers  [2; 3] he extends
Lorentz invariance\index{Lorentz invariance} on all the forces of nature, including gravitational; he discovers
equations of the relativistic 
mechanics\index{equations of relativistic mechanics}
\index{Poincar\'{e}'s equations of mechanics}; he discovers fundamental invariant
\[
c^2t^2-x^2-y^2-z^2,
\]
which determines the geometry of space-time.

H.\,Poincar\'e\index{Poincar\'e} approach
is transparent and contemporary though it is realized almost one hundred years ago. How is it possible not to understand
this after reading Poincar\'e works [2; 3]?\index{Poincar\'e}

In the article (1905) {\sf ``On the electrodynamics of moving bodies''} (\S\,3)
A.\,Einstein\index{Einstein} took
the relation (3.12) as the initial equation in
searching for the function $\tau$. But hence one can na\-tu\-ral\-ly obtain
nothing, but
Lorentz's\index{Lorentz} 
``{\bf local time}''.\index{local time}
We write the equation obtained
by him in the form
\[
\tau=\f{a}{1-\ds\f{v^2}{c^2}}
\left[\left(1-\f{v^2}{c^2}\right)T-\f{v}{c^2}x\right],
\]
Where
{\bf \textit{a}}
is an unknown function depending only on the velo-\break city
$v$.

Hence it is seen that this expression differs from the
Lorentz\index{Lorentz} 
``{\bf local time}''\index{local time}
(3.6) only by a factor depending on the velocity $v$ and which
is not determined by condition (3.12). It is strange to see that 
A.\,Einstein\index{Einstein} knows that this is Lorentz 
''local time'',\index{local time}
but he does not refer to the author. Such a treatment is not an exception for him.

Further, for a beam of light leaving
the source at the time moment $\tau=0$ in the direction of increasing
$\xi$ values,
Einstein\index{Einstein} 
writes:
\[
\xi=c\tau
\]
or
$$
\xi=\f{ac}{1-\ds\f{v^2}{c^2}}
\left[\left(1-\f{v^2}{c^2}\right)T-\f{v}{c^2}x\right].\eqno(\beta)
$$
He further finds
$$
x=(c-v)T.\eqno(\delta)
$$
Substituting this value of $T$ into the equation for $\xi$,
Einstein\index{Einstein} 
obtains
\[
\xi=\f{a}{1-\ds\f{v^2}{c^2}}x.
\]
Since, as it will be seen further from
Einstein's\index{Einstein} 
article, the quantity
{\bf \textit{a}}
is given as follows
\[
a=\sqrt{1-\f{v^2}{c^2}},
\]
then, with account of this expression, we obtain:
\[
\xi=\f{x}{\sqrt{1-\ds\f{v^2}{c^2}}}.
\]
Substituting, instead of $x$, its value (3.3),
$$
x=X-vT,\eqno(\nu)
$$
Einstein\index{Einstein} 
obtains for $\xi$ an expression of the form:
\[
\xi=\f{X-vT}{\sqrt{1-\ds\f{v^2}{c^2}}},
\]
which he namely considers as the
Lorentz\index{Lorentz} 
transformation\index{Lorentz transformations}
for $\xi$,
{\bf implying that
{\bf \textit{X}}
and
{\bf \textit{T}}
are arbitrary and
independent}. However, this is not so. He does not take into account,
that according to ($\delta$) and ($\nu$), there exists the equality
\[
X-vT=(c-v)T,
\]
hence it follows that
\[
X=cT.
\]

Hence it follows, that
Einstein\index{Einstein} 
obtained the
Lorentz\index{Lorentz} 
trans\-for\-ma\-tions for
$\xi$ only for the partial case of $X=cT$:
\[
\xi=X\sqrt{\f{1-\ds\f{\,v\,}{\,c\,}}{1+\ds\f{\,v\,}{\,c\,}}}.
\]
This can be directly verified, if in formula ($\beta$) for $\xi$ one
sub\-sti\-tu\-tes, instead of the value of $T$ from formula ($\delta$), as done
by Einstein,\index{Einstein} 
the value of $x$ from the same formula.
Then we obtain:
\[
\xi=\f{a}{1+\ds\f{\,v\,}{\,c\,}}X,\quad X=cT.
\]
Taking into account the expression for
{\bf \textit{a}},
we again arrive at
the formula found
earlier,
\[
\xi=X\sqrt{\f{1-\ds\f{\,v\,}{\,c\,}}{1+\ds\f{\,v\,}{\,c\,}}},\quad X=cT.
\]

But further in the text of the article A.\,Einstein\index{Einstein} exploits the general
form of Lorentz\index{Lorentz}  transformations without any comments. 
A.\,Einstein\index{Einstein}  has not observed that the principle of
relativity\index{relativity principle} together with electrodynamics obligatory requires a construction of
four-dimensional physical quantities, in accordance with the 
Lorentz\index{Lorentz} group\index{group, Lorentz}\index{group}. As
a result this requires  presence of the group invariants testifying to the pseudo-Euclidean geometry of
space-time\index{pseudo-Euclidean geometry of space-time}. Just due to this Einstein has not succeeded in finding
relativistic equations of mechanics\index{Poincar\'{e}'s equations of mechanics}, because he has not discovered
the law of transformation for Lorentz\index{Lorentz} 
force\index{Lorentz force}. He also has not understood that energy and
momentum of a particle constitute a unified quantity and that they transform under Lorentz\index{Lorentz}
transformations\index{Lorentz transformations} in the same way as $ct,x,y,z$. It should be especially emphasized
that Einstein,\index{Einstein} in his  work of 1905, in contrast to 
Poincar\'e\index{Poincar\'e}, has not extended Lorentz transformations onto all forces of nature, for example,
onto gravitation. He wrote later that \textit {``in the framework of special relativity theory there is no place
for a satisfactory theory of gravitation''.} But as it is shown in [5] this statement is not correct.

Owing to the
Maxwell-Lorentz\index{Lorentz} 
\index{Maxwell} 
equations,\index{Maxwell-Lorentz equations}
the relativity prin\-ciple\index{relativity principle}
for
inertial reference systems\index{inertial reference systems}
led
Poincar\'{e}~\cite{3}\index{Poincar\'e} 
and, sub\-se\-quen\-tly,
Minkowski\index{Minkowski} 
[4] to discovering the pseu\-do-Euc\-li\-de\-an ge\-o\-met\-ry of space-time.
Precisely for this, we indebted to
Poincar\'{e}\index{Poincar\'e} 
and
Min\-kow\-ski.\index{Minkowski} 
In 1908
H.\,Minkowski,\index{Minkowski} 
addressing the 80-th mee\-ting of German naturalists
and doctors in Co\-logne, noted [4]:

\begin{quote}
{\it\hspace*{5mm}``The views of space and time which I wish to lay
before you have sprung from the soil
of experimental physics, and therein lies their strength. 
They are radical. Henceforth space by itself, 
and time by itself, are doomed to fade away into mere
shadows, and  only a kind of union
of the two will preserve an independent reality''.}
\end{quote}

Therefore, the essence of special relativity theory consists in the
following ({\bf it is a postulate}): {\bf all physical processes proceed
in four-dimensional\index{four-dimensional space-time}
space-time $(ct, \vec x)$, the geometry of which is
pseudo-Euclidean\index{pseudo-Euclidean geometry of space-time}
and is determined by the
interval} (3.23).

The consequences of this postulate are energy-momentum and angular momentum conservation laws, the existence of
inertial reference systems\index{inertial reference systems}, the relativity principle\index{relativity principle}
for all physical phenomena, the Lorentz transformations, the constancy of the velocity of light in Galilean
coordinates of an inertial system, the retardation of time, the Lorentz contraction\index{Lorentz
contraction}\index{Lorentz}, the opportunity to use non-inertial reference
systems\index{non-inertial reference systems}, 
the ``clock paradox''\index{``clock paradox''}, the Tho\-mas
precession\index{Thomas}, the Sagnac effect\index{Sagnac} 
and so on. On the base of this postulate and the quantum ideas a set of fundamental conclusions
was obtained and the quantum field theory was constructed.

By centennial of the theory of relativity it is high time to make clear that  constancy of the light velocity in
all inertial systems of reference is not a fundamental statement of the theory of relativity.

Thus, investigation of electromagnetic phenomena to\-ge\-ther with
Poincar\'{e}'s\index{Poincar\'e} 
relativity principle\index{relativity principle}
resulted in the uni\-fi\-cation of space
and time in a unique four-dimensional con\-ti\-nu\-um of events and permitted
to establish the
pseudo-Eucli\-dean\index{pseudo-Euclidean geometry of space-time}
geometry of this con\-ti\-nu\-um. Such a
four-dimensional\index{four-dimensional space-time}
space-time is homogeneous and isotropic.

These properties of space-time  provide validity
 of
fundamental
con\-ser\-vation laws\index{fundamental conservation laws}
of
energy, momentum and angular momentum  in a closed
physical system.
{\bf The pseudo-Euclidean\index{pseudo-Euclidean geometry of space-time} geometry of space-time reflects the general
dy\-na\-mi\-cal properties of matter, which make it universal}.
Investigation of various forms of matter, of its laws of motion is at
the same time investigation of space and time. Although the actual
structure of space-time has been revealed to us as a result of studying
matter (electrodynamics), we sometimes speak of space as of an arena,
in which some or other phenomena take place.
Here, we will make no mistake,
if we remember that this arena does not exist by itself, without
matter. Some\-times it is said that space-time
(Minkowski\index{Minkowski} 
space)\index{Minkowski space}
is
given a priori, since its structure does not change under the influence
of matter. Such an invariability of
Minkowski\index{Minkowski} 
space\index{Minkowski space}
arises owing to its
{\bf universality} for all physical fields, so the impression is thus
created that it exists as if independently of matter. Probably just due to a vagueness of the essence of special
relativity theory for him
A.\,Einstein\index{Einstein} arrived at the conclusion that {\it
``within
special relativity theory there is no place for a satisfactory theory of gravity''.}

In Einstein's\index{Einstein} 
general relativity theory, special relativity the\-ory is
certainly not satisfied, it is considered a limit case. In 1955
A.\,Einstein\index{Einstein} 
wrote:
\begin{quote}
{\it\hspace*{5mm}$\ll$An essential achievement of general relativity theory consists in that it has saved physics
from the necessity of introducing an ``inertial reference 
system''\index{inertial reference systems} (or
``iner\-ti\-al reference systems'')$\gg$.}
\end{quote}

However, even now, there exists absolutely no experimental or ob\-ser\-va\-ti\-o\-nal
fact that could testify to the violation of spe\-ci\-al relativity theory.
For this reason no renunciation, to what\-ever extent, of its rigorous and
precise application in  studies of gravitational phe\-no\-me\-na, also, can be
justified. Especially  taking into account that all known gravitational
effects are explained precisely within the framework of special relativity
theory~\cite{5}. {\bf Renunciation of special relativity theory leads to
renunciation of the fundamental con\-ser\-va\-tion laws of energy, momentum and
angular momentum.} Thus, having adopted the hypothesis that all natural
phenomena proceed in
pseudo-Euclidean\index{pseudo-Euclidean geometry of space-time}
space-time, we automatically
comply with all the requirements of
fundamental conservation laws\index{fundamental conservation laws}
and
confirm the {\bf existence of
inertial reference systems.}\index{inertial reference systems}

The space-time continuum, determined by the interval (3.23) can be described in arbitrary coordinates, also. {\bf
In transition to arbitrary coordinates, the geometry of 
four-dimensional\index{four-dimensional space-time}
space-time does not change. However, three-di\-men\-si\-onal space will no longer be Euclidean\index{Euclid}
in arbitrary co\-or\-di\-na\-tes.} To simplify our writing we shall, instead of variables $T,X,Y,Z$,
introduce the variables $X^\nu, \nu=0,1,2,3, X^0=cT$. We now perform transition from the variables $X^\nu$ to the
arbitrary variables $x^\nu$ with the aid of the transformations \be
X^\nu=f^\nu (x^\sigma).
\ee These transformation generally lead to a non-inertial reference 
system.\index{non-inertial reference systems}
Calculating the differentials \be
dX^\nu =\f{\pa f^\nu}{\pa x^\lambda}dx^\lambda
\ee
(here and further summation is performed from 0 to 3 over identical
indices $\lambda$) and substituting them into (3.23) we obtain an
expression for the interval
in the non-inertial reference system\index{non-inertial reference systems} \be
(d\sigma)^2=\gamma_{\mu\lambda}(x)dx^\mu dx^\lambda.
\ee Here, $\gamma_{\mu\lambda}(x)$ is the metric 
tensor\index{metric tensor of space} of
four-dimensional\index{four-dimensional space-time} space-time, it is given as follows \be
\gamma_{\mu\lambda}(x)=\sum_{\nu=0}^3 \varepsilon^\nu \f{\pa f^\nu}{\pa x^\mu}\cdot\f{\pa f^\nu}{\pa x^\lambda},
\quad \varepsilon^\nu=(1,-1,-1,-1).
\ee
Expression (3.32) is invariant with respect to arbitrary co\-or\-di\-nate
transformations. Expression (3.33) represents the general form of
the pseudo-Euclidean metric.

The difference between a metric of the form (3.23) from the metric (3.32)
is usually, in accordance with
Einstein's\index{Einstein} 
ideas, at\-tri\-bu\-ted to the
existence of the gravitational field. But this is incorrect. No
gravitational field is present in a metric of the form (3.32). Ideas of
accelerated reference systems in
Minkowski\index{Minkowski} 
space\index{Minkowski space}
have played an important
heuristic role in Einstein\index{Einstein} reflections on the problem of gravitation.
They contributed to his arriving
at the idea of describing the gravitational field with the aid of the
metric tensor\index{metric tensor of space}
of
Riemannian\index{Riemann} 
space, and for this reason
Einstein\index{Einstein} 
tried to
retain them, although they have nothing to do with the gravitational field.
Precisely such circumstances prevented him from revealing the essence of
special relativity the\-o\-ry. From a formal, mathematical, point of view
Einstein\index{Einstein} 
highly ap\-pre\-ci\-a\-ted
Minkowski's\index{Minkowski} 
work, but he never penetrated the
pro\-found physical essence of
Minkowski's\index{Minkowski} 
work, even though the ar\-ticle
dealt with a most important dis\-co\-ve\-ry in physics --- the {\bf dis\-co\-ve\-ry of
the pseudo-Euclidean structure of space and time}.

Einstein\index{Einstein} 
considered special relativity theory only related to an
interval
of the form (3.23), while ascribing (3.32) to general re\-la\-ti\-vi\-ty
theory. Regretfully, such a point of view still prevails in textbooks
and monographs expounding relativity theory.

Consider a certain non-inertial reference 
system\index{non-inertial reference systems} where the metric
tensor\index{metric tensor of space} of  space-time is given as $\gamma_{\mu\lambda}(x)$. It is, then, readily
shown that there exists an infinite number of reference systems, in which the interval (3.32)\index{interval} is
as follows
\be
(d\sigma)^2=\gamma_{\mu\lambda}(x^\prime)
dx^{\prime\mu}dx^{\prime\lambda}.
\ee A partial case of such transformations is represented by the 
Lorentz\index{Lorentz} transformations,\index{Lorentz transformations} which relate one inertial re\-fe\-rence
system\index{inertial reference systems} to another. We see that the transformations of co\-or\-di\-nates, which
leave the metric form-invariant, result in that physical phenomena proceeding in such reference systems at
identical conditions can never permit to distinguish one reference system from another. Hence, one can give a more
general formulation of the {\bf relativity principle},\index{relativity principle} which not only concerns {\bf
inertial reference systems},\index{inertial reference systems} 
but {\bf non-inertial ones}~\cite{6},\index{non-inertial reference systems} as well:
\begin{quote}{\it\hspace*{5mm}``Whatever physical reference
system (inertial\index{inertial reference systems} 
or non-inertial)\index{non-inertial reference systems} we
choose, it is always possible to point to an infinite set of other reference systems, such as all physical
phenomena proceed there exactly like in the initial reference system, so we have no, and cannot have any,
experimental means to distinguish, namely in which reference system of this infinite set we are''.}
\end{quote}

It must be noted that, though the metric tensor\index{metric tensor of space} $\gamma_{\mu\lambda}(x)$ in
(3.33) de\-pends on  coordinates, nevertheless the space
remains pseudo-Euclidean.\index{pseudo-Euclidean geometry of space-time} Although this is evident, it must be
pointed out, since even in 1933 A.\,Einstein\index{Einstein} wrote the ab\-so\-lute
opposite~\cite{7}:
\begin{quote}
{\it\hspace*{5mm}$\ll$In the special theory of relativity, as Minkowski had shown, this metric was а quasi-Euclidean one,\break i.\,e., the square of the ``length'' $ds$ of а line element was a certain quadratic function of the 
differentials of the coordinates.\\
\hspace*{5mm}If other coordinates are introduced by means of а non-linear
transformation, $ds^2$ remains a homogeneous function of the
differentials of the coordinates, but the coefficients of this 
function ($g_{\mu\nu}$) cease to be constant and become certain functions of the coordinates. In mathematical terms this means that physical 
(four-dimensional)\index{four-dimensional space-time} space has a 
Riemannian\index{Riemann} metric$\gg$.}
\end{quote}

This is, naturally, {\bf wrong}, since it is impossible to transform pseudo-Euclidean\index{pseudo-Euclidean
geometry of space-time} geometry into Riemannian\index{Riemann} 
geometry\index{Riemannian geometry} by applying the transformations of coordinates (3.30). Such a statement by
A.\,Einstein\index{Einstein} had profound physical roots. 
Ein\-stein\index{Einstein}
was convinced that the pseudo-Euclidean metric in ar\-bit\-ra\-ry coordinates,
$\gamma_{\mu\lambda}(x)$, describes the gravitational field, also. These ideas, put forward by
Einstein,\index{Einstein} restricted the framework of special relativity theory and in
such form be\-came part of the material expounded in textbooks and mo\-no\-graphs, which had hindered
comprehension of the essence of relativity theory.

Thus, for example, Academician L.I.\,Mandel'stam,\index{Mandel'stam} in his
lectures on relativity theory~\cite{8}, especially noted:
\begin{quote}
{\it\hspace*{5mm}``What actually happens, how an accelerated moving clock
shows time and why it slows down or does the opposite cannot be answered
by special relativity theory, because it absolutely does not deal with
the issue of accelerated moving reference systems''.}
\end{quote}

The physical sources of such a limited understanding of special relativity theory origin from
A.\,Einstein.\index{Einstein} Let us pre\-sent a number of his statements concerning special re\-la\-ti\-vi\-ty theory. In 1913 he wrote~\cite{9}:
\begin{quote}
{\it\hspace*{5mm}``In the case of the customary theory of  relativity 
only linear orthogonal substitutions are permissible''.}
\end{quote}
In the next article of the same year he writes~\cite{10}:
\begin{quote}
{\it\hspace*{5mm}``While in the original theory of relativity
the independent of the physical equations from
the special choice of the reference system is based on
the postulation of the fundamental in\-va\-ri\-ant 
$ds^2=\underset{i}\varSigma dx_i^2$,
we are concerned with constructing 
a theory in which the most general line element of the  form
\[ds^2=\underset{i,k}\varSigma g_{ik}dx^i dx^k\]
plays the role of the fundamental invariant''.}
\end{quote}
Later, in 1930,
A.\,Einstein wrote~\cite{11}:\index{Einstein} 
\begin{quote}{\it\hspace*{5mm} ``In the special theory of relativity  
those coordinate changes (by transformation) are permitted for
which also in new coordinate system
the quantity  $ds^2$ (fundamental invariant) 
equals the sum of squares of the coordinate.
Such trans\-for\-ma\-tions are called
Lo\-rentz\index{Lorentz} 
transformation''.}\index{Lorentz transformations}
\end{quote}

Although
Einstein,\index{Einstein} 
here, takes advantage of the invariant (in\-ter\-val)
discovered by
Poincar\'{e},\index{Poincar\'e} 
he understands it only in a limited (strictly
diagonal) sense. For
A.\,Einstein\index{Einstein}
it was difficult to see that the
Lorentz\index{Lorentz} 
transformations\index{Lorentz transformations}
and the relativity of time concealed a fun\-da\-men\-tal
fact: space and time form a unique four-dimensional con\-ti\-nu\-um with
pseudo-Euclidean\index{pseudo-Euclidean geometry of space-time}
geometry, determined by the
interval
\be
ds^2=\gamma_{\mu\nu}(x)dx^\mu dx^\nu,\qquad \det(\gamma_{\mu\nu})=\gamma<0,
\ee
with the metric tensor\index{metric tensor of space}
$\gamma_{\mu\nu}(x)$, for which the
Riemannian\index{Riemann} 
cur\-va\-ture\index{Riemannian curvature tensor}
tensor equals zero. But, precisely, the existence of the
four-di\-men\-si\-o\-nal space of events with a pseudo-Euclidean metric permitted
to establish that a number of vector quan\-ti\-ti\-es in
Euclidean\index{Euclid}
three-dimensional space are at the same time components of four-di\-men\-si\-o\-nal
quantities together with certain scalars in
Euclidean\index{Euclid}
space.

This was performed by
H.\,Poincar\'{e}\index{Poincar\'e} 
and further developed by
H.\,Minkowski.\index{Minkowski} 
Very often, without understanding the es\-sence of theory,
some people write that
Minkowski\index{Minkowski} 
allegedly gave the ge\-o\-met\-ri\-cal
interpretation of relativity theory. This is not true.
\textbf{On the
basis of the
group\index{group}\index{group, Poincar\'e}
discovered by
Poincar\'{e},\index{Poincar\'e} 
H.\,Poincar\'{e}\index{Poincar\'e} 
and
H.\,Minkowski\index{Minkowski} 
revealed the pseu\-do-Euc\-li\-de\-an geometry of\break space-time, which
is precisely the essence of special relativity theory.}

In 1909
H.\,Minkowski\index{Minkowski}
wrote about this in the article {\sf ``Space and
time''}:\index{Minkowski space}
\begin{quote}{\it\hspace*{5mm}``Neither
Einstein,\index{Einstein} nor Lorentz\index{Lorentz} 
dealt with the con\-cept of space, maybe because in the case of the afor\-e\-men\-ti\-o\-ned special
transformation, under which the $x^\prime, t^\prime$  plane coincides with the $x, t$ plane, it may be understood
that the $x$ axis of space retains its position. The attempt to thus evade the concept of space could have indeed
been regarded as a certain impudence of the mathematical thought. But after making this step, surely unavoidable
for true comprehension of the $G_c$ group\index{group}
 ({\rm the Lorentz\index{Lorentz} 
group.} --- A.L.),\index{group}\index{group, Lorentz} the term ``{\bf \textit{relativity postulate}}''
for requiring invariance with respect to the $G_c$ group\index{group} seems to me too insipid. Since the meaning
of the postulate reduces to that in phenomena we only have the four-dimensional world in space and time, but that
the projections of this world onto space and time can be taken with a certain arbitrariness, I would rather give
this statement the title ``{\bf \textit{postulate of the absolute world}}'' or, to be short, world postulate''.}
\end{quote}

It is surprising, but in H.\,Minkowski's\index{Minkowski} work there is no reference to
the articles [2] and [3] by H.\,Poincar\'{e},\index{Poincar\'e} although it just gives the
details of what had already been presented in refs. [2] and [3]. However, by the brilliant exposition before a
broad audience of naturalists it attracted general attention. In 1913, in Germany, a col\-lec\-tion of articles on
relativity by H.\,A.\,Lorentz,\index{Lorentz}  A.\,Einstein,\index{Einstein}
H.\,Minkowski\index{Minkowski}  was published. The fundamental works
[2] and [3] by H.\,Poincar\'{e}\index{Poincar\'e} were not in\-clu\-ded in the collection.
In the comments by A.\,Sommerfeld\index{Sommerfeld} to
Min\-kow\-ski's\index{Minkowski}  work Poincar\'{e}\index{Poincar\'e} 
is only mentioned in relation to par\-ti\-cu\-lars. Such hushing up of the fundamental works of
H.\,Poincar\'{e}\index{Poincar\'e} in relativity theory is difficult to understand.

E.\,Whittaker\index{Whittaker} was the first who came to the conclusion of
the decisive contribution of H.\,Poincar\'e\index{Poincar\'e} to this problem when studying
the history of creation of the special relativity theory, 50 years ago. His monograph caused a remarkably angry
reaction of some authors. But {\bf E.\,Whittaker\index{Whittaker} was mainly right. H.\,Poincar\'e\index{Poincar\'e}  really created the special
theory of relativity grounding upon  the
Lorentz work of 1904 and gave to this theory a general character by
extending it onto all physical phenomena.} Instead of a more thorough study and comparison of 
Einstein's\index{Einstein} 1905 work and Poincar\'{e}'s\index{Poincar\'e} papers (it is 
the only way of objective study of the problem) the way of complete rejection of  Whittaker's\index{Whittaker} conclusions was chosen. So, the idea that the theory of relativity was
created independently and exclusively by A.\,Einstein\index{Einstein} was propagated in
literature without detail investigations. This was also my view up to the middle of 80-s until I had read the articles
by H.\,Poincar\'e\index{Poincar\'e} and A.\,Einstein\index{Einstein}.

\newpage
\markboth{thesection\hspace{1em}}{}
\section{The relativity of time and the contraction of length}

Consider the course of time in two inertial reference 
systems,\index{inertial reference systems} one of which will
be considered to be at rest, while another one will move with respect to the first one with a velocity $v$.
According to the relativity principle,\index{relativity principle} the change in time shown by the clocks (for a
given time scale) in both reference systems is the same. Therefore, the both count {\bf their own
physical}\index{physical time} time in the same manner. If the clock in the moving reference system is at rest,
then its interval in this system of reference is \be
d\sigma^2=c^2dt^{\prime 2},
\ee
$t^\prime$ is the time shown by the clock in this reference system.

Since this clock moves relative to the other reference system with
the velocity $v$, the same
interval,
but now in the reference system
at rest will be
\be
d\sigma^2=c^2dt^2\Bigl(1-\f{v^2}{c^2}\Bigr),
\ee
here $t$ is the time shown by the clock at rest in this reference
system, and
\be
v^2=\Biggl(\f{dx}{dt}\Biggr)^2+\Biggl(\f{dy}{dt}\Biggr)^2+
\Biggl(\f{dz}{dt}\Biggr)^2.
\ee

From relations (4.1) and (4.2) we find the relationship between
the {\bf time durations} in these
inertial reference systems\index{inertial reference systems}
in
the description of the {\bf physical phenomenon}
\be
dt^\prime=dt\sqrt{1-\f{v^2}{c^2}}.
\ee

One often reads that a retardation of moving clocks takes place. It is wrong, because such a statement contradicts
the principle of relativity.\index{relativity principle} The clock rate in all inertial reference
systems\index{inertial reference systems} does not change. The clocks equally measure physical 
time\index{physical time} of their own inertial system of 
reference.\index{inertial reference systems} This is not a change of the
clock rate but a change of a physical process duration. The duration of a local physical process according to the
clock of this inertial system\index{inertial reference systems} or a clock in other inertial system is in general
different. It is minimal in the system where the process is localized in one spatial point. Precisely this meaning
is implied in saying about the  {\bf retardation of time}. 
\markboth{thesection\hspace{1em}}{4. The relativity of
time and the contraction of length}

Integrating this relation, we obtain
\be
\bigtriangleup t^\prime =\bigtriangleup t
\sqrt{1-\f{v^2}{c^2}}.
\ee
This expression is a consequence of the existence of the fundamental
invariant (3.22). The ``time dilation'' (4.5) was considered as
early as in 1900 by J.\,Larmor.\index{Larmor}
As noted by
W.\,Pauli,\index{Pauli} 
``\textbf{it received its first clear statement only
by Einstein}''\index{Einstein} 
from the
``Lorentz\index{Lorentz} 
trans\-for\-mation''.\index{Lorentz transformations}

\markboth{thesection\hspace{1em}}{4. The relativity of time and the contraction of length}
We shall apply this equality
to an elementary particle with a lifetime
at rest equal to $\tau_0$. From (4.5) after setting
$\bigtriangleup t^\prime=\tau_0$, we find the lifetime of the moving
particle
\be
\bigtriangleup t=\f{\tau_0}{\sqrt{1-\ds\f{v^2}{c^2}}}.
\ee
Precisely owing to this effect it turns out to be possible to transport
beams of high energy particles in vacuum over quite large distances from
the accelerator to the experimental de\-vices, although their lifetime in
the state of rest is very small.

In the case considered above we dealt with a time-like interval\index{interval} $d\sigma^2>0$. We shall now
consider another example, when the interval\index{interval} between the events is space-like, $d\sigma^2<0$. Again
we consider two such inertial reference sys\-tems.\index{inertial reference systems} Consider measurement, in a
moving reference system, of the length of a rod that is at rest in another reference system. We first determine
the method for measuring the length of a moving rod. Consider an observer in the moving reference system, who
records the ends of the rod, $X_1^\prime$ and $X_2^\prime$, at the {\bf same moment of time} \be
T_1^\prime=T_2^\prime,\,
\ee this permits to reduce the interval $S_{12}^2$\index{interval} in the moving reference system to the spatial
part only \be
S_{12}^2=-\left(X_2^\prime-X_1^\prime\right)^2=-\ell^2.
\ee Thus, in our method of determining the length of a moving rod, it is rather natural to consider the quantity
$\ell$ as its length.

The same interval\index{interval} in the reference system at rest, where the rod is in the state of rest, is given
as follows \be
S_{12}^2=c^2(T_2-T_1)^2-(X_2-X_1)^2.
\ee

But, in accordance with the
Lorentz\index{Lorentz} 
transformations\index{Lorentz transformations}
we have
\be
T_2^\prime-T_1^\prime=\gamma\Bigl[(T_2-T_1)-
\f{v}{c^2}(X_2-X_1)\Bigr],
\ee
whence for our case (4.7) we find
\be
T_2-T_1=\f{v}{c^2}(X_2-X_1)=\f{v}{c^2}\ell_0\,,
\ee $\ell_0$ is length of the rod in the reference system at rest. Substituting this expression into (4.9) we
obtain \be
S_{12}^2=-\ell_0^2\Bigl(1-\f{v^2}{c^2}\Bigr).
\ee
Comparing (4.8) and (4.12) we find
\be
\ell=\ell_0\,\sqrt{1-\f{v^2}{c^2}}.
\ee
From relations (4.7) and (4.11) we see, \textbf{that events that are
si\-mul\-ta\-ne\-ous in one
inertial reference system\index{inertial reference systems}
will not be
simultaneous in another
inertial reference system,\index{inertial reference systems}
so the
notion of
simultaneity\index{simultaneity}
is relative}.
\textbf{Relativity of time} is a straightforward consequence of the
\textbf{definition of simultaneity}\index{simultaneity}
for different spatial points of
inertial reference system\index{inertial reference systems}
by means of a light signal.
The con\-traction (4.13)
is a consequence of the relative nature of
si\-mul\-ta\-nei\-ty,\index{simultaneity}
or
to be more precise, of the {\bf existence of the fun\-da\-men\-tal
invariant} (3.22).

Length contraction (4.13), as a hypothesis to explain the negative result of
the Michelson-Morley\index{Michelson}\index{Morley} experiment,
was initially suggested by G.F.\,FitzGerald\index{FitzGerald} in 1889.
Later, in 1892, the same hypothesis was formulated by
H.A.\,Lorentz.\index{Lorentz}

Thus, we have established that, in accordance with special re\-la\-ti\-vi\-ty theory, the time
interval\index{interval} between events for a local object and the length of a rod, given the method of
mea\-su\-re\-ment of (4.7), are relative. They depend on the choice of the inertial reference
system.\index{inertial reference systems} Only the interval\index{interval} between events has an absolute sense.
It must be especially noted that contraction of the length of a rod (4.13) is determined not only by the
pseudo-Euclidean\index{pseudo-Euclidean geometry of space-time} structure of space-time, but also by our {\bf
method of measuring length}, so contraction, unlike the slowing down of time (4.5), does not have such essential
 physical significance. This is due to the slowing down of time being related to a {\bf local object}, and
such objects exist in Nature, they are described by the time-like interval $d\sigma^2
>0$;\index{interval} consequently, a causal relationship is realized, here. Con\-traction of length is related to
different points in space and is, therefore, described by the space-like interval\index{interval} $d\sigma^2 <0$,
when no causal relationship is present.

Let us return to the issue of Lorentz\index{Lorentz} contraction,\index{Lorentz
contraction} de\-ter\-mi\-ned by formula (4.13). We saw that in the case considered above, when the rod is at rest
in the unprimed inertial re\-fe\-rence 
sys\-tem\index{inertial reference systems} and has a length $\ell_0$, for
all observers in other inertial reference systems there occurs, given the adopted method of measuring length
(4.7), contraction, and the length will be determined by formula (4.13). It is quite evident that here nothing
happens to the rod. Some authors call this contraction effect kinematical, since the rod undergoes no deformation,
here. And they are right in this case, and there is no reason for criticizing them. However, it must be noted that
this kinematics is a consequence of the 
pseudo-Euclidean\index{pseudo-Euclidean geometry of space-time} structure
of space, which reflects the general dynamic pro\-per\-ties of matter --- the conservation laws.

Back in 1905
H.\,Poincar\'{e}\index{Poincar\'e} 
wrote the following about this situation:
\begin{quote}
{\it\hspace*{5mm}``If we were to accept the relativity 
principle,\index{relativity principle} then we would find a
common constant in the law of gravity and in electromagnetic laws, the velocity of light. Precisely in the same
way, we would also encounter it in all the other forces of whatever origin, which can be only explained from two
points of view: either eve\-ry\-thing existing in the world is of electromagnetic origin, or this property, that
is, so to say, common to all phy\-si\-cal phenomena, is nothing more, than an external ap\-pe\-a\-rance, something
related to the methods of our measurements. How do we perform our measurements? Earlier we would have answered as
follows: by car\-rying bodies, considered solid and unchangeable, one to the place of the other; but in modern
theory, taking into account the 
Lorentz\index{Lorentz} contraction,\index{Lorentz contraction} this is no longer correct. Ac\-cor\-ding to this theory, {\bf \textit{two segments are, by
definition, equal, if light covers them in the same time}} (\rm singled out by me. --- {\it A.L.})}''.
\end{quote}
A totally different situation arises in the case of motion with ac\-ce\-le\-ra\-tion. If, for instance, the rod,
that is at rest in the unprimed inertial reference system and has a length $\ell_0$, starts moving with
acceleration along its length so that {\bf both of its ends start moving simultaneously}, then in the reference
system, related to the rod, its length will increase according to the law
\[
L=\f{\ell_0}{\sqrt{1-\ds\f{v^2(t)}{c^2}}},
\]
or, if formula (12.3) is taken into account, then one can express the velocity $v(t)$ via the acceleration $a$ and
obtain the ex\-pression
\[
L=\ell_0\sqrt{1+\f{a^2t^2}{c^2}}.
\]
of events being the same in the unprimed inertial reference system and in the reference system moving with
ac\-ce\-le\-ra\-tion $a$, in which the rod is at rest. This means that the rod undergoes rupture strains.

Earlier we found the Lorentz\index{Lorentz} 
transformations\index{Lorentz transformations} for the case, when the motion of one reference system with respect to another inertial reference
system proceeded with a constant velocity along the $X$ axis. Now, consider the general case, when the motion
takes place with a velocity $\vec v$ in an arbitrary direction \be
\vec r=\vec R-\vec v\,T.
\ee Transformation (4.14) provides the transition to the inertial reference system the origin of which moves with a
constant velocity $\vec v$ related to the initial reference system.

Let us decompose vectors $\vec R,\vec R^\prime$  in the initial 
Galilean\index{Galilei} 
reference system along the direction of the velocity $\vec v$ and along the direction perpendicular to the velocity $\vec
v$: \be \vec R=\f{\vec v}{\vert\vec v\vert}R_\parallel +\vec R_\perp\,, \qquad \vec R^\prime=
\f{\vec v}{\vert\vec v\vert}R_\parallel^\prime +\vec R_\perp^\prime\,.
\ee On the basis of the Lorentz\index{Lorentz} transformations
(3.1),\index{Lorentz transformations} one may expect that only the longitudinal quantities will be changed, while the
transverse quantities remain without change. \be R_\parallel^\prime=\gamma(R_\parallel -vT),\quad
T^\prime=\gamma\,\Biggl(T-\f{v}{c^2}R_\parallel\Biggr)\,,
\quad \vec R_\perp^\prime=\vec R_\perp\,.
\ee
In accordance with (4.15) we find
\be
R_\parallel=\f{(\vec v\vec R)}{v}\,,\quad
\vec R_\perp=\vec R-\f
{\vec v(\vec v \vec R)}{v^2}\,.
\ee
Substituting (4.17) into (4.16) and, then, into (4.15) we obtain
\be
\vec R^\prime=\vec R+(\gamma -1)\f
{(\vec v \vec R)}{v^2}\vec v-\gamma\,\vec v\,T\,,
\ee
\be
T^\prime=\gamma\,\Biggl(T-\f
{(\vec v\vec R)}{c^2}\Biggr)\,.
\ee In obtaining formulae (4.18) and (4.19) we have considered that under the general transformation (4.14) only
the component $\vec R$ of the vector along the velocity $\vec v$ changes, in accordance with the
Lorentz\index{Lorentz} transformations (3.1),\index{Lorentz transformations} while
the transverse component remains un\-chan\-ged.

Let us verify that this assertion is correct. To this end we shall
take as a starting point the invariant (3.22). Substituting (4.14) into
(3.22) we obtain
\be
J=c^2T^2-(\vec r +\vec v\,T)^2=c^2T^2\left(1-\f{v^2}{c^2}\right)
-2(\vec v\,\vec r)T-\vec r^{\,2}.
\ee In  the invariant $J$ we single out the time-like part \be J=c^2\left[\f{T}{\gamma}-\gamma\f{(\vec v\,\vec
r)}{c^2}\right]^2
-\vec r^{\,2}-\f{\gamma^2}{c^2}(\vec v\,\vec r)^2.
\ee
Our goal is to find such new variables $T^\prime$ and $\vec R^\prime$,
in which this expression can be written in a diagonal form
\be
J=c^2T^{\prime\,2}-\vec R^{\prime\,2}.
\ee From comparison of (4.21) and (4.22) we find time $T^\prime$ in the moving reference system: \be
T^\prime =\f{T}{\gamma}-\f{\gamma}{c^2}(\vec v\,\vec r).
\ee
Expressing the right-hand part in (4.23) via the variables $T,\,\vec R$,
we obtain
\be
T^\prime =\gamma\left(T-\f{\vec v\,\vec R}{c^2}\right).
\ee We also express the space-like part of the  invariant $J$ in terms of  variables  $T,\,\vec R$ \be \vec
r^{\,2}+\f{\gamma^2}{c^2}(\vec v\,\vec r)^2 =\vec R^2+\f{\gamma^2}{c^2}(\vec v\,\vec R)^2
-2\gamma^2(\vec v\,\vec R)T+\gamma^2 v^2T^2.
\ee
One can readily verify that the first two terms in (4.25) can be written
in the form
\be
\vec R^2+\f{\gamma^2}{c^2}(\vec v\,\vec R)^2
=\left[\vec R+(\gamma -1)
\f{\vec v(\vec v\,\vec R)}{v^2}\right]^2,
\ee
and, consequently, expression (4.25) assumes the form
\ba
&&\vec r^{\,2}+\ds\f{\gamma^2}{c^2}(\vec v\,\vec r)^2
=\left[\vec R+(\gamma -1)
\ds\f{\vec v(\vec v\,\vec R)}{v^2} \right]^2-\nonumber\\
\label{4.27}\\
&&-2\gamma^2(\vec v\,\vec R)T+v^2\gamma^2 T^2.\nonumber
\ea
The right-hand part of (4.27) can be written as
\be
\vec r^{\,2}+\f{\gamma^2}{c^2}(\vec v\,\vec r\,)^2
=\left[\vec R+(\gamma -1)
\f{\vec v(\vec v\,\vec R)}{v^2}-\gamma\vec v\, T \right]^2.
\ee Thus, the space-like part of  the invariant $J$ assume a diagonal form \be
\vec r^{\,2}+\f{\gamma^2}{c^2}(\vec v\,\vec r\,)^2=(\vec R^\prime)^2,
\ee
where
\be
\vec R^\prime =\vec R +(\gamma -1)
\f{\vec v(\vec v\,\vec R)}{v^2}-\gamma\vec v\, T.
\ee
Formulae (4.24) and (4.30) coincide with formulae (4.18) and (4.19);
this testifies to our assumption, made earlier in the course of their
derivation, being correct.

It should be especially emphasized that we have derived above the {\bf general formulae} relating coordinates \(
(T, \vec R) \) of   the initial inertial reference system\index{inertial reference systems} to  coordinates \(
(T^\prime,  \vec R^\prime) \)  of   the  reference system moving with constant velocity \(  \vec v \)   relative
to the first system. We have used the form-invariance of invariant (3.22) and {\bf identical transformations}
only. If by means of transformations (4.24) and (4.30) we go from an inertial reference system S  to a system $S^\prime$ and later to a system 
$S^{\prime\prime}$, then after these two subsequent transformations  we will
get transformation which will be different from transformations (4.24) and (4.30) by a rotation in 3-dimensional
space. It means that transformations (4.24) and (4.30)  do not form a subgroup of the Lorentz group\index{group,
Lorentz}\index{group}\index{Lorentz}. The rotation mentioned above reduces the axes of the reference system S to the same orientation as axes of the
system $S^{\prime\prime}$. 
Thomas'\index{Thomas}effect which will be considered in Section 14 is caused just by this
circumstance.

The general derivation of  Lorentz transformations from the relativity principle, Galilean principle of inertia
and the wavefront equation was done by Academician 
V.A.\,Fock\index{Fock} in his monograph [12, Appendix A]. His analysis
demonstrates that it is impossible to derive Lorentz transformations from the two Einstein\index{Einstein} postulates only (see
p.~\pageref{cyte:postulates} of this book).

\newpage
\markboth{thesection\hspace{1em}}{}
\vspace*{20mm}
\section{Adding velocities}

Differentiating the
Lorentz\index{Lorentz} 
transformations (3.1)\index{Lorentz transformations}
with respect to the
variable $T$, we obtain the formulae
relating velocities\index{relating velocities}
\be
u_x^\prime=\f{u_x-v}{1-\ds\f{u_x v}{c^2}}\,,\;
u_y^\prime=u_y\ds\f{\sqrt{1-\ds\f{v^2}{c^2}}}{1-\ds\f{u_x v}{c^2}}\,,\;
u_z^\prime=u_z\ds\f{\sqrt{1-\ds\f{v^2}{c^2}}}{1-
\ds\f{u_x v}{c^2}}.
\ee
Here
\be
u_x=\f{dX}{dT},\quad u_y=\f{dY}{dT},\quad
u_z=\f{dZ}{dT},
\ee
\be
u_x^\prime=\f{dX^\prime}{dT^\prime}\,,\quad
u_y^\prime=\f{dY^\prime}{dT^\prime}\,,
\quad u_z^\prime=\f{dZ^\prime}{dT^\prime}.
\ee
In deriving (5.1) we made use of the formula
\be
\f{dT^\prime}{dT}=\gamma\Bigl(1-\f{v}{c^2}\,u_x\Bigr).
\ee
It is possible, in a similar way, to obtain general formulae, also,
if one takes advantage of expressions (4.18), (4.19):
\be
\vec u^{\,\prime}=\f{\ds\vec u+(\gamma-1)
\f{(\vec v\vec u)}{v^2}\vec v-
\gamma\vec v}{\gamma\Biggl[1-\ds\f{(\vec v\vec u)}{c^2}\Biggr]}\,,\qquad
\f{dT^\prime}{dT}=\gamma\,\Biggl[1-\ds\f{(\vec v\vec u)}{c^2}
\Biggr]\,.
\ee We will further (Section 16) see, that the velocity space is the Lo\-ba\-chev\-sky
\index{Lobachevsky}
space.\index{Lobachevsky space}

\newpage
\markboth{thesection\hspace{1em}}{}
\section{Elements of vector and tensor analysis in
Min\-kow\-ski space}

All physical quantities must be defined in  a way,  to have their physical meaning  independent on the choice of
reference system.

Consider a certain reference system $x^\nu, \nu=0,1,2,3$ given in
four-dimensional\index{four-dimensional space-time}
Minkowski\index{Minkowski} 
space.\index{Minkowski space}
Instead of this reference system,
it is possible to choose another, defined by the ex\-pression
\be
x^{\prime\nu}=f^\nu(x^\sigma).
\ee We shall consider functions $f^\nu$ as continuous and differentiable.

If at any point  Jacobian
\index{Jacobian} of the transformation
\be J=\det \left|\f{\pa f^\nu}
{\pa x^\sigma}\right|
\ee
differs from zero, then under this condition the variables $x^{\prime\nu}$
will be independent, and, consequently, the initial variables $x^\nu$ may
unambiguously be expressed in terms of the new ones $x^{\prime\nu}$
\be
x^\alpha=\varphi^\alpha(x^{\prime\sigma}).
\ee

Physical quantities must not depend on the choice of re\-fe\-rence system,
 therefore, it should be possible to express them in terms of
geometrical objects. The simplest ge\-o\-met\-ri\-cal object is  scalar $\phi (x)$, which under transition to new
variables transforms as follows: \be
\phi^\prime (x^\prime)=\phi\Bigl[x(x^\prime)\Bigr].
\ee The gradient of scalar function $\phi (x)$ transforms by the rule for differentiating composite functions \be
\f{\pa \phi^\prime (x^\prime)} {\pa x^\prime{^\nu}}= \f{\pa \phi}{\pa x^\sigma}\cdot
\f{\pa x^\sigma}{\pa x^\prime{^\nu}},
\ee
here, and below, summation is performed over identical in\-di\-ces $\sigma$
from 0 to 3.
\markboth{thesection\hspace{1em}}{6. Elements of vector and tensor analysis \ldots}

A set of functions that transforms, under coordinate trans\-for\-ma\-tion,
by the rule (6.5) is called a
{\bf covariant vector}\index{covariant vector}
\be
A^{\prime}_\nu (x^\prime)=
\f{\pa x^\sigma}{\pa x^{\prime \nu}}
A_\sigma (x).
\ee
Accordingly, a quantity $B_{\mu \nu}$, that transforms by the
rule
\be
B^\prime_{\mu \nu}(x^\prime)=
\f{\pa x^\sigma}{\pa x^\prime{^\mu}}
\cdot\f{\pa x^\lambda}{\pa x^{\prime \nu}}
B_{\sigma \lambda},
\ee
is called a
covariant tensor\index{covariant tensor}
of the second rank, and so on.

Let us now pass to another group\index{group} of geometrical objects. Con\-si\-der the  transformation of the
differentials of coordinates: \be dx^\prime{^\nu}=\f{\pa x^{\prime\nu}}
{\pa x^\sigma}dx^\sigma.
\ee

A set of functions that transforms, under coordinate trans\-for\-ma\-tions, by the rule (6.8) is called a {\bf
contravariant vector}:\index{contravariant vector} \be A^{\prime \nu}(x^\prime)= \f{\pa x^\prime {^\nu}}
{\pa x^\sigma}A^\sigma (x).
\ee
Accordingly, a quantity $B^{\mu \nu}$, that transforms by the rule
\be
B^{\prime\mu\nu}(x^\prime)=
\f{\pa x^{\prime\mu}}{\pa x^\sigma}
\cdot\f{\pa x^{\prime\nu}}{\pa x^\lambda}
B^{\sigma\lambda}(x),
\ee
has been termed a
contravariant tensor\index{contravariant tensor}
of the second rank, and so on.

From the transformational properties of a vector or a tensor it follows, that, if all its components are zero in
one reference system, then they are zero, also, in any other reference system. Note, that  \textbf{coordinates}
{\mathversion{bold}\( x^\nu \)}
\textbf{do not form a vector}, while the differential $dx^\nu$ is a vector. The coordinates $x^\nu$ only form a
vector with respect to linear transformations.

Now, we calculate the quantity
$A_\sigma^\prime (x^\prime)B^{\prime\sigma}(x^\prime)$
\be
A_\sigma^\prime (x^\prime)B^{\prime\sigma}(x^\prime)=
\f{\pa x^\mu}{\pa x^{\prime\sigma}}
\cdot\f{\pa x^{\prime\sigma}}
{\pa x^\lambda}A_\mu (x)B^\lambda (x)\,,
\ee but, it is easy to see that
\be \f{\pa x^\mu} {\pa x^\prime{^\sigma}}\cdot \f{\pa x^\prime{^\sigma}} {\pa x^\lambda}=\delta_\lambda^\mu=
\begin{cases} 0,  & \text{for $\mu \ne \lambda$} \\
1,  & \text{for $\mu =\lambda$.}
\end{cases}
\ee
The symbol $\delta_\lambda^\mu$ is a mixed tensor of the second rank and
is known as the
Kronecker\index{Kronecker} 
symbol.\index{Kronecker symbol}

Taking into account (6.12) in expression (6.11) we find
\be
A_\sigma^\prime (x^\prime)B^{\prime\sigma} (x^\prime)=
A_\lambda (x)B^\lambda (x)\,.
\ee
Hence it is evident that this {\bf quantity is a scalar, it is
usually called an invariant}.

In writing expression (3.32) we actually dealt with the fun\-da\-men\-tal
invariant
\be
d\sigma^2=\gamma_{\mu \lambda}(x)dx^\mu dx^\lambda\,,
\quad \det(\gamma_{\mu\nu})=\gamma<0.
\ee The existence of the metric tensor\index{metric tensor of space} 
of Minkowski\index{Minkowski} 
space,\index{Minkowski space} that has the general form (3.33), permits to raise and to lower indices
of vector and tensor quantities, for example: \be A_\nu=\gamma_{\nu \lambda}(x)A^\lambda\,,\quad
A^\lambda=\gamma^{\lambda \sigma}A_\sigma\,,\quad
A_\nu A^\nu=\gamma_{\lambda \sigma}A^\lambda A^\sigma\,.
\ee
\be
\gamma_{\mu \lambda}\,\gamma^{\lambda \nu}=
\delta_\mu^\nu\,.
\ee
Tensors can be added and subtracted, for example,
\be
C_{\mu\nu\sigma}^{\alpha\beta}=A_{\mu\nu\sigma}^{\alpha\beta}
\pm B_{\mu\nu\sigma}^{\alpha\beta}\,.
\ee
They can also be multiplied, independently of their structure
\be
C^{\alpha\beta\lambda}_{\mu\nu\sigma\rho}=
A_{\mu\nu\sigma}^{\alpha\beta}\cdot
B_\rho^\lambda\,.
\ee
Here it is necessary to observe both the order of multipliers
and the order of indices.

Transformations (6.9) form a
group.\index{group}
Consider
\be
A^{\prime\nu} (x^\prime)=
\f{\pa x^{\prime\nu}}{\pa x^\sigma}
A^\sigma (x)\,,\quad
A^{\prime\prime\mu}=\f{\pa x^{\prime\prime\mu}}
{\pa x^{\prime\lambda}}
A^{\prime\lambda} (x^\prime)\,,
\ee
hence we have
\be
A^{\prime\prime\mu} (x^{\prime\prime})=
\f{\pa x^{\prime\prime\mu}}
{\pa x^{\prime\lambda}}\cdot
\f{\pa x^{\prime\lambda}}
{\pa x^{\sigma}}
A^\sigma (x)=
\f{\pa x^{\prime\prime\mu}}
{\pa x^\sigma}A^\sigma (x)\,.
\ee Note that tensor calculus does not depend on the metric properties of space. It is persisted, for example, in
Riemannian\index{Riemann} geometry,\index{Riemannian geometry}
 where the group\index{group} of motion of space-time is absent, in the general case. On the other hand, the
group\index{group} of general coordinate transformations (6.19--6.20) is fully persisted, since it is
in\-de\-pen\-dent of the metric pro\-per\-ti\-es of space, but it has no any physical meaning.

\newpage
\markboth{thesection\hspace{1em}}{}
\section{Lorentz group}

\textbf{H.Poincar\'{e}\index{Poincar\'e} discovered that the 
Lorentz\index{Lorentz}
transformations,\index{Lorentz transformations} to\-ge\-ther with all space rotations,
form a group.}\index{group} Consider, for example, \be x^\prime=\gamma_1 (x-v_1t),\quad t^\prime=
\gamma_1 \left(t-\f{v_1}{c^2}x\right)\,,
\ee
\be
x^{\prime\prime}=\gamma_2 (x^\prime-v_2t^\prime),\quad t^{\prime\prime}=
\gamma_2 \left(t^\prime-\f{v_2}{c^2}x^\prime\right)\,.
\ee
Substituting (7.1) into (7.2) we obtain
\be
x^{\prime\prime}=\gamma_1\gamma_2\left(1+\f{v_1 v_2}{c^2}
\right)x-\gamma_1\gamma_2(v_1+v_2)t\,,
\ee
\be
t^{\prime\prime}=\gamma_1\gamma_2\left(1+\f{v_1 v_2}{c^2}
\right)t-\gamma_1\gamma_2 \left( \f{v_1+v_2}{c^2}\right)x\,,
\ee
But since
\be
x^{\prime\prime}=\gamma (x-vt)\,,\quad t^{\prime\prime}=
\gamma \left(t-\f{v}{c^2}x \right)\,.
\ee
From comparison of (7.3) and (7.4) with (7.5) we obtain
\be
\gamma=\gamma_1\gamma_2\left(1+\f{v_1 v_2}{c^2}\right)\,,\quad
\gamma v=\gamma_1\gamma_2(v_1+v_2)\,.
\ee
From relations (7.6) we find
\be
v=\f{v_1+v_2}{\ds1+\f{v_1 v_2}{c^2}}\,.
\ee
It is readily verified that
\be
\gamma=\f{1}{\sqrt{\ds1-\f{v^2}{c^2}}}=\gamma_1\gamma_2
\left(1+\ds\f{v_1 v_2}{c^2}\right)\,.
\ee

\markboth{thesection\hspace{1em}}{7. Lorentz group} 
Thus, we have established that transition from the
re\-fe\-rence system $x^\nu$ to the reference system $x^{\prime\nu}$ and, sub\-se\-quent\-ly, to the reference
system $x^{\prime\prime\nu}$ is equivalent to direct transition from the reference system $x^\nu$ to the reference
sys\-tem $x^{\prime\prime\nu}$. Precisely in this case, it can be said that the Lorentz\index{Lorentz}
transformations\index{Lorentz transformations} form a 
group.\index{group}\index{group, Lorentz} {\bf Poincar\'{e}\index{Poincar\'e} discovered~\cite{2} this group\index{group}
and named it the Lorentz\index{Lorentz} 
group.\index{group}\index{group, Lorentz}
He found the group\index{group} generators and constructed the 
Lie\index{Lie}
algebra of the Lorentz\index{Lorentz} group.\index{group}\index{group, Lorentz}
Poincar\'{e}\index{Poincar\'e} was the first to establish that, for universal invariance
of the laws of Nature with respect to the Lorentz\index{Lorentz} 
trans\-for\-mations\index{Lorentz transformations} to hold valid, it is necessary for the physical fields and for
other dynamical and kinematical characteristics to form a set of quan\-ti\-ti\-es transforming under the
Lorentz\index{Lorentz} transformations\index{Lorentz transformations} in
ac\-cor\-dance with the group,\index{group} or, to be more precise, in accordance with one of the
re\-pre\-sen\-ta\-tions of the Lorentz\index{Lorentz} 
group}.\index{group}\index{group, Lorentz}

Several general words about a
group.\index{group}
A group\index{group}
is a set of elements
$A, B, C\ldots $ for which the operation of multiplication is defined.
Elements may be of any nature.
The product of any two elements of
a group\index{group}
yields an element of the same
group.\index{group}
In the case of a
group,\index{group}
multiplication must have
 the following properties.

1. The law of associativity\\
\[
(AB)C=A(BC)\,.
\]

2. A group\index{group}
contains a unit element $E$
\[
AE=A\,.
\]

3. Each element of a
group\index{group}
has its inverse element
\[
AB=E\,,\quad B=A^{-1}\,.
\]

Transformations of the Lorentz\index{Lorentz} group\index{group, Lorentz}\index{group}
can be given in matrix form
\[
X^\prime =AX\,,
\]
where
\[
A=\begin{pmatrix} \gamma&-\f{v}{c}\gamma\\ -\f{v}{c}\gamma&\gamma\end{pmatrix}\,,\; X=\begin{pmatrix}x\\
x_0\end{pmatrix}\,,\; X^\prime=\begin{pmatrix}x^\prime\cr x_0^\prime\end{pmatrix}\,,\;
\]
\[
x_0^\prime=ct^\prime\,,\;x_0=ct\,.
\]
It is readily verified, that the set of all
Lorentz\index{Lorentz} 
trans\-for\-mations\index{Lorentz transformations}
satisfies all the listed requirements of a
group.\index{group}

Coordinate transformations which preserve the form of the metric tensor  form the {\bf
group\index{group}
of motions of the space}.
In particular the Lorentz\index{Lorentz} group\index{group, Lorentz}\index{group}
is such a group.\index{group}

\newpage
\markboth{thesection\hspace{1em}}{}
\section{Invariance of  Maxwell-Lorentz equations}

The
Maxwell-Lorentz\index{Lorentz} 
\index{Maxwell} equations\index{Maxwell-Lorentz equations}
in an inertial reference sys\-tem, said
to be ``at rest``, have the form
\ba
&&\rot \vec H=\ds\f{4\pi}{c}\rho\vec v+\f{\,1\,}{\,c\,}\cdot
\f{\pa \vec E}{\pa t}\,,\;
\rot \vec E=-\f{\,1\,}{\,c\,}\cdot\f{\pa \vec H}{\pa t}\,,\nonumber\\%
\label{8.1}\\[-3mm]
&&\qquad\qquad{\rm div} \vec E=4\pi\rho\,,\quad
{\rm div} \vec H=0\,,\nonumber\\[5mm]
&&\qquad\qquad\quad\vec f=\rho\vec E+\f{\,1\,}{\,c\,}\rho\left[\vec v, \vec H \right].
\ea
The second term in the right-hand part of the first equation of (8.1) is precisely that small term --- the {\bf
displacement current},\index{displacement current} introduced 
by Maxwell\index{Maxwell} 
in the equations of electrodynamics. Namely it was mentioned in Section 3.  Since the divergence of a
curl is zero, from the first and third equations of (8.1) follows the 
conservation\index{conservation law of current} law of current \be \f{\pa \rho}{\pa t}+{\rm div} \vec j=0\,,\quad
\vec j=\rho \vec v\,.
\ee
As one sees from (8.3), the displacement current permitted to achieve
accordance between the equations of elec\-tro\-dy\-na\-mics and the con\-ser\-va\-tion
law of electric charge. To make the fourth equation from (8.1) be satisfied
identically we re\-pre\-sent $\vec H$ in the form
\be
\vec H=\rot \vec A\,.
\ee
Thus, we have introduced the vector
potential\index{potential}
$\vec A$. Substituting
ex\-pres\-sion (8.4) into the second equation from (8.1) we obtain
\be
\rot \left(\vec E+ \f{\,1\,}{\,c\,}\cdot
\f{\pa \vec A}{\pa t}\right)=0\,.
\ee
To have equation (8.5) satisfied identically
the expression in brackets
must
be  gradient of
some function $\phi$
\be
\vec E=-\f{\,1\,}{\,c\,}\cdot\f{\pa \vec A}{\pa t}
-\grad\, \phi\,.
\ee

\markboth{thesection\hspace{1em}}{8. Invariance of Maxwell-Lorentz equations} Thus, we have introduced the notion
of scalar potential $\phi$.\index{potential} For given values of $\vec E$ and $\vec H$ the potentials $\phi$ and
$\vec A$,\index{potential} as we shall see below (Section 10), are determined ambiguously. So by choosing them to
provide for the validity of the L.\,Lorenz\index{Lorenz} 
condition\index{L.\,Lorenz condition} \be \f{\,1\,}{\,c\,}\cdot\f{\pa \phi}{\pa t}
+{\rm div} \vec A=0\,,
\ee
from equations (8.1), with account of formulae
\[
\ds {\rm div \grad} \phi=\nabla^2 \phi,\;
{\rm \rot \rot} \vec A={\rm \grad div}
\vec A-\nabla^2 \vec A,\;
\]
\[
\nabla^2 \phi=\f{\pa^2 \phi}{\pa x^2}
+\f{\pa^2 \phi}{\pa y^2}
+\f{\pa^2 \phi}{\pa z^2},
\]
and relation (8.7) as well, we find  equations for  po\-ten\-ti\-als $\phi$ and $\vec A$\index{potential} in the
following form: \be \Box\vec A=\f{4\pi}{c}\vec j\,,\quad
\Box\phi=4\pi\rho\,.
\ee

For the equation of charge conservation\index{equations of charge conservation} to be form-\-in\-va\-ri\-ant with
respect to the Lorentz\index{Lorentz} 
transformations\index{Lorentz transformations} it is necessary that the density $\rho$ and current be components of the contravariant
vector $S^\nu$\index{contravariant vector} \be
S^\nu=(c\rho, \vec j)=(S^0, \vec S)\,,\quad \vec j=\rho \vec v\,.
\ee
The contravariant vector $S^\nu$\index{contravariant vector}
transforms under the
Lorentz\index{Lorentz} 
trans\-for\-ma\-tions in the same way as $(ct\,,\vec x)$. The equation (8.3)
of charge conservation assume the form
\be
\f{\pa S^\nu}{\pa x^\nu}=0\,,
\ee summation is performed over identical indices $\nu$.  Taking into account Eq.~(8.9) we rewrite Eqs.~(8.8) as
follows \be \Box\vec A=\f{4\pi}{c}\vec S\,,\quad
\Box\phi=\f{4\pi}{c}S^0\,.
\ee For these equations not to alter their form under Lorentz\index{Lorentz} trans\-for\-mations,\index{Lorentz transformations} it is necessary that the scalar and vector potentials be
components\index{potential} of a contravariant 
vector $A^\nu$\index{contravariant vector} \be
A^\nu=(A^0, \vec A)=(\phi, \vec A)\,.
\ee Since, as we showed earlier, the operator $\Box$ does not alter its form under the Lorentz\index{Lorentz}
transformations,\index{Lorentz transformations} Eqs.~(8.11) at any inertial system of
reference\index{inertial reference systems} will have the following form \be
\Box A^\nu=\f{4\pi}{c}S^\nu\,,\quad \nu=0, 1, 2, 3.
\ee
\textbf{The vectors}
{\mathversion{bold}\( S^\nu \)}
\textbf{and}
{\mathversion{bold}\( A^\nu \)}
\textbf{were first introduced by 
Henri Poin\-car\'e}~\cite{3}.\index{Poincar\'e} 

Unification of $\phi$ and $\vec A$ into the
four-vector $A^\nu$\index{four-vector}
is
necessary, since, as the right-hand part of (8.13) represents the
vector $S^\nu$, then the left-hand part must also transform like a
vector. Hence, it directly follows that, if in a certain inertial
reference system only an electric field exists, then in any other
reference system there will be found, together with the electric field,
a magnetic field, also, owing to $A^\nu$ transforming like a vector.
This is an immediate consequence of validity of the relativity
principle\index{relativity principle}
for electromagnetic phenomena.

The
Lorentz\index{Lorentz} 
transformations\index{Lorentz transformations}
for the vector $S^\nu$ have the same form,
as in the case of the vector $(ct, \vec x)$
\be
S_x^\prime=\gamma\left(S_x-\f{u}{c}S_0 \right)\,,\quad
S^{\prime 0}=\gamma\left(S^0-\f{u}{c}S_x \right).
\ee
Taking into account the components of the vector $S^\nu$ (8.9),
we find
\be
\rho^\prime=\gamma\rho\left(1-\f{u}{c^2}v_x\right)\,,\;
\rho^\prime v_x^\prime=\gamma\rho (v_x-u).
\ee
Here
\be
\gamma=\f{1}{\sqrt{1-\ds\f{u^2}{c^2}}}\,,
\ee
where $u$ is the velocity of the reference system.

The transformations for the components $S_y, S_z$ have the form
\be
\rho^\prime v_y^\prime=\rho v_y\,,\quad\rho^\prime v_z^\prime=
\rho v_z.
\ee {\bf All these formulae were first obtained 
by H.\,Poincar\'{e}}~\cite{2}.\index{Poincar\'e} 
From these the formulae for velocity\index{relating velocities} 
addition follow \be
v_x^\prime=\f{v_x-u}{1-\ds\f{uv_x}{c^2}},\; v_y^\prime=v_y\f{\sqrt{\ds1-\f{u^2}{c^2}}}{\ds1-\f{uv_x}{c^2}},\;
v_z^\prime=v_z\f{\sqrt{\ds1-\f{u^2}{c^2}}}{\ds1-\f{uv_x}{c^2}}.
\ee
We now introduce the covariant
vector $S_\nu$\index{covariant vector}
\be
S_\nu=\gamma_{\nu\lambda}S^\lambda.
\ee Taking into account that $\gamma_{\nu\lambda}=(1,-1,-1,-1)$, we obtain from (8.19) \be
S_0=S^0,\quad S_i=-S^i,\quad i=1,2,3.
\ee
Now compose the invariant
\be
S_\nu S^\nu=c^2 \rho^2 \left(1-\f{v^2}{c^2}\right)
=c^2 \rho_0^2,
\ee
here $\rho_0$ is the charge density in the reference system, where
the charge is at rest. Hence we have
\be
\rho_0=\rho\sqrt{1-\f{v^2}{c^2}}.
\ee {\bf H.\,Minkowski\index{Minkowski} introduced  antisymmetric tensor}
{\mathversion{bold}\( F_{\mu\nu} \)}
\be
F_{\mu\nu}=\f{\pa A_\nu}{\pa x^\mu}-
\f{\pa A_\mu}{\pa x^\nu},\quad \nu=0,1,2,3,
\ee
which automatically satisfies the equation
\be
\f{\pa F_{\mu\nu}}{\pa x^\sigma}+
\f{\pa F_{\nu\sigma}}{\pa x^\mu}+
\f{\pa F_{\sigma\mu}}{\pa x^\nu}=0.
\ee

Since $\vec H=\rot\vec A,\, \vec E=\ds -\grad\,\phi-\f{\,1\,}{\,c\,}
\f{\pa \vec A}{\pa t}$, the following equations are easily verified
\ba
&&-H_x=F_{23}\,,\;-H_y=F_{31}\,,\;-H_z=F_{12}\,,\nonumber\\
\label{8.25}\\
&&-E_x=F_{10}\,,\;-E_y=F_{20}\,,\;-E_z=F_{30}\,.\nonumber
\ea
The set of equations (8.24) is equivalent to the set of
Maxwell\index{Maxwell} 
equations
\be
\rot \vec E=-\f{\,1\,}{\,c\,}\cdot\f{\pa \vec H}{\pa t}\,,\;
{\rm div} \vec H=0\,.
\ee
With the aid of the tensor $F^{\mu\nu}$, the set of equations (8.13)
can be written as follows:
\be
\f{\pa F^{\mu\nu}}{\pa x^\nu}=
-\f{4\pi}{c}S^\mu\,.
\ee
The tensor $F^{\mu\nu}$ is related to the field components $\vec E$
and $\vec H$ by the following relations:
\ba
&&-H_x=F^{23}\,,\;-H_y=F^{31}\,,\;-H_z=F^{12}\,,\nonumber\\
\label{8.28}\\
&&E_x=F^{10}\,,\;E_y=F^{20}\,,\;E_z=F^{30}\,.\nonumber \ea All this can be presented in the following table form,
where first index \( \mu =0, 1, 2, 3 \) numerates lines, and second  \( \nu \) --- columns
\[
F_{\mu\nu}=\left( \begin{array}{cccc}
0 & E_x & E_y & E_z \\
-E_x & 0 & -H_z & H_y \\
-E_y & H_z & 0 & -H_x \\
-E_z & -H_y & H_x & 0
\end{array} \right),\quad
\]
\[
F^{\mu\nu}=\left( \begin{array}{cccc}
0 & -E_x & -E_y & -E_z \\
E_x & 0 & -H_z & H_y \\
E_y & H_z & 0 & -H_x \\
E_z & -H_y & H_x & 0
\end{array} \right).
\]

Hence it is seen that the quantities $\vec E$ and $\vec H$ change under
the
Lorentz\index{Lorentz} 
transformations\index{Lorentz transformations}
like individual components of the tensor
$F^{\mu\nu}$. Neither
Lorentz,\index{Lorentz} 
nor
Einstein\index{Einstein} 
established this, so they,
 did not succeed in demonstrating the in\-va\-ri\-ance of the
Max\-well-Lorentz\index{Lorentz} 
\index{Maxwell} 
equations with respect to the
Lorentz\index{Lorentz} 
transformations\index{Lorentz transformations}
neither in space without charges, nor in space with charges.

\textbf{We emphasize that the identical appearance of equations in two systems of coordinates under
Lorentz\index{Lorentz} transformations\index{Lorentz transformations} still does not mean their form-invariance
under these transformations. To prove the form-invariance of equations we are to ascertain that
Lorentz\index{Lorentz} transformations\index{Lorentz transformations} form a group\index{group} and field
variables (for example, {\mathversion{bold}\( \vec E \)} and {\mathversion{bold}\( \vec H \)}) transform according
to some representation of this group}.

Taking into account the relationship between the com\-po\-nents of the
tensor $F^{\mu\nu}$ and the components of the electric and magnetic
fields, it is possible to obtain the transformation law for the
com\-po\-nents of the electric field
\ba
&&E_x^\prime=E_x,\;E_y^\prime=\gamma\left(\ds E_y-\f{\,u\,}{\,c\,}H_z \right),\nonumber\\
\label{8.29}\\
&&E_z^\prime=\gamma\left(\ds E_z+\f{\,u\,}{\,c\,}H_y \right),\nonumber \ea and for the components of the magnetic field
\ba
&&H_x^\prime=H_x,\;H_y^\prime=\gamma\left(\ds H_y+\f{\,u\,}{\,c\,}E_z \right),\nonumber\\
\label{8.30}\\
&&H_z^\prime=\gamma\left(\ds H_z-\f{\,u\,}{\,c\,}E_y \right).\nonumber \ea

{\bf These formulae were first discovered by Lorentz,\index{Lorentz} however,
neither he, nor, later, Einstein\index{Einstein} established their group\index{group}
nature. This was first done by H.\,Poincar\'{e},\index{Poincar\'e} who discovered the
trans\-for\-mation law for the scalar and vector 
potentials}~\cite{3}.\index{potential} Since $\phi$ and $\vec A$
transform like $(ct, \vec x)$, H.\,Poincar\'{e}\index{Poincar\'e} has found, with the aid
of formulae (8.4) and (8.6), the procedure of calculation  for the quantities $\vec E$ and $\vec H$ under
transition to any other  inertial reference system.

From the formulae for transforming the electric and mag\-ne\-tic fields it follows that, if, for example, in a
reference system $K^\prime$ the magnetic field is zero, then in another reference system it already differs from
zero and equals \be H_y=-\f{\,u\,}{\,c\,}E_z,\,H_z=\f{\,u\,}{\,c\,}E_y,\, {\textrm or}\;
\vec H=\f{\,1\,}{\,c\,}\left[\vec u, \vec E \right].
\ee
From the field components it is possible to construct two invariants
with respect to the
Lorentz\index{Lorentz} 
transformations.\index{Lorentz transformations}
\be
E^2-H^2,\quad(\vec E\vec H).
\ee
{\bf These invariants of the electromagnetic
field\index{electromagnetic field invariants}
were first dis\-co\-ve\-red
by H.\,Poincar\'{e}}~\cite{3}.\index{Poincar\'e} 

The invariants (8.32) can be expressed via  an\-ti\-sym\-met\-ric tensor of the electromagnetic field $F^{\mu\nu}$
\be E^2-H^2=\f{\,1\,}{\,2\,}F_{\mu\nu}F^{\mu\nu},\;\vec E\vec H=-\f{\,1\,}{\,4\,}
{F}_{\mu\nu}\overset{\!\!\!\!\!\ast}{F^{\mu\nu}}
\ee
here
\be
\overset{\!\!\!\!\!\ast}{F^{\mu\nu}}=-\f{\,1\,}{\,2\,}
\varepsilon^{\mu\nu\sigma\lambda}F_{\sigma\lambda},
\ee
$\varepsilon^{\mu\nu\sigma\lambda}$ is the
Levi-Civita\index{Levi-Civita} 
tensor,
$\varepsilon^{0123}=1$, transposition of any two indices alters the sign
of the
Levi-Civita\index{Levi-Civita} 
tensor.\index{Levi-Civita tensor}

In accordance with the second invariant (8.32), the fields $\vec E$ and $\vec H$, that are reciprocally orthogonal
in one reference system, persist this property in any other reference system. If in reference system $K$ the
fields $\vec E$ and $\vec H$ are orthogonal, but not equal, it is always possible to find such a reference system,
in which the field is either purely electric or purely magnetic, depending on the sign of the first invariant from
(8.33).

Now let us consider the derivation of the \textbf{Poynting\index{Poynting} equation\index{Poynting equation}}
(1884). To do so we multiply both parts of first equation from Eqs.~(8.1) by vector \( \vec E, \) and both parts
of second equation  from Eqs.~(8.1) --- by vector \( \vec H \); then we subtract the results and obtain
\[
\f{1}{4\pi}\left(\vec E\f{\pa \vec E}{\pa t} +\vec H\f{\pa \vec H}{\pa t}\right) =-\rho \vec v \vec E -\f{c}{4\pi}
\left(\vec H \rot \vec E- \vec E \rot \vec H\right).
\]
By using the following formula from vector analysis
\[
{\rm div}[\vec a,\,\vec b\,]=\vec b \rot \vec a-\vec a \rot \vec b,
\]
we obtain the \textbf{Poynting\index{Poynting} equation\index{Poynting equation}}
\[
\f{\pa}{\pa t}\left(\f{E^2+H^2}{8\pi}\right) =-\rho \vec v \vec E-{\rm div}\vec S,
\]
where
\[
\vec S=\f{c}{4\pi}[\vec E\, \vec H]
\]
is called the \textbf{Poynting\index{Poynting} vector\index{Poynting vector}}. After integration of the
Poynting\index{Poynting} equation\index{Poynting equation} over volume \( V \) and using Gauss\index{Gauss}
theorem\index{Gauss theorem} we get
\[
\f{\pa}{\pa t}\int\limits_V\f{E^2+H^2}{8\pi}dV =-\int\limits_V \rho \vec v \vec E dV -\oint\limits_\varSigma \vec
S d\vec \sigma.
\]
The term standing in l.h.s. determines a change of electromagnetic energy in volume \( V \) at a unit of time. The
first term in r.h.s. characterizes work done by electric field on charges in volume \( V. \) The second term in
r.h.s. determines the  \textbf{energy flow of electromagnetic field} through surface \( \varSigma, \) bounding
volume \( V. \)

Formulation of the energy conservation law with help of the \textbf{notion of the energy flow} was
first proposed by N.\,A.\,Umov.\index{Umov} The notion of the  \textbf{energy flow} has become one of the most
important in physics. With help of  the \textbf{Poynting\index{Poynting} equation\index{Poynting equation}} it
is possible to prove the  \textbf{uniqueness theorem} in the following formulation
 (see: \textit{I.\,E.\,Tamm}\index{Tamm} \textsf{Foundations of the theory of electricity.} Moscow: ``Nauka'', 1976
 (in Russian),
pp.~428-429):
\begin{quote}
\hspace*{5mm}\textit{``\ldots electromagnetic field at any 
instant of time\break \( t_1>0 \) and at any point of volume
\( V \), bounded by an arbitrary closed surface  \( S \) is uniquely determined by Maxwell
equations,\index{Maxwell} if initial values for electromagnetic vectors \( \vec E \) and \( \vec H \) are
prescribed in all this part of space at time \( t=0 \) and if also \textbf{for one of these vectors} (for example,
\( \vec E \)) \textbf{boundary values of its tangential components on surface \( S \) are given during the whole
time interval from \( t=0 \) to \( t=t_1 \).}\\
\hspace*{5mm}Let us suppose the opposite, i.\,e. suppose there are two different systems of solutions of Maxwell\index{Maxwell}
equations \( \vec E^\prime, \vec H^\prime \) and \( \vec E^{\prime\prime}, \vec H^{\prime\prime} \), satisfying
the same initial and boundary conditions. Due to linear character of the field equations the difference of these
solutions  \( \vec E^{\prime\prime\prime}=\vec E^\prime-\vec E^{\prime\prime} \) and \( \vec
H^{\prime\prime\prime}=\vec H^\prime-\vec H^{\prime\prime} \) should also satisfy Maxwell\index{Maxwell} equations under
the following additional conditions:\\
\hspace*{5mm}a) \( \vec E^{extra} =0 \),\\
\hspace*{5mm}b) at time \( t=0 \) in each point of the volume \( V \): \( \vec E^{\prime\prime\prime}=0, \vec
H^{\prime\prime\prime}=0 \) (because at \( t=0\;\vec E^\prime, \vec E^{\prime\prime} \) and\( \vec H^{\prime},
\vec H^{\prime\prime} \) have, as supposed, equal given values),\\
\hspace*{5mm}c) during the whole time interval from \( t=0 \) to \( t=t_1 \) in all points of the surface  $S$
tangential components of vector $\vec E^{\prime\prime\prime}$ or vector
$\vec H^{\prime\prime\prime}$ are equal to zero (by the same reason).\\
\hspace*{5mm}Let us apply the Poynting\index{Poynting} theorem (which is a consequence of Maxwell\index{Maxwell} equations) to
this field \( \vec E^{\prime\prime\prime}, \vec H^{\prime\prime\prime} \)  and put work of extraneous forces  \( P
\) equal to zero. The surface integral which enters the Poynting equation\index{Poynting equation} is equal to
zero during the whole time interval from \( t=0 \) to \( t=t_1 \), because from Eq.~(\( c \)) it follows that on
surface \( S \)
\[
S_n=[\vec E^{\prime\prime\prime} \vec H^{\prime\prime\prime}]_n=0;
\]
therefore, at any time during this interval we get
\[
\f{\pa W^{\prime\prime\prime}}{\pa t} =-\int\limits_V \f{\vec j^{\,'''\,2}}{\lambda}\,dV.\, \footnote{\( \vec
j^{\,'''}=\lambda \vec E^{\,'''}\)}
\]
\hspace*{5mm}As the integrand is strictly positive, we have
\[
\f{\pa W^{\prime\prime\prime}}{\pa t}\leq 0,
\]
i.\,e. field energy \( W^{\prime\prime\prime} \) may decrease or stay constant.  But at \( t=0 \), according to
Eq.~(\( b \)), energy \( W^{\prime\prime\prime} \) of field \( \vec E^{\prime\prime\prime}, \vec
H^{\prime\prime\prime} \) is equal to zero. It also can not become negative, therefore during the whole interval
considered   $0\leq t\leq t_1$ the energy
\[
W^{\prime\prime\prime}=\f{1}{8\pi}\int\limits_V (\vec E^{\prime\prime\prime\,2}+\vec H^{\prime\prime\prime\,2})dV
\]
should stay equal to zero. This may take place only if \( \vec E^{\prime\prime\prime} \) and \( \vec
H^{\prime\prime\prime} \) stay equal to zero at all points of the volume \( V \). Therefore, the two systems of
solutions of the initial problem \( \vec E^{\prime}, \vec H^{\prime} \) and \( \vec E^{\prime\prime}, \vec
H^{\prime\prime} \), supposed by us to be different, are necessarily identical. So the uniqueness theorem is proved.\\
\hspace*{5mm}It is easy to get convinced that in case of the whole infinite space the fixing of field vectors
values on the bounding surface \( S \) may be replaced by putting the following conditions at infinity:
\[
ER^2\;\textrm{and}\;HR^2\;\textrm{at}\;R\rightarrow \infty\; \textrm{stay finite}.
\]
\hspace*{5mm}Indeed, it follows from these conditions that the integral of the Poynting\index{Poynting}
vector\index{Poynting vector} over an infinitely distant surface is occurred to be zero.  This fact enables us to
prove applicability of the above inequality to the whole infinite space, starting from the Poynting\index{Poynting}
equation\index{Poynting equation}.  Also uniqueness of solutions for the
field equations follows from this
inequality''.}
\end{quote}

{\bf For consistency with the
relativity principle\index{relativity principle}
for all elec\-tro\-mag\-ne\-tic
phenomena, besides the requirement that the
Max\-well-Lorentz\index{Lorentz} 
\index{Maxwell} 
equations
remain unaltered under the
Lorentz\index{Lorentz} 
trans\-for\-ma\-tions, it is necessary that
the equations of motion of charged particles under the influence of the
Lorentz\index{Lorentz} 
force\index{Lorentz force}
remain unaltered, also}.

All the aforementioned was only performed in  works~\cite{2,3} 
by H.\,Poincar\'{e}.\index{Poincar\'e} 
The invariability of physical equations in all inertial reference systems is just what signifies the
identity of physical phenomena in these reference systems under identical conditions. Pre\-ci\-se\-ly for this
reason, all {\bf natural standards} are {\bf identical} in all inertial reference systems. Hence, for instance,
follows the {\bf equality} of the $NaCl$ crystal lattice con\-stants taken to be at rest in two inertial reference
systems moving with respect to each other. This is just the essence of the relativity principle.\index{relativity
principle} The relativity principle\index{relativity principle} was understood exactly in this way in classical
mechanics, also. Therefore, one can only be surprised at what Academician V.\,L.\,Ginzburg\index{Ginzburg}
writes in the same article (see this edition, the footnote on page
\pageref{wash15}):\label{wash2}
\begin{quote}{\it\hspace*{5mm}``I  add that, having reread now
(70 years after they were published!) the works of Lorentz\index{Lorentz} and
Poincar\'{e},\index{Poincar\'e} I have been only able with difficulty and knowing the
result beforehand (which is known to extremely fa\-ci\-li\-tate apprehension) to understand why invariance of the
equations of electrodynamics with respect to the Lorentz\index{Lorentz} 
transformations,\index{Lorentz transformations} demonstrated in those works, could at the time be considered as
evidence for validity of the relativity principle''.}\index{relativity principle}
\end{quote}
Though  A.\,Einstein\index{Einstein} wrote in 1948
\begin{quote}
{\it\hspace*{5mm}``With the aid of the Lorentz\index{Lorentz} 
transformation\index{Lorentz transformations} the spe\-ci\-al relativity principle\index{relativity principle} can
be formulated as follows: the laws of Nature are invariant with respect to the Lorentz\index{Lorentz} 
transformation\index{Lorentz transformations} (i.\,e. a law of Nature must not change, if it
would be referred to a new inertial reference system obtained with the aid of  Lorentz\index{Lorentz} 
transformation\index{Lorentz transformations} for $x,y,z,t$)''.}
\end{quote}

Now, compare the above with that written 
by H.\,Poincar\'{e}\index{Poincar\'e} in 1905:
\begin{quote}
{\it ``\ldots If it is possible to give general trans\-la\-ti\-onal motion to a whole system without any visible
changes taking place in phenomena, this means that the equations of the electromagnetic field will not
change as a result of certain transformations, which we shall 
call {\bf \textit{Lorentz\index{Lorentz}
transformations}};\index{Lorentz transformations} two systems, one at rest, and another
undergoing translational mo\-tion represent, therefore, an exact image of each other''.}
\end{quote}

We see that classical works require attentive reading, not to
mention contemplation.

We shall now establish the law for transformation\index{Lorentz transformations} of the Lorentz\index{Lorentz} force\index{Lorentz force} under transformation from one inertial reference
sys\-tem to another. The equations of motion will be established in Section 9. The expression for the
Lorentz\index{Lorentz} force,\index{Lorentz force} referred to unit volume, will,
in reference system $K$, have the form (8.2) \be \vec f=\rho\vec E+\rho\f{\,1\,}{\,c\,}
\left[\vec v, \vec H\right]\,.
\ee
Then, in reference system $K^\prime$ we must have a similar ex\-pression
\be
\vec f^\prime=\rho^\prime\vec E^\prime+\rho^\prime\f{\,1\,}{\,c\,}
\left[\vec u^\prime, \vec H^\prime\right]\,.
\ee Replacing all the quantities by their values (8.15), (8.17), (8.29), (8.30) and (8.35), we obtain \be
f_x^\prime = \gamma\left(f_x-\f{\,u\,}{\,c\,}f\right),\quad
f^\prime = \gamma\left(f-\f{\,u\,}{\,c\,}f_x\right),
\ee
\be
f_y^\prime =f_y,\quad f_z^\prime = f_z,
\ee
here by $f$ we denote the expression
\be
f=\f{\,1\,}{\,c\,}\left(\vec v \vec f\right).
\ee

{\bf These formulae were first found by Poincar\'{e}}.\index{Poincar\'e} We see that
scalar $f$ and vector $\vec f$ transform like components of $(x^0, \vec x)$. Now let us establish the law for the
transformation of a force referred to unit charge 
\be \vec F=\vec E+\f{1}{c}\left[\vec v, \vec H\right],
\quad \vec F=\f{\,\vec f\,}{\,\rho\,},\quad F=\f{\,f\,}{\,\rho\,}.
\ee Making use of (8.37), (8.38) and (8.39), we find \be F_x^\prime =\gamma\f{\rho}{\rho^\prime}
\left(F_x-\f{\,u\,}{\,c\,}F\right),\; F^\prime =\gamma\f{\rho}{\rho^\prime}
\left(F-\f{\,u\,}{\,c\,}F_x\right),
\ee
\be
F_y^\prime =\f{\rho}{\rho^\prime}F_y,\quad
F_z^\prime =\f{\rho}{\rho^\prime}F_z.
\ee
On the basis of (8.15) we have
\be
\f{\rho}{\rho^\prime}=
\f{\sqrt{1-\ds\f{u^2}{c^2}}}{1-\ds\f{u}{c^2}v_x}\,.
\ee To simplify (8.43) we shall derive an identity. Consider a certain inertial reference system $K$, in which
there are two bodies with  four-velocities $U_1^\nu$ and $U_2^\nu$ (see (9.1))  respectively
\be U_1^\nu
=\left(\gamma_1, \f{\vec v_1}{c}\gamma_1 \right),\;
U_2^\nu =\left(\gamma_2, \f{\vec v_2}{c}\gamma_2 \right);
\ee then, in the reference system $K^\prime$, in which the first body is at rest, their
four-velocities\index{four-velocity} will be \be U_1^{\prime\nu}=(1,0),\quad U_2^{\prime\nu}= \left(\gamma^\prime,
\f{\vec v^\prime}{c}
\gamma^\prime \right).
\ee
Since the product of
four-vectors\index{four-vector}
is an invariant, we obtain
\be
\gamma^\prime =\gamma_1 \gamma_2
\left(1-\f{\vec v_1 \vec v_2}{c^2}\right).
\ee Setting in this expression $\vec v_2=\vec v$,\, and\, $\vec v_1=\vec u$ (velocity along the $x$ axis), we find
\be \f{\sqrt{1-\ds\f{u^2}{c^2}}\, \sqrt{1-\ds\f{v^2}{c^2}}}{1-\ds\f{uv_x}{c^2}}=
\sqrt{1-\f{(v^\prime)^2}{c^2}}.
\ee On the basis of (8.43) and (8.47) we obtain \be \f{\rho}{\rho^\prime}=\f{\sqrt{1-\ds\f{(v^\prime)^2}{c^2}}}
{\sqrt{1-\ds\f{v^2}{c^2}}},
\ee where $\vec v^\prime$ is the charge's velocity in the reference system $K^\prime$. Sub\-sti\-tu\-ting (8.48)
into (8.41) and (8.42) we obtain the four-force\index{four-force} $R^\nu$ determined by the expression \be
R=\f{F}{\sqrt{1-\ds\f{v^2}{c^2}}},\;
\vec R=\f{\vec F}{\sqrt{1-\ds\f{v^2}{c^2}}},
\ee which transforms under the Lorentz\index{Lorentz} 
transformations\index{Lorentz transformations} like $(ct, \vec x)$ \ba &&R_x^\prime
=\gamma\left(R_x-\ds\f{\,u\,}{\,c\,}R\right),\;
R^\prime =\gamma\left(R-\ds\f{\,u\,}{\,c\,}R_x\right),\nonumber\\
\label{8.50}\\
&&R_y^\prime =R_y,\quad R_z^\prime =R_z\,.\nonumber
\ea
{\bf Such a four-vector\index{four-vector}
 of force was first introduced by
Poincar\'{e}}\break\cite{2,3}.\index{Poincar\'e} 

With the aid of formulae (8.28) and (8.9) the
Lorentz\index{Lorentz} 
force
(8.35) can be written as
\be
f^\nu =\f{\,1\,}{\,c\,}F^{\nu\mu}S_\mu;
\ee
similarly, for the
four-vector\index{four-vector}
of force $R^\nu$ we have
\be
R^\nu =F^{\nu\mu}U_\mu.
\ee
Now let us calculate the energy-momentum tensor of the electromagnetic field. By means of Eqs. (8.51) and (8.27) we
obtain
\be
f_\nu=-\frac{1}{4\pi}[\partial_\alpha(F_{\nu\mu}F^{\mu\alpha})-F^{\mu\alpha}\partial_\alpha F_{\nu\mu}].
\ee
With the help of identity (8.24) the second term may be written as follows
$$
F^{\mu\alpha}\partial_\alpha F_{\nu\mu}=-\frac{\,1\,}{\,4\,}\partial_\nu F^{\mu\beta}F^{\mu\beta}.
$$
Taking into account this equation we get
\be
f_\nu=-\partial_\alpha T^\alpha_\nu,
\ee
where $T^\alpha_\nu$ is energy-momentum tensor of  the electromagnetic field
$$
T^\alpha_\nu=\frac{1}{4\pi}F_{\nu\mu}F^{\mu\alpha}+\frac{1}{16\pi}\delta^\alpha_\nu F^{\mu\beta}F_{\mu\beta},
$$
or in symmetric form \be
T^{\alpha\sigma}=-\frac{1}{4\pi}F^{\alpha\mu}F^{\sigma\rho}\gamma_{\mu\rho}+\frac{1}{16\pi}\gamma^{\alpha\sigma}
F^{\mu\beta}F_{\mu\beta}. \ee For more details see Section 15 p.~\pageref{e-m-tensor}.

The components of energy-momentum tensor may be expressed through $\vec E$ and $\vec H$ as follows
$$
T^{00}=\frac{1}{8\pi}(E^2+H^2),
$$
$$
cT^{0i}=S^i=\frac{c}{4\pi}[\vec E \vec H]_i,
$$
$$
T^{ik}= - \frac{1}{4\pi} \left( E_iE_k+H_iH_k-\frac{1}{2}\delta_{ik}(E^2+H^2) \right).
$$
From Eq.~(8.54) by integrating it over the whole space we get
$$
\vec F=\int dV \vec f=-\frac{d}{dt}\int\frac{1}{4\pi c}[\vec E \vec H]dV.
$$
This result coincides with the expression obtained 
by H.\,Poincar\'e\index{Poincar\'e} 
(see Section 9, p.~\pageref{wash6})

Thus, the entire set of
Maxwell-Lorentz\index{Lorentz} 
\index{Maxwell} 
equations\index{Maxwell-Lorentz equations}
is written
via vectors and tensors of
four-dimensional\index{four-dimensional space-time}
space-time.
{\bf The
Lo\-rentz\index{Lorentz} group,\index{group}\index{group, Lorentz}
that was discovered on the basis of
studies of elec\-tro\-mag\-ne\-tic phenomena, was extended by
H.\,Poincar\'{e}\index{Poincar\'e} 
{\rm [2, 3]} to all physical phenomena.}

In ref.~\cite{3} developing Lorentz\index{Lorentz}  ideas he wrote:
\begin{quote}
{\it ``\ldots All forces, of whatever origin they may be, behave,
owing to the
Lorentz\index{Lorentz} 
transformations\index{Lorentz transformations}
(and, con\-seq\-uently, owing to
translational motion) precisely like electromagnetic forces''.}
\end{quote}

H.\,Poincar\'{e}\index{Poincar\'e} wrote:
\begin{quote}
{\it\hspace*{5mm}``The principle of physical relativity may serve us in defining space. It gives us, so to say, a
new instrument for measurement. Let me explain. How can a solid body serve us in measuring or, to be more correct,
in constructing space? The point is the following: by transferring a solid body from one place to another, we thus
notice that it can, first, be applied to one figure and, then, to another, and we agree to consider these figures
equal. This convention gave rise to geometry\ldots Geometry is nothing, but a doctrine  on the reciprocal
relationships between these transformations or, to use mathematical language, a doctrine on the structure of the
group\index{group} composed by these transformations, i.\,e. the group\index{group} of motions
of solid bodies.\\
\hspace*{5mm}Now, take another group,\index{group} the group\index{group} of trans\-for\-ma\-tions, that do not
alter our differential equations. We obtain a new way of determining the equality of two figures. We no longer
say: two figures are equal, when one and the same solid body can be applied both to one figure and to the other.
We will say: two figures are equal, when one and the same mechanical system, sufficiently distant from
neighbouring that it can be considered isolated, being first accommodated so that its material points reproduce
the first figure, and then so that they reproduce the second figure, be\-ha\-ves in the second case like in the
first. Do these two views differ
from each other in essence? No\ldots\\
\hspace*{5mm}A solid body is much the same mechanical system as any
other. The only difference between our previous and new definitions
of space consists in that the latter is broader, allowing the solid
body to be replaced by any other mechanical system. Moreover, our
new con\-ven\-ti\-o\-nal agreement not only defines space, but time, also.
It explains to us, what are two simultaneous mo\-ments, what are
two equal
intervals\index{interval}
of time, or what is an
interval\index{interval}
of time twice
greater than another
interval''.}\index{interval}
\end{quote}

Further he notes:
\begin{quote}
{\it\hspace*{5mm}``Just transformations of the ``Lorentz\index{Lorentz} 
group\index{group, Lorentz}\index{group}'' do not alter differential equations of dynamics. If we suppose that our
system is referred not to axes at rest, but to axes in translational motion, then we have to admit, that all bodies
are deformed. For example, a sphere is transformed to an ellipsoid which smallest axis coincides with the
direction of translational motion of coordinate axes. In this case the time itself is experienced profound
changes. Let us consider two observers, the first is connected to axes at rest, the second --- to moving axes, but
both consider themselves at rest. We observe that not only the geometric object treated as a sphere by first
observer will be looked liked an ellipsoid for the second observer, but also two events treated as simultaneous by
the first will not be simultaneous for the second.''.}
\end{quote}

All the above formulated by H.\,Poincar\'{e}'s\index{Poincar\'e} (not mentioning the content
of his articles [2, 3]\,) completely contradicts to A.\,Einstein\index{Einstein} words
written in his letter to professor Zangger\index{Zangger} (Director of Law Medicine
Institute of Zurich University) 16.11.1911, that 
H.\,Poin\-ca\-r\'e\index{Poincar\'e} {\it
``has taken up a position of  unfounded denial (of the theory of relativity) and  has revealed insufficient
understanding of the new situation at all''.} (B.\,Hoffmann\index{Hoffmann} {\sf
``А.\,Einstein''},\index{Einstein} Moscow: Progress, 1984, p. 84 (in Russian)).

If one reflects upon H.\,Poincar\'e's\index{Poincar\'e} words, one can immediately perceive
the depth of his penetration into the es\-sence of physical 
relativity\index{relativity principle} and the
relationship between ge\-o\-met\-ry and group.\index{group} Pre\-ci\-sely in this way, starting from the
invariability of the Maxwell-Lorentz\index{Lorentz}\index{Maxwell} 
equations\index{Maxwell-Lorentz equations} under  the Lorentz\index{Lorentz} 
group\index{group}\index{group, Lorentz} trans\-for\-mations, which provided for
con\-sis\-ten\-cy with the principle of physical re\-la\-ti\-vi\-ty, 
H.\,Poincar\'{e}\index{Poincar\'e}
discovered the ge\-o\-met\-ry of space-time, de\-ter\-mi\-ned by the invariant (3.22).

Such space-time possesses the properties of ho\-mo\-ge\-nei\-ty and isotropy.
It reflects the existence in Nature of the fun\-da\-men\-tal conservation
laws\index{fundamental conservation laws}
of energy, momentum and angular momentum for a closed system. Thus,
the ``new convention'' is not arbitrary, it is based on the fundamental
laws\index{fundamental conservation laws}
of Nature.

Now let us quote one striking statement by Hermann\,Weyl\index{Weyl}. It is written in
his book {\sf ``Raum. Zeit. Materie''} appeared in 1918: 
\begin{quote}
{\it\hspace*{5mm}``The solution of Einstein\index{Einstein} ({\rm here is
the reference to the 1905 paper by A.\,Einstein\index{Einstein}} --- A.L.), which at one stroke overcomes
all difficulties, is then this: {\bf \textit{the world is a four-dimensional
affine space whose metrical structure is determined by a non-definite
quadratic form\\
\[
Q(\vec{x})=(\vec{x}\vec{x})
\]
with has one negative and three positive dimensions}}''.}
\end{quote}
Then he writes:
\begin{quote}
{\it``$(\overrightarrow{OA},\overrightarrow{OA})=-x_0^2+x_1^2+x_2^2+x_3^2\,,$\\
in which the $x_i$'s are the co-ordinates of $A$''.}
\end{quote}

But all this mentioned by H.\,Weyl\index{Weyl} was discovered by
H.\,Poincar\'e\index{Poincar\'e} (see articles [2, 3]\,), and not by
A.\,Einstein.\index{Einstein}  Nonetheless H.\,Weyl\index{Weyl} does not
see this and even more, he writes in his footnote:
\begin{quote}
{\it\hspace*{5mm}``Two almost simultaneously appeared works by\break {\bf \textit{H.\,Lorentz}}\index{Lorentz}
and {\bf \textit{H.\,Poincar\'e}},\index{Poincar\'e} are
closely related to it ({\rm the article by A.\,Einstein\index{Einstein} of 1905} ---
A.L.). They are not so clear and complete in presenting principal issues as Einstein's\index{Einstein}
article is.}
\end{quote}
Then references to works by Lorentz\index{Lorentz} 
and Poincar\'e\index{Poincar\'e} are given. Very strange logic. H.\,Weyl\index{Weyl} has
exactly formulated the solution, ``{\it which at one stroke
overcomes all difficulties}'', but namely this
\textbf{is contained in articles by H.\,Poincar\'{e}\index{Poincar\'e} [2, 3], and not in
Einstein's} ones. It is surprising how he has not seen this during his reading the Poincar\'{e}\index{Poincar\'e}  articles, because, as he
mentions correctly, the essence of the theory of relativity is namely this. All the main consequences of it follow
trivially from this, including the definition of the simultaneity\index{simultaneity} concept for different space
points by means of the light signal, introduced by 
H.\,Poincar\'{e}\index{Poincar\'e} in his
articles published in 1898, 1900 and 1904.

What a clearness and completeness of presentation of the principal issues is additionally necessary for
Weyl\index{Weyl} when he  himself has demonstrated 
what ``\textit{at one stroke
overcomes all difficulties.}'' H.\,Weyl\index{Weyl} should better be more attentive in
reading and more accurate in citing literature.

Above we have convinced ourselves that the symmetric set of equations of electrodynamics, (8.1), (8.2), which is
invariant with respect to coordinate three-dimensional orthogonal trans\-for\-mations, at the same time turned out
to be invariant, also, with respect to Lorentz\index{Lorentz} 
transformations\index{Lorentz transformations} in 
four-dimensional\index{four-dimensional space-time} space-time.
This became po\-ssible due to a number of vector quantities of 
Euclidean\index{Euclid} space
become, together with certain scalar quantities of the same space, components of four-dimensional quantities. At
the same time, some vector quantities, such as, for example,
$\vec E, \vec H$, are de\-ri\-vatives of the com\-po\-nents of four-dimensional quantities, which is the evidence
that they are com\-po\-nents of a tensor of the second rank in 
Min\-kow\-ski\index{Minkowski} 
space.\index{Minkowski space} The latter leads to the result that such concepts as {\it electric and
magnetic field strengths} are not absolute.

\newpage
\markboth{thesection\hspace{1em}}{}
\section{Poincar\'{e}'s relativistic mechanics}

In Section 3 we saw that the requirement of fulfilment
 of the relativity
principle\index{relativity principle} for electrodynamics leads to transition from one inertial reference system
to another, moving with respect to the first along the $x$ axis with a velocity $v$, being realized not by
Galilean\index{Galilei} transformations (2.5),\index{Galilean transformations} but by
 Lorentz\index{Lorentz} transformations\index{Lorentz transformations} (3.1).
Hence it follows, of necessity, that the equations of mechanics must be changed to make them form-invariant with
respect to the Lorentz\index{Lorentz} 
transformations.\index{Lorentz transformations} Since space and time are four-dimensional, the physical quan\-ti\-ti\-es described by vectors
will have four components. The sole four-vector\index{four-vector} describing a point-like body has the form \be
U^\nu=\f{dx^\nu}{d\sigma}.
\ee Here the interval $d\sigma$ in Galilean\index{Galilei} 
coordinates\index{Galilean (Cartesian) coordinates} is as follows \be
(d\sigma)^2=c^2 dt^2\left(1-\f{v^2}{c^2}\right).
\ee
Substituting the expression for $d\sigma$ into (9.1) we obtain
\be
U^0=\gamma,\quad
U^i=\gamma\,\f{v^i}{c}\,,\quad
v^i=\f{dx^i}{dt}\,,\quad
i=1,2,3.
\ee {\bf This four-vector\index{four-vector} of velocity was first introduced by Poincar\'{e}} 
[3].\index{Poincar\'e} 

We now introduce the
four-vector\index{four-vector}
of momentum
\be
P^\nu=mcU^\nu\,
\ee where $m$ is rest mass of a point-like body.

The relativistic
equations\index{Poincar\'{e}'s equations of mechanics}
of mechanics can intuitively be written
in the form
\be
mc^2\,\f{dU^\nu}{d\sigma}=F^\nu\,,
\ee
here $F^\nu$ is the
four-vector\index{four-vector}
of force, which is still to be
ex\-pres\-sed via the ordinary
Newtonian\index{Newton} 
force $\vec f$. It is
readily verified that the
four-force\index{four-force}
is orthogonal to the
four-velocity,\index{four-velocity}
i.\,e.
\[
F^\nu U_\nu=0.
\]

\markboth{thesection\hspace{1em}}{9. Poincar\'{e}'s relativistic mechanics}
On the basis of (9.2) and (9.3) equation (9.5) can be writ\-ten
in the form
\be
\f{d}{dt}\left(\f{m\vec v}{\sqrt{1-\ds\f{v^2}{c^2}}}\right)=
\vec F\,\sqrt{1-\f{v^2}{c^2}},
\ee
\be
\f{d}{dt}\left(\f{mc}{\sqrt{1-\ds\f{v^2}{c^2}}}\right)=
F^0\,\sqrt{1-\f{v^2}{c^2}}.
\ee
Since from the
correspondence
principle at small velocities equa\-tion
(9.6) should coincide with
Newton's\index{Newton} 
equation, it is natural to
define $\vec F$ as follows:
\be
\vec F=\f{\vec f}{\sqrt{\ds 1-\f{v^2}{c^2}}},
\ee
here $\vec f$ is the usual three-dimensional force.

Now let us verify, that equation (9.7) is a consequence of equation (9.6). Multiplying equation (9.6) by  the velocity
$\vec v$ and differentiating with respect to time, we obtain \be \f{m}{\left(\ds 1-\f{v^2}{c^2}\right)^{3/2}}\cdot
\left(\vec v\,\f{d\vec v}{dt} \right)=
\vec f\,\vec v.
\ee
On the other hand, upon differentiation with respect to time,
equation (9.7) assumes the form
\be
\f{m}{\left(\ds 1-\f{v^2}{c^2}\right)^{3/2}}\cdot
\left(\vec v\,\f{d\vec v}{dt} \right)=
cF^0\,\sqrt{\ds 1-\f{v^2}{c^2}}.
\ee
Comparing (9.9) and (9.10), we find
\be
F^0=\f{\left(\ds\f{\vec v}{c}\vec f\right)}
{\sqrt{\ds 1-\f{v^2}{c^2}}}.
\ee On the basis of relations (9.8) and (9.11) the equations of 
relativistic\index{equations of relativistic mechanics} mechanics assume the form \be \f{d}{dt}\left(\f{m\vec v}{\sqrt{\ds 1-\f{v^2}{c^2}}}\right)=
\vec f,
\ee
\be
\f{d}{dt}\left(\f{mc^2}{\sqrt{\ds 1-\f{v^2}{c^2}}}\right)=
\vec f\vec v.
\ee {\bf These equations\index{Poincar\'{e}'s equations of mechanics} were first obtained by
H.\,Poincar\'{e}}~\cite{3}\index{Poincar\'e}. Equation (9.13) relates the change in
particle energy  and the work done per unit time.

Having obtained these
equations,\index{Poincar\'{e}'s equations of mechanics}
Poincar\'{e}\index{Poincar\'e} 
applied them for
explaining the anomalies in the movement of Mercury. In this
connection he wrote:
\begin{quote}
{\it\hspace*{5mm} ``Thus, the {\bf \textit{new mechanics}} is still on unsteady soil. So we are to wished it new
confirmations. Let us see what astronomical observations give us in this connection. The velocities of planets
are, doubtless, relatively very small, but, on the other hand, as\-tro\-no\-mi\-cal observations exhibit a high
degree of precision and extend over long intervals\index{interval} of time. Small actions can, apparently, add up
to such an extent, that they acquire values permitting to be estimated. The only effect, with respect to which one
could expect it to be noticeable is the one we actually see: I mean the perturbations of the fastest of all
planets --- Mercury. It indeed shows such anomalies in its motion that can still not be explained by celestial
mechanics. The shift of its perihelion is much more significant than calculated on the basis of classical theory.
Much effort has been applied with the aim of explaining these de\-vi\-a\-tions \ldots The new mechanics somewhat
corrects the error in the theory of Mercury's motion lowering it to $32^{\prime\prime}$, but does not achieve
total accordance between the ob\-ser\-vation and calculation. This result, is, thus, not in favour of the {\bf
\textit{new mechanics}}, but at any rate, it also is not against it. The new
doctrine
does
not contradict astronomical observations directly''.}
\end{quote}

One can see here, how careful H.\,Poincar\'{e}\index{Poincar\'e} was in his estimation of
results. This was quite understandable, since the theory was still under development, and therefore attentive and
multiple ex\-pe\-ri\-men\-tal tests of its conclusions were required. It turned out that these equations were
valid only when gravity was neglected. Later A.\,Einstein\index{Einstein} explained the
anomaly in the motion of Mercury on the basis of general relativity theory, in which gravity is a consequence of
the  curvature of space-time. But to explain the anomaly in the motion of Mercury 
Einstein\index{Einstein} actually had to renounce special relativity theory and, as a consequence, the
fundamental con\-ser\-vation laws\index{fundamental conservation laws} of energy-momentum and of angular momentum.

From equations (9.12) it follows that the equations of clas\-si\-cal mechanics are valid only when the velocity
$v$ is small as compared with the velocity of light. It is just the approximate character of the equations of
classical me\-cha\-nics that has led to the origination of the 
Galilean\index{Galilei} 
transformations,\index{Galilean transformations} that leave the equations of mechanics unchanged in all inertial
reference systems.

In three-dimensional form the momentum and energy have the form
\be
\vec p=\gamma m\vec v, \quad E=p^0 c=\gamma m c^2.
\ee From (9.12) and (9.13) it follows that for a closed system energy and momentum are conserved. \textbf{As we
see from for\-mu\-la {\textrm{(9.14),}} energy $E$ is not an invariant. It has been and remains to be  an
invariant only with respect to three-di\-men\-si\-o\-nal coordinate trans\-for\-ma\-tions, and at the same time it
is the zeroth component of the four-dimensional mo\-men\-tum vector in 
Minkowski.\index{Minkowski}\index{Minkowski space}}

As an example let us calculate the energy of a system of two particles
{\bf \textit{a}}
and
{\bf \textit{b}}
in two different systems of reference. To proceed so let us consider the
invariant
\[
{\cal V}=(p_a+p_b)^2.
\]
In the system of reference where one particle is at rest,
\[
\vec p_a=0\,,
\]
we have
\[
{\cal V}=2mE+2m^2c^2.
\]
Here we take masses for particle
{\bf \textit{a}}
and for particle
{\bf \textit{b}}
as equal.
The same invariant is
\[
{\cal V}=(p_a+p_b)^2=4\f{{\cal E}^2}{c^2},
\]
when estimated at the reference system where the center of mass is at rest
\[
\vec p_a+\vec p_b=0,
\]
and $\cal E$ is a particle energy  calculated in this system of reference.

When comparing these expressions we get a connection between the energies in these two reference systems:
\[
E=2\f{{\cal E}^2}{mc^2}-mc^2.
\]

The collision energy of two particles is used with most efficiency in case when the center of mass of the two
particles is at rest in laboratory system of reference. Just this situation is realized in colliders. There is no
loss of energy for the center of mass motion.

One who has felt the four-dimensionality of space-time, could have seen immediately that energy and momentum are
combined in the four-momentum.\index{four-momentum} Moreover, he would have understood that in the case of a
closed system they obey the  energy and momentum conservation law.

In  1905 A.\,Einstein\index{Einstein} has proposed really existent 
quanta of the light energy
$\hbar\omega$ to explain the photo-effect. If he would understand in deep the existence and  meaning of the group, and
so the requirement of relativity principle that physical quantities should be four-dimensional, then he could
introduce for light the quantum of  momentum in line with the quantum
of energy. Moreover  that time it was already
proved experimentally (P.\,N. Lebedev\index{Lebedev}, 1901) that the light was carrying not only the energy, but
also the momentum and so it was exerting pressure on solid bodies. 
But A.\,Einstein\index{Einstein} has not done this. 
The  momentum of the quantum
of light has been introduced by J.\,Stark\index{Stark} in 1909. He took it
into account in the momentum conservation law. So the quantum of 
light,  the {\bf photon}, has appeared (as a particle).

Energy and momentum according to (9.4) transform as follows under Lorentz  transformations
$$
p'_x=\gamma\left(p_x-v\frac{E}{c^2} \right),\quad p'_y=p_y,\quad p'_z=p_z,\quad E'=\gamma(E-vp_x).
$$
A monochromatic plane light wave is characterized by frequency $\omega$ and wave vector $\vec K=\ds\frac{\,\omega\,}{\,c\,}\vec n$.
Together they are components of four-dimensional wave vector
$$
K^\nu=\left(\frac{\,\omega\,}{\,c\,},\frac{\,\omega\,}{\,c\,}\vec n \right).
$$
Square of this four-dimensional  wave vector is zero due to the wave equation
$$
K^\nu K_\nu=0.
$$
The meaning of this fact is that the rest mass is zero.

The frequency $\omega$ and the wave vector $\vec K$ transform under Lorentz transformations in the same way as $ct$,
$\vec x$, i.\,e. as follows
\ba
&&\omega^\prime=\omega\gamma\left(1-\f{\,v\,}{\,c\,}n_x\right), \nonumber\\
&&\omega^\prime n_x^\prime=\omega\gamma\left(n_x-\f{\,v\,}{\,c\,}\right),\nonumber\\
&&\omega^\prime n_y^\prime=\omega n_y,\quad \omega^\prime n_z^\prime=\omega n_z.\nonumber \ea Just the same
formulae stay valid for \textbf{photon} which rest mass is zero. The vector of four-momentum\index{four-momentum}
of photon is as follows
\[
p^\nu=\left(\f{\hbar\omega}{c}, \hbar\vec K\right),
\]
where \( \hbar \) is the Planck constant.\index{Planck}

It follows from the above that energy \( E \) and frequency \( \omega \) transforms in the same way. Formulae
given above explain 
\textbf{Doppler effect}\index{Doppler effect},\index{Doppler} i.\,e. the change of the light frequency when it is
emitted by a moving source. The \textbf{Doppler effect}\index{Doppler} takes place also when the direction of movement of the light
source is perpendicular to the direction of observation ($n_x^\prime=0$). So far as
\[
\omega=\omega^\prime\gamma\left(1+\f{\,v\,}{\,c\,}n_x^\prime\right),
\]
we obtain for the transverse 
 \textbf{Doppler effect}\index{Doppler effect}\index{Doppler}
the following result
\[
\omega^\prime=\omega\sqrt{1-v/c^2}.
\]
This effect is small enough in comparison with the longitudinal one. From the above formulae it is also possible
to determine how the direction of light beam changes under transformation to another inertial reference
system\index{inertial reference systems}
\[
n_x^\prime=\f{n_x-\ds\f{\,v\,}{\,c\,}}{1-\ds\f{\,v\,}{\,c\,}n_x}.
\]
This formula shows the effect of \textbf{aberration}.
\index{aberration of light} We will return to this subject in Section
\( 16 \).

The covariant vector\index{covariant vector}
of four-velocity\index{four-velocity}
is
$U_\nu=U^\sigma \gamma_{\sigma \nu}$, but since in
Galilean\index{Galilei} 
coordinates\index{Galilean (Cartesian) coordinates}
$\gamma_{\sigma\nu}=(1,-1,-1,-1)$, we obtain
\be
U_\nu=(U^0,\,-U^i).
\ee
Taking into account
(9.1) and (9.15) it is possible to compose the
invariant
\be
U_\nu U^\nu=(U^0)^2-(\vec U)^2=1,
\ee
which by virtue of the definition of the
four-vector $U^\nu$\index{four-vector}
will be
unity. This is readily verified, if the values determined by formulae
(9.3) are substituted into (9.16). Thus, we have
\be
p_\nu p^\nu=(mc)^2,\; \textrm{or}\; E=c\,\sqrt{\vec p\,^2+m^2c^2}.
\ee In formula (9.17) we have retained for energy only the positive sign, however the negative sign of energy also
has sense. It turns out to be significant in the case of unification of relativity theory and quantum ideas. This
led Dirac\index{Dirac} to predicting the particle (po\-si\-tron) with the
mass of the electron and positive charge, equal to the electron charge. Then the ideas arose of ``elementary''
particles creation in the process of interaction, of the physical vacuum\index{physical vacuum}, of the
antiparticles\index{antiparticles} (V.\,Ambartzumyan,\index{Ambartzumyan} D.\,Ivanenko,\index{Ivanenko}  E.\,Fermi\index{Fermi}
). It has opened the possibility of transformation of the colliding particles \textbf{kinetic
energy} to the  \textbf{material substance} possessing \textbf{rest mass}. So the need to construct accelerators
for high energies to study microcosm's mysteries has arisen.

On the basis of (9.14) equation (9.12) assumes the form
\be
\f{d}{dt}\left(\f{E}{c^2}\vec v\right)=\vec f,\;
{\rm or}\; \f{E}{c^2}\cdot\f{d\vec v}{dt}=
\vec f-\f{\vec v}{c^2}\cdot\f{dE}{dt}.
\ee From (9.18) it follows that the acceleration of a body, de\-ter\-mi\-ned by the expression $\ds\f{d\vec
v}{dt}$ does not coincide in direction with the acting force $\vec f$. From the equations of
Poincar\'{e}'s\index{Poincar\'e} 
relativistic\index{equations of relativistic mechanics}
mechanics we have on the basis of (9.17), for a body in a state of rest
\[
E_0=mc^2,
\]
where $E_0$ is the energy, $m$ is the mass of the body at rest.

From (9.17) it is evident, that mass
\textit{m}
is an invariant. This relation is a direct consequence of pseudo-Euclidean structure of the space-time geometry.
The connection between energy and mass\index{inert mass}  first arose in relation to the inert property of the
electromagnetic radiation. Formula $E=m{c^2}$ for radiation had been found for the first time in the article by
H.\,Poincar\'e in 1900 in clear and exact form. \label{Poincare_expression}

Let us quote some extractions from the article by H.\,Poincar\'e published in 1900 ``Lorentz theory and principle
of equality of action and reaction'' (put into modern notations by V.\,A.\,Petrov\index{Petrov}):
\begin{quote}
{\it\hspace*{5mm}``First of all let us shortly remind  the derivation  proving that the principle of equality of
action and reaction is no more valid in the Lorentz\index{Lorentz} theory,
at least when it is applied to the matter.\\
\hspace*{5mm} We shall search for the resultant of all ponderomotive forces applied to all electrons located
inside  a definite volume.
This resultant is given by the following integral
\[
\vec F=\int \rho dV \left(\f{\,1\,}{\,c\,}[\vec v, \vec H]+\vec E \right),
\]
where integration is over elements \( dV \) of the considered volume,
and  \( \vec v \) is the electron velocity.\\
\hspace*{5mm}Due to the following equations
\ba
&&\f{4\pi}{c}\rho\vec v=-\f{\,1\,}{\,c\,}\cdot\f{\pa \vec E}{\pa t}+\rot \vec H,\nonumber\\
&&4\pi\rho={\rm div} \vec E,\nonumber
\ea
and by adding and subtracting the expression \( \ds \f{1}{8\pi}\nabla H^2, \)
I can write the following formula
\[
\vec F=\sum_1^4 \vec F_i,
\]
\vspace*{-5mm}
where
\ba
&&\vec F_1=\f{1}{4\pi c}\int dV \left[\vec H \f{\pa \vec E}{\pa t}\right], \nonumber\\
&&\vec F_2=\f{1}{4\pi}\int dV(\vec H \nabla)\vec H,  \nonumber\\
&&\vec F_3=-\f{1}{8\pi}\int dV \nabla H^2, \nonumber\\
&&\vec F_4=\f{1}{4\pi}\int dV \vec E ({\rm div} \vec E).  \nonumber
\ea
Integration by parts gives the following
\ba
&&\vec F_2=\f{1}{4\pi}\int d\sigma \vec H (\vec n \vec H)
-\f{1}{4\pi}\int dV \vec H ({\rm div} \vec H),  \nonumber\\
&&\vec F_3=-\f{1}{8\pi}\int d\sigma \vec n H^2, \nonumber
\ea
where integrals are taken over all elements \( d\sigma \) of the surface bounding the volume considered, and
where \( \vec n \) denotes the normal vector to this element.
Taking into account
\[
{\rm div} \vec H=0,
\]
it is possible to write the following
$$
\vec F_2+\vec F_3=
\f{1}{8\pi}\int d\sigma\Bigl(2\vec H (\vec n \vec H)-\vec n H^2\Bigr).\eqno(A)
$$
Now let us transform expression \( \vec F_4. \) Integration by parts gives the following
\[
\vec F_4=\f{1}{4\pi}\int d\sigma \vec E (\vec n \vec E)
-\f{1}{4\pi}\int dV (\vec E \nabla)\vec E.
\]
Let us denote two integrals from r.h.s. as \( \vec F_4^\prime \) and \( \vec F_4^{\prime\prime} \), then
\[
\vec F_4=\vec F_4^\prime -\vec F_4^{\prime\prime}.
\]
Accounting for the following equations
\[
[\nabla \vec E]=-\f{\,1\,}{\,c\,}\cdot \f{\pa \vec H}{\pa t},
\]
we can obtain the following formula
\[
\vec F_4^{\prime\prime}=\vec Y +\vec Z,
\]
where
\ba
&&\vec Y=\f{1}{8\pi}\int dV\nabla E^2,\nonumber\\
&&\vec Z=\f{1}{4\pi c}\int dV\left[\vec E\,\f{\pa \vec H}{\pa t}\right].\nonumber
\ea
As a result we find that
\ba
&&\vec Y=\f{1}{8\pi}\int d\sigma \vec n E^2,\nonumber\\
&&\vec F_1-\vec Z=\f{d}{dt}\int\f{dV}{4\pi c}[\vec H\,\vec E].\nonumber
\ea
At last we get the following
\[
\vec F=\f{d}{dt}\int \f{dV}{4\pi c}
[\vec H\,\vec E]+(\vec F_2+\vec F_3)+(\vec F_4^\prime -\vec Y),
\]
where \( (\vec F_2+\vec F_3) \) is given by Eq.~( A), whereas
\[
\vec F_4^\prime -\vec Y=\f{1}{8\pi}\int d\sigma
\Bigl(2\vec E (\vec n \vec E)-\vec n E^2\Bigr).
\]
Term \( (\vec F_2+\vec F_3) \) represents the pressure experienced by different  elements \( d\sigma \)
of the surface bounding the volume considered. It is straightforward to see that this pressure is nothing
else, but the Maxwell\index{Maxwell}
\textbf{magnetic pressure} 
introduced by this scientist in  well-known theory.
Similarly, term \( (\vec F_4^\prime-\vec Y) \)
represents action of the Maxwell\index{Maxwell}
electrostatic pressure. 
In the absence of the first term,
\[
\f{d}{dt}\int dV\f{1}{4\pi c}[\vec H\,\vec E]\,,
\]
the ponderomotive force would be nothing else, but a result of 
the Maxwell\index{Maxwell} pressures.
If our integrals are extended on the whole space, then forces \( \vec F_2, \vec F_3, \vec F_4^\prime \)
and \( \vec Y \) disappear, and the rest is simply\label{wash6}
\[
\vec F=\f{d}{dt}\int \f{dV}{4\pi c}[\vec H\,\vec E].
\]

If we denote as \( M \) the mass of one of particles considered, and as \( \vec v \)  its velocity, then we will
have in case when the  principle of equality of action and reaction is valid the following:
\[
\sum M\vec v={\rm const}.\;\footnote{The matter only is considered  here. --- {\sl A.\,L.}}
\]
Just the opposite, we will have:
\[
\sum M\vec v-\int \f{dV}{4\pi c}[\vec H\,\vec E]={\rm const}.
\]
Let us notice that
\[
-\f{c}{4\pi}[\vec H\,\vec E]
\]
is the Poynting vector\index{Poynting vector} of radiation.

If we put
\[
J=\f{1}{8\pi}(H^2+E^2),
\]
then the Poynting\index{Poynting}  equation\index{Poynting equation} gives the following
$$
\int \ds\f{dJ}{dt}dV
=-\int d\sigma\ds \f{c}{4\pi}\vec n [\vec H\,\vec E]-
\int dV \rho (\vec v \vec E).\eqno(B)
$$
The first integral in the r.h.s., as well known, is the amount of electromagnetic energy flowing into the
considered volume through the surface and the second term is the amount of electromagnetic energy created
in the volume by means of transformation from other species of energy.\\
\hspace*{5mm}We may treat the electromagnetic energy as a fictitious fluid with density \( J \) which is
distributed in space according to the Poynting laws.\index{Poynting} It is only necessary to admit that this fluid
is not indestructible, and it is decreasing over value \( \rho dV\vec E \vec v\) in volume element \( dV \)   in a
unit of time (or that an equal and opposite in sign amount of it is created, if this expression is negative). This
does not allow us to get a full analogy with the real fluid for our fictitious one. The amount of this fluid which
flows through a unit square surface oriented perpendicular to the axis 
$i,$  at a unit of time is equal to the
following
\[
JU_i,
\]
where \( U_i \) are corresponding components of the fluid velocity.

\hspace*{5mm}Comparing this to the Poynting\index{Poynting}  formulae, we obtain
\[
J\vec U=\f{c}{4\pi}[\vec E\,\vec H];
\]
so our formulae take the following form
$$
\sum M\vec v+\int dV \f{J\vec U}{c^2}={\rm const}.\;\footnote{ In Eq. 
($C$) the second term in the l.h.s.
determines the total momentum of the electromagnetic radiation. Just here the concept of  \textit{radiation
momentum density} arises
\[
\vec g=\f{J}{c^2}\vec U,
\]
and also the concept of \textit{mass density of the electromagnetic field}
\[
m=\f{J}{c^2},
\]
where \( J \) is the electromagnetic energy density.
It is also easy to see from here that radiation energy density
\[
\vec S=\f{c}{4\pi}[\vec E\,\vec H]
\]
is related to the momentum density
\[
\vec g=\f{\vec S}{c^2}.
\]
So the notions of local \textit{energy} and \textit{momentum} appeared. All this was firstly obtained by
H.\,Poincar\'e.\index{Poincar\'e} Later these items were discussed in the Planck work\index{Planck} (Phys.
Zeitschr. 1908. {\bf 9.} S. 828) --- {\sl A.\,L.}}\eqno(C)
$$
They demonstrate that the momentum of substance plus the momentum of our fictitious fluid is given by a constant
vector.

\hspace*{5mm}In standard mechanics one concludes from the constancy of the momentum that the motion of the mass
center is rectilinear and uniform. But here we have no right to conclude that the center of mass of the system
composed of the substance and our fictitious fluid is moving rectilinearly and uniformly. This is due to the fact
that this fluid is not indestructible.\\
\hspace*{5mm}The position of the mass center depends on  value of the following integral
\[
\int \vec x JdV,
\]
which is taken over the whole space. The derivative of this integral is as follows
\[
\int \vec x\f{dJ}{dt}dV=-\int \vec x{\rm div}(J\vec U)dV-
\int \rho \vec x(\vec E \vec v)dV.
\]
But the first integral of the r.h.s. after integration transforms to the following expression
\[
\int J\vec UdV
\]
or
\[
\Bigl(\vec C-\sum M\vec v\Bigr)\,c^2,
\]
when we denote by \( \vec C \) the constant sum of vectors from Eq.~(\( C \)).\\
\hspace*{5mm}Let us denote by \( M_0 \) the total mass of substance, 
by $\vec R_0$ the coordinates of its center of mass, by $M_1$  the total mass of fictitious fluid, by $\vec R_1$  its center of mass, by $M_2$ the
total mass of the system (substance + fictitious fluid), by  $\vec R_2$ 
its center of mass, then we have
\[
M_2=M_0+M_1,\quad M_2\vec R_2=M_0\vec R_0+M_1\vec R_1,
\]
\[
\int \vec x\ds\f{J}{c^2}dV=M_1\vec R_1\,.\footnote{ H.\,Poincar\'e\index{Poincar\'e} also exploits in this formula the
concept of the \textit{mass density of the electromagnetic field} introduced by him earlier. --- {\sl A.\,L.}}
\]
Then we come to the following equation
$$
\f{d}{dt}(M_2\vec R_2)=\vec C-\int \vec x \f{\rho (\vec v\vec E)}{c^2}dV.\eqno(D)
$$
Eq.~(\( D \)) may be expressed in standard terms as follows. If the electromagnetic energy is created or
annihilated nowhere, then the last term disappears, whereas the center of mass of the system formed of the matter and
electromagnetic energy (treated as a fictitious fluid) has a rectilinear and uniform motion''.}
\end{quote}

Then H.\,Poincar\'e\index{Poincar\'e}
writes:
\begin{quote}
{\it\hspace*{5mm}``So, the electromagnetic energy behaves as a fluid having inertia from our point of view. And we
have to conclude that if some device producing electromagnetic energy will send it by means of radiation in a
definite direction, then this device must experience a recoil, as a cannon which fire a shot. Of course, this
recoil will be absent if the device radiates energy iso\-tro\-pi\-cally in all directions; just opposite, it will be
present when this symmetry is absent and when the energy is emitted in 
a single direction. This is just the
same as this proceeds, for example, for the H.\,Hertz\index{Hertz} emitter situated in a parabolic mirror. It is
easy to estimate numerically the value of this recoil. If the device has mass 1~\textit{kg}, and if it sends three
billion Joules in a single direction with the light velocity, then the
 velocity due to recoil is equal to  1~\textit{sm/sec}''.}
\end{quote}
When determining the velocity of recoil H.\,Poincar\'e\index{Poincar\'e}
again exploits the formula
\[
M=\f{E}{c^2}.
\]
In  \S\,7 of article [3] H.\,Poincar\'e\index{Poincar\'e} derives equations of  relativistic
mechanics\index{equations of relativistic mechanics}.
\index{Poincar\'{e}'s equations of mechanics} If we change the
system of units in this paragraph from \( M=1, c=1 \) to Gaussian system of units, then it is easy to see that
\textbf{inert\index{inert mass} mass of a body} is also determined by formula:
\[
M=\f{E}{c^2}.
\]
Therefore, it follows from works by H.\,Poincar\'e\index{Poincar\'e}
that the \textbf{inert\index{inert mass}
mass} both of \textbf{substance}, and of \textbf{radiation} is determined by their energy.
All this has been a consequence of the electrodynamics and the relativistic mechanics.

In 1905 Einstein\index{Einstein} has published the article
\textsf {``Does the inertia of a body depend on
the energy contained in it?''}. Max Jammer\index{Jammer} wrote on this article in his book
 \textsf{``The concept of mass in classical and modern physics''} 
 (Harvard University Press, 1961.):
\begin{quote}
\hspace*{5mm}{\Large\guillemotleft}\textit{It is generally said that ``the theorem of 
inertia of energy in its full generality was stated
by Einstein\index{Einstein} (1905)'' (Max Born.\index{Born} {\sf``Atomic physics''. Blackie, London, Glasgow ed.~6, p.~55}). 
The article referred  to is Einstein's\index{Einstein} paper, {\sf
``Does the inertia of a body depend upon its energy content?''}. On the 
basis of the 
Maxwell-Hertz\index{Maxwell}\index{Hertz} equations of the electromagnetic
field  Einstein\index{Einstein} contended that ``if a body gives off the
energy $E$ in the form of radiation, its mass diminishes by $E/c^2$''.
Generalizing this result for all energy transformations Einstein\index{Einstein} concludes: ``The mass of a body is a measure of its energy content''.\\
\hspace*{5mm}It is a curious incident in the history of
scientific thought that Einstein's\index{Einstein} own derivation of formula $E=mc^2$, as published in 
his article  in the {\sf ``Annalen der Physik'',} was basically fallacious. In fact, what for the layman is
known as ``the most famous mathematical formula ever projected'' in science
({\sf William Cahn\index{Cahn}. ``Ein\-stein, a pictorial biography''. New York: Citadel. 1955. P.~26}) was
but the result of {\textsf{petitio prin\-ci\-pii}}, the conclusion of
begging the question{\Large\guillemotright}}.\\
\hspace*{5mm}\textit{``The logical illegitimacy of Einstein's\index{Einstein} derivation has been shown by
Ives\index{Ives} {\sf (Journal of the Op\-ti\-cal Society of America. 1952. {\bf 42}, pp.~540-543)}}''.
\end{quote}

Let us consider shortly Einstein's\index{Einstein} article
of 1905 \textsf {``Does the inertia of a body depend on
the energy contained in it?''}
Einstein\index{Einstein} writes:
\begin{quote}
{\it \hspace*{5mm}``Let there be a body at rest in the system $(x, y, z)$, whose energy, referred to the system $(x, y, z)$, is $ E_0$. The energy of 
the body with respect to the system $(\zeta, \eta, \varsigma)$, which 
is moving with velocity $v$ as above, shall be $H_0$.\\ 
\hspace*{5mm} Let this body simultaneously
emit plane waves of light of energy $L/2$ (measured
relative to $(x, y, z)$) in a direction forming an angle $\varphi$ with the
 $x$-axis and an equal amount of light in the opposite direction. 
 All the while, the body shall stay at rest with respect to the system
$(x, y, z)$. This process must satisfy the energy principle, and this 
must be true (according to the  principle 
of relativity)\index{relativity principle} 
with respect to both coordinate systems. If $E_1$ and $H_1$ denote the energy
of the body after the emission of light, as measured relative to systems 
$(x, y, z)$ and $(\zeta, \eta, \varsigma)$, 
respectively, we obtain, using the relation indicated above,
\[
E_0=E_1+\left[\ds\f{\,L\,}{2}+\ds\f{\,L\,}{2}\right],
\]

$$
\begin{array}{l}
H_0=H_1+\left[\ds\f{\,L\,}{2}\cdot\f{1-\ds\f{\,v\,}{V}\cos\varphi}{\sqrt{1-\left[\ds\f{v}{V}\right]^2}}
+\f{\,L\,}{2}\cdot\f{1+\ds\f{\,v\,}{V}\cos\varphi}{\sqrt{1-\left[\ds\f{v}{V}\right]^2}}\right]=
\\[10mm]
=H_1+\ds\f{L}{\sqrt{1-\left[\ds\f{v}{V}\right]^2}}.
\end{array}
$$

Subtracting, we get from these equations
$$
\!\!\!\!\!\!\!\!\!\!(H_0-E_0)-(H_1-E_1)=
L\,\Biggl\{\ds\f{1}{\sqrt{1-\left[\ds\f{v}{V}\right]^2}}-1\Biggr\}".\eqno(N)
$$}
\end{quote}

A.\,Einstein\index{Einstein}
tries to get all the following just from this relation. Let us make an elementary analysis of the equation derived
by him. According to the theory of relativity
\[
H_0=\ds\f{E_0}{\sqrt{1-\ds\f{v^2}{c^2}}},\quad
H_1=\ds\f{E_1}{\sqrt{1-\ds\f{v^2}{c^2}}}.
\]
Einstein\index{Einstein}  seemingly did not take into account such formulae. It follows then  that
\[
H_0-E_0=E_0\Biggl(\ds\f{1}{\sqrt{1-\ds\f{v^2}{c^2}}}-1\Biggr),\quad
H_1-E_1=E_1\Biggl(\ds\f{1}{\sqrt{1-\ds\f{v^2}{c^2}}}-1\Biggr),
\]
and consequently the l.h.s. of the Einstein\index{Einstein} equation is equal to the following
\[
(H_0-E_0)-(H_1-E_1)=
(E_0-E_1)\Biggl(\ds\f{1}{\sqrt{1-\ds\f{v^2}{c^2}}}-1\Biggr);
\]
then Eq. (\( N \)) takes an apparent form
\[
E_0-E_1=L.
\]
Therefore, it is impossible to get something more substantial from the initial Einstein\index{Einstein} equation $(N)$. In this work A.\,Einstein\index{Einstein} has not succeeded in discovering neither physical arguments, nor
a method of calculation to prove that formula
\[
M=\f{E}{c^2}
\]
is valid at least for radiation. So, the critics given by Ives\index{Ives} on the A.\,Einstein\index{Einstein}
work is correct. In 1906  Einstein\index{Einstein} once more returns to this subject, but his work reproduces
the Poincar\'e\index{Poincar\'e} results of 1900, as he notes himself.

Later,
Planck\index{Planck} 
in 1907 and
Langevin\index{Langevin} 
in 1913 revealed, on this basis,
the role of internal interaction energy
(binding energy),\index{binding energy}
which  \textbf{led
to the mass defect,\index{mass defect}
providing conditions for possible energy
release, for example,  in fission and fusion of atomic  nuclei}.
The relativistic mechanics has become an engineering discipline. Accelerators of elementary particles are
constructed with the help of it.

``Disproofs'' of the special theory of relativity  appearing sometimes are  related to unclear and inexact
presentation of its basics in many textbooks.
Often its meaning is deeply hidden by plenty of minor or even needless details presented. The special theory of
relativity is strikingly simple in its basics, almost as Euclidean 
geometry\index{Euclidean geometry}\index{Euclid}.

\noindent
{\it On the transformations of force}

According to (9.8) and (9.11) the
four-force\index{four-force}
is
\be
F^\nu=\left(\gamma\,\f{\vec v\vec f}{c},\,\gamma\vec f \right),
\ee
\be
F_\nu=\left(\gamma\,\f{\vec v\vec f}{c},\,-\gamma\vec f \right).
\ee As we noted above, the force as a four-vector\index{four-vector} transforms like the quantities $ct$ and $x$,
so, \be f_x^\prime=\f{\gamma}{\gamma^\prime}\gamma_1 \left(f_x-\beta
\f{\,\vec v\,}{\,c\,}\vec f\right),\quad (\vec
v^{\,\prime} \vec f^{\,\prime})= \f{\gamma}{\gamma^\prime}\gamma_1
(\vec v\vec f-c\beta f_x),
\ee
\be
f_y^\prime =\f{\gamma}{\gamma^\prime}f_y,\quad
f_z^\prime =\f{\gamma}{\gamma^\prime}f_z.
\ee
Here
\be
\beta=\f{\,u\,}{\,c\,},\quad \gamma_1=\f{1}{\sqrt{1-\ds\f{u^2}{c^2}}},
\ee
$u$ is the velocity along the $x$ axis.

Consider two particles in the unprimed inertial reference system with the four-velocities\index{four-velocity} \be
U_1^\nu=\left(\gamma, \gamma\f{\,\vec v\,}{\,c\,}\right),\;
U_2^\nu=\left(\gamma_1, \gamma_1\f{\,\vec u\,}{\,c\,}\right).
\ee Then, in the inertial reference system, in which the second particle is at rest, we  have the following
expressions for the respective four-vectors:\index{four-vector}
\[
U_1^{\prime\nu}=\left(\gamma^\prime, \gamma^\prime\f{\vec v^\prime}{c}\right),\;
U_2^{\prime\nu}=(1,0).
\]
Hence, on the basis of invariance of  expression $U_{1\sigma}U_2^\sigma$ we have the following equality: \be
\gamma^\prime =\gamma\gamma_1\left(1-\f{\vec v\vec u}{c^2}\right).
\ee
Thus, we obtain
\be
\f{\gamma}{\gamma^\prime}=
\f{1}{\gamma_1\left(1-\ds\f{\vec v\vec u}{c^2}\right)}.
\ee
In our case, when the velocity $\vec u$ is directed along
the $x$ axis, we have
\be
\f{\gamma}{\gamma^\prime}=
\f{1}{\gamma_1\left(1-\ds\f{v_x u}{c^2}\right)}.
\ee
Substituting (9.27) into (9.21) and (9.22) we obtain
\be
f_x^\prime =\f{f_x-\ds\beta\,\f{\,\vec v\,}{\,c\,}\,\vec f}{1-\ds\beta\,
\f{v_x}{c}},\,
f_y^\prime =\f{f_y\sqrt{1-\ds\beta^2}}{1-\ds\beta\,
\f{v_x}{c}},\,
f_z^\prime =\f{f_z\sqrt{1-\ds\beta^2}}{1-\ds\beta\,
\f{v_x}{c}},
\ee
\be
\vec f\,^\prime\vec v\,^\prime=
\f{(\vec f\,\vec v)-\beta\, c\,f_x}{\ds 1-\beta\f{v_x}{c}}.
\ee

Hence it is evident that, if the force $\vec f$ in a certain inertial Ga\-li\-le\-an
\index{Galilei} 
re\-fe\-rence system is zero, it is, then, zero in any other inertial re\-fe\-rence system, also. This
means that, if the law\index{law of inertia} of inertia is valid in one inertial re\-fe\-rence system, then it is
also obeyed in any other inertial reference system. Moreover, the conclusion concerning the force is not only
valid for an inertial reference sys\-tem, but also for any ac\-ce\-le\-ra\-ted (non-inertial) reference
system.\index{non-inertial reference systems} {\bf Force cannot arise as a result of coordinate transformations}.
If mo\-tion by iner\-tia in an inertial reference system proceeds along a straight line, then in a non-inertial
reference system\index{non-inertial reference systems} free motion will proceed along the geodesic
line,\index{geodesic line} which in these coordinates will no longer be a straight line.

In classical mechanics the force $\vec f$ is the same in all inertial
reference systems, in relativistic mechanics this is no longer so,
the components of force, in this case, vary in accordance with (9.28).

Let us, now, dwell upon a general comment concerning inertial reference systems. Inertial reference systems being
equitable signifies that, if we create in each reference system identical conditions for the evolution of matter,
then we, na\-tu\-rally, should have the same description of a phenomenon in each reference system, in other words,
we will not be able to single out any one of the inertial reference systems. But, if we have provided some
conditions for the motion of matter in one  inertial  reference system, then, in describing what goes on in this
reference system by observing  from any other inertial reference system, we will already obtain another picture.
This does not violate the equality of inertial reference systems, since in this case the initial re\-fe\-rence
system has been singled out by the actual formulation of the problem.
{\bf Precisely such a si\-tu\-ation arises,
when we consider the Universe.\index{Universe} In this case, there exists a unique, physically singled out inertial
reference system in the Universe\index{Universe}, which is de\-ter\-mi\-ned by the dis\-tri\-bution of matter.
Time in this reference system 
will have a special status
as compared with  other iner\-tial reference systems. This time
could be termed the ``true time'' of the Universe.\index{Universe} As an example of 
such 
a special reference system
one could choose a reference system, in which the relict elec\-tro\-mag\-ne\-tic radiation is ho\-mo\-ge\-neous
and isotropic} (see ref.[5]\,).

From the above exposition, especially from Sections 3, 5, 7, 8, and 9 it is evident that Henri
Poincar\'{e}\index{Poincar\'e} discovered all the essentials that make up the content of the
special theory of relativity. Any person, who has graduated from a University in theoretical physics and who has
attentively read at least two of his articles {\sf ``On the dynamics of the electron''}, may verify this.

There exist, also, other points of view: {\it ``Poincar\'{e}\index{Poincar\'e} did not
make the decisive step''} (de Broglie),\index{Broglie} 
{\it ``Poincar\'{e}\index{Poincar\'e} was, most likely, quite close to creating the STR, but did not arrive at the end. One
can only guess why this happened.''} (V.\,L.\,Ginzburg).\index{Ginzburg}  But
these statements characterize their authors' own level of understanding the problem, instead of
H.\,Poincar\'{e}'s\index{Poincar\'e} outstanding achievements in 
the theory of relativity. What
is surprising is that the authors show no trace of doubt in considering their own incomprehension, or the
difficulty they had in understanding, as a criterion in evaluating the outstanding studies performed by
Poincar\'{e}.\index{Poincar\'e} In this case there is no need to ``guess''. It is only
necessary to read the works by Poincar\'e~\cite{2,3}\index{Poincar\'e} and to think.

Professor
A.\,Pais\index{Pais} 
wrote the following in his book 
{\sf ``Subtle is the Lord: the science and the life of Albert
Einstein''}, Oxford University Press, 1982:\index{Einstein} 
\begin{quote}
{\it``It is evident that as late as 1909 Poincar\'{e}\index{Poincar\'e} did not
know \textbf{that the contraction of rods is a consequence of the two Einstein's\index{Einstein}
postulates}. ({\rm singled out by me} --- {\sl A.\,L.})  
Poincar\'{e}\index{Poincar\'e}  therefore did not understand one of the 
most basic traits of special relativity''.}
\end{quote}
We right away note that the underlined statement is wrong. But
about this later.

From everything that
A.\,Pais\index{Pais} 
has written it clearly fol\-lows that he
himself {\bf did not understand} the fundamentals of special 
re\-la\-ti\-vi\-ty.
Let me explain.
Poincar\'{e}\index{Poincar\'e} 
demonstrated the invariability of
the
Maxwell-Lorentz\index{Lorentz} 
\index{Maxwell} 
equations\index{Maxwell-Lorentz equations}
with respect to the
Lorentz\index{Lorentz} 
trans\-for\-mations,
which was consistent with the
relativity principle,\index{relativity principle}
for\-mu\-la\-ted by
Poincar\'{e}\index{Poincar\'e} 
in 1904 for all natural physical phenomena. As we already
noted,
H.\,Poincar\'{e}\index{Poincar\'e} 
discovered the fundamental in\-va\-ri\-ant (3.22)
\[
J=c^2T^2-X^2-Y^2-Z^2\,,
\]
that establishes the geometry of space-time. Namely hence it fol\-lows, that the light velocity being constant is
a particular con\-se\-quence of this formula, when the invariant $J$ is zero.
A.\,Pais\index{Pais}  had to understand that the Lorentz\index{Lorentz}
contraction\index{Lorentz contraction} is related to negative $J$, i.\,e. to a space-like
value of $J$, not equal to zero. As to the slowing down of time, it is related to positive $J$, i.e time-like $J$,
but certainly not equal to zero. Thus, from the above it is clear, \textbf{that contraction of the dimensions of
rods is not a consequence of the two Einstein's\index{Einstein} postulates only}.
Such is the result of a superficial know\-ledge of the relativity theory foundations.

So with such a knowledge of material A.\,Pais\index{Pais} had tried to prove on
the pages of his book that H.\,Poincar\'e\index{Poincar\'e} had not made the decisive step
to create the theory of relativity! He,a physicist, ``reinforced'' his view on the contribution of
H.\,Poincar\'e\index{Poincar\'e} by the decision of the Paris Session of the French Philosophical
Society in 1922.

So simple it is! The philosophers have met and made a decision whereas they probably have not studied works by
Poincar\'e\index{Poincar\'e} on the theory of relativity at all. But 
such a study required
a corresponding professional level. I doubt whether their professional level had been higher than one by 
A.\,Pais\index{Pais} in this field. We should say that  
A.\,Pais\index{Pais}
was an outstanding scientist irrespective to this criticism and he made a lot of remarkable
investigations.

As to the Lorentz\index{Lorentz} contraction,\index{Lorentz contraction} in the
article [3] (\S\,6 {\sf ``The con\-traction of electrons''}) 
H.\,Poincar\'{e}\index{Poincar\'e} 
deals with this issue in detail, making use of the Lorentz\index{Lorentz} transformations.\index{Lorentz transformations} All this is clearly pre\-sen\-ted in article [3].
Precisely unification of relativity and the Maxwell-Lorentz\index{Lorentz} \index{Maxwell} electrodynamics permitted Poincar\'{e}\index{Poincar\'e} to formulate in articles [2] and [3] the foundation of the theory
of relativity. As to the postulate
concerning the constancy\index{principle of constancy of velocity of light} of the velocity of light, it proved to
be just a simple heuristic device, but not a fundamental of the theory. It is a con\-se\-quence of the requirement
that electrodynamical phenomena, des\-cri\-bed by 
the Maxwell-Lorentz\index{Lorentz}\index{Maxwell} 
equations\index{Maxwell-Lorentz equations} in
Galilean\index{Galilei} co\-or\-di\-nates,\index{Galilean (Cartesian) coordinates} be
consistent with the relativity principle.\index{relativity principle}

A.\,Pais,\index{Pais} mentioning the group character of Lorentz\index{Lorentz}
transformations\index{Lorentz transformations}, writes (see p.~130 of the book cited above):
\begin{quote}
{\hspace*{5mm} \it``He did, of course, not 
know that a few weeks earlier someone ({\rm A.\,Einstein
is understood.}\index{Einstein} --- A.L.{\sl )} 
else had independently noted the group properties of
Lorentz\index{Lorentz} 
transformations \ldots\index{Lorentz transformations}''}
\end{quote}
But all this is \textbf{absolutely incorrect}. Article [2] by H.\,Poincar\'e,\index{Poincar\'e} appeared in
\textsf{``Comptes Rendus''} on June, 5, 1905, whereas the article by A.\,Einstein\index{Einstein} had been sent to
publisher on June, 30, 1905.

H.\,Poincar\'e,\index{Poincar\'e} discovered the group\index{group} and named it as \textbf{Lorentz group\index{group, Lorentz}}.\index{Lorentz} He wrote in article [2]:
\begin{quote}
{\it\hspace*{5mm}``All these transformations together with all rotations should form a group''.}\index{group}
\end{quote}
In articles [2; 3] by H.\,Poincar\'e,\index{Poincar\'e} the group properties are widely used for constructing
four-dimensional physical quantities, providing the invariance of electrodynamics equations  under the\break
Lorentz\index{Lorentz} group\index{group, Lorentz}. While in the article by A.\,Einstein\index{Einstein} only the
following is told:
\begin{quote}
{\it\hspace*{5mm}``\ldots from this we see that such parallel 
transformations form a group\index{group} --- as they indeed must''.}
\end{quote}
There is no any other word on the group\index{group} in the Einstein\index{Einstein} article. From here his
misunderstanding that electrodynamic quantities should be transformed according to the group\index{group} in
order
to provide the invariance of equations required by relativity 
principle\index{relativity principle}  follows
naturally. But all this leads to the consequence that some physical quantities become four-dimensional, for
example, current density, potentials, momentum,\index{four-momentum} and force.\index{four-force}

Striking ``discoveries'' are made by certain historians near sci\-ence.
Here, follows, for example, one ``masterpiece'' of such a cre\-a\-tive
activity.
S.\,Goldberg\index{Goldberg}
wrote the following in his article
({\sf ``The British Journal for the History of Science''}. 1970. Vol.~V,
No.~17, p.~73):
\begin{quote}
{\it\hspace*{5mm}
``Poincar\'{e}\index{Poincar\'e} had retained the notion
of ab\-so\-lute space\index{absolute space} in his work,
whether or not that space was accessible to ob\-ser\-vation''.}
\end{quote}
\begin{quote}
{\it ``There was in Poincar\'{e}'s\index{Poincar\'e} 
mind a {\bf\textit{preferred}} frame of
re\-fe\-rence in which the velocity of light was 
{\bf\textit{really}} a constant, and only one such frame''.}
\end{quote}
S.\,Goldberg\index{Goldberg} 
attributes all this to
Poincar\'{e}\index{Poincar\'e} 
without any grounds
what\-so\-e\-ver. Thus, back in 1902, in the book {\sf ``Science and Hypothesis''},
Poincar\'{e}\index{Poincar\'e} 
wrote:
\begin{quote}
{\it\hspace*{5mm}``Absolute space does not exist. We only perceive
relative motions''.}\index{absolute space}
\end{quote}
\begin{quote}
{\it\hspace*{5mm}``Absolute time does not exist''.}\index{absolute time}
\end{quote}
In 1904 Poincar\'{e}\index{Poincar\'e} formulated the 
principle\index{relativity principle} of relativity for all phy\-si\-cal phenomena (see Section 3 p.~\pageref{wash}) and in 1905
es\-tab\-li\-shed that, in accordance with the relativity principle,\index{relativity principle} the equations of
the elec\-tro\-mag\-netic field remain the same in all inertial reference systems, owing to the 
Lorentz\index{Lorentz} trans\-for\-mations.\index{Lorentz transformations}

Thus the equality and constancy\index{principle of constancy of velocity of light} of the velocity of light is
pro\-vi\-ded for  any inertial reference system. All this is ex\-po\-un\-ded in the articles by H.\,Poincar\'{e}
[2, 3],\index{Poincar\'e} which should have been studied carefully by
S.\,Goldberg\index{Goldberg} before writing about an opinion of
Poincar\'{e}.\index{Poincar\'e} 

\textbf{In evaluating works [2] and [3], as well as the early works of H.\,Poincar\'{e}\index{Poincar\'e}
in physics}, it is necessary to proceed only from their content, comparing it with contemporary
ideas, but not to be guided by outside statements on the issue, even made by well-known scientists, contemporaries of
Poincar\'{e},\index{Poincar\'e} since the level of many of them was insufficient to fully
apprehend what Poincar\'{e}\index{Poincar\'e} has written. At the time his personality was
espe\-ci\-ally manifest in that for him physical problems and their adequate mathematical formulation joined
naturally and com\-po\-sed a single whole. Na\-me\-ly for this reason, his creations are exact and modern even
after a hundred years. H.\,Poincar\'{e}\index{Poincar\'e} was one of those rare
researchers, to whom natural sciences and mathematics are their proper sur\-ro\-un\-dings. The young people of
today, who are prepared in theoretical phy\-sics, can readily perceive this, if only they, at least, read
Poincar\'{e}'s\index{Poincar\'e} works [2] and [3]. What concerns the statements by
Professor A.\,Pais\index{Pais} and Doctor S.\,Goldberg,\index{Goldberg} 
we once more en\-co\-un\-ter, what we saw earlier is a clear attempt to attribute their own
incomprehension to the author.

Some authors wishing to stress the preceding character 
of\break H.\,Poincar\'{e}'s\index{Poincar\'e} 
articles
[2], [3] on  relativity  give  two following quotes from the book 
of W.\,Pauli\index{Pauli}
{\sf ``Theory of Relativity''}  written by him in young age in 1921:
\begin{quote}
{\it\hspace*{5mm}``Is was Einstein,\index{Einstein}
finally, who in a way completed the basic formulation
of this new discipline ''.}
\end{quote}
\begin{quote}
{\it\hspace*{5mm}``It includes not only all the
essential results contained in the other two papers,
but shows an entirely novel, and much more profound, 
understanding of the whole problem''.}
\end{quote}
Below we will give a quotation from
W.\,Pauli\index{Pauli} related to the same subject, but written later, in 1955.

To the first Pauli\index{Pauli} quotation it should be said that no further completion
of works [2], [3] by H.\,Poincar\'{e}\index{Poincar\'e} is required. All the main
results which contain the full content of the theory of relativity are formulated there and in the most definite
form.

What about the second statement by Pauli,\index{Pauli} the case is just opposite.
It is sufficient to compare the content of the Poincar\'{e}\index{Poincar\'e}  and Einstein\index{Einstein} works to conclude that articles [2], [3] by
Poincar\'e\index{Poincar\'e} 
contain not only all the main content of the article by
Einstein\index{Einstein} of
1905 (moreover Poincar\'e\index{Poincar\'e} has formulated everything definitely in
contrast to Einstein\index{Einstein}), but also contain main parts of the later work
by Minkowski.\index{Minkowski} 
What about words by Pauli\index{Pauli}  on
 ``\textit{deep understanding of the whole problem}'',
it is just present in articles [2], [3] by
Poincar\'e\index{Poincar\'e}.
For example:
\begin{quote}
{\it\hspace*{5mm}``All forces behave in the same way as electromagnetic forces irrespective of their origin. This
is  due to Lorentz\index{Lorentz} transformations\index{Lorentz transformations}
(and consequently due to translational motion)''.}
\end{quote}
In other words Lorentz\index{Lorentz} invariance is universal.
All the above in full can be said about gravitational 
forces\index{forces of gravity}.

Further, Poincar\'e\index{Poincar\'e} discovered 
pseudo-Euclidean\index{pseudo-Euclidean geometry of space-time} geometry of space-time, revealed the four-dimensionality of physical quantities. He
constructed the equations of relativistic 
mechanics\index{Poincar\'{e}'s equations of mechanics}, predicted
existence of the gravitational waves,\index{gravitational waves} propagating with the velocity of light. Then,
what else ``\textit{deep understanding of the whole problem}'' may be spoken about?

There is a surprising statement by L.\,de Broglie\index{Broglie} made in
1954:
\begin{quote}
{\it\hspace*{5mm}``A bit more and it  would be H.\,Poincar\'e\index{Poincar\'e}, and not A.\,Einstein,\index{Einstein}
who first built the theory of relativity in its full generality and that would  deliver to French science the
honor of this discovery.\ldots But Poincar\'e\index{Poincar\'e} has not made the decisive step and left to
Einstein\index{Einstein} the honor to uncover all the consequences following from the principle of
relativity\index{relativity principle}, and in particular, by means of a deep analysis of measurements of
length and time, to discover the real physical nature of relation between space and time maintained by the
principle of relativity\index{relativity principle}''.}
\end{quote}

\textbf{In fact all is just opposite} to the L.\,de Broglie\index{Broglie} writings.
H.\,Poin\-car\'e\index{Poincar\'e} gave detailed analysis of time measurements already in his article of 1898
\textsf{``The measurement of time''}, in particular, by means of a light signal. Later, in articles of 1900 and 1904
he describes a procedure for \textbf{determination of simultaneity}\index{simultaneity} at different points of
space by means of a light signal in a moving inertial system of 
reference\index{inertial reference systems},
and therefore reveals the physical meaning of \textbf{local\index{local time} time} by  Lorentz.\index{Lorentz} In
1904 in article [1] he was the first who formulated the principle of 
relativity\index{relativity principle} for
all physical phenomena. In 1905 being based on the Lorentz\index{Lorentz}  paper H.\,Poincar\'e\index{Poincar\'e} has
discovered the Lorentz group\index{group, Lorentz} in articles
 [2; 3] and on this ground proved  invariance of Maxwell-Lorentz equations \index{Maxwell-Lorentz equations}
under Lorentz
 transformations\index{Lorentz transformations} in full agreement with the relativity principle\index{relativity principle}.
H.\,Poincar\'e\index{Poincar\'e} extrapolated the   Lorentz 
group\index{group, Lorentz} on all physical forces.
Therefore the Lorentz invariance\index{Lorentz invariance} became universal and valid also for gravitational
phenomena. In article [3], being based on the 
Lorentz\index{group, Lorentz} group\index{group}
H.\,Poincar\'e\index{Poincar\'e} introduced 
pseudo-Euclidean\index{pseudo-Euclidean geometry of space-time}
space-time geometry. So, the homogeneous and isotropic space-time arose which was defined by the \textbf{invariant}
\[
c^2t^2-x^2-y^2-z^2.
\]
It was developed in relativity of
 \textit{time} and  \textit{length} concepts, in symmetry of physical laws, in conservation laws, in existence
 of the limiting velocity for material bodies, in four-dimensionality of physical quantities.
The connection between space and time was determined in full by the structure of geometry. There is no such a deep
insight into the essence of the problem  in the article by A.\,Einstein\index{Einstein}. Following these ideas
H.\,Poincar\'e\index{Poincar\'e} discovered 
equations\index{Poincar\'{e}'s equations of mechanics} of relativistic
mechanics\index{equations of relativistic mechanics} and predicted existence of gravitational
waves\index{gravitational waves} propagating with velocity of light. Therefore H.\,Poin\-ca\-r\'e\index{Poincar\'e}
deduced all the most general consequences from the principle of relativity. \textbf{There is no an  idea from
the 1905 work by A.\,Einstein\index{Einstein} which has not been present in articles by
H.\,Poin\-ca\-r\'e\index{Poincar\'e}}. The  work by A.\,Einstein\index{Einstein} is rather elementary in  realization of ideas. Though in fact the realization of ideas required high level of analysis. In
H.\,Poincar\'{e}'s\index{Poincar\'e} works [2; 3] there is not only a high level  analysis and realization, but they
contain also much new which is not contained in the article by A.\,Einstein\index{Einstein} and which has
determined further development of the theory of relativity. How Louis de Broglie\index{Broglie} has not seen all this
when reading the Poincar\'e\index{Poincar\'e} articles? Compare writings by Louis de Broglie\index{Broglie} to
writings by W.\,Pauli\index{Pauli} of 1955 (see present edition, p.~\pageref{wash7}).

It is quite evident, \textbf{that Louis de Broglie\index{Broglie} has not gained an understanding of the essence of
the problem as a matter of fact}. Though being the Director of the Henri Poincar\'e\index{Poincar\'e} Institute he
had to do so.

Being based upon opinions by Louis de Broglie\index{Broglie} 
Academician V.\,L.\,Ginzburg\index{Ginzburg} writes:
\begin{quote}
{\it\hspace*{5mm}``As we see, the position of L.\,de Broglie\index{Broglie}, referring to the memory of
H.\,Poincar\'e\index{Poincar\'e} with a deep respect and with a maximal
 kindness, should be considered as one more testimony that the main author of the SRT is A.\,Einstein''\index{Einstein}}.
\end{quote}
All this is strange. One would think everything is simple here: if your qualification admits you, then take the
article by A.\,Einstein\index{Einstein} of 1905 and the articles by H.\,Poincar\'e\index{Poincar\'e}, compare them
and  \textbf{all will be clear}. Just this will be considered in details in further Sections. What about the
quotation of  L.\,de Broglie\index{Broglie}, it clearly demonstrates 
his superficial knowledge of the works by
H.\,Poincar\'e\index{Poincar\'e}.

P.\,A.\,M.\,Dirac\index{Dirac} wrote in 1979 (Proceedings of the 1979 Einstein Centennial Symposium: Some Strangeness in the Proportion. Addison-Wesley MA 1980. P.~111.):
\begin{quote}
{\it\hspace*{5mm}``In one respect Einstein went far beyond Lorentz and 
Poincar\'{e}\index{Poincar\'e}
and the others, and that was in asserting that the 
Lorentz\index{Lorentz} transformation\index{Lorentz transformations}
would apply to the whole of physics and not merely to phenomena based on
electrodynamics. Any other physical forces that may be introduced in the
future will have to conform to 
Lorentz\index{Lorentz} transformations,\index{Lorentz transformations} 
which is going
far beyond what the people who were working with electrodynamics were
thinking about''.}
\end{quote}
But just relating to this H.\,Poincar\'e\index{Poincar\'e} wrote in
 1905-1906 in articles [2; 3]:
\begin{quote}
\textit{``\ldots All forces, despite of the nature they may have, behave according to Lorentz\index{Lorentz}
transformations\index{Lorentz transformations} (and consequently, according to translational motion) just in the
same way  as electromagnetic forces''.}
\end{quote}
Comparing the quotation from Poincar\'e\index{Poincar\'e} with the words by Dirac,\index{Dirac} it is easy to get
convinced, that all this considered by Dirac as the achievement by Einstein\index{Einstein} is contained in full
in article [2] by Poincar\'e\index{Poincar\'e}. Therefore the quoted statement by Dirac\index{Dirac}: \textit{``In
one respect Einstein\index{Einstein} went far beyond \ldots Poincar\'e''}\index{Poincar\'e} is
simply incorrect. Poincar\'e\index{Poincar\'e} was the first who extrapolated Lorentz transformations onto any
forces of nature, including gravitational ones.

The following, for example, is what Richard
Feynman\index{Feynman} 
wrote (see his book {\sf The Character of Physical law}. BBC, 1965):
\begin{quote}
{\it\hspace*{5mm}``It was Poincar\'{e}'s\index{Poincar\'e} 
suggestion to make this analysis of
what you can do to the equations and leave them alone.
It was Poincar\'{e}'s\index{Poincar\'e} attitude to pay
attention to the symmetries of physical laws''.}
\end{quote}

In 1955, in connection with the 50-th anniversary of re\-la\-ti\-vi\-ty
theory W.\,Pauli\index{Pauli} wrote:\label{wash7}
\begin{quote}
{\it\hspace*{5mm}``
Both Einstein
\index{Einstein} and Poincar\'{e},
\index{Poincar\'e} took their stand
on the preparatory work of 
H.A.\,Lorentz\index{Lorentz}, who 
had already come quite
close to  the result, 
without however quite reaching it.
In the agreement between the
results of the methods followed independently of each other by 
Einstein\index{Einstein} and
Poincar\'{e}\index{Poincar\'e} I 
discern a deeper significance
of a harmony between the mathematical
method and analysis by means of conceptual
experiments (Gedankenexperimente), which rests on general features
of physical experience''}.
\end{quote}
Compare this quotation from W.\,Pauli\index{Pauli} with words by L.\,de\,Broglie\index{Broglie} of 1954.

The articles~\cite{2,3} by Henri
Poincar\'{e}\index{Poincar\'e} 
are extremely modern
both in content and form and in the exactness of exposition. Truly,
they are pearls of theoretical physics.

Now let us return to words by Academician V.L.\,Ginzburg\index{Ginzburg} (see
this edition, p.~\pageref{wash2}), further he says about the principle of relativity\index{relativity
principle}:\label{wash3}
\begin{quote}
{\it``\ldots Moreover,
Lorentz\index{Lorentz} 
and Poincar\'e\index{Poincar\'e} 
interpreted this principle only as a statement on impossibility to register the uniform motion of a body relative
to ether''.}\index{ether}
\end{quote}

This is absolutely incorrect in relation to Poincar\'e.\index{Poincar\'e} 
Let me explain. This principle in Poincar\'e\index{Poincar\'e} formulation is as follows [1]:
\begin{quote}
{\it\hspace*{5mm}``The principle of relativity,\index{relativity principle} according to which the laws for physical
phenomena should be the same both for observer at rest and for observer in uniform motion, i.\,e. we have no any method
to determine whether we participate in such motion or not and we cannot have such a method in principle.''.}
\end{quote}
There is no term ``ether''\index{ether} in this formulation of the relativity principle. Therefore the statement by
V.L.\,Ginzburg\index{Ginzburg} is a simple misunderstanding. Let us present
some trivial explanations in this connection. It follows from the formulation of  the relativity principle
\index{relativity principle} that an observer performing a translational uniform motion can move with any constant
velocity and so there is an infinite set of equitable reference systems with the same  laws for physical
phenomena. This set of equitable reference systems includes also a system of reference taken as a system of
rest.

Then V.\,L.\,Ginzburg\index{Ginzburg} 
continues:
\begin{quote}
{``\ldots\it It is possible to go from  above to consideration of all inertial systems of 
reference\index{inertial reference systems} as completely equitable (this is the modern treatment of the relativity
principle\index{relativity principle}) without special efforts only in case {\bf \textit{if we understand Lorentz
transformations\index{Lorentz}  as transformations corresponding to transition to
the moving reference system}} (\textrm{emphasized by me.} --- {\sl A.\,L.})''.}
\end{quote}
To have in mind that Poincar\'e\index{Poincar\'e} 
has not understood that Lorentz\index{Lorentz} 
transformations\index{Lorentz transformations}
correspond to transition from the ``rest'' system of reference to the moving one is also a misunderstanding.
This trivially follows from the Lorentz\index{Lorentz} 
 transformations\index{Lorentz transformations}.

From the Lorentz\index{Lorentz} transformations\index{Lorentz transformations}
Лоренца\\[-1mm]
\[
x^\prime=\gamma(x-\varepsilon t)
\]\\[-1mm]
it follows that the origin of the new system of reference\\[-1mm]
\[
x^\prime=0,\;y^\prime=0,\;z^\prime=0
\]\\[-1mm]
moves along axis $x$ with velocity $\varepsilon$:\\[-1mm]
\[
x=\varepsilon t
\]\\[-1mm]
in relation to another system of reference. Therefore, Lorentz\index{Lorentz} transformations\index{Lorentz transformations} connect variables ($t, x, y, z$) referring to one system of
reference with variables ($t^\prime, x^\prime, y^\prime, z^\prime$) referring to another system moving uniformly
and straightforwardly with velocity  $\varepsilon$ along axis $x$ relatively to the first system. The
Lorentz\index{Lorentz}  transformations\index{Lorentz transformations} has taken
place of the Galilean\index{Galilei} transformations speaking figuratively.

Let us consider in more detail the statement by V.\,L.\,Ginzburg.\index{Ginzburg} He notes that
{``\it if one understands Lorentz\index{Lorentz} 
transformations\index{Lorentz transformations}
as transformations corresponding to transition to a moving system of reference'',} then
``\textit{it is possible without  special efforts''}
to go on to
``\textit{the treatment of all inertial systems of reference as completely equitable (this is the modern
treatment of the relativity principle\index{relativity principle})}''.

But it is not so. This is not enough for the fulfilment of requirements of the principle of
relativity\index{relativity principle}. It is necessary to prove (and this is the most important) that the
Lorentz\index{Lorentz} transformations\index{Lorentz transformations} together
with the spatial rotations form {\bf the group}. But we are obliged for this solely to Poincar\'e\index{Poincar\'e}. Only after discovering \textbf{the group}\index{group} it is possible to say that
all physical equations stay untouchable at any inertial reference system\index{inertial reference systems}. Then
all the corresponding physical characteristics transform exactly according to
 {\bf the group}.\index{group}
Just this provides the fulfilment of requirements of the relativity principle\index{relativity principle}.

In connection with the quotation  from Ginzburg\index{Ginzburg} (see this
edition, p.~\pageref{wash3}) we will give some comments. Let us admit that the principle of
relativity\index{relativity principle} is treated as a statement of impossibility to register a uniform
translational motion of a body relative to the ether.\index{ether} What follows from here?  First, from here it follows
directly that the physical equations are the same, both in the ether\index{ether} system of reference and in any other
reference system, moving with constant velocity relative to the ether\index{ether} system. The invariableness of equations is
provided by the Lorentz\index{Lorentz} 
transformations\index{Lorentz transformations}. Second, as the Lorentz\index{Lorentz} 
transformations\index{Lorentz transformations} \textbf{form a group},\index{group} it is impossible to prefer one
system of reference to another. The ether\index{ether} system of reference will be a member of this totality of equitable
inertial systems\index{inertial reference systems}. Therefore it will lose the meaning of the fixed system of
reference. But this leads to the fact that \textbf{the ether}\index{ether} in the 
Lorentz\index{Lorentz} sense \textbf{disappears}. \vspace{-1mm}

Very often in order to stress that Poincar\'e\index{Poincar\'e} has not created the theory
of relativity one cites his words:
\begin{quote}
{\it\hspace*{5mm}``The importance of this subject ought me to return to this again; the results obtained by me
are in correspondence with those of Lorentz\index{Lorentz} in all the most
important points. I only tried to modify slightly  and enlarge them.''.}
\end{quote}

One usually concludes from this that Poincar\'e\index{Poincar\'e} has exactly followed
 Lorentz\index{Lorentz} views.
But Lorentz\index{Lorentz}, as he notes himself, has not established the
relativity principle\index{relativity principle} for electrodynamics. So, one concludes that also
Poincar\'e\index{Poincar\'e} has not made this decisive step. {\bf But this is incorrect}.
Those authors who write so have not read Poincar\'{e}\index{Poincar\'e} articles [2, 3]
carefully. Let us give some more explanations. H.\,Poincar\'e\index{Poincar\'e} writes in
his article [2]:
\begin{quote}
{\it\hspace*{5mm}``The idea by Lorentz\index{Lorentz} is that
electromagnetic field equations are invariant under some 
transformations\index{Lorentz transformations}
(which I will call by name of H.A.\,Lorentz)\index{Lorentz} 
of the following form\ldots}''.
\end{quote}
Poincar\'e\index{Poincar\'e} writes: ``{\bf the idea 
by Lorentz}'',\index{Lorentz}
but Lorentz\index{Lorentz} never wrote so before
Poincar\'{e}.\index{Poincar\'e}  Here Poincar\'{e}\index{Poincar\'e} 
has formulated his own fundamental idea, but ascribed it to 
Lorentz.\index{Lorentz} 
He always appreciated and celebrated extremely high anybody who gave a stimulus to his thought, a joy
of creation, probably as nobody else. He was absolutely deprived of personal priority sentiments. But descendants
are obliged to  restore truth and pay duty to the creator.

In the same article (see this edition, the footnote on p.~\pageref{wash15}) Academician V.L.\,Ginzburg\index{Ginzburg}  writes:
\begin{quote}
{\it\hspace*{5mm}``One can suspect that Poincar\'e\index{Poincar\'e} has not
 estimated the Einstein\index{Einstein} contribution as a very substantial one, and
  maybe he even has believed that he ``has made everything himself''.
But that's just the point that we are trying to guess about the Poincar\'e\index{Poincar\'e} feelings from his silence and not from some claims told by him.\,''}.
\end{quote}
One may readily find out what Poincar\'e\index{Poincar\'e} has done in the theory of
relativity: for a theoretical physicist it is enough to read his articles [2, 3]. Therefore it is not necessary
``to guess'' about the Poincar\'e\index{Poincar\'e} feelings in order to answer the
question: what he really has done. Academician V.\,L.\,Ginzburg\index{Ginzburg} usually cites writings by
W.\,Pauli\index{Pauli} of 1921, but surprisingly does not cite writings  by W.\,Pauli\index{Pauli} of  1955. Some
people for some reason want to see only A.\,Einstein\index{Einstein} treated as the creator of special theory of
relativity.  But we should follow facts and only them.

Now let us consider words by professor Pais\index{Pais}  written in 
the same book at p.~169.
\begin{quote}
{\it``Why did Poincar\'{e}\index{Poincar\'e} 
not mention Einstein\index{Einstein} in his G\"{o}ttingen lectures? Why 
is there no paper by Poincar\'{e}\index{Poincar\'e} 
in which Einstein\index{Einstein} and relativity 
are linked? It is inconceivable that Poincar\'e\index{Poincar\'e}
would have studied  Einstein's\index{Einstein} papers of 1905  
without understanding them. It is impossible
that in 1909 (the year he spoke at G\"{o}ttingen) he 
would never have heard of 
Einstein's\index{Einstein} activities in this area. 
Shall I write of petulance or professional envy?''.}
\end{quote}
There is a unique answer to these questions. After reading the articles and books published by
Poincar\'{e}\index{Poincar\'e} up to 1905 it is easy to get convinced that there has been
nothing new for Poincar\'e in the Einstein\index{Einstein} article. Being based on his
own previous works and on Lorentz\index{Lorentz} investigations
Poincar\'e\index{Poincar\'e} formulated all the main content of the special theory of
relativity, discovered the laws of relativistic mechanics, extended 
Lorentz\index{Lorentz} 
transformations\index{Lorentz transformations} to all the forces of nature. But all this he ascribed
to  {\bf the Great destructor} H.A.\,Lorentz,\index{Lorentz} because just his
article of 1904 provided a stimulus for Poincar\'e\index{Poincar\'e} thought. This was his
usual practice. It is strange that professor Pais\index{Pais} addresses questions only
to Poincar\'e\index{Poincar\'e}, and not to Einstein.\index{Einstein}  How Einstein\index{Einstein} decided to submit his paper on electrodynamics of moving body if he knew
papers by Lorentz\index{Lorentz}  of ten years ago only and papers by
Poincar\'e\index{Poincar\'e} of five years ago only? What prevented 
Einstein\index{Einstein} from acquaintance with reviews published 
in the journal \textsf{``Beibl\"{a}tter Annalen
der Physik''}, if he himself prepared many reviews for this journal?
21 reviews by Einstein\index{Einstein} were published there in 1905.

The journal \textsf{``Beibl\"{a}tter Annalen der Physik''} was printed in 
Lei\-pzig in separate
issues. 24 issues were published in a year. The review of 
the Lorentz\index{Lorentz} article which appeared in the journal \textsf{``Versl. K. Ak. van Wet.''} (1904. \textbf{12} (8). S.
986--1009) was published in 4th issue of 1905. This review contained Lorentz transformations also.

 A review by Einstein on the article by M.\,Ponsot\index{Ponsot} from 
 the May issue of the French journal
 \textsf{``Comptes Rendus''} 1905. \textbf{140}.  S. 1176--1179 was published in the 18th issue of 1905.
The same issue (S. 1171--1173)  contains article by P.\,Langevin\index{Langevin} \textsf{``On impossibility to
register the translational motion of Earth by physical experiments''.} In this article
P.\,Langevin\index{Langevin} refers to the articles by Lorentz\index{Lorentz} of 1904 and Larmor\index{Larmor} of 1900.

Why Einstein\index{Einstein}  never refers to articles [2, 3] by
Poincar\'e\index{Poincar\'e}?
By the way, he wrote a lot of articles on the theory of relativity during the next  50 years. What
 personal qualities explain this?
How is it possible not to refer to articles, if they are published earlier and if you exploit
ideas and concepts from them?

Academicians
V.\,L.\,Ginzburg\index{Ginzburg} 
and Ya.\,B.\,Zel'dovich\index{Zel'dovich} 
wrote in 1967 (see ``Zel'dovich --- known and unknown (in the
recollections of his friends, colleagues, students).
Moscow: ``Nauka'', 1993, p.~88):
\begin{quote}
{\it\hspace*{5mm}``For example, despite how much a person would do himself, he could not pretend to have
a priority, if later it will be clear that the same result has been obtained earlier by other persons''.}
\end{quote}
This is a quite right view. We are to follow it. Ideas and results should be referred to that person who has
discovered them first.

How strange the fate happened to be, if one can say that, of the works by Henri 
Poincar\'{e},\index{Poincar\'e} {\sf ``On the dynamics of the electron''}, published in 1905-1906. These outstanding
papers by H.\,Poincar\'e\index{Poincar\'e} have become a peculiar source from which ideas
and methods were drawn and then published without references to the author. When references to these articles were
done, they always had nothing to do with the essence. All those discovered and introduced by
Poincar\'{e},\index{Poincar\'e} in articles [2; 3] can be easily found in one or another
form in articles by other authors published later.

M.\,Planck\index{Planck} wrote in article of 1906 \textsf{``The relativity principle\index{relativity principle}
and the general equations of mechanics''}:
\begin{quote}
{\it\hspace*{5mm} ``\textbf{The relativity principle\index{relativity principle}}, suggested by
Lorentz\index{Lorentz} and in more general formulation by Einstein\index{Einstein} means\ldots''}
\end{quote}
But after all this is incorrect. The relativity 
principle\index{relativity principle} was first formulated in
general form by Poincar\'{e},\index{Poincar\'e} in 1904. Then M.\,Planck\index{Planck}
derives equations of relativistic mechanics\index{equations of relativistic mechanics}, but there are no
references to the Poincar\'{e}\index{Poincar\'e} article [3], though the equations of relativistic
mechanics\index{equations of relativistic mechanics} have been derived in it earlier. If ever
M.\,Planck\index{Planck} has not been informed on 
the Poincar\'{e}\index{Poincar\'e} work
that time, he could refer to it later. But such a reference to article [3] did not appear also later. Articles by
Poincar\'{e},\index{Poincar\'e}  [2; 3] did not appear also in the German collection
devoted to the theory of relativity. How one could explain all this?

B.\,Hoffmann\index{Hoffmann} (Proceedings of the 1979 Einstein Centennial Symposium: Some Strangeness in the Proportion. Addison-Wesley MA 1980. P.~111):
\begin{quote}
{\it\hspace*{5mm} ``I am wondering whether people would have discovered the special  theory of relativity without Einstein. It is true that 
Poincar\'{e}\index{Poincar\'e} had all  the mathematics and somewhat more than Einstein had in his 1905 paper,  but in Poincar\'{e}'s\index{Poincar\'e} work there was always the implication that there was a rest system - something at rest in the ether\index{ether} --- and so you get the  impression 
that Poincar\'{e}\index{Poincar\'e} and any followers would have said, yes, if  something is moving relative to the ether, it is contracted. But, of  course, people who believe this would think that our stationary rods were  expanded, instead of contracted, and Poincar\'{e}\index{Poincar\'e} would have had one clock  going slower, but the other going faster. This reciprocity was a very
subtle point, and it is quite likely that people might never have realized
that it was a reciprocal relationship''.}
\end{quote}

All this is inaccurate or follows from misunderstanding of the SRT basics. First, the SRT
has already been discovered by Poincar\'{e}\index{Poincar\'e}  in articles [2; 3] according to the principle of
relativity\index{relativity principle} formulated by Poincar\'{e}\index{Poincar\'e} in 1904 for all physical
phenomena. In accordance with the principle of 
relativity\index{relativity principle} physical equations are the
same in all inertial reference systems\index{inertial reference systems}. All inertial reference
systems\index{inertial reference systems} are equitable, and so the existence of a rest system of reference is
excluded. From this it follows that the reversibility is realized here. Second, Poincar\'{e}\index{Poincar\'e}
discovered the Lorentz\index{Lorentz} group\index{group}
\index{group, Lorentz} and the existence of the inverse
element follows from here, consequently, the  reversibility follows from existence of the group. Third, in the SRT
constructed by Poincar\'{e}\index{Poincar\'e} really this fact --- ``the reversible nature of this connection is a
very subtle point'' --- is a trivial consequence, so  writing  ``that people would never recognize this'' is an
intention of the author to see the problem there where it is absent. Moreover, it is absurdly to ascribe his own
misunderstanding to Poincar\'{e}\index{Poincar\'e}.

It is surprising to read a quotation from A.\,Einstein\index{Einstein} given by G.\,Holton\index{Holton} (Proceedings of the 1979 Einstein Centennial Symposium: Some Strangeness in the Proportion. Addison-Wesley MA 1980. P.~111):
\begin{quote}
{\it\hspace*{5mm} ``Einstein himself said that not Poincar\'{e} or 
Lorentz\index{Lorentz}  
but Langevin\index{Langevin}  might have developed the special theory of relativity''.}
\end{quote}
If we trust
G.\,Holton\index{Holton}, then we see that  A.\,Einstein\index{Einstein}
without any doubt thought that it was exclusively he who discovered the special theory of relativity. Was it possible
that he did not read the Poincar\'{e}\index{Poincar\'e} papers [2; 3] where all the main content of the
special theory of relativity was given in the extremely definite and general form?
Therefore it is rather strange even such an appearance of this statement from  A.\,Einstein\index{Einstein}.
But if we admit that  A.\,Einstein\index{Einstein} really has not read 
Poincar\'{e}\index{Poincar\'e}
articles [2; 3] during next fifty years, then this is also surprising. How this could be connected with the
``punctilious honesty of
Einstein''\index{Einstein} as a scientist which is 
expressively described  by G.\,Holton\index{Holton}?

\textbf{The suppression of Poincar\'{e}\index{Poincar\'e} articles  [2; 3] continued all the twentieth century. The
opinion was created that the special theory of relativity is created 
by A.\,Einstein\index{Einstein} alone.} This
is written in textbooks, including those used at school, in monographs, in science popular books, in encyclopedia.
German physicists as distinct from French physicists have made a lot of efforts in order to arrange the situation
when  A.\,Einstein\index{Einstein}  alone was considered as the creator of the special theory of relativity, and
this scientific achievement as a fruit of German science. But fortunately ``manuscripts do not burn''.  Articles
[2; 3] clearly demonstrate the fundamental contribution 
by Poincar\'{e}\index{Poincar\'e} to the discovering of the
special theory of relativity. All the following done in this direction are applications and developments of his
ideas and methods.

In 1913, a collection of the works of
Lorentz,\index{Lorentz} 
Einstein\index{Einstein} and
Minkowski\index{Minkowski} 
in the special relativity theory was published in
Germany. But the fundamental works by
H.\,Poincar\'{e}\index{Poincar\'e} 
were not included
in the collection. How this could be explained?

In 1911 the French physicist Paul
Langevin\index{Langevin} 
published two ar\-ticles on the
relativity theory: {\sf ``Evolution of the concept of space and time''};
{\sf ``Time, space and causality in modern physics''}. But in these
articles
H.\,Poincar\'{e}\index{Poincar\'e} 
is not even mentioned, although they deal
with the
relativity principle,\index{relativity principle}
the
Lorentz\index{Lorentz} 
group,\index{group}\index{group, Lorentz}
space and time,
determined by the
interval.\index{interval}
In 1920 in the article by
P.\,Langevin\index{Langevin}
{\sf ``The historical de\-ve\-lo\-pment
of the relativity principle''}\index{relativity principle}
H.\,Poincar\'{e}\index{Poincar\'e} 
is also not men\-ti\-o\-ned.
How could P.\,Langevin\index{Langevin} do that?

In 1935 a collection 
{\sf ``The relativity principle''},\index{relativity principle} edited by pro\-fe\-ssors
V.\,K.\,Frederix\index{Frederix} and D.\,D.\,Ivanenko\index{Ivanenko}
was published, which for the first time contained works in the relativity theory of
Lorentz,\index{Lorentz}  Poin\-ca\-r\'{e},\index{Poincar\'e} 
Einstein\index{Einstein} and Minkowski.\index{Minkowski} 
However, the first work by H.\,Poin\-ca\-r\'{e},\index{Poincar\'e} {\sf ``On the dynamics of
the electron''} happened not to be included. And only in 1973, in the collection {\sf ``The relativity
principle''}\index{relativity principle} (with an introductory article by corresponding member of the USSR Academy
of Sciences Professor D.\,I.\,Blokhintsev;\index{Blokhintsev} the collection was
compiled by Professor A.\,A.\,Tyapkin),\index{Tyapkin} the works of\linebreak
H.\,Poincar\'{e}\index{Poincar\'e} in relativity theory were presented most completely,
which permitted many people to appreciate the decisive contribution made by Poincar\'{e}\index{Poincar\'e}
in the creation of special relativity theory. Somewhat later, Academician
V.\,A.\,Matveev\index{Matveev} and me decided to rewrite the for\-mu\-lae in
the articles by H.\,Poincar\'{e}\index{Poincar\'e} {\sf ``On the dynamics of the
electron''} in modern notations, so as to facilitate studying these articles.

In 1984, to the 130-th anniversary of H.\,Poincar\'{e}\index{Poincar\'e} his articles {\sf
``On the dynamics of the electron''} together with comments were published by the Publishing Department of the
Joint Institute for Nuclear Research (Dubna), and later, in 1987 they were published by the Publishing Department of the
M.V.\,Lomonosov Moscow State Uni\-ver\-si\-ty.

Henri Poincar\'{e}\index{Poincar\'e} is one of the most rare personalities in the
his\-to\-ry of science. A greatest mathematician, specialist in mechanics, the\-o\-re\-ti\-cal physicist; his
fundamental works have left a most bril\-li\-ant imprint in many fields of modern science. He, moreover, possessed
the rare gift of profound vision of science as a whole. In the beginning of the past century (1902--1912) several
books by Poincar\'{e}\index{Poincar\'e}  were published: {\sf ``Science and
hypothesis''}; {\sf ``The value of science''}; {\sf ``Science and method''}; {\sf ``Recent thoughts''}. Some of
them were nearly at once translated into Russian. These books are marvellous both in content and in
the free, ex\-tre\-me\-ly brilliant and illustrative manner of presentation. They have not become obsolete, and
for everyone, who studies mathematics, physics, mechanics, philosophy, it would be extremely useful to become
familiarized with them. It is quite regretful that for various reasons they were not republished for a long time.
And only owing to the persistent efforts of Academician 
L.\,S.Pontryagin\index{Pontryagin} 
 they have been republished and become available to present-day readers
 in Russia.

We also would like to note that some interesting books devoted
to various aspects and ``non-orthodoxal'' views of the history
of the relativity theory were published recently in the West [12].
\newpage
\markboth{thesection\hspace{1em}}{}
\section{The principle of stationary action in electrodynamics}

Many equations of theoretical physics are obtained from requiring the functional, termed action, to achieve an
extremum. Earlier (Section 2), the principle of least action was applied in mechanics, resulting in the
Lagrange\index{Lagrange} equations.\index{Lagrangian equations} We must in the case
of elec\-tro\-dy\-na\-mics, also, compose action so as to have its variation with respect to the fields leading to
the Maxwell-Lorentz\index{Lorentz}\index{Maxwell} 
equations.\index{Maxwell-Lorentz equations}

Action is constructed with the aid of scalars composed of functions of the field and current. We introduce tensor
of the electromagnetic field \be F_{\mu\nu}=\f{\pa A_\nu}{\pa x^\mu}-
\f{\pa A_\mu}{\pa x^\nu},
\ee which by  construction satisfies the equation \be \f{\pa F_{\mu\nu}}{\pa x^\sigma}+ \f{\pa F_{\nu\sigma}}{\pa
x^\mu}+
\f{\pa F_{\sigma\mu}}{\pa x^\nu}=0,
\ee that is equivalent to the Maxwell-Lorentz\index{Lorentz}\index{Maxwell}
equations (8.26).\index{Maxwell-Lorentz equations} We  need  further the two  simplest
invariants only\be
A_\nu S^\nu,\;F_{\lambda\nu}F^{\lambda\nu}.
\ee
Here $S^\nu$ is the
four-vector\index{four-vector}
of current (8.9).

The sought action will have the form
\be
S=\f{1}{c}\int Ld\Omega,
\ee
$L$ is the density of the
Lagrangian\index{Lagrange} 
function,\index{Lagrangian density}
equal to
\be
L=-\f{\,1\,}{\,c\,}A_\nu S^\nu -\f{1}{16\pi}F_{\lambda\sigma}
F^{\lambda\sigma},\; d\Omega=dx^0 dx^1 dx^2 dx^3.
\ee In seeking for the field equations we shall only vary the field po\-ten\-ti\-als in the action functional,
considering the field sources $S^\nu$ as given. \markboth{thesection\hspace{1em}}{10. The principle of stationary action \ldots}

Then
\be
\delta S=-\f{\,1\,}{\,c\,}\int\left[\f{\,1\,}{\,c\,}S^\nu\delta A_\nu+
\f{1}{8\pi}F^{\lambda\sigma}\delta F_{\lambda\sigma}
\right]d\Omega=0.
\ee
Since the variations commute with differentiation, we obtain
\be
F^{\lambda\sigma}\left(\f{\pa}{\pa x^\lambda}\delta A_\sigma-
\f{\pa}{\pa x^\sigma}\delta A_\lambda
\right)=-2F^{\sigma\lambda}\f{\pa}{\pa x^\lambda}\delta A_\sigma.
\ee
Substituting (10.7) into (10.6) we find
\be
\delta S=-\f{\,1\,}{\,c\,}\int\left[\f{\,1\,}{\,c\,}S^\nu \delta A_\nu-
\f{1}{4\pi}F^{\sigma\lambda}\f{\pa}
{\pa x^\lambda}\delta A_\sigma \right]d\Omega=0.
\ee
Integrating in the second term by parts and taking into account
that the variations of
potentials\index{potential}
at the initial and final moments
of time are zero, while the field vanishes at infinity, we obtain
\be
\delta S=-\f{\,1\,}{\,c\,}\int\left[\f{\,1\,}{\,c\,}S^\sigma+
\f{1}{4\pi}\cdot\f{\pa F^{\sigma\lambda}}
{\pa x^\lambda}\right]\delta A_\sigma d\Omega=0.
\ee
Hence, owing to the arbitrariness of $\delta A_\sigma$, we find
\be
\f{\pa F^{\sigma\lambda}}{\pa x^\lambda}=
-\f{4\pi}{c}S^\sigma.
\ee
Thus, our choice of density of the
Lagrangian\index{Lagrange} 
function (10.5)
is justified, since we have obtained exactly the second pair of
Maxwell-Lorentz\index{Lorentz}
\index{Maxwell}
equations\index{Maxwell-Lorentz equations}
\be
\rot \vec H=\f{4\pi}{c}\vec S+
\f{\,1\,}{\,c\,}\cdot\f{\pa \vec E}{\pa t},\;
{\rm div}\vec E=4\pi\rho.
\ee One must bear in mind, that the choice of density\index{Lagrangian density} of the 
Lagrangian\index{Lagrange} function in the action functional is not unambiguous, however, it is readily
verified that adding to the density of the Lagrangian\index{Lagrange} function an
additional term in the form of the four-dimensional divergence of a vector does not influence the form of the
field equations. The Maxwell-Lorentz\index{Lorentz}\index{Maxwell} 
equations (10.2), (10.10)\index{Maxwell-Lorentz equations} are invariant with res\-pect to
gauge transformations of the potentials,\index{potential} \be A_\sigma^\prime=A_\sigma+\f
{\pa f}{\pa x^\sigma},
\ee
here $f$ is an arbitrary function.

The density of the
Lagrangian (10.5)\index{Lagrangian density}
we have constructed is not
invariant under transformations (10.12). On the basis of the
conservation law of current\index{conservation law of current}
 $S^\nu$ (8.10), it only varies by a divergence,
\be
L^\prime=L-\f{\,1\,}{\,c\,}\cdot
\f{\pa}{\pa x^\nu}(fS^\nu),
\ee
which has not effect on the field equations.

From the point of view of classical electrodynamics the po\-ten\-ti\-al $A^\nu$ has no physical sense, since only
the Lorentz\index{Lorentz} force\index{Lorentz force} acts on the charge, and it
is expressed via the field strength $\vec E,\,\vec H$. However, in quantum mechanics this is no longer so. It
turns out to be that the vector-potential\index{vector-potential} does act on the electron in a certain situation.
This is the Aharonov-Bohm\index{Aharonov}\index{Bohm} effect. It
was observed in 1960. The experiment was carried out as follows: a long narrow solenoid was used, the magnetic
field outside the solenoid was zero, nevertheless, the motion of electrons outside the solenoid was influenced.
The effect is explained by the solenoid violating the simple con\-nec\-ted\-ness of space-time, which gave rise to
the influence of the potential $A^\nu$,\index{potential} as it should be  in quantum gauge theory.

We shall now find the equations of motion for charged particles in an electromagnetic field. To obtain them it is
necessary to com\-po\-se an action with a part related to the particles and, also, the already known to us part
containing the field interaction with particles. Since for a particle having charge \( e \) the following equations
are valid  \be \rho=e\delta (\vec r-\vec r_0),\;
j^i=e\f{dx^i}{dt}\delta (\vec r-\vec r_0),
\ee
we have
\be
-\f{1}{c^2}\int S^\nu A_\nu d\Omega =
-\f{\,e\,}{\,c\,}\int A_\nu dx^\nu.
\ee
The action for a particle in an electromagnetic field is
\be
S=-mc\int d\sigma -\f{\,e\,}{\,c\,}\int A_\nu dx^\nu.
\ee
Varying over the particle coordinates, we obtain
\be
\!\!\!\delta S=
-\ds\int \left(mcU_\nu d\delta x^\nu+
\f{\,e\,}{\,c\,}A_\nu d\delta x^\nu +
\f{\,e\,}{\,c\,}\delta A_\nu dx^\nu\right)=0.
\ee
Integrating by parts in the first two terms and setting the variations
of coordinates to zero at the ends, we obtain
\be
\delta S=\int \left(mcdU_\nu \delta x^\nu+
\f{\,e\,}{\,c\,}dA_\nu \delta x^\nu-
\f{\,e\,}{\,c\,}\delta A_\nu dx^\nu\right)=0.
\ee
With account of the obvious relations
\be
dA_\nu=\f{\pa A_\nu}{\pa x^\lambda}dx^\lambda,\;
\delta A_\nu=\f{\pa A_\nu}{\pa x^\lambda}\delta x^\lambda,
\ee
expression (10.18) assumes the form
\be
\!\!\delta S=\int \left[mc\f{dU_\nu}{d\sigma}-
\f{\,e\,}{\,c\,}\left(\f{\pa A_\lambda}{\pa x^\nu}-
\f{\pa A_\nu}{\pa x^\lambda}\right)U^\lambda
\right]d\sigma\delta x^\nu=0,
\ee hence, due to arbitrariness of variation $\delta x^\nu$ being arbitrary, we have \be mc^2\f{dU_\nu}{d\sigma}=
eF_{\nu\lambda}U^\lambda,
\ee
or
\be
mc^2\f{dU^\nu}{d\sigma}=eF^{\nu\lambda}U_\lambda.
\ee
In three-dimensional form (10.22) assumes the form
\be
mc\f{d}{dt}\f{1}{\sqrt{\ds 1-\f{v^2}{c^2}}}=
\f{\,e\,}{\,c\,}\,(\vec v\,\vec E)\,,
\ee
\be
\f{d}{dt}\left(\f{m\vec v}{\sqrt{\ds 1-\f{v^2}{c^2}}}\right)=
e\vec E+\f{\,e\,}{\,c\,}\,[\vec v\,,\vec H\,].
\ee
Let us calculate the energy loss for an electron moving with acceleration. In case of electron velocity small in
comparison to the velocity of light the radiation energy loss is given by the following formula due to
Larmor\index{Larmor}\index{Larmor formula}:
\be
-\f{\pa E}{\pa t}=\f{2e^2}{3c^3}\left(\f{d\vec v}{dt}\right)^2.
\ee
In the  system of reference where the electron is at rest this formula takes the form
\be
-\f{\pa E}{\pa t}=\f{2e^2}{3c^3}\left(\f{d\vec v}{dt}\right)_0^2,
\ee
where acceleration is calculated in the given system of reference.

In the given system of reference the total momentum radiated is zero due to the symmetry of radiation:
\be
-\f{\pa p^i}{\pa t}=0.
\ee In order to find the formula for radiation energy loss of a charge with high velocity it is necessary to apply
the Lorentz\index{Lorentz} group\index{group}\index{group, Lorentz}, according to which it is easy to make the
transition  from one system of reference to another. To do so we consider the acceleration
four-vector\index{four-vector}, which is as follows according to (9.5) \be
a^\nu=c^2\f{dU^\nu}{d\sigma}=c\f{dU^\nu}{d\tau}.
\ee By means of this relation and also formulae (9.3) we get \be a^0=\gamma^4\left(\f{\vec v}{c}\f{d\vec v}{dt}
\right),\quad \vec a=\gamma^2\f{d\vec v}{dt}+\vec v\f{\gamma^4}{c^2}
\left(\vec v\f{d\vec v}{dt}\right). 
\ee
Using (10.29) we find  invariant
\be
(a^0)^2-(\vec a)^2=-\gamma^6\Bigl\{\Bigl(\f{d\vec v}{dt}\Bigr)^2
-\left[\f{\vec v}{c},\,\f{d\vec v}{dt}\right]^2\Bigr\}. 
\ee
In the rest system of reference we have
\be
\gamma^6\Bigl\{\Bigl(\f{d\vec v}{dt}\Bigr)^2
-\left[\f{\,\vec v\,}{\,c\,},\,\f{d\vec v}{dt}\right]^2\Bigr\}
=\Bigl(\f{d\vec v}{dt}\Bigr)_0^2. 
\ee Let us now write formulae (10.26) and (10.27) in the covariant form \be
-\f{\pa p^\nu}{\pa \tau}=\f{2e^2}{3c^4}\left(\f{d\vec v}{dt}\right)_0^2U^\nu.
\ee
Substituting now (10.31) into this relation we obtain
\be
-\f{\pa E}{\pa t}=\f{2e^2}{3c^3}\gamma^6\Bigl\{\Bigl(\f{d\vec v}{dt}\Bigr)^2
-\left[\f{\,\vec v\,}{\,c\,},\,\f{d\vec v}{dt}\right]^2\Bigr\},
\ee
\be
-\f{\pa \vec p}{\pa t}=\f{2e^2}{3c^5}\vec v\gamma^6\Bigl\{\Bigl(\f{d\vec v}{dt}\Bigr)^2
-\left[\f{\,\vec v\,}{\,c\,},\,\f{d\vec v}{dt}\right]^2\Bigr\}.
\ee
Formula (10.33) has been derived first by Li\'enard\index{Li\'enard} in 1898.

The equations of motion (10.22) in an external electromagnetic field do not account for the reaction  of
radiation. Therefore, these equations are valid only for the motion of a charged particle in weak fields. In 1938
Dirac\index{Dirac} took into account the reaction 
forces,\index{forces of reaction} and this led to the equation 
\ba
&&mc^2\ds\f{dU^\nu}{d\sigma}=eF^{\nu\lambda}U_\lambda+\nonumber\\
\label{10.35}\\
&&+\ds\f{\,2\,}{\,3\,}e^2\left[\ds\f{d^2U^\nu}{d\sigma^2}
+U^\nu\cdot\left(\ds\f{dU_\mu}{d\sigma}\cdot
\ds\f{dU^\mu}{d\sigma}\right)\right],\nonumber
\ea
called the
Dirac-Lorentz\index{Dirac} 
\index{Lorentz} 
equation.\index{Dirac-Lorentz equation}

Let us apply these formulae to motion of an ultra-relativistic charge with mass $m$ in a strong constant uniform
magnetic field $H$. We admit that the circular charge motion is determined by the Lorentz\index{Lorentz}
force\index{Lorentz force} only. So we neglect by influence of the force of
reaction\index{forces of reaction} on the motion. Let us write equations
 (10.35) in the form of Eqs.~(9.12), (9.13):
\be
\f{d}{dt}(m\gamma \vec v)
=\f{\,e\,}{\,c\,}\cdot\left[\vec v,\vec H\right]+\vec f_R,
\ee
\be
\vec f_R=\f{2e^2}{3\gamma}\cdot
\left[\f{d^2\vec U}{d\sigma^2}+\vec U\cdot
\left(\f{dU_\nu}{d\sigma}\cdot\f{dU^\nu}{d\sigma}\right)\right]\,,
\ee
\be
\f{dE}{dt}=\f{2e^2c}{3\gamma^2}
\left[\vec U\cdot\f{d^2\vec U}{d\sigma^2}
+U^2\cdot\left(\f{dU_\nu}{d\sigma}\cdot\f{dU^\nu}{d\sigma}\right)\right]\,,
\ee
where $E$ is the energy of the particle.

As in our approximation the equations of motion are the following \be mc\cdot\f{d\vec U}{d\sigma}
=\f{\,e\,}{\,c\,}\cdot\left[\vec U,\,\vec H\right]\,,\quad
\left(\vec U\cdot\vec H\right)=0,
\ee
it follows from here that
\be
U^2\cdot\left(\f{d\vec U}{d\sigma}\right)^2
=\left(\f{eH}{mc^2}\right)^2\cdot U^4,
\ee
\be
\vec U\cdot\f{d^2\vec U}{d\sigma^2}
=-\left(\f{eH}{mc^2}\right)^2\cdot U^2,
\ee where $U$ is the length of vector $\vec U$. For ultra-relativistic particles \be
U\simeq\f{E}{mc^2}.
\ee As $U^2\gg1$ we can neglect first term  (10.41) in comparison to second one (10.40) in expression (10.38). In
our approximation we  can also neglect by the following term \be
U^2\cdot\left(\f{dU^0}{d\sigma}\right)^2\,,
\ee in second term (10.38) due to its smallness  in comparison with (10.40). Expression (10.38) after taking into
account  (10.40) and (10.42) is as follows \be -\f{dE}{dt}=\f{\,2\,}{\,3\,}\cdot
\f{e^4H^2E^2}{m^4c^7}\,.
\ee 
With regard to the fact that  for the motion of a charge over circle of radius $R$ the following equation
takes place \be
H=\f{E}{eR}\,;
\ee
we can rewrite formula (10.44) in the following form
\be
-\f{dE}{dt}=\f{2e^2c}{3R^2}\cdot\left(\f{E}{mc^2}\right)^4\,.
\ee If the energy of electrons and the value of the magnetic field are large enough, then energy losses for
synchrotron radiation become rather substantial. Synchrotron radiation is widely used in biology and medicine, in
production of integral schemes an so on. Special storage rings for generation of the intense X-rays are
constructed
 (see more details in:
Ya.\,P.\,Terletsky,\index{Terletsky}  Yu.\,P.\,Rybakov\index{Rybakov} 
{\sf ``Electrodynamics''}. Moscow: ``Vysshaja Shkola'', 1980 (in Russian)).

\newpage
\markboth{thesection\hspace{1em}}{}
\section{Inertial motion of a test body. Covariant
differentiation}

In an arbitrary reference system the
interval\index{interval}
is known to have
the form
\be
d\sigma^2=\gamma_{\mu\nu}(x)dx^\mu dx^\nu,\;
\det(\gamma_{\mu\nu})=\gamma<0.
\ee The pseudo-Euclidean metric $\gamma_{\mu\nu}$ is determined by expression (3.33). Precisely for this metric
the Riemannian\index{Riemann} 
curvature\index{Riemannian curvature tensor} tensor is zero. The action for a free moving point-like body of mass $m$ has the form \be
S=-mc\int d\sigma.
\ee
Owing to the principle of stationary action, we have
\be
\delta S=-mc\int \delta (d\sigma)=0,
\ee
\ba
&&\delta(d\sigma^2)=2d\sigma\delta(d\sigma)=\delta
(\gamma_{\mu\nu}(x)dx^\mu dx^\nu)=\nonumber\\
\label{11.4}\\
&&=\ds\f{\pa \gamma_{\mu\nu}}{\pa x^\lambda}
\delta x^\lambda dx^\mu dx^\nu+2\gamma_{\mu\nu}
dx^\mu \delta(dx^\nu).\nonumber
\ea
Since
\be
\delta (dx^\nu)=d(\delta x^\nu),
\ee
from expression (11.4) we find
\be
\delta (d\sigma)=\f{\,1\,}{\,2\,}\cdot\f
{\pa \gamma_{\mu\nu}}{\pa x^\lambda}
U^\mu dx^\nu \delta x^\lambda+
\gamma_{\mu\nu}U^\mu d(\delta x^\nu),
\ee
here
\be
U^\mu=\f{dx^\mu}{d\sigma}.
\ee
Substituting (11.6) into (11.3) we obtain
\be
\!\!\!\!\!\!\!\!\!\delta S=
-\ds mc\int \left[\f{\,1\,}{\,2\,}\cdot
\f{\pa \gamma_{\mu\nu}}{\pa x^\lambda}
U^\mu U^\nu \delta x^\lambda+
\gamma_{\mu\nu}U^\mu\f
{d(\delta x^\nu)}{d\sigma}\right]d\sigma=0.
\ee
\markboth{thesection\hspace{1em}}{11. Inertial motion of a test body.
Covariant
differentiation} Since \be \gamma_{\mu\nu}U^\mu
\f{d(\delta x^\nu)}{d\sigma}= \f{d}{d\sigma}(\gamma_{\mu\nu}U^\mu \delta x^\nu)- \delta x^\nu \f{d}{d\sigma}
(\gamma_{\mu\nu}U^\mu),
\ee
then, with account of the variations at the boundary of the
region being zero, we find
\ba
&&\!\!\!\!\!\!\!\delta S=\ds -mc\int \Biggl[\f{\,1\,}{\,2\,}\cdot\f
{\pa \gamma_{\mu\nu}}{\pa x^\lambda}
U^\mu U^\nu -\gamma_{\mu\lambda}\f
{dU^\mu}{d\sigma}-\Biggr.\nonumber\\
\label{11.10}\\
&&\!\!\!\!\!\!\!\ds -\f {\pa \gamma_{\mu\lambda}}{\pa x^\alpha} U^\mu U^\alpha\Biggl.\Biggr]d\sigma \delta
x^\lambda =0.\nonumber \ea We represent the last term in (11.10) as \be \f{\pa \gamma_{\mu\lambda}}{\pa x^\alpha}
U^\mu U^\alpha=\f{\,1\,}{\,2\,}\left(\f {\pa \gamma_{\mu\lambda}}{\pa x^\alpha}+\f {\pa \gamma_{\alpha\lambda}}{\pa
x^\mu}\right)
U^\mu U^\alpha.
\ee
With account of (11.11), expression (11.10) assumes the form
\ba
&&\ds\int \Biggl[\f{\,1\,}{\,2\,}\left(\f
{\pa \gamma_{\mu\lambda}}{\pa x^\alpha}+\f
{\pa \gamma_{\alpha\lambda}}{\pa x^\mu}-
\f{\pa \gamma_{\mu\alpha}}{\pa x^\lambda}\right)\Biggr.
U^\mu U^\alpha+\nonumber\\
\label{11.12}\\
&&\ds +\gamma_{\mu\lambda}\f
{dU^\mu}{d\sigma}\Biggl.\Biggr]d\sigma\delta x^\lambda=0.\nonumber
\ea
Since the factors $\delta x^\lambda$ are arbitrary, we find
\be
\gamma_{\mu\lambda}\f{dU^\mu}{d\sigma}+\f{\,1\,}{\,2\,}\left(\f
{\pa \gamma_{\mu\lambda}}{\pa x^\alpha}+\f
{\pa \gamma_{\alpha\lambda}}{\pa x^\mu}-
\f{\pa \gamma_{\mu\alpha}}{\pa x^\lambda}\right)
U^\mu U^\alpha=0.
\ee
Multiplying (11.13) by $\gamma^{\lambda\nu}$, we obtain
\be
\f{dU^\nu}{d\sigma}+{\mit\Gamma}_{\mu\alpha}^\nu U^\mu U^\alpha=0,
\ee
here ${\mit\Gamma}_{\mu\alpha}^\nu$ is the
Christoffel\index{Christoffel} 
symbol\index{Christoffel symbol}
\be
{\mit\Gamma}_{\mu\alpha}^\nu=\f{\,1\,}{\,2\,}\gamma^{\nu\lambda}
(\pa_\alpha \gamma_{\mu\lambda}+\pa_\mu \gamma_{\alpha\lambda}-
\pa_\lambda \gamma_{\mu\alpha}).
\ee We see that inertial motion of any test body, independently of its mass, proceeds along the geodesic
line,\index{geodesic line} determined by equa\-tion (11.14). It is absolutely evident that in arbitrary
coordinates the geodesic lines\index{geodesic line} could not be treated as  direct lines, this is  confirmed by
nonlinear dependence of spatial coordinates $x^i(i=1,2,\break 3)$ on time variable $x^0$. Motion along a geodesic line
(11.14)\index{geodesic line} in Min\-kow\-ski\index{Minkowski} 
space\index{Minkowski space} is a free motion. {\bf Thus, forces of inertia cannot cause any deformation by themselves. Under their
influence free mo\-tion takes place. The situation changes, when there are forces of reaction,\index{forces of reaction} which counteract the forces of inertia. In this case deformation is unavoidable}. In weightlessness, in
a satellite, de\-for\-mation does not exist, because, owing to the gravitational field being homogeneous, in each
element of the volume of a body com\-pen\-sation of the force of 
gravity\index{forces of gravity} by the forces of
inertia\index{forces of inertia} takes place. The forces of 
gravity\index{forces of gravity} and the forces of
inertia are volume forces.

Physical forces are four-vectors\index{four-vector} in Minkowski\index{Minkowski} space.\index{Minkowski space} But the forces of inertia are not, since they can be rendered equal to zero by
transition to an inertial reference system in Minkowski\index{Minkowski} space.\index{Minkowski space}

Now we shall dwell upon the issue of covariant differentiation.
In Cartesian coordinates $x^\lambda$ ordinary differentiation, for
example, of a vector $A^\nu$ results in a tensor quantity
\be
\f{\pa A^\nu}{\pa x^\lambda}=
B_\lambda^\nu
\ee
with respect to linear transformations. In arbitrary coordinates
$y^\lambda$ this property is not conserved, and, therefore, the
quantity $\pa A^\nu/\pa y^\lambda$ will
no longer be a tensor.

It is necessary to introduce the
covariant derivative,\index{covariant derivative}
which
will provide for differentiation of a tensor yielding a tensor, again.
This will permit us to easily render covariant any physical equations.
{\bf Covariance is not a physical, but a mathematical, requirement}.

Earlier (see 6.13) we saw that of two vectors $A^\nu, B_\nu$ it is
possible to construct an invariant
\be
A^\nu (x)B_\nu (x).
\ee
We shall consider an invariant of a particular form
\be
A_\lambda (x)U^\lambda (x),
\ee
where
\be
U^\lambda=\f{dx^\lambda}{d\sigma},
\ee
fulfils Eq. (11.14).

Differentiating (11.18) with respect to $d\sigma$, we also obtain
an invariant (a scalar)
\[
\f{d}{d\sigma}(A_\lambda U^\lambda)=
\f{dA_\lambda}{d\sigma}U^\lambda+
A_\nu\f{dU^\nu}{d\sigma}.
\]
Substituting expression (11.14) into the right-hand part,
we find
\[
\f{d}{d\sigma}(A_\lambda U^\lambda)=
\f{\pa A_\lambda}{\pa x^\alpha}U^\alpha U^\lambda-
{\mit\Gamma}_{\alpha\lambda}^\nu U^\alpha U^\lambda A_\nu, \textrm{i.\,e.}
\]
\be
\f{d}{d\sigma}(A_\lambda U^\lambda)=
\left(\f{\pa A_\lambda}{\pa x^\alpha}-
{\mit\Gamma}_{\alpha\lambda}^\nu A_\nu\right)U^\alpha U^\lambda.
\ee
Since (11.20) is an invariant, $U^\lambda$ is a vector, hence it
follows that the quantity
\[
\f{\pa A_\lambda}{\pa x^\alpha}-
{\mit\Gamma}_{\alpha\lambda}^\nu A_\nu
\]
{\bf is a
covariant tensor\index{covariant tensor}
of the second
rank}~{\mathversion{bold}\( A_{\lambda ;\alpha} \)}
\be
A_{\lambda ;\alpha}=\f{DA_\lambda}{\pa x^\alpha}=
\f{\pa A_\lambda}{\pa x^\alpha}-
{\mit\Gamma}_{\alpha\lambda}^\nu A_\nu.
\ee Here and further the semicolon denotes covariant differentiation.

Thus, we have defined the
covariant derivative\index{covariant derivative}
of the covariant
vector $A_\nu$. Now, we shall define the
covariant derivative\index{covariant derivative}
of
the contravariant vector $A^\nu$.\index{contravariant vector}
 To this end we write the same
invariant as
\[
\begin{array}{l}
\ds\f{d}{d\sigma}(A^\mu U^\nu \gamma_{\mu\nu})=
\f{\pa A^\mu}{\pa x^\lambda}U^\nu U^\lambda\gamma_{\mu\nu}+
\\[5mm]
\ds +A^\mu \gamma_{\mu\nu}\f{dU^\nu}{d\sigma}+
A^\mu U^\nu U^\lambda \f{\pa \gamma_{\mu\nu}}{\pa x^\lambda}.
\end{array}
\]

Substituting expression (11.14) into
the right-hand part, we ob\-tain
\ba
&&\!\!\!\!\!\!\ds\f{d}{d\sigma}(A_\nu U^\nu)=\nonumber\\
\label{11.22}\\
&&\!\!\!\!\!\!=\ds\left[\gamma_{\mu\nu}\f{\pa A^\mu}{\pa x^\lambda}-
A^\mu \gamma_{\mu\alpha}{\mit\Gamma}_{\nu\lambda}^\alpha+
A^\mu\f{\pa \gamma_{\mu\nu}}{\pa x^\lambda}\right]
U^\nu U^\lambda.\nonumber
\ea
Taking into account definition (11.15) we find
\be
A^\mu \pa_\lambda \gamma_{\mu\nu}-
A^\mu \gamma_{\mu\alpha}{\mit\Gamma}_{\nu\lambda}^\alpha=
\f{\,1\,}{\,2\,}A^\mu (\pa_\lambda \gamma_{\mu\nu}+
\pa_\mu \gamma_{\nu\lambda}-
\pa_\nu \gamma_{\mu\lambda}).
\ee
Substituting this expression into (11.22), and applying expression
$U_\alpha \gamma^{\alpha\nu}$, instead of $U^\nu$, we obtain
\be
\f{d}{d\sigma}(A_\nu U^\nu)=
\left[\f{\pa A^\alpha}{\pa x^\lambda}+
{\mit\Gamma}_{\mu\lambda}^\alpha A^\mu\right]
U^\lambda U_\alpha.
\ee
Since (11.24) is an invariant (a scalar), and  $U^\nu$ is
a vector, from (11.24) it follows that the first factor in
the right-hand part is a tensor.

Hence it follows, \textbf{that the
covariant derivative\index{covariant derivative}
of the
contravariant vector}
{\mathversion{bold}\( A^\nu \)}
\textbf{is}
\be
A_{;\lambda}^\alpha=\f{DA^\alpha}{\pa x^\lambda}=
\f{\pa A^\alpha}{\pa x^\lambda}+
{\mit\Gamma}_{\mu\lambda}^\alpha A^\mu.
\ee
Making use of formulae (11.21) and (11.25) one can also obtain
covariant derivatives\index{covariant derivative}
of a tensor of the second rank.
\be
A_{\mu\nu;\lambda}=\f{\pa A_{\mu\nu}}{\pa x^\lambda}-
{\mit\Gamma}_{\lambda\mu}^\alpha A_{\alpha\nu}-
{\mit\Gamma}_{\lambda\nu}^\alpha A_{\mu\alpha},
\ee
\be
A_{;\lambda}^{\mu\nu}=\f{\pa A^{\mu\nu}}{\pa x^\lambda}+
{\mit\Gamma}_{\lambda\alpha}^\mu A^{\alpha\nu}+
{\mit\Gamma}_{\lambda\alpha}^\nu A^{\mu\alpha},
\ee
\be
A_{\rho;\sigma}^{\nu}=\f{\pa A_\rho^\nu}{\pa x^\sigma}-
{\mit\Gamma}_{\rho\sigma}^\lambda A_\lambda^\nu+
{\mit\Gamma}_{\sigma\lambda}^\nu A_\rho^\lambda.
\ee
We see, that the rules established for (11.21) and (11.25) are
applied independently for each index of the tensor. Precisely
in this way, one can obtain the
covariant derivative\index{covariant derivative}
of a
tensor of any rank.

With the aid of expression (11.26) it is easy to show that
the covariant derivative\index{covariant derivative}
of a metric tensor\index{metric tensor of space}
is zero,
\be
\gamma_{\mu\nu;\rho}\equiv 0.
\ee
Applying the technique of covariant differentiation, one can
readily write the
equations of relativistic\index{Poincar\'{e}'s equations of mechanics} mechanics and of
electrodynamics in arbitrary coordinates of
Minkowski\index{Minkowski} 
space.\index{Minkowski space}

Thus, substituting the
covariant derivative\index{covariant derivative}
for the ordinary
one in (9.5) we find the equation of
relativistic\index{equations of relativistic mechanics}
mechanics in
arbitrary coordinates
\be
mc^2\f{DU^\nu}{d\sigma}=mc^2\left(\f{dU^\nu}{d\sigma}+
{\mit\Gamma}_{\mu\lambda}^\nu U^\mu U^\lambda\right)=
F^\nu,
\ee
here
\be
{\mit\Gamma}_{\mu\lambda}^\nu=\f{\,1\,}{\,2\,}\gamma^{\nu\rho}
(\pa_\mu \gamma_{\lambda\rho}+
\pa_\lambda \gamma_{\mu\rho}-
\pa_\rho \gamma_{\mu\lambda}).
\ee
In a similar way, it is also possible to write the
Maxwell-Lorentz\index{Lorentz} 
\index{Maxwell} 
equations\index{Maxwell-Lorentz equations}
in arbitrary coordinates. To this end,
it is necessary to substitute
covariant derivatives\index{covariant derivative}
for ordinary
derivatives in equations (8.24) and (8.27),
\be
D_\sigma F_{\mu\nu}+D_\mu F_{\nu\sigma}+
D_\nu F_{\sigma\mu}=0,
\ee
\be
D_\nu F^{\mu\nu}=-\f{4\pi}{c}S^\mu.
\ee
One can readily verify, that the following equalities hold valid:
\be
F_{\mu\nu}=D_\mu A_\nu-D_\nu A_\mu=
\pa_\mu A_\nu -\pa_\nu A_\mu,
\ee
\be
D_\sigma F_{\mu\nu}+D_\mu F_{\nu\sigma}+
D_\nu F_{\sigma\mu}=\pa_\sigma F_{\mu\nu}+
\pa_\mu F_{\nu\sigma}+\pa_\nu F_{\sigma\mu}.
\ee
On the basis of (11.31) we find
\be
{\mit\Gamma}_{\mu\nu}^\nu=\f{\,1\,}{\,2\,}\gamma^{\nu\rho}
\pa_\mu \gamma_{\nu\rho}.
\ee
But, since the following equalities hold valid:
\be
\f{\,1\,}{\,\gamma\,}\cdot\f{\pa \gamma}{\pa x^\mu}=
\gamma^{\nu\rho}\pa_\mu \gamma_{\rho\nu},\;
\f{\pa \sqrt{-\gamma}}{\pa \gamma_{\mu\nu}}=
\f{\,1\,}{\,2\,}\sqrt{-\gamma}\,\gamma^{\mu\nu},
\ee
[here $\gamma=\det (\gamma_{\mu\nu})<0$], we obtain
\be
{\mit\Gamma}_{\mu\nu}^\nu=\f{1}{2\gamma}\cdot
\f{\pa \gamma}{\pa x^\mu}=
\pa_\mu \ln \sqrt{-\gamma}\,.
\ee
Making use of (11.27), we find
\be
D_\nu F^{\mu\nu}=\pa_\nu F^{\mu\nu}+
{\mit\Gamma}_{\nu\alpha}^\mu F^{\alpha\nu}+
{\mit\Gamma}_{\alpha\nu}^\nu F^{\mu\alpha}.
\ee
The second term in (11.39) equals zero, owing to the
tensor $F^{\alpha\nu}$ being antisymmetric. On the basis
of (11.38), expression (11.39) can be written as
\be
D_\nu (\sqrt{-\gamma}\, F^{\mu\nu})=\pa_\nu (\sqrt{-\gamma}\,
F^{\mu\nu}).
\ee
Thus, equation (11.33) assumes the form
\be
\f{1}{\sqrt{-\gamma}}\pa_\nu
(\sqrt{-\gamma}\,F^{\mu\nu})=-\f{4\pi}{c}S^\mu.
\ee

The equations of motion of charged particles can be obtained by
substituting
covariant derivatives\index{covariant derivative}
for the ordinary derivatives
in (10.22)
\be
mc^2\f{DU^\nu}{d\sigma}=
eF^{\nu\lambda}U_\lambda.
\ee

Thus, we have established that transition in
Minkowski\index{Minkowski} 
space\index{Minkowski space}
from Galilean\index{Galilei} 
coordinates\index{Galilean (Cartesian) coordinates}
in an inertial reference system to
arbitrary coordinates is a simple mathematical procedure, if
covariant dif\-fe\-ren\-ti\-ation has been defined.

{\bf The property of covariance of the equations has nothing
to do with the
relativity principle.\index{relativity principle}
This has long ago been
clarified by
V.\,A.\,Fock}~\cite{13}.\index{Fock} 

\textbf{Therefore, no ``general
relativity prin\-ci\-ple'',\index{relativity principle}
as a physical prin\-ci\-ple, exists}.

\newpage
\markboth{thesection\hspace{1em}}{}
\section{Relativistic motion with constant acceleration.
The clock paradox. Sagnac effect}

Relativistic motion with constant acceleration is a motion under the influence of a force $\vec f$, that is
constant in value and direction. Ac\-cor\-ding to (9.12) we have \be \f{d}{dt}\left(\f{\vec
v}{\sqrt{1-\ds\f{v^2}{c^2}}}\right)=
\f{\vec f}{m}=\vec a.
\ee
Integrating equation (12.1) over time, we obtain
\be
\f{\vec v}{\sqrt{1-\ds\f{v^2}{c^2}}}=\vec at+\vec v_0.
\ee
Setting the constant $\vec v_0$ to zero, which corresponds to
 zero initial velocity, we find after squaring
\be
\f{1}{1-\ds\f{v^2}{c^2}}=1+\f{a^2t^2}{c^2}.
\ee
Taking into account this expression in (12.2), we obtain
\be
\vec v=\f{d\vec r}{dt}=\f
{\vec a t}{\sqrt{1+\ds\f{a^2t^2}{c^2}}}.
\ee
Integrating this equation, we find
\be
\vec r=\vec r_0+\f{\vec a c^2}{a^2}
\left[\sqrt{1+\f{a^2t^2}{c^2}}-1\right].
\ee
Since the interval\index{interval}
$ds$ is
\be
ds=cdt\sqrt{1-\f{v^2}{c^2}},
\ee the proper time\index{proper time} $d\tau$ for a moving test body is \be d\tau=\f{ds}{c}=dt
\sqrt{1-\f{v^2}{c^2}}.
\ee
Taking account of (12.3), from equation (12.7) we find
the total proper time $\tau$\index{proper time}
\be
\tau=t_0+\f{\,c\,}{\,a\,}\ln
\left[\f{at}{c}+\sqrt{1+\f{a^2t^2}{c^2}}\;\right].
\ee
From this formula it follows that, as time $t$ increases in
an inertial reference system, the
proper time\index{proper time}
for a moving body
flows slowly, according to a logarithmic law. We considered
the motion of a body with acceleration $\vec a$ with respect to an
inertial reference system in
Galilean\index{Galilei} 
coordinates.\index{Galilean (Cartesian) coordinates}
\markboth{thesection\hspace{1em}}{12. Relativistic motion with constant acceleration \ldots}

Now consider a reference system moving with constant ac\-ce\-le\-ration. Let
the inertial and moving reference systems have coordinate axes oriented
in the same way, and let one of them be moving with respect to the
other along the $x$ axis. Then, if one considers their origins to
have coincided at $t=0$, from expression (12.5) one obtains the law of
motion of the origin of the reference system moving relativistically
with constant acceleration,
\be
x_0=\f{c^2}{a}
\left[\sqrt{1+\f{a^2t^2}{c^2}}-1 \right].
\ee
Therefore, the formula for coordinate transformation, when
transition is performed from the inertial reference system
$(X,T)$ to the reference system $(x,t)$ moving relativistically
with constant acceleration, will have the form
\be
x=X-x_0=X-\f{c^2}{a}
\left[\sqrt{1+\f{a^2T^2}{c^2}}-1\right].
\ee The transformation of time can be set arbitrarily. Let it be the same in both reference systems \be
t=T.
\ee
In the case of transformations (12.10) and (12.11) the
interval $d\sigma$\index{interval}
assumes the form
\be
d\sigma^2=\f{c^2dt^2}{1+\ds\f{a^2t^2}{c^2}}-
\f{2a\,t\,dtdx}{\sqrt{1+\ds\f{a^2t^2}{c^2}}}-
dx^2-dY^2-dZ^2.
\ee

\vspace*{5mm}
\noindent
{\it We shall now proceed to deal with the
``clock paradox''}.\index{``clock paradox''}

Consider two reference systems. If two observers, who are in these reference systems, compared their clocks at
moment $t=0$, and then departed from each other, and after some period of time they again met at one point in
space, what time will their clocks show? The answer to this question is the solution of the so-called ``clock
paradox''.\index{``clock paradox''} However, two observers, who are in different inertial reference systems, after
they have compared their clocks at one and the same point of space, will never be able to meet in the future at
any other point of space, because to do so, at least, one of them would have to interrupt his inertial motion and
for some time go over to a non-inertial reference 
system.\index{non-inertial reference systems} In scientific
literature, and in textbooks, as well, it is often written that the answer to this question cannot be given within
the framework of special relativity theory.

This is, naturally, wrong, the issue is resolved precisely within the framework of special relativity theory. The
point is that re\-fe\-rence systems moving with acceleration in 
pseudo-Euclidean\index{pseudo-Euclidean geometry of space-time} ge\-o\-met\-ry, contrary to A.\,Einstein's\index{Einstein}  point of view,
have nothing to do with the gravitational field, and for this reason general relativity theory is not required for
explaining the ``clock paradox''.\index{``clock paradox''}

Let us illustrate this statement by a concrete computation. Sup\-po\-se
we have two identical (ideal) clocks at one and the same point of
an inertial reference system. Consider their readings to coincide
at the initial moment $T=0$. Let one of these clocks always be at
rest at the initial point and, thus, be inertial. Under the
influence of an applied force, at moment $t=0$, the other clock
starts to move relativistically with a constant acceleration
$a$ along the $x$ axis, and continues moving thus till the moment
of time $t=T_1$, shown by the clock at rest. Further, the influence
of the force on the second clock ceases, and during the time
interval\index{interval}
$T_1\leq t\leq T_1+T_2$ it moves with constant velocity.
After that a decelerating force is applied to it, and under the
influence of this force it starts moving relativistically with
constant acceleration $-a$ and continues to move thus till the
moment of time $t=2T_1+T_2$, as a result of which its velocity
with respect to the first clock turns zero. Then, the entire
cycle is reversed, and the second clock arrives at the same
point, at which the first clock is.

We shall calculate the difference in the readings of these clocks
in the inertial reference system, in which the first clock is at
rest. By virtue of the symmetry of the problem (four segments of
motion with constant acceleration and two segments of uniform
rectilinear motion), the reading of the clock at rest, by the moment
the two clocks meet, will be
\be
T=4T_1+2T_2.
\ee
For the second clock
\be
T^\prime =4T_1^\prime +2T_2^\prime.
\ee Here $T_1^\prime$  is the time interval\index{interval} between the moment when the second clock started to
accelerate and the moment when the acceleration ceased, measured by the moving clock. $T_2^\prime$ is the
interval\index{interval} of the second clock's proper time\index{proper time} between the first and second
accelerations, during which second clock's motion is uniform and rectilinear.

In an inertial reference system, the
interval\index{interval}
for a moving body is
\be
ds=cdt\sqrt{1-\f{v^2(t)}{c^2}}.
\ee
Therefore
\be
T_1^\prime=\int\limits_0^{T_1}\sqrt{1-\f{v^2(t)}{c^2}}\,dt.
\ee
On the basis of (12.3) we obtain
\be
T_1^\prime=\int\limits_0^{T_1}\f{dt}{\sqrt{1+\ds\f{a^2t^2}{c^2}}}.
\ee
Hence we find
\be
T_1^\prime=\f{\,c\,}{\,a\,}\ln\left(
\f{aT_1}{c}+\sqrt{1+\f{a^2T_1^2}{c^2}}\;\right).
\ee

The motion of the second clock during the
interval\index{interval}
of time
\[
T_1\leq t\leq T_1+T_2
\]
due to Eq.~(12.4) proceeds with the velocity \be
v=\f{aT_1}{\sqrt{1+\ds\f{a^2T_1^2}{c^2}}},
\ee and, therefore, in accordance with (12.15) we obtain \be
T_2^\prime =\f{T_2}{\sqrt{1+\ds\f{a^2T_1^2}{c^2}}}.
\ee
Consequently, by the moment the two clocks meet the reading of the
second clock will be
\be
T^\prime=\f{4c}{a}\ln\left(
\f{aT_1}{c}+\sqrt
{1+\f{a^2T_1^2}{c^2}}\;\right)+
\f{2T_2}{\sqrt{1+\ds\f{a^2T_1^2}{c^2}}}.
\ee
Subtracting (12.13) from (12.21), we find
\ba
&&\!\!\!\!\!\!\!\!\!\!\!\!\!\!\!\!\!\!\!\Delta T=T^\prime -T=\ds
\f{4c}{a}\ln\left(\f{aT_1}{c}+
\sqrt{1+\ds\f{a^2T_1^2}{c^2}}\;\right)-\nonumber\\
\label{12.22}\\
&&\!\!\!\!\!\!\!\!\!\!\!\!\!\!\!\!\!\!\!-4T_1+2T_2\left[\ds\f{1}{\ds\sqrt{1+
\f{a^2T_1^2}{c^2}}}-1\right].\nonumber
\ea
It can be verified that for any $a>0, T_1>0, T_2>0$
the quantity $\Delta T$ is negative. This means that at the
moment the clocks meet the reading of the second clock will
be less than the reading of the first clock.

Now consider the same process in the reference system, where the
second clock is always at rest. This reference system is not inertial,
since part of the time the second clock moves with a con\-stant
acceleration with respect to the inertial reference system re\-la\-ted
to the first clock, while the remaining part of time its motion is
uniform. At the first stage the second clock moves with constant
acceleration, according to the law (12.9)
\[
x_0=\f{c^2}{a}\left[\sqrt{1+\f{a^2t^2}{c^2}}-1\right].
\]
Therefore, at this segment of the journey, the interval\index{interval} in the non-inertial reference
system,\index{non-inertial reference systems} according to (12.12) has the form \be
d\sigma^2=\f{c^2dt^2}{1+\ds\f{a^2t^2}{c^2}}- \f{2a\,t\,dxdt}{\sqrt{1+\ds\f{a^2t^2}{c^2}}}-
dx^2-dY^2-dZ^2.
\ee In this reference system the second clock is at rest at point $x=0$, while the first clock moves along the
geodesic line\index{geodesic line} determined by Eqs.~(11.14) \be
\f{dU^\nu}{d\sigma}+{\mit\Gamma}_{\alpha\beta}^\nu
U^\alpha U^\beta=0,\quad\nu=0,1,2,3.
\ee
Of these four equations only three are independent, since
 the fol\-low\-ing relation is always valid:
\be
\gamma_{\mu\nu} U^\mu U^\nu =1,\;
U^\nu=\f{dx^\nu}{d\sigma}.
\ee
From expression (12.23) we find
\be
\gamma_{00}=\f{1}{1+\ds\f{a^2t^2}{c^2}},\quad
\gamma_{01}=-\f{at}{c\,\sqrt{1+\ds\f{a^2t^2}{c^2}}},\quad
\gamma_{11}=-1.
\ee From Eq.~(12.26) and the following equation
\[
\gamma^{\mu\nu}\cdot\gamma_{\nu\lambda}=\delta_\lambda^\mu\,,
\]
we find
\[
\gamma^{00}=1,\quad\gamma^{01}=-\f{at}{c\,\sqrt{1+\ds\f{a^2t^2}{c^2}}}\,,\quad
\gamma^{11}=-\f{1}{1+\ds\f{a^2t^2}{c^2}}\,.
\]
By means of these formulae and also Eqs.~(11.31) and (12.26) it is easy to see that there is only one nonzero
Christoffel\index{Christoffel}  symbol\index{Christoffel symbol}
\[
{\mit\Gamma}_{00}^1=\f{1}{c^2\,\left(1+\ds\f{a^2t^2}{c^2}\right)^{3/2}}\,.
\]

We do not have to resolve equation (12.24), we shall only take
advantage of relation (12.25)
\be
\gamma_{00}(U^0)^2+2\gamma_{01}U^0 U^1 -(U^1)^2=1.
\ee Taking into account (12.26), from equation (12.27) we find a partial solution \be
U^1=-\f{at}{c\,\sqrt{1+\ds\f{a^2t^2}{c^2}}}\,,\quad U^0=1\,,
\ee which as easy to check satisfies also Eqs.~(12.24). From  (12.28) it follows \be
\f{dx^1}{dt}=-\f{at}{\sqrt{1+\ds\f{a^2t^2}{c^2}}}.
\ee
Resolving this equation with the initial conditions
$x(0)\!\!\!=\!\!\!0,\linebreak \dot x(0)=0$, we obtain
\be
x=\f{c^2}{a}\left[1-\sqrt{1+\f{a^2t^2}{c^2}}\;\right].
\ee

Thus, we have everything necessary for determining the re\-a\-dings
of both clocks by the final moment of the first stage in their motion.
The proper time $d\tau$\index{proper time}
of the first clock at this stage of motion,
by virtue of (12.29), coincides with the time $dT$ of the inertial
reference system
\be
d\tau=\f{ds}{c}=dT,
\ee
therefore, by the end of this stage of the journey the reading
$\tau_1$ of the first clock will be $T_1$
\be
\tau_1=T_1.
\ee Since the second clock is at rest with respect to the non-inertial reference system,\index{non-inertial
reference systems} its proper time\index{proper time} can be determined from expression \be
d\tau^\prime=\sqrt{\gamma_{00}}\,dt.
\ee
Since the first stage of the journey occupies the
interval\index{interval}
$0\leq t\leq T_1$ of inertial time, then at the end of this segment the
reading $\tau_1^\prime$ of the second clock will be
\be
\tau_1^\prime=\int\limits_0^{T_1}\sqrt{\gamma_{00}}\,dt=
\f{\,c\,}{\,a\,}\ln\left[\f{aT_1}{c}+
\sqrt{1+\f{a^2T_1^2}{c^2}}\;\right].
\ee
At the end of the first stage of the journey, upon reaching
the velocity
\be
v=\f{aT_1}{\sqrt{1+\ds\f{a^2T_1^2}{c^2}}},
\ee
the action of the accelerating force ceases, this means
that the re\-fe\-rence system related to the second clock will
be inertial. The
interval\index{interval}
in this reference system, in accordance
with (12.23) will, by the moment $T_1$, have the form
\be
d\sigma^2=c^2\left(1-\f{v^2}{c^2}\right)dt^2-
2vdxdt-dx^2-dY^2-dZ^2,
\ee
here
\be
v=\f{aT_1}{\sqrt{1+\ds\f{a^2T_1^2}{c^2}}}.
\ee

Taking advantage, for the metric (12.36), of the identity \be \gamma_{\mu\nu}U^\mu U^\nu =1,\;
U^\nu =\f{dx^\nu}{d\sigma},
\ee
we find
\be
\f{dx^1}{dt}=-v.
\ee
Taking into account (12.39) in (12.36), we obtain
\be
d\tau=\f{d\sigma}{c}=dt,
\ee
i.\,e. the time, shown by the first clock at this stage,
coincides with the time $T_2$
\be
\tau_2=T_2.
\ee
Since the second clock is at rest, its reading of its
proper time\index{proper time}
is
\be
d\tau^\prime =\sqrt{\gamma_{00}}\,dt.
\ee
Hence follows
\be
\tau_2^\prime =\int\limits_{T_1}^{T_1+T_2}\sqrt{\gamma_{00}}\,dt=
\f{T_2}{\sqrt{1+\ds\f{a^2T_1^2}{c^2}}}.
\ee

Owing to the symmetry of the problem, the information obtained
is sufficient for determining the readings of the clocks at the
moment they meet. Indeed, the reading of the first clock $\tau$,
determined in the reference system, related to the second
clock, is
\be
\tau=4\tau_1+2\tau_2,
\ee
which on the basis of (12.32) and (12.41) gives
\be
\tau=4T_1+2T_2.
\ee
The reading of the second clock $\tau^\prime$, determined in the
same reference system, where the second clock is at rest, is
\be
\tau^\prime=4\tau_1^\prime+2\tau_2^\prime,
\ee
which on the basis of (13.34) and (13.43) gives
\be
\tau^\prime=\f{4c}{a}\ln\left[\f{aT_1}{c}+
\sqrt{1+\f{a^2T_1^2}{c^2}}\;\right]+
\f{2T_2}{\sqrt{1+\ds\f{a^2T_1^2}{c^2}}}.
\ee
Subtracting from (12.47) expression (12.45), we obtain
\ba
&&\!\!\!\!\!\!\!\!\!\!\!\!\!\!\!\!\!\!\!\Delta\tau=\ds\tau^\prime-\tau=\f{4c}{a}\ln
\left[\f{aT_1}{c}+\sqrt{1+\ds\f{a^2T_1^2}{c^2}}\;\right]-\nonumber\\
\label{12.48}\\
&&\!\!\!\!\!\!\!\!\!\!\!\!\!\!\!\!\!\!\!-4T_1+2T_2\left[\ds\f{1}{\sqrt{1+\ds\f{a^2T_1^2}{c^2}}}-1\right].\nonumber
\ea Comparing (12.22) and (12.48) we see, that the computation per\-for\-med in the inertial reference system,
where the first clock is at rest, yields the same result as the computation performed in the non-inertial
reference system\index{non-inertial reference systems} related to the second clock.

Thus,
\be
\Delta\tau=\Delta T<0.
\ee Hence it follows that no paradox exists, since the reference system related to the first clock is inertial,
while the reference system, in which the second clock is at rest, 
is non-inertial.\index{non-inertial reference systems}

Precisely for this reason, the slowing down of the second clock,
as compared to the first clock, is an absolute effect and does
not depend on the choice of reference system, in which this
effect is computed.

The arguments concerning the relativity of motion, which were used previously, in this case cannot be applied,
since the reference systems are not equitable. Qualitatively, the slowing down of the second clock, as compared to
the first, can be explained as follows. It is known, that in arbitrary coordinates the free motion of a test body
proceeds along a geodesic line,\index{geodesic line} i.\,e. the extremal line, which in
pseudo-Euclidean\index{pseudo-Euclidean geometry of space-time} space is the maximum distance between two points,
if on the entire line, joining these points, the quantity $d\sigma^2$ is positive. In the case, when we choose an
inertial reference system in Galilean\index{Galilei} 
coordinates,\index{Galilean (Cartesian) coordinates} related to the first clock, this means that the first clock describes a geodesic
line,\index{geodesic line} while the second clock, owing to the influence of the force, moves along a line
differing from the geodesic, and, therefore, slows down. The same happens, also, when the reference system is
related to the second clock. In the case of transition to this reference system, the interval\index{interval}
somewhat changes its form. In this case the first clock again describes a geodesic line\index{geodesic line} in an
altered metric, while the second clock is at rest, and, consequently, do not describe a geodesic
line\index{geodesic line} and, therefore, slow down.

We have considered the influence of accelerated motion on the
readings of clocks and have showed their slowing down. But this
effect concerns not only clocks, but all physical, or, to be more
general, all natural phenomena. On this basis, interstellar flights
become fantastically fascinating. Back in 1911, Paul
Langevin\index{Langevin} 
discussed in an article~\cite{14} the voyage of a human being at high
velocities, close to the velocity of light, subsequently returning
to the Earth. In principle, this is possible, but it still remains only
a fantasy.

Let us now pay attention to the Sagnac\index{Sagnac} effect
(see more details in: Uspekhi Fiz. Nauk. 1988. Vol. 156,
issue 1,  pp.~137-143. In collaboration 
with {\it Yu.\,V.\,Chugreev}).\index{Chugreev}
As is well known, the Sagnac\index{Sagnac} 
effect in line with the Michelson\index{Michelson} 
experiment is one of the basic experiments of the theory of relativity. But till now it is possible
to read  incorrect explanations of this effect with the help of signals propagating faster than
light or with the help of general relativity (see in more detail below). So we consider it as
necessary to stress once more purely special relativistic nature of Sagnac effect.\index{Sagnac} 

Let us at first describe the Sagnac experiment.\index{Sagnac} 
There are mirrors situated at the angles of a quadrangle on a disk. The angles of their reciprocal
disposition are such that the beam from a monochromatic source after reflections over these mirrors
passes a closed circle and returns to the source. With the help of a semi-transparent plate it is
possible to divide the beam coming from  a source into two beams moving in opposite directions over
this closed circle.
\vspace{-1mm}

Sagnac\index{Sagnac} has discovered that if the disk is subjected
to rotation, then the beam with the direction of its round coinciding with the direction of rotation
will come back to the source later than the beam with opposite round, resulting in a shift of the
interference picture on  the photographic plate.
After interchanging the direction of rotation the interference bands shift in opposite direction.

What explanation was given to this effect?
Sagnac\index{Sagnac} 
himself has obtained  a theoretical value for the magnitude of the effect by purely classical
addition of the light velocity\index{relating velocities} with the linear velocity of rotation for the
beam moving oppositely to rotation and corresponding subtraction for the beam moving in the direction
of rotation.
The discrepancy of this result with the experiment was of percent order.

This explanation of the experimental results  remained later more or less invariable or even became obscure.
As a typical example we  present a related quotation from ``Optics'' 
by A.\,Sommer\-feld\index{Sommerfeld}:
\begin{quote}
{\it\hspace*{5mm} ``The negative result of Michelson's\index{Michelson} 
experiment has, of course, 
no bearing on the problem of the 
propagation of light in {\bf\textit{rotating}} media.
To discuss this problem one must use not the 
special but rather the general theory of relativity with its additional 
terms which correspond to the mechanical
centrifugal centrifugal forces. However, in view of the fact 
that in the following experiments 
({\rm by Sagnac\index{Sagnac} and others.} --- {\sl A.L.}) only velocities $v\ll c$ occur and only
first order effects in $v/c$ are important,
relativity theory can be dispensed with entirely
and the computations can be carried out classically''.}
\end{quote}

We will see below that the explanation of the Sagnac\index{Sagnac} effect
lies in full competence of the special theory of relativity and neither general theory of relativity nor
super-luminal velocities are not required as well as any 
other additional postulates. We will consider in detail how to
calculate the time difference between arrivals of the two beams to the source in the inertial rest system of
reference. We will also do that in the rotating with the disk non-inertial reference 
system.\index{non-inertial reference systems} The results of calculations will coincide as should be expected. For simplicity of calculations
we will consider the motion of light in a light guide over circular trajectory  which corresponds to the case of
infinite number of mirrors in the Sagnac\index{Sagnac} experiment.

We begin with the case of inertial system of 
reference.\index{inertial reference systems}
Let us express the interval\index{interval} in cylindrical coordinates:
\be
ds^2 = c^2dt^2-dr^2-r^2d{\phi}^2-dz^2.
\ee Let as it has been told before light beams move in  plane $z=0$ over circle of radius $r=r_0={\textrm const}$.
The interval\index{interval} is exactly equal to zero for light, so we obtain the following \be
\f{d\phi_{\pm}(t)}{dt}=\pm\f{c}{r_0}.
\ee The beam moving in the direction of rotation is marked by index ``$+$'', and the beam moving in opposite
direction is marked by ``$-$''.

With account for the initial conditions $\phi_{+}(0)=0,\,\phi_{-}(0)=2\pi$
we find the law of angle $\phi_{\pm}$ dependence of the two beams on time $t$:
\ba
&&\phi_{+}(t)=\ds\f{c}{r_0}t,\nonumber\\
\label{12.52}\\
&&\phi_{-}(t)=2\pi-\ds\f{c}{r_0}t.\nonumber
\ea
The beams will meet at time $t_1$, when $\phi_{+}(t_1)=\phi_{-}(t_1)$.
Substituting (12.52) we obtain
\[
\phi_{+}(t_1)=\phi_{-}(t_1)=\pi.
\]

Then taking time $t_1$ as the initial time and repeating our argumentation we will find that the next meeting of
beams will take place just at that spatial point where they have been emitted, i.\,e. at point with coordinates
 $\phi=0,
r=r_0,\,z=0$.

We emphasize that this result does not depend on the angular velocity of rotation of the system of reference
which is the rest system for the source and mirrors.

The law of dependence of the angular coordinate of the source by definition is as follows (for initial
condition $\phi_s (0)=0)$:
\be
\phi_s(t)=\omega t.
\ee

Therefore, the meeting of the source with  ``+''-beam will take place at time moment $t_{+}$ determined by
condition $\phi_s(t_{+})=\break\phi_{+}(t_{+})-2\pi$, i.\,e. \be
t_{+}=\f{2\pi}{(c/r_0)-\omega},
\ee and with ``$-$''-beam --- at time moment $t_{-}$ determined by condition $\phi_s(t_{-})=\phi_{-}(t_{-})$: \be
t_{-}=\f{2\pi}{(c/r_0)+\omega}\,.
\ee It may seem from the form of Eqs.~(12.54), (12.55) that the velocity of light is here anisotropic and is
different from {\bf \textit{c}}.
But this is incorrect. The light velocity is the same for both beams and it is equal to
{\bf \textit{c}},
and the different time of return to the source  is explained by the fact that the source has moved over some
distance during the time of beams propagation (``+''-beam has travelled over larger distance).

Let us now find the interval of proper time\index{proper time}
between arrivals of the two beams for an observer sitting on the source. By definition it is equal to
\be
\Delta=\f{\,1\,}{\,c\,}\!\int\limits_{s(t_{-})}^{s(t_{+})}\!ds=
\f{\,1\,}{\,c\,}\int\limits_{t_{-}}^{t_{+}}\f{ds}{dt}dt,
\ee
where $s$ is the interval.
As a value of interval\index{interval} after using (12.53) we get
\[
ds^2=c^2dt^2-r_0^2d\phi^2=c^2dt^2\left(1-\f{r_0^2\omega^2}{c^2}\right),
\]
where $\omega^2r_0^2/c^2<1$.

Substituting this into Eq. (12.56) we will find exact value of the Sagnac effect\index{Sagnac}
\footnote{In calculation for the realistic Sagnac effect, when the light beam trajectory
is a polygonal line, it is necessary to take into account the centrifuge deformation due to centrifugal forces.}:
\be
\Delta=\left(1-\f{r_0^2\omega^2}{c^2}\right)^{1/2}\!\!\!\!\!\!\!\cdot(t_{+}-t_{-})
=\f{4\pi\omega r_0^2}{c^2[1-(r_0^2\omega^2/c^2)]^{1/2}}.
\ee

Let us note that in deriving Eq.~(12.57) we used only absolute concepts of events of beams meeting (with each
other and with the source), and not the concept of the light velocity relative to the rotating reference system.

Let us consider now the same physical process of propagation of beams over circle towards each other in rotating
with angular velocity $\omega$ non-inertial system of reference. In order to find out the form of
interval\index{interval} in this system we will make a coordinate transformation:
\ba
&&\phi_{new}=\phi_{old}-\omega t_{old},\nonumber\\
&&t_{new}=t_{old},\nonumber\\[-5mm]
\label{12.58}\\
&&r_{new}=r_{old},\nonumber\\
&&z_{new}=z_{old}.\nonumber
\ea
In new coordinates $t_{new},\,r_{new},\,\phi_{new},\,z_{new}$ we  obtain
 (after lowing index ``new'' for simplicity)
interval\index{interval} in the following form
\ba
&&\!\!\!\!\!\!\!\!\!\!\!ds^2=\left(1-\ds\f{\omega^2r^2}{c^2}\right)c^2dt^2
-\ds\f{2\omega r^2}{c}d\phi cdt-\nonumber\\
\label{12.59}\\
&&\!\!\!\!\!\!\!\!\!\!\!-dr^2-r^2d\phi^2-dz^2.\nonumber \ea Let us note that time $t$ in this expression is the
coordinate time\index{coordinate time} for the rotating system of reference.

After accounting for initial conditions $\phi_{+}(0)=0,\,\phi_{-}(0)=2\pi$ we get:
\ba
&&\phi_{+}(t)=\ds\f{ct}{r_0}\left(1-\ds\f{\omega r_0}{c}\right),\nonumber\\
\label{12.60}\\
&&\phi_{-}(t)=2\pi-\ds\f{ct}{r_0}\left(1+\ds\f{\omega r_0}{c}\right).\nonumber
\ea
the first meeting of beams will happen at time $t_1$, when
 $\phi_{+}(t_1)=\phi_{-}(t_1)$, i.\,e. when angular variable will be equal to  $\phi_1=\break\pi [1-(\omega r_0/c)]$.
After analogous reasoning we conclude that the second meeting of beams will happen ``at angle''
\be
\phi_2=2\pi\left(1-\f{\omega r_0}{c}\right),
\ee
i.\,e. at  angular distance $2\pi r_0\omega/c$ from the source. The dependence of source angular coordinate
is trivial $\phi_s={\rm const}=0$.

The moment of coordinate time\index{coordinate time} $t_{+}$ corresponding to meeting of  ``+''-beam with the
source could be found, as before, from relation $\phi_s(t_{+})=0=\phi_{+}(t_{+})-2\pi$: \be
t_{+}=\f{2\pi r_0}{c-\omega r_0},
\ee
and similarly we find moment $t_{-}$:
\be
t_{-}=\f{2\pi r_0}{c+\omega r_0}.
\ee

The proper time\index{proper time} interval\index{interval} between two events of coming the beams into the point
where the source is disposed can be calculated with the help of definition (12.56) and interval (12.59):
\[
\begin{array}{l}
\Delta=\ds\f{\,1\,}{\,c\,}\int\limits_{t_{-}}^{t_{+}}\ds\f{ds}{dt}dt
=\left(1-\ds\f{\omega^2 r_0^2}{c^2}\right)^{1/2}\!\!\!\!\!\!\!\cdot(t_{+}-t_{-})=
\\[5mm]
=\ds\f{4\pi\omega r_0^2}{c^2[1-(r_0^2\omega^2/c^2)]^{1/2}},
\end{array}
\]
i.\,e. we come to the same expression (12.57).

Therefore we demonstrated that for explanation of the Sagnac 
effect\index{Sagnac} 
one does not need neither modify the special theory of relativity, nor use super-luminal velocities,
nor  apply to the general theory of relativity. 
On only has 
to strictly follow the special theory of
relativity.

\newpage
\markboth{thesection\hspace{1em}}{}
\section{Concerning the limiting velocity}

The interval\index{interval} for 
pseudo-Euclidean\index{pseudo-Euclidean geometry of space-time} geometry, in
arbitrary co\-or\-di\-na\-tes, has, in accordance with (3.32) and (3.33) the following general form: \be
d\sigma^2=\gamma_{\mu\nu}(x)dx^\mu dx^\nu,\;
\gamma=\det(\gamma_{\mu\nu})<0.
\ee
The metric tensor\index{metric tensor of space}
$\gamma_{\mu\nu}$ equals
\be
\gamma_{\mu\lambda}(x)=\sum_{\nu=0}^3 \varepsilon^\nu
\f{\pa f^\nu}{\pa x^\mu}\cdot\f{\pa f^\nu}{\pa x^\lambda},
\quad \varepsilon^\nu=(1,-1,-1,-1).
\ee
Here $f^\nu$ are four arbitrary continuous functions with
continuous derivatives, that relate
Galilean\index{Galilei} 
coordinates\index{Galilean (Cartesian) coordinates}
with
the arbitrary $x^\lambda$.

Depending on the sign of $d\sigma^2$, events can be identified
as time-like
\be
d\sigma^2>0,
\ee space-like \be
d\sigma^2<0,
\ee and isotropic \be
d\sigma^2=0.
\ee
Such a division of
intervals\index{interval}
is absolute, it does not depend
on the choice of reference system.

For a time-like interval\index{interval} $d\sigma^2>0$ there always exists an inertial reference system, in which
it is only determined by time
\[
d\sigma^2=c^2dT^2.
\]
For a space-like interval\index{interval} $d\sigma^2<0$ there can always be found an inertial reference system, in
which it is determined by the distance between infinitesimally close points
\[
d\sigma^2=-d\ell^2,\;d\ell^2=dx^2+dy^2+dz^2.
\]
These assertions are also valid in the case of a finite interval $\sigma$.\index{interval}
\markboth{thesection\hspace{1em}}{13. Concerning the limiting velocity}
\newpage

Any two events, related to a given body, are described by a time-like interval.\index{interval} An isotropic
interval\index{interval} corresponds to a field with\-out rest mass. Let us see, what conclusions result from an
isotropic interval\index{interval} \be \gamma_{\mu\nu}dx^\mu dx^\nu=\gamma_{00}(dx^0)^2+2\gamma_{0i}dx^0 dx^i+
\gamma_{ik}dx^i dx^k =0.
\ee We single out in (14.6) the time-like part \be c^2\left[\sqrt{\gamma_{00}}\,dt+\f
{\gamma_{0i}dx^i}{c\sqrt{\gamma_{00}}}\,\right]^2- \left[-\gamma_{ik}+\f{\gamma_{0i}\gamma_{0k}}
{\gamma_{00}}\right]dx^idx^k=0.
\ee
The quantity
\be
d\tau=\sqrt{\gamma_{00}}\,dt+\f
{\gamma_{0i}dx^i}{c\sqrt{\gamma_{00}}}=
\f{\,1\,}{\,c\,}\left(\f{\gamma_{0\lambda}dx^\lambda}
{\sqrt{\gamma_{00}}}\right)
\ee is to be considered as physical time,\index{physical time} which, as we shall see below, is independent of the
choice of time variable. In the general case (non-inertial reference systems)\index{non-inertial reference
systems} the quantity $d\tau$ is not a total dif\-fe\-ren\-ti\-al, since the following conditions will not be
satisfied: \ba &&\ds\f{\pa}{\pa x^i}(\sqrt{\gamma_{00}})= \f{\,1\,}{\,c\,}\f{\pa}{\pa t}
\left(\f{\gamma_{0i}}{\sqrt{\gamma_{00}}}\right),\nonumber\\
\label{13.9}\\
&&\ds\f{\pa}{\pa x^k} \left(\f{\gamma_{0i}}{\sqrt{\gamma_{00}}}\right)= \f{\pa}{\pa x^i}
\left(\f{\gamma_{0k}}{\sqrt{\gamma_{00}}}\right).\nonumber \ea The second term in (13.7) is nothing, but the
square distance between two infinitesimally close points of {\bf three-dimensional space}, which is independent of
the choice of coordinates in this space: \be
d\ell^2=\chi_{ik}dx^idx^k,
\ee
here the metric tensor\index{metric tensor of space}
of three-dimensional space, $\chi_{ik}$, is
\be
\chi_{ik}=-\gamma_{ik}+\f{\gamma_{0i}\gamma_{0k}}{\gamma_{00}}.
\ee
With account of (13.8) and (13.10), from expression (13.7) we find
\be
\f{d\ell}{d\tau}=c.
\ee

The quantities $d\ell$ and $d\tau$ are of local character. In
this case the
{\bf concept of simultaneity}\index{simultaneity}
loses {\bf sense}
for events at different sites, because it is impossible to
{\bf synchronize clocks} with the aid of a light signal, since
it depends on the synchronization path. From (13.12) it follows,
that the field at each point of
Minkowski\index{Minkowski} 
space,\index{Minkowski space}
in accordance
with the local characteristics of $d\ell$ and $d\tau$, have a
velocity equal to the
electrodynamic\index{electrodynamic constant}
constant $c$. This is the
li\-mi\-ting velocity, that is not achievable for particles with rest
mass, since for them
\[
d\sigma^2>0.
\]
This inequality is the causality condition. The causality principle
is not contained in the
Maxwell-Lorentz\index{Lorentz} 
\index{Maxwell} 
equations.\index{Maxwell-Lorentz equations}
It is imposed as
a natural complementary condition. In 1909
H.\,Minkowski\index{Minkowski} 
for\-mu\-la\-ted
it as the principal axiom as follows:
\begin{quote}
{\it\hspace*{5mm} ``{\bf \textit{A substance, found at any world point, given
the appropriate definition of space and time}} ({\rm i.\,e. given the
corresponding choice of reference system in
Min\-kow\-ski\index{Minkowski} 
space.} --- A.L.)\index{Minkowski space}
{can be considered to be at rest}. The axiom expresses the idea, that
at each world point the expression
\[
c^2dt^2-dx^2-dy^2-dz^2
\]
is always positive or, in other words, that any velocity $v$ is always less than
{\bf \textit{c}}''.}
\end{quote}

H.\,Poincar\'e\index{Poincar\'e} has demonstrated the deep physical meaning of the limiting velocity
in his article  [1] published in 1904 even before his fundamental works  [2; 3].
He wrote:
\begin{quote}
{\it\hspace*{5mm}$\ll$If all these results would be confirmed there will arise an absolutely new mechanics. It will
be characterized mainly by the fact that neither velocity could exceed the velocity of light,\footnote{\it Because
bodies would oppose to the forces trying to accelerate their motion by means of the increasing inertia, and this
inertia would become infinite in approaching the velocity of light.} as the temperature could not drop below the
absolute zero. Also no any observable velocity could exceed the light velocity for any observer performing a
translational motion but not suspecting about it. There would be a contradiction here if we will not remember that
this observer uses another clock than the observer at rest. Really he uses the clock showing ``the local
time''\index{local time}$\gg$.}
\end{quote}

Just these thoughts  by H.\,Poincar\'e\index{Poincar\'e} and his principle of relativity\index{relativity
principle} were reported by him in a talk given at The Congress of Art and Science in Sent-Louis
(in September of 1904) and they found their realization in articles [2; 3]. They underlie the
work by A.\,Einstein\index{Einstein} of 1905.

Signal from one object to another can only be transferred by
means of a material substance; from the aforementioned it is clear,
that $c$ is the {\bf maximum velocity for transferring interaction or
in\-for\-mation}. Since particles, corresponding to the electromagnetic
field, --- photons --- are usually considered to be 
massless, the quantity
{\bf \textit{c}}
is identified with the velocity
of light. The existence of a maximum velocity is a direct consequence
of the pseudo-Euclidean\index{pseudo-Euclidean geometry of space-time} geometry of space-time.

If we choose the function $f^\nu$ in (13.2) by a special way as follows
\be
f^0(x^\lambda),\;f^i(x^k),
\ee
then, owing such transformation, we do not leave the
inertial re\-fe\-rence system.\index{inertial reference systems}

In this case the
metric tensor\index{metric tensor of space}
$\gamma_{\mu\nu}$, in
accordance with (13.2) and (13.13), assumes the form
\be
\gamma_{00}=\left(\f{\pa f^0}{\pa x^0}\right)^2,\quad
\gamma_{0i}=\f{\pa f^0}{\pa x^0}\cdot
\f{\pa f^0}{\pa x^i},
\ee
\be
\gamma_{ik}=\f{\pa f^0}{\pa x^i}\cdot
\f{\pa f^0}{\pa x^k}-\sum_{\ell=1}^3
\f{\pa f^\ell}{\pa x^i}\cdot
\f{\pa f^\ell}{\pa x^k}.
\ee Substituting the values  for the metric coefficients $\gamma_{00}$, $\gamma_{0i}$ from (13.14) into (13.8) we
obtain, with account for (3.30) and (13.13), \be d\tau=\f{1}{c}\left(\f{\pa f^0}{\pa x^\nu}dx^\nu\right)=
\f{1}{c}df^0=\f{1}{c}dX^0.
\ee
We see, that proper time,\index{proper time}
in this case, is a total differential,
since our reference system is inertial. Substituting (13.14)
and (13.15) into (13.11), we obtain
\be
\chi_{ik}=\sum_{n=1}^3\f{\pa f^n}{\pa x^i}\cdot
\f{\pa f^n}{\pa x^k}.
\ee Hence, with account for (3.30) and (13.13), we find \be
d\ell^2=\chi_{ik}dx^i dx^k=\sum_{n=1}^3(df^n)^2=\sum_{n=1}^3(dX^n)^2.
\ee

In an inertial reference system, ambiguity exists in the co\-or\-di\-nate description of
 Minkowski\index{Minkowski} space,\index{Minkowski space} depending on the choice of func\-tions (13.13). This
is the reason for arbitrariness in adopting an {\bf agreement} concerning simultaneity\index{simultaneity} at
different points of space. All such agreements are conventional. However, this ambiguity and, consequently, the
arbitrariness in reaching an agreement do not influence the physical quantities. Eqs.~(13.16) and (13.18) show
that in an inertial reference system the physical quantities of time (13.8) and distance (13.10) do not de\-pend
on the choice of agreement concerning simultaneity.\index{simultaneity} Let me clarify. In formulae (13.16) and
(13.18), given any choice of func\-tions (13.13), there only 
arise {\bf Galilean\index{Galilei} 
coordinates}\index{Galilean (Cartesian) coordinates} {\mathversion{bold}\( X^0, X^n \)}
of
Min\-kow\-ski\index{Minkowski} 
space,\index{Minkowski space}
that correspond to the invariant (3.22). This is
pre\-ci\-se\-ly what removes, in the physical quantities of time (13.8) and
distance (13.10), arbitrariness in the choice of a conventional
ag\-ree\-ment concerning
simultaneity.\index{simultaneity}
Moreover, no physical quan\-ti\-ti\-es
can, in principle, depend on the choice of this agreement on
si\-mul\-ta\-ne\-i\-ty. And if someone has written, or writes, the opposite,
this only testifies to that person's {\bf incomprehension} of the
essence of relativity theory. One must distinguish between
coordinate\index{coordinate quantities}
quan\-ti\-ti\-es and physical quantities. For details con\-cer\-ning this issue
see ref. [6].

Let us demonstrate a particular special example of the simultaneity\index{simultaneity} convention. Let the
synchronization of clocks\index{synchronization of clocks} in different spatial points is provided by the light
signal having velocity {\mathversion{bold}\( c_1 \)} in the direction parallel to the positive semi-axis $X$, and
having velocity {\mathversion{bold}\( c_2 \)} in the  direction of the negative semi-axis $X$. Then  the signal sent
from point  \textit{A} at the moment of time  $t_A$ will arrive  to point  $B$ at time  $t_B$ which is given as
follows
$$
t_B=t_A+\f{X_{AB}}{c_1}.\eqno(M)
$$
The reflected signal will arrive at point $A$ at time $t_A^\prime$
\[
t_A^\prime=t_B+\f{X_{AB}}{c_2}.
\]
After substituting into this expression value $t_B$, determined by  formula
({\it M})
we get
\[
t_A^\prime-t_A=X_{AB}\left(\f{1}{c_1}+\f{1}{c_2}\right).
\]
From here it follows
\[
X_{AB}=\f{c_1c_2}{c_1+c_2}(t_A^\prime -t_A).
\]
Applying this expression to Eq.~({\it M}) we find
\[
t_B=t_A+\f{c_2}{c_1+c_2}(t_A^\prime -t_A).
\]
So we come to the synchronization proposed by Reichenbach\index{Reichenbach} (see his book: ``\textsf{The philosophy of space \& time}''. 
Dover Publications, Inc. New York. 1958, p.~127):
\[
t_B=t_A+\varepsilon (t_A^\prime -t_A),\quad 0<\varepsilon <1.
\]

The conditional convention on the synchronization of 
clocks\index{synchronization of clocks} and therefore on
simultaneity\index{simultaneity} at different spatial points accepted by us corresponds to the choice of
interval\index{interval} in inertial reference 
system\index{inertial reference systems} in the following form:
$$
\begin{array}{l}
d\sigma^2=(dx^0)^2-\ds\f{c(c_2-c_1)}{c_1c_2}dx^0dx-
\\
-\ds\f{c^2}{c_1c_2}(dx)^2-(dy)^2-(dz)^2.
\end{array}\label{nnK}\eqno(K)
$$
Here we deal with coordinate time\index{coordinate time}
$t=x^0/c$ and other coordinate values.

Metric coefficients of interval\index{interval} ({\it K}) are as follows:
$$
\begin{array}{l}
\gamma_{00}=1,\quad \gamma_{01}=-\ds\f{c(c_2-c_1)}{2c_1c_2},
\\
\gamma_{11}=-\ds\f{c^2}{c_1c_2},\quad \gamma_{22}=-1,\quad \gamma_{33}=-1.
\end{array}\label{nnL}\eqno(L)
$$
With the help of Eqs.~(13.14), (13.15), and also $(L)$, we obtain transformation functions (13.13) for our case:
\vspace*{-2mm}
\[
\begin{array}{l}
f^0=X^0=x^0-\ds\f{x}{2}\cdot\f{c(c_2-c_1)}{c_1c_2},
\\
f^1=X=x\ds\f{c(c_1+c_2)}{2c_1c_2},
\\[3mm]
f^2=Y=y,\quad f^3=Z=z.
\end{array}
\]
Deriving from the above the inverse transformation functions, calculating with them differentials $dx^0,\,dx$ and
then substituting them into
({\it K}),
we find
$$
d\sigma^2=(dX^0)^2-(dX)^2-(dY)^2-(dZ)^2.\eqno(H)
$$
Therefore,
\textbf{the physical time}\index{physical time}
$d\tau$ in our example is given as follows:
\[
\begin{array}{l}
d\tau=dt-\ds\f{dx}{2}\cdot\ds\f{c_2-c_1}{c_1c_2},
\\[5mm]
dX^0=cd\tau,
\end{array}
\]
and it does not depend on the choice of functions (13.13), because it is completely determined by
interval\index{interval}
(\textit{H}) only.
Any change in coordinate values like (13.13) leads only to changing of the connection between
\textbf{the physical time}\index{physical time} and coordinate values.

To any conditional convention on the simultaneity\index{simultaneity} there will  correspond a definite choice of
the coordinate system in an inertial system of 
reference\index{inertial reference systems} of the
Minkowski\index{Minkowski}  space\index{Minkowski space}. Therefore \textbf{a
conditional convention on the simultaneity}\index{simultaneity} is nothing more than {\bf a definite choice of the
coordinate system} in an inertial system of 
reference\index{inertial reference systems}  of the
Minkowski\index{Minkowski} space\index{Minkowski space}.

An important contribution into understanding of some fundamental questions of the theory of relativity related to
the definition of simultaneity\index{simultaneity} in different spatial points was provided by Professor
A.\,A.\,Tyapkin\index{Tyapkin} (Uspekhi Fiz. Nauk. 1972. Vol.~106, issue 4.)

Now let us return to the analysis of physical time $d\tau$.\index{physical time} Quantity $d\tau$ characterizes
physical time,\index{physical time} which is independent on the choice of coordinate time.\index{coordinate time}
Indeed, let us introduce new variable $x^{\prime 0}$, such that \be x^{\prime 0}=x^{\prime 0}(x^0,x^i),\;
x^{\prime i}=x^{\prime i}(x^k).
\ee
Then due to tensorial character of  $\gamma_{\mu\nu}$ transformation
\[
\gamma_{\mu\nu}^\prime=\gamma_{\alpha\beta}
\f{\pa x^\alpha}{\pa x^{\prime\mu}}\cdot
\f{\pa x^\beta}{\pa x^{\prime\nu}},
\]
we will obtain for our case
\be
\gamma_{00}^\prime=\gamma_{00}\left(\f{\pa x^0}
{\pa x^{\prime 0}}\right)^2,\quad
\gamma_{0\lambda}^\prime=\gamma_{0\beta}\f{\pa x^0}
{\pa x^{\prime 0}}\cdot\f{\pa x^\beta}
{\pa x^{\prime\lambda}}\,;
\ee
similarly
\be
dx^{\prime\lambda}=\f{\pa x^{\prime\lambda}}
{\pa x^\sigma}dx^\sigma.
\ee
Exploiting the Kronecker\index{Kronecker} delta symbol\index{Kronecker symbol}
\be
\f{\pa x^\beta}{\pa x^{\prime\lambda}}\cdot
\f{\pa x^{\prime\lambda}}
{\pa x^\sigma}=\delta_\sigma^\beta,
\ee
we get
\be
cd\tau=\f{\gamma_{0\lambda}^\prime dx^{\prime\lambda}}
{\sqrt{\gamma_{00}^\prime}}=
\f{\gamma_{0\sigma}dx^\sigma}
{\sqrt{\gamma_{00}}}.
\ee
We can see that physical time $d\tau$\index{physical time}
does not depend on the choice of the coordinate system in an inertial system of reference\index{inertial reference
systems} of the Minkowski\index{Minkowski} space\index{Minkowski space}.

Physical time\index{physical time} determines the flow of time in a physical process, however, the quantity
$d\tau$ exhibits local character in a non-inertial reference 
system,\index{non-inertial reference systems} since
it is not a total differential and therefore {\bf no variable $\tau$ exists}.

In this case, there exists no unique physical time with lines orthogonal to three-dimensional space. In a
non-inertial reference system\index{non-inertial reference systems} the interval\index{interval} $d\sigma$ is
expressed via the physical quantities $d\tau, d\ell$ as follows:
\[
d\sigma^2=c^2d\tau^2-d\ell^2.
\]
There {\bf exist no} variables $\tau, \ell$  in this case. Here {\bf coordinate quantities} arise which permit to
describe any effects in space and time in non-inertial system of reference.

In an inertial reference system $d\tau$ coincides, in
Galilean\index{Galilei} 
co\-or\-di\-na\-tes,\index{Galilean (Cartesian) coordinates}
with the differential $dt$, so in
Minkowski\index{Minkowski} 
space\index{Minkowski space}
one can introduce unique time $t$. It will be physical. Introduction
of si\-mul\-ta\-nei\-ty for all the points of three-dimensional space
is a con\-se\-quence of the
pseudo-Euclidean\index{pseudo-Euclidean geometry of space-time}
geometry of the
four-dimensional\index{four-dimensional space-time}
space of events.

One can only speak of the velocity of light being constant, the same in all directions, and identical with the
electrodynamic\index{electrodynamic constant} constant $c$ in an {\bf inertial reference system in
Galilean\index{Galilei} coordinates}.\index{Galilean (Cartesian) coordinates} In an
inertial reference system, in any other admissible coordinates, the velocity of light will be the same, if time is
defined in ac\-cor\-dance with formula (13.8) and distance by formula (13.10). In a
non-inertial\index{non-inertial reference systems} reference system the 
electrodynamic\index{electrodynamic constant} constant $c$ is only expressed via the local quantities $d\tau, d\ell$. There  exist no variables $\tau,
\ell$  in this case.

It is often written that the principle of 
constancy\index{principle of constancy of velocity of light} of the
velocity of light underlies special relativity theory. This is wrong. \textbf{No principle of
constancy\index{principle of constancy of velocity of light} of the velocity of light exists as a first physical
principle,} because this principle is a simple consequence of the Poincar\'e\index{Poincar\'e} relativity
principle\index{relativity principle} for all the nature phenomena. It is enough to apply it to the emission of a
spherical electromagnetic wave to get convinced that the velocity of light at any inertial reference
system\index{inertial reference systems} is equal to electrodynamic constant {\bf
\textit{c}}.\index{electrodynamic constant} Therefore, 
this pro\-po\-sition, 
having only secondary role,
as we
already noted (see Sections 3 and 9), does not underlie relativity theory. Precisely in the same way, the
syn\-chro\-ni\-zation\index{synchronization of clocks} of clocks at different points of space, also, has a limited
sense, since it is possible only in inertial reference systems. \vspace*{-1mm} One cannot perform transition to
accelerated reference systems on the basis of the principle of the constancy
of the velocity of light, because the
concept of simultaneity\index{simultaneity} loses sense, since the syn\-chro\-ni\-zation of
clocks\index{synchronization of clocks} at different points in space depends on the synchronization path. The need
to describe effects by means of coordinate quantities arises.

We, now, define the
coordinate velocity of light\index{coordinate velocity of light}
\be
v^i=\f{dx^i}{dt}=v\ell^i,
\ee
here $\ell^i$ is a unit vector satisfying the condition
\be
\chi_{ik}\ell^i\ell^k=1.
\ee
With account for formulae (13.8), (13.10) and (13.25)
expression (13.12) assumes the following form
\be
\f{v}{\sqrt{\gamma_{00}}+\ds\f{\,v\,}{\,c\,}\cdot\f{\gamma_{0i}\,\ell^i}
{\sqrt{\gamma_{00}}}}=c.
\ee
Hence one finds the coordinate velocity\index{coordinate velocity of light}
\be
v=c\cdot\f{\sqrt{\gamma_{00}}}
{1-\ds\f{\gamma_{0i}\,\ell^i}{\sqrt{\gamma_{00}}}}.
\ee

In the general case, the coordinate
velocity\index{coordinate velocity of light}
varies, both
in value and in direction. It can take any value satisfying
the condition
\be
0<v<\infty.
\ee
In Galilean\index{Galilei} 
coordinates\index{Galilean (Cartesian) coordinates}
of an inertial reference system
coordinate velocity coincides with physical velocity.

In an arbitrary non-inertial reference system, for describing physical pro\-ces\-ses it is possible to introduce
unique coordinate time\index{coordinate time} throughout space in many ways. In this case, the
synchronization\index{synchronization of clocks} of clocks at different points in space must  be performed with
the aid of coordinate velocity. In non-inertial systems it is necessary to use {\bf coordinate
quantities}\index{coordinate quantities} in order to describe physical processes because in this case {\bf
physical quantities are determined only locally}.

\newpage
\markboth{thesection\hspace{1em}}{}
\section{Thomas precession}

\noindent
Consider a particle with its own angular momentum (spin) $S^\nu$.
In a reference system, where the particle is at rest, its
four-vector\index{four-vector}
of angular momentum (spin) has the components $(0, \vec J)$. In any
arbitrary inertial reference system we have the relation
\be
S^\nu U_\nu =0.
\ee
When a force $\vec f$ without torque acts on the particle,
the following relation should  be valid
\be
\f{dS^\nu}{d\tau}=ZU^\nu,
\ee
here $U^\nu$ is the
four-vector\index{four-vector}
of velocity; $\tau$ is
proper time,\index{proper time}
\be
d\tau=dt\f{\,1\,}{\,\gamma\,}\,.
\ee
If the velocity $U^i$ is not zero, then the quantity $Z$ can
be determined from the relation
\be
\f{d}{d\tau}(S^\nu U_\nu)=
\f{dS^\nu}{d\tau}U_\nu+
\f{dU_\nu}{d\tau}S^\nu=0.
\ee
Substituting (14.2) into (14.4), we obtain
\be
Z=-\left(S_\mu\f{dU^\mu}{d\tau}\right),
\ee
the covariant vector $S_\mu$\index{covariant vector}
has the components
\be
S_\mu=(S^0,-S^1,-S^2,-S^3).
\ee With account for Eq.~(14.5) the equation of motion for the spin vector (14.2)\index{vector, spin} assumes the
form \be
\f{dS^\nu}{d\tau}=-\left(S_\mu\f{dU^\mu}{d\tau}\right)U^\nu.
\ee \markboth{thesection\hspace{1em}}{14. Thomas precession}

Our further goal will be to try to provide the details of these
equations making use of the
Lorentz\index{Lorentz} 
transformations.\index{Lorentz transformations}
Consider a
particle of spin $\vec J$ moving with a velocity $\vec v$ in a
laboratory inertial reference system. In this case, the inertial
laboratory reference system will move with respect to the inertial
reference system, in which the particle is at rest, with a velocity
$-\vec v$. Applying the
Lorentz\index{Lorentz} 
transformations (4.18) and (4.19)\index{Lorentz transformations}
and taking the sign of the velocity into account, we obtain
\be
S^0=\gamma\f{(\vec v\vec J)}{c},\;
\vec S=\vec J+\f{\gamma -1}{v^2}\vec v(\vec v\vec J).
\ee
The four-vectors $\ds U^\mu,\index{four-vector}
\f{dU^\mu}{d\tau}$
have the following components:
\be
U^\mu=\left(\gamma, \gamma\f{\,\vec v\,}{\,c\,}\right),\;
\f{dU^\mu}{d\tau}=\left(
\f{d\gamma}{d\tau}, \f{\,\gamma\,}{\,c\,}\cdot
\f{d\vec v}{d\tau}+\f{\,\vec v\,}{\,c\,}\cdot
\f{d\gamma}{d\tau}\right).
\ee
Applying (14.6), (14.8) and (14.9), we obtain
\ba
&&\!\!\!\!\!\!\!\!\!\!\ds\left(S_\mu\f{dU^\mu}{d\tau}\right)=
\gamma\f{(\vec v\vec J)}{c}\cdot\f{d\gamma}{d\tau}-\nonumber\\
\label{14.10}\\
&&\!\!\!\!\!\!\!\!\!\!-\ds\left(\f{\,\gamma\,}{\,c\,}\cdot\f{d\vec v}{d\tau}+
\f{\,\vec v\,}{\,c\,}\cdot\f{d\gamma}{d\tau}\right)\cdot
\left(\vec J+\f{\gamma-1}{v^2}\vec v(\vec v\vec J)\right).\nonumber
\ea

Computations in the right-hand part of expression (14.10) will only
leave terms obtained by multiplication of the first term in brackets
and the two terms in the second pair of brackets, while all other
terms mutually cancel out
\be
\left(S_\mu\f{dU^\mu}{d\tau}\right)=-\f{\,\gamma\,}{\,c\,}
\left\{\left(\vec J\,\f{d\vec v}{d\tau}\right)+
\f{\gamma -1}{v^2}(\vec v\vec J)
\left(\vec v\f{d\vec v}{d\tau}\right)\right\}.
\ee Making use of (14.8) and (14.11), we write equation (14.7) se\-pa\-ra\-te\-ly for the zeroth component of the
four-vector\index{four-vector} of spin\index{vector, spin} $S^\nu$ and for its vector part, \be
\f{d}{d\tau}\left\{\gamma (\vec v\vec J)\right\}= \gamma^2\left\{\left(\vec J\,\f{d\vec v}{d\tau}\right)+
\f{\gamma -1}{v^2}(\vec v\vec J)
\left(\vec v\f{d\vec v}{d\tau}\right)\right\},
\ee
\ba
&&\!\!\!\!\!\!\!\!\!\ds\f{d}{d\tau}\left\{\vec J+\f{\gamma -1}{v^2}\vec v
(\vec v\vec J)\right\}=\nonumber\\
\label{14.13}\\
&&\!\!\!\!\!\!\!\!\!=\ds\f{\gamma^2}{c^2}\vec v
\left\{\left(\vec J\,\f{d\vec v}{d\tau}\right)+
\f{\gamma -1}{v^2}(\vec v\vec J)
\left(\vec v\f{d\vec v}{d\tau}\right)\right\}.\nonumber
\ea
From equations (14.12) and (14.13) we find
\be
\f{d}{d\tau}\left\{\vec J+\f{\gamma -1}{v^2}\vec v\,
(\vec v\vec J)\right\}-
\f{\vec v}{c^2}\f{d}{d\tau}
\left\{\gamma\, (\vec v\vec J)\right\}=0.
\ee
From equation (14.12) we find
\be
\gamma^2\f{(\gamma -1)}{v^2}(\vec v\vec J)
\left(\vec v\,\f{d\vec v}{d\tau}\right)=
\f{d}{d\tau}\left\{\gamma\, (\vec v\vec J)\right\}-
\gamma^2\left(\vec J\,\f{d\vec v}{d\tau}\right).
\ee Now we write the first term of equation (14.14) in expanded form \ba
&&\!\!\!\!\!\!\!\!\!\!\!\!\!\!\!\!\!\!\ds\f{d\vec J}{d\tau}+\f{\gamma^4}{c^4(1+\gamma)^2}
\vec v(\vec v\vec J)\left(\vec v\f{d\vec v}{d\tau}\right)+\nonumber\\
\label{14.16}\\
&&\!\!\!\!\!\!\!\!\!\!\!\!\!\!\!\!\!\!+\ds\f{\gamma\vec v}{c^2(1+\gamma)}\f{d}{d\tau}
\left\{\gamma (\vec v\vec J)\right\}+
\ds\f{\gamma^2}{c^2(1+\gamma)}(\vec v\vec J)
\f{d\vec v}{d\tau}.\nonumber
\ea
In computation we took into account the equalities
\be
\f{\gamma -1}{v^2}=\f{\gamma^2}{c^2(1+\gamma)},\;
\f{d\gamma}{d\tau}=\f{\gamma^3}{c^2}
\left(\vec v\f{d\vec v}{d\tau}\right).
\ee
The second term in (14.16) can be transformed, taking advantage
of (14.15), to the form
\ba
&&\!\!\!\!\!\!\!\!\!\!\!\!\!\!\!\!\!\!\ds\f{\gamma^4}{c^4(1+\gamma)^2}
\vec v(\vec v\vec J)\left(\vec v\f{d\vec v}{d\tau}\right)=\nonumber\\
\label{14.18}\\
&&\!\!\!\!\!\!\!\!\!\!\!\!\!\!\!\!\!\!=\ds\f{\vec v}{c^2(1+\gamma)}
\left[\f{d}{d\tau}\left\{\gamma (\vec v\vec J)\right\}-
\gamma^2\left(\vec J\,\f{d\vec v}{d\tau}\right)
\right].\nonumber
\ea
Applying (14.18) we see, that the second term together with the
third term in (14.16) can be reduced to the form
\be
\f{\vec v}{c^2}\cdot\f{d}{d\tau}\left\{\gamma (\vec v\vec J)\right\}-
\f{\gamma^2}{c^2(1+\gamma)}\vec v
\left(\vec J\f{d\vec v}{d\tau}\right).
\ee
With account of (14.16) and (14.19) equation (14.14) is reduced
to the following form:
\be
\f{d\vec J}{d\tau}+\f{\gamma^2}{c^2(1+\gamma)}
\left\{\f{d\vec v}{d\tau}(\vec v\vec J)-
\vec v\left(\vec J\f{d\vec v}{d\tau}\right)\right\}=0.
\ee
Using the formula
\be
\Bigl[\vec a\,[\vec b,\vec c\,]\Bigr]=\vec b\,(\vec a\vec c\,)-\vec c\,(\vec a\,\vec b\,),
\ee
and choosing the vectors
\be
\vec a=\vec J,\;\vec b=\f{d\vec v}{d\tau},\;\vec c=\vec v,
\ee
equation (14.20) is reduced to the form
\be
\f{d\vec J}{d\tau}=\Bigl[\vec \Omega \vec J\Bigr],
\ee
here
\be
\vec \Omega=-\f{\gamma -1}{v^2}
\left[\vec v,\f{d\vec v}{d\tau}\right].
\ee

{\bf When the particle moves along a curvilinear trajectory, the
spin vector\index{vector, spin}
{\mathversion{bold}\( \vec J \)}
undergoes precession around the direction {\mathversion{bold}\( \vec \Omega \)}
with angular velocity {\mathversion{bold}\( |\vec \Omega|. \)}
This effect was first discovered by\linebreak
Thomas}~\cite{15}.\index{Thomas} 

The equation of
relativistic\index{equations of relativistic mechanics}
mechanics (9.12) can be written in the
form
\be
m\f{d\vec v}{d\tau}=\vec f-\f{\vec v}{c^2}(\vec v\vec f).
\ee
With account of this equation, expression (14.24) assumes the form
\be
\vec \Omega=-\f{\gamma -1}{mv^2}[\vec v, \vec f\,].
\ee

Thus, a force without torque, by virtue of the 
pseudo-Euclidean\index{pseudo-Euclidean geometry of space-time}
structure of space-time, gives rise to the precession of spin, if its action results in curvilinear motion in the
given inertial reference system. In the case, when the force is directed, in a certain reference system, along the
velocity of the particle, no precession of the spin occurs. But parallelism of the vectors of force $\vec f$ and
of velocity $\vec v$ is violated, when transition is performed from one inertial reference system to another.
Therefore, the effect of precession, equal to zero for an observer in one inertial reference system, will differ
from zero for an observer in some other inertial reference  system.

\newpage
\markboth{thesection\hspace{1em}}{}
\section{The equations of motion and conservation laws
in classical field theory}

Earlier we saw that, with the aid of the Lagrangian approach, it is possible to construct all the
Maxwell-Lorentz\index{Lorentz}\index{Maxwell} 
equations.\index{Maxwell-Lorentz equations} This approach possesses an explicit general covariant character. It
per\-mits to obtain field equations of motion and conservation laws in a general form without explicit
concretization of the Lagrangian density function.\index{Lagrangian density} In this approach each physical field
is described by a one- or mul\-ti-\-com\-po\-nent function of coordinates and time, called the field function (or
field variable). As field variables, quan\-ti\-ti\-es are chosen that transform with respect to one of
 the linear representations of the Lorentz\index{Lorentz} 
group,\index{group}\index{group, Lorentz} for example, scalar, spinor, vector, or even tensor. Apart the field
variables, an important role is attributed, also, to the metric tensor of space-time, which determines the
geometry for the physical field, as well as the choice of one or another coordinate system, in which the
description of physical processes is performed. The choice of co\-or\-di\-nate system is, at the same time, a
choice of reference system. Naturally, not every choice of coordinate system alters the reference system. Any
trans\-for\-ma\-tions in a given reference system of the form \ba
&&x^{\prime\, 0}=f^0(x^0, x^1, x^2, x^3),\nonumber\\
\label{15.1}\\
&&x^{\prime\, i}=f^i(x^1, x^2, x^3),\nonumber
\ea
always leave us in this reference system. Any other choice of
coordinate system will necessarily lead to a change in reference
system. The choice of coordinate system is made from the class of
admissible coordinates,
\be
\gamma_{00}>0,\;\gamma_{ik}dx^i dx^k<0,\;
\det|\gamma_{\mu\nu}|=\gamma <0.
\ee
\markboth{thesection\hspace{1em}}{15. The equations of motion and
conservation laws  \ldots}
The starting-point of the Lagrangian\index{Lagrangian} formalism is
construction of the action function. Usually, the expression determining
the action function is written as follows
\be
S=\f{1}{c}\int\limits_\Omega L(x^0, x^1, x^2, x^3)dx^0 dx^1 dx^2 dx^3,
\ee
where integration is performed over a certain arbitrary four-di\-men\-si\-o\-nal
region of space-time. Since the action must be invariant, the Lagrangian\index{Lagrangian}
density function\index{Lagrangian density}
is the density of a scalar of weight $+1$. The density of
a scalar of weight $+1$ is the product of a scalar function and the
quantity $\sqrt{-\gamma}$. The choice of
Lagrangian\index{Lagrangian} density\index{Lagrangian density}
is performed
in accordance with a number of requirements. One of them is that the
lagrangian density\index{Lagrangian density}
must be real.

Thus, the Lagrangian density\index{Lagrangian density}\index{Lagrangian}
may be constructed with the aid of
the fields studied, $\varphi$, the
metric tensor\index{metric tensor of space}
$\gamma_{\mu\nu}$,
and partial derivatives with respect to the coordinates,
\be
L=L(\varphi_A, \pa_\mu \varphi_A, \ldots, \gamma_{\mu\nu},
\pa_\lambda \gamma_{\mu\nu}).
\ee
For simplicity we shall assume, that the system we are dealing with
consists of a real vector field. We shall consider the field
Lagrangian\index{Lagrangian}
not to contain derivatives of orders higher, than the first. This
restriction results in all our field equations being equations of the
second order,
\be
L=L(A^\nu, \pa_\lambda A^\nu, \gamma_{\mu\nu},
\pa_\lambda \gamma_{\mu\nu}).
\ee
Note, {\bf that, if the
Lagrangian\index{Lagrangian}
has been constructed, the theory
is defined}. We find the field equations from the least action principle.
\be
\delta S=\f{1}{c}\int
\limits_\Omega
d^4 x\delta L=0.
\ee
The variation $\delta L$ is
\be
\delta L=\f{\pa L}{\pa A_\lambda}\delta A_\lambda+
\f{\pa L}{\pa (\pa_\nu A_\lambda)}\delta
(\pa_\nu A_\lambda),\;{\rm or}
\ee
\be
\delta L=\f{\delta L}{\delta A_\lambda}\delta A_\lambda+
\pa_\nu \left[\f{\pa L}{\pa (\pa_\nu A_\lambda)}
\delta A_\lambda \right].
\ee
Here we have denoted
Euler's\index{Euler}
variational derivative\index{variational derivative}
by
\be
\f{\delta L}{\delta A_\lambda}=\f{\pa L}{\pa A_\lambda}-
\pa_\nu \left(\f{\pa L}{\pa (\pa_\nu A_\lambda)}
\right).
\ee
In obtaining expression (15.8) we took into account, that
\be
\delta (\pa_\nu A_\lambda)=\pa_\nu (\delta A_\lambda).
\ee
Substituting (15.8) into (15.6) and applying the
Gauss\index{Gauss} 
theorem\index{Gauss theorem}
we obtain
\[
\delta S=\f{\,1\,}{\,c\,}\int
\limits_\Omega d\Omega
d^4 x
\left(\f{\delta L}{\delta A_\lambda}\right)\delta A_\lambda +
\f{\,1\,}{\,c\,}\int\limits_\Sigma ds_\nu \left[\f{\pa L}{\pa (\pa_\nu A_\lambda)}
\delta A_\lambda \right].
\]
Since the field variation at the boundary $\Sigma$ is zero, we have
\be
\delta S=\f{\,1\,}{\,c\,}\int
\limits_\Omega d\Omega
d^4 x
\left(\f{\delta L}{\delta A_\lambda}\right)\delta A_\lambda =0.
\ee
Owing to the variations $\delta A_\nu$ being arbitrary, we obtain,
with the aid of the main lemma of variational calculus, the equation
for the field
\be
\f{\delta L}{\delta A_\lambda}=\f{\pa L}{\pa A_\lambda}-
\pa_\nu \left(\f{\pa L}{\pa (\pa_\nu A_\lambda)}
\right)=0.
\ee

{\bf We see, that if the
Lagrangian\index{Lagrangian}
has been found, then the theory has
been defined}. Besides field equations, the
Lagrangian\index{Lagrange} 
method\index{Lagrangian method}
provides
the possibility, also, to obtain
differential\index{differential conservation laws}
con\-ser\-vation laws:
{\bf strong} and {\bf weak}. {\bf A strong conservation law} is a
differential relation, that holds valid by virtue of the invariance of
action under the transformation of coordinates. Weak conservation laws
are obtained from strong laws, if the field equation (15.12) is taken
into account in them.

It must be especially stressed that, in the general case, strong
differential conservation laws do not establish the conservation of
anything, neither local, nor global. For our case the action has
the form
\be
S=\f{\,1\,}{\,c\,}\int\limits_\Omega d^4 x
L(A_\lambda, \pa_\nu A_\lambda, \gamma_{\mu\nu},
\pa_\lambda \gamma_{\mu\nu}).
\ee
Now, we shall perform infinitesimal transformation of the
co\-or\-di\-nates,
\be
x^{\prime\,\nu}=x^\nu +\delta x^\nu,
\ee
here $\delta x^\nu$ is an infinitesimal
four-vector.\index{four-vector}

Since action is a scalar, then in this transformation it remains
unaltered, and, consequently,
\be
\delta_c S=\f{\,1\,}{\,c\,}\int\limits _{\Omega^\prime}d^4 x^\prime L^\prime (x^\prime)
-\f{\,1\,}{\,c\,}\int\limits_\Omega d^4 x L(x)=0,
\ee
where
\[
L^\prime (x^\prime)=L^\prime \Bigl(A^\prime_\lambda,
\pa_\nu^\prime A_\lambda^\prime (x^\prime), \gamma_{\mu\nu}^\prime (x^\prime),
\pa_\lambda^\prime \gamma_{\mu\nu}^\prime (x^\prime)\Bigr).
\]
The first term in (15.15) can be written as
\be
\int\limits_{\Omega^\prime} d^4 x^\prime L^\prime (x^\prime)=
\int\limits_\Omega J d^4 x L^\prime (x^\prime),
\ee
where the
Jacobian\index{Jacobian}
of the transformation
\be
J=\f{\pa (x^{\prime\,0}, x^{\prime\,1},
x^{\prime\,2}, x^{\prime\,3})}
{\pa (x^0, x^1, x^2, x^3)}=
\det \left|\f{\pa x^{\prime \nu}}{\pa x^\lambda}\right|.
\ee
In the case of transformation (15.14)
the Jacobian\index{Jacobian}
has the form
\be
J=1+\pa_\lambda \delta x^\lambda.
\ee
Expanding $L^\prime (x^\prime)$ into a
Taylor\index{Taylor} 
series, we have
\be
L^\prime (x^\prime)=L^\prime (x)+\delta x^\lambda
\f{\pa L}{\pa x^\lambda}.
\ee
Taking (15.16), (15.18) and (15.19) into account, we rewrite
va\-ri\-a\-tion (15.15) as
\be
\delta_c S=\f{\,1\,}{\,c\,}\int\limits_\Omega d^4 x \left[\delta_L L(x)+
\f{\pa}{\pa x^\lambda}
\Bigl(\delta x^\lambda L(x)\Bigr)\right]=0;
\ee
here we have denoted
$$
\delta_L L(x)=L^\prime (x)-L(x).
$$
This variation is usually called the
Lie\index{Lie}
variation.\index{variation, Lie}
It
commutes with partial differentiation
\be
\delta_L \pa_\nu =\pa_\nu \delta_L.
\ee
The Lie\index{Lie}
variation\index{variation, Lie}
of the
Lagrangian\index{Lagrangian} 
density function\index{Lagrangian density}
is
\ba
&&\!\!\!\!\!\!\!\!\!\!\ds\delta_L L(x)=\f{\pa L}{\pa A_\lambda}\delta_L A_\lambda+
\f{\pa L}{\pa(\pa_\nu A_\lambda)}\delta_L\pa_\nu A_\lambda+\nonumber\\
\label{15.23}\\
&&\!\!\!\!\!\!\!\!\!\!+\ds\f{\pa L}{\pa \gamma_{\mu\nu}}\delta_L \gamma_{\mu\nu}+
\f{\pa L}{\pa (\pa_\lambda \gamma_{\mu\nu})}
\delta_L \pa_\lambda \gamma_{\mu\nu}.\nonumber
\ea
The following identity
\be
\delta_L L(x)+
\f{\pa}{\pa x^\lambda}
\bigl(\delta x^\lambda L(x)\bigr)=0,
\ee is a consequence of Eq.~(15.20) due to  arbitrariness of volume \( \Omega \). It was obtained by
D.\,Hilbert\index{Hilbert} in 1915.

Upon performing elementary transformations, we obtain
\ba
&&\!\!\!\!\!\!\!\!\!\!\!\!\!\!\!\!\!\!\!\!\!\delta_c S=\nonumber\\
\label{15.24}\\
&&\!\!\!\!\!\!\!\!\!\!\!\!\!\!\!\!\!\!\!\!\!=\ds\f{\,1\,}{\,c\,}\int\limits_\Omega d^4 x
\left[\f{\delta L}{\delta A_\lambda}\delta_L A_\lambda+
\f{\delta L}{\delta \gamma_{\mu\nu}}\delta_L\gamma_{\mu\nu}+
D_\lambda J^\lambda\right]=0,\nonumber
\ea
here
\[
\f{\delta L}{\delta \gamma_{\mu\nu}}=
\f{\pa L}{\pa \gamma_{\mu\nu}}-\pa_\sigma
\left(\f{\pa L}{\pa (\pa_\sigma \gamma_{\mu\nu})}\right),
\]
\be
J^\nu=L\delta x^\nu+\f{\pa L}
{\pa(\pa_\nu A_\lambda)}\delta_L A_\lambda+
\f{\pa L}
{\pa(\pa_\nu \gamma_{\lambda\mu})}\delta_L \gamma_{\lambda\mu}.
\ee Since $J^\nu$ is the density of a vector of weight $+1$, then, in accordance with (11.25) and (11.28), we find
\be
\pa_\nu J^\nu=D_\nu J^\nu,
\ee
where $D^\nu$ is a
covariant derivative\index{covariant derivative}
in pseudo-Euclidean\index{pseudo-Euclidean geometry of space-time}  space-time.
It must be pointed out, that the variations $\delta_L A_\lambda,
\delta_L \gamma_{\mu\nu}$ originate from the coordinate transformation
(15.14), so they can, therefore, be expressed via the components
$\delta x^\lambda$.

Let us find the
Lie\index{Lie}
variation\index{variation, Lie}
of field variables, that is due to
coordinate transformation. According to the transformation law of
the vector $A_\lambda$
\[
A_\lambda^\prime (x^\prime)=A_\nu (x)
\f{\pa x^\nu}{\pa x^{\prime \lambda}},
\]
we have
\be
A_\lambda^\prime (x+\delta x)=A_\lambda (x)-
A_\nu (x)\f{\pa \delta x^\nu}{\pa x^\lambda}.
\ee
Expanding the quantity $A_\lambda^\prime (x+\delta x)$ in a
Taylor\index{Taylor} 
series, we find
\be
A_\lambda^\prime (x+\delta x)=A_\lambda^\prime (x)+
\f{\pa A_\lambda}{\pa x^\nu}\delta x^\nu.
\ee
Substituting (15.28) into (15.27) we obtain
\be
\delta_L A_\lambda (x)=-\delta x^\nu\f{\pa A_\lambda}{\pa x^\nu}
-A_\nu (x)\f{\pa \delta x^\nu}{\pa x^\lambda},
\ee
or, in covariant form
\be
\delta_L A_\lambda (x)=-\delta x^\nu D_\nu A_\lambda -A_\nu D_\lambda \delta x^\nu.
\ee

Now let us find the
Lie\index{Lie}
variation\index{variation, Lie}
of the
metric tensor\index{metric tensor of space}
$\gamma_{\mu\nu}$ from the transformation law
\[
\gamma_{\mu\nu}^\prime (x^\prime)=
\f{\pa x^\lambda}{\pa x^{\prime \mu}}\cdot
\f{\pa x^\sigma}{\pa x^{\prime \nu}}
\gamma_{\lambda\sigma}(x)
\]
we obtain
\be
\gamma_{\mu\nu}^\prime (x+\delta x)=
\gamma_{\mu\nu}-\gamma_{\mu\sigma}\pa_\nu \delta x^\sigma-
\gamma_{\nu\sigma}\pa_\mu \delta x^\sigma,
\ee
hence we find
\be
\delta_L \gamma_{\mu\nu}=-
\gamma_{\mu\sigma}\pa_\nu \delta x^\sigma-
\gamma_{\nu\sigma}\pa_\mu \delta x^\sigma-
\delta x^\sigma \pa_\sigma \gamma_{\mu\nu}.
\ee
Taking into account the equality
\be
\pa_\sigma \gamma_{\mu\nu}=\gamma_{\mu\lambda}
\Gamma_{\nu\sigma}^\lambda+\gamma_{\nu\lambda}
\Gamma_{\mu\sigma}^\lambda,
\ee we write expression (15.32) through covariant derivatives,\index{covariant derivative} \be \delta_L
\gamma_{\mu\nu}=- \gamma_{\mu\sigma}D_\nu \delta x^\sigma-
\gamma_{\nu\sigma}D_\mu \delta x^\sigma.
\ee
Substituting expressions (15.30) and (15.34) into
action (15.24), we obtain
\ba
&&\!\!\!\!\!\!\!\!\!\!\!\!\!\!\!\!\!\!\!\!\!\delta_c S=\ds\f{\,1\,}{\,c\,}\int\limits_\Omega d^4 x
\Biggl[-\delta x^\lambda \f{\delta L}{\delta A_\nu}D_\lambda A_\nu-
\f{\delta L}{\delta A_\lambda}A_\nu D_\lambda \delta x^\nu-\Biggr.\nonumber\\
\label{15.35}\\
&&\!\!\!\!\!\!\!\!\!\!\!\!\!\!\!\!\!\!\!\!\!-(\gamma_{\mu\sigma}D_\nu \delta x^\sigma+\gamma_{\nu\sigma}D_\mu
 \delta x^\sigma)
\ds\f{\delta L}{\delta \gamma_{\mu\nu}}+D_\nu J^\nu \Biggl.\Biggr]=0.\nonumber
\ea
We introduce the following notation:
\be
T^{\mu\nu}=-2\f{\delta L}{\delta \gamma_{\mu\nu}}.
\ee
As we will further see, {\bf this quantity, first introduced by
Hilbert,\index{Hilbert} 
is the tensor
density\index{density of  Hilbert energy-momentum tensor}
of the field energy-mo\-men\-tum}.

Integrating by parts in expression (15.35) we obtain
\ba
&&\!\!\!\!\!\!\!\!\!\!\!\!\!\!\!\delta_c S=\ds\f{\,1\,}{\,c\,}\int\limits_\Omega d^4 x \Biggl\{-\delta x^\lambda
\biggl[\f{\delta L}{\delta A_\nu}D_\lambda A_\nu-\biggr.\Biggr.\nonumber\\
&&\!\!\!\!\!\!\!\!\!\!\!\!\!\!\!-D_\nu \left(\f{\delta L}{\delta A_\nu}A_\lambda\right)+\biggr.\Biggr.
D_\nu\left(T^{\mu\nu}\gamma_{\mu\lambda}\right)\biggr]+\Biggr.\label{15.37}\\
&&\!\!\!\!\!\!\!\!\!\!\!\!\!\!\!+D_\nu\biggl(J^\nu -\f{\delta L}{\delta A_\nu}
A_\lambda \delta x^\lambda +T^{\mu\nu}
\gamma_{\mu\sigma}\delta x^\sigma \biggr)\Biggl.\Biggr\}=0.\nonumber
\ea
Substituting into expression (15.25) for the density of vector $J^\nu$
the values of variations
$\delta_L A_\lambda (x), \delta_L \gamma_{\mu\nu} (x)$, in accordance
with for\-mu\-lae (15.30) and (15.34), and grouping the terms at
$\delta x^\nu$ and $D_\lambda \delta x^\nu$, we obtain
\be
J^\nu -\f{\delta L}{\delta A_\nu}A_\lambda \delta x^\lambda=-
\tau_\sigma^\nu \delta x^\sigma -\sigma_\mu^{\nu\lambda}
D_\lambda \delta x^\mu,
\ee here we denote \be \tau_\sigma^\nu=-L \delta_\sigma^\nu+ \f{\pa L}{\pa (\pa_\nu A_\lambda)}D_\sigma A_\lambda+
\f{\delta L}{\delta A_\nu}A_\sigma.
\ee
This quantity is usually {\bf called the density of the
canonical\index{density of canonical energy-momentum tensor}
energy-mo\-men\-tum tensor}, while the quantity
\be
\sigma^{\nu\lambda}_\mu =2\f{\pa L}
{\pa (\pa_\nu \gamma_{\sigma\lambda})}\gamma_{\sigma\mu}+
\f{\pa L}{\pa (\pa_\nu A_\lambda)}A_\mu
\ee
is called the
spin tensor density.\index{density of spin tensor}

If function \( L \) depends only on \( \gamma^{\mu\nu}, A_\mu, \pa_\nu A_\mu \), then quantity \(
\sigma_\mu^{\nu\lambda} \) according to Eq.~(15.40)  may be written as follows
\begin{gather}
\sigma_\mu^{\nu\lambda}=
\left(\f{\pa L}{\pa (\pa_\nu A_\lambda)}\right)A_\mu.\tag{15.40a}%
\end{gather}

On the basis of (15.38) we represent the covariant divergence in
(15.37) as
\ba
&&\!\!\!\!\!\!\!\!\!\!\!\!\!\!\!D_\nu\biggl(J^\nu -\f{\delta L}{\delta A_\nu}
A_\lambda \delta x^\lambda +T^\nu_\sigma \delta x^\sigma \biggr)=\nonumber\\
&&\!\!\!\!\!\!\!\!\!\!\!\!\!\!\!-\delta x^\sigma\biggl[D_\nu T_\sigma^\nu- D_\nu \tau_\sigma^\nu\biggr]+
D_\nu (\delta x^\lambda)\times\label{15.41}\\
&&\!\!\!\!\!\!\!\!\!\!\!\!\!\!\!\times\biggl[T_\lambda^\nu -\tau_\lambda^\nu \biggr.
-D_\mu \sigma_\lambda^{\mu\nu}\biggr]-
\sigma_\mu^{\nu\lambda}D_\nu D_\lambda \delta x^\mu.\nonumber
\ea
Taking advantage of this expression, the variation of action (15.37)
can be written in the form
\ba
&&\!\!\!\!\!\!\!\!\!\!\!\!\!\!\!\delta_c S=\ds\f{\,1\,}{\,c\,}\int\limits_\Omega d^4 x \Biggl[-\delta x^\lambda
\biggl(\f{\delta L}{\delta A_\nu}D_\lambda A_\nu-\biggr.\nonumber\\
&&\!\!\!\!\!\!\!\!\!\!\!\!\!\!\!-D_\nu \left(\f{\delta L}{\delta A_\nu}A_\lambda\right)+
D_\nu \tau_\lambda^\nu\biggr)+D_\nu (\delta x^\lambda)\times\label{15.42}\\
&&\!\!\!\!\!\!\!\!\!\!\!\!\!\!\!\times\left(T_\lambda^\nu -\tau_\lambda^\nu-
D_\mu \sigma_\lambda^{\mu\nu}\right)-
\sigma_\mu^{\nu\lambda} D_\nu D_\lambda \delta x^\mu \Biggl.\Biggr]=0.\nonumber
\ea
Since the integration volume $\Omega$ is arbitrary, it hence
follows that the integrand function is zero.
\ba
&&\!\!\!\!\!\!\!\!\!\!\!\!\!\!\!\!\!\!\!\!\!\!\!\!\!\!-\delta x^\lambda \Biggl(\f{\delta L}{\delta A_\nu}D_\lambda A_\nu
-D_\nu\biggl(\f{\delta L}
{\delta A_\nu}A_\lambda \biggr)+D_\nu \tau_\lambda^\nu\Biggr)+\nonumber\\
\label{15.43}\\
&&\!\!\!\!\!\!\!\!\!\!\!\!\!\!\!\!\!\!\!\!\!\!\!\!\!\!+\biggl(T_\lambda^\nu-
\tau_\lambda^\nu-D_\mu\sigma_\lambda^{\mu\nu}\biggr)
D_\nu \delta x^\lambda -\sigma_\mu^{\nu\lambda}D_\nu D_\lambda \delta x^\mu =0.\nonumber \ea This expression turns
to zero for arbitrary $\delta x^\lambda$ independently of the choice of coordinate system. Precisely this permits
to readily establish that the tensor $\sigma_\mu^{\nu\lambda}$ is antisymmetric with respect to $\nu, \lambda$.
Due to antisymmetry of quantity \( \sigma_\mu^{\nu\lambda} \) in upper indices \( \nu \), \( \lambda \) we get
from Eq.~(15.40а) the following
\[
\left(\f{\pa L}{\pa (\pa_\nu A_\lambda)}
+\f{\pa L}{\pa (\pa_\lambda A_\nu)}\right)=0.
\]
It follows from the above that function \( L \) depends on derivatives in this case as follows
\[
L(F_{\nu\lambda}),
\]
где
\[
F_{\nu\lambda}=D_\nu A_\lambda - D_\lambda A_\nu.
\]
This result was obtained by D.\,Hilbert\index{Hilbert} in 1915. Of course, this does not exclude an explicit dependence of
\( L \) on variable \( A_\nu \).

By virtue of the tensor transformation law, if it becomes zero in one coordinate system, then it is equal to zero in
any other coordinate system. Hence the identities follow: \be D_\nu \tau_\lambda^\nu+\f{\delta L}{\delta
A_\nu}D_\lambda A_\nu
-D_\nu\left(\f{\delta L}{\delta A_\nu}A_\lambda \right)=0.
\ee
\be
T_\lambda^\nu -\tau_\lambda^\nu -D_\mu \sigma_\lambda^{\mu\nu}=0,\;
\sigma_\mu^{\nu\lambda}=-\sigma_\mu^{\lambda\nu}.
\ee
As to the last term in (15.43), it should become zero owing to the
quantities $\sigma_\mu^{\nu\lambda}$ being antisymmetric with respect to
the upper indices. From the antisymmetry of the spin tensor follows
\be
D_\nu T_\lambda^\nu=D_\nu \tau_\lambda^\nu.
\ee
Identities (15.44) and (15.45) are called strong conservation laws,
they are obeyed by virtue of action being invariant under coordinate
transformations. Applying relation (15.46), expression (15.44) can be
written in the form
\ba
&&D_\nu T_\lambda^\nu+\f{\delta L}{\delta A_\nu}F_{\lambda\nu}
-A_\lambda D_\nu \left(\f{\delta L}{\delta A_\nu} \right)=0,\nonumber\\
\label{15.47}\\
&&F_{\lambda\nu}=D_\lambda A_\nu -D_\nu A_\lambda.\nonumber
\ea
If we take into account the field equations (15.12), we will obtain
\be
D_\nu T_\lambda^\nu=0,\;T_\lambda^\nu-\tau_\lambda^\nu=
D_\mu \sigma_\lambda^{\mu\nu},
\ee
here the quantity $\tau_\lambda^\nu$ equals
\be
\tau_\lambda^\nu=-L\delta_\lambda^\nu+
\f{\pa L}{\pa (\pa_\nu A_\mu)}
D_\lambda A_\mu.
\ee
The existence of a weak conservation law of the symmetric
energy-momentum tensor provides for conservation of the field
angular momentum tensor. By defining the angular momentum
tensor in
Galilean\index{Galilei}
coordinates\index{Galilean (Cartesian) coordinates}
of an inertial reference system
\be
M^{\mu\nu\lambda}=x^\nu T^{\mu\lambda}-x^\mu T^{\nu\lambda},
\ee
it is easy, with the aid of (15.48), to establish that
\be
\pa_\lambda M^{\mu\nu\lambda}=0.
\ee

The weak conservation laws we have obtained for the energy-mo\-men\-tum
tensor and for the angular momentum tensor do not yet testify in favour
of the conservation of energy-momentum or angular momentum for a closed
system.

{\bf The existence of integral conservation 
laws\index{integral conservation laws} for a closed sys\-tem is due to
the properties of space-time, namely, to the exis\-tence of the group\index{group} of space-time motions. The
existence of the Poincar\'{e}\index{Poincar\'e}  group\index{group} {\rm (}the
Lorentz\index{Lorentz}  group\index{group}\index{group, Lorentz} together with the
group\index{group} of translations{\rm )} for pseudo-Euclidean\index{pseudo-Euclidean geometry of space-time}
space provides for the existence of the conservation laws of energy, momentum and angular momentum for a closed
system} [6]. The group\index{group} of space-time motion provides  form-invariance of the metric
tensor\index{metric tensor of space} $\gamma_{\mu\nu}$ of 
Minkowski\index{Minkowski} 
space.\index{Minkowski space}

Let us consider this in more detail.
The density of substance energy-momentum 
tensor\index{Hilbert energy-momentum tensor density}
according to Eq.~(15.36) is the following
\be
T^{\mu\nu}=-2\f{\delta L}{\delta \gamma_{\mu\nu}},
\ee
\[
\f{\delta L}{\delta\gamma_{\mu\nu}}
=\f{\pa L}{\pa \gamma_{\mu\nu}}-\pa_\sigma
\left(\f{\pa L}{\pa\gamma_{\mu\nu,\sigma}}\right).
\]
This tensor density\index{Hilbert energy-momentum tensor density} satisfies
тензора удовлетворяет соотношению Eq.~(15.48)
\be
D_\nu T^{\mu\nu}=0,
\ee
that may be written as follows
\be
\pa_\nu T_\mu^\nu+\f{1}{2}T_{\sigma\nu}\pa_\mu g^{\sigma\nu}=0.
\ee
In general case Eq.~(15.53) could not be written as an equality of an ordinary divergence  to zero, and so it
does not demonstrate any conservation law.
But an expression of the form
\be
D_\nu A^\nu,
\ee
where \( A^\nu \) is an arbitrary vector, is easy to convert into a divergence form even in the Riemannian space.

From Eq.~(11.25) one has
\be
D_\lambda A^\lambda=\pa_\lambda A^\lambda
+{\mit\Gamma}_{\mu\lambda}^\lambda A^\mu.
\ee
By means of Eq.~(11.38) one obtains
\be
D_\lambda (\sqrt{-\gamma}\,A^\lambda)
=\pa_\lambda (\sqrt{-\gamma}\,A^\lambda).
\ee Let us exploit this below. Multiply the energy-momentum 
density\index{Hilbert energy-momentum tensor density}
onto vector
 \( \eta_\nu \)
\be
T^{\mu\nu}\eta_\nu.
\ee
According to Eq.~(15.57) we obtain
\be
D_\mu (T^{\mu\nu}\eta_\nu)
=\pa_\mu (T^{\mu\nu}\eta_\nu).
\ee
Quantity (15.58) already is a vector density in our case.
Therefore we should not substitute \( \sqrt{-\gamma} \) into Eq.~(15.59).
We rewrite Eq.~(15.59) in the following form
\be
\f{\,1\,}{\,2\,}T^{\mu\nu}(D_\mu\eta_\nu+D_\nu\eta_\mu)
=\pa_\mu (T^{\mu\nu}\eta_\nu).
\ee
After integration of Eq.~(15.60) over volume containing the substance we get
\be
\f{\,1\,}{\,2\,}\int\limits_V dV T^{\mu\nu}(D_\mu \eta_\nu +D_\nu \eta_\mu)
=\f{\pa}{\pa x^0} \int\limits_V (T^{\nu 0}\eta_\nu)dV.
\ee
If vector \( \eta_\nu \) fulfils the Killing\index{Killing} 
equation\index{Killing equation}
\be
D_\mu \eta_\nu +D_\nu \eta_\mu =0,
\ee
then we have  integral of motion
\be
\int\limits_V T^{\nu\, 0}\eta_\nu dV={\rm const}.
\ee
We have already derived Eq.~(15.34):
\be
\delta_L \gamma_{\mu\nu}
=-(D_\nu \delta x_\mu +D_\mu \delta x_\nu).
\ee
From Eqs.~(15.62) it follows that if they are fulfilled, then the metric is form-invariant
\be
\delta_L \gamma_{\mu\nu}=0.
\ee In case of pseudo-Euclidean (Minkowski\index{Minkowski} space\index{Minkowski space})
geometry\index{pseudo-Euclidean geometry of space-time}\break Eqs.~(15.62) may be written in a Galilean
(Cartesian)
coordinate system\index{Galilean (Cartesian) coordinates}: \be
\pa_\mu \eta_\nu +\pa_\nu \eta_\mu =0.
\ee
This equation has the following general solution
\be
\eta_\nu=a_\nu +\omega_{\nu\sigma} x^\sigma,\quad \omega_{\nu\sigma}
=-\omega_{\sigma\nu},
\ee
containing ten arbitrary parameters \( a_\nu, \omega_{\mu\nu}\).
This means that there are ten independent Killing\index{Killing} 
vectors\index{Killing vector}, and so
there are ten integrals of motion.
Taking
\be
\eta_\nu=a_\nu
\ee
and substituting this to Eq.~(15.63), one finds four integrals of motion:
\be
P^\nu=\f{\,1\,}{\,c\,}\int\limits_V T^{\nu\,0}dV={\rm const}.
\ee
Here \( P^0 \) is the system energy, and \( P^i \) is the momentum of the system.
Taking  Killing\index{Killing} vector\index{Killing vector} in the following form
\be
\eta_\nu=\omega_{\nu\sigma}x^\sigma
\ee
and substituting it in the initial expression (15.63), one gets the following expression
for the angular momentum tensor:
\be
P^{\sigma\nu}=\f{\,1\,}{\,c\,}\int (T^{\nu\,0}x^\sigma
-T^{\sigma\,0}x^\nu)dV.
\ee Quantities \( P^{i\,0} \) are center of mass integrals of motion, and
 \( P^{i\,k} \) are angular momentum integrals of motion.

In correspondence with Eq.~(15.50) we introduce the following quantity
\be
P^{\sigma\nu\lambda}=\f{\,1\,}{\,c\,}\int (T^{\nu\lambda}x^\sigma
-T^{\sigma\lambda}x^\nu)dV,
\ee
where
\be
M^{\sigma\nu\lambda}=T^{\nu\lambda}x^\sigma
-T^{\sigma\lambda}x^\nu
\ee
is tensor density, satisfying the following condition
\be
\pa_\lambda M^{\sigma\nu\lambda}=0.
\ee Therefore, we have been convinced, by deriving Eqs.~(15.69) and (15.71), that all these ten integrals of
motion arise on the base of pseudo-Euclidean geometry of 
space-time\index{pseudo-Euclidean geometry of space-time}.
Namely this geometry possesses 
ten independent Killing\index{Killing} 
vectors.\index{Killing vector} There may be also ten
Killing\index{Killing} vectors\index{Killing vector} in a Riemannian space, but only in case of a constant curvature
space [6].

Note that conservation laws are automatically satisfied for an arbitrary scalar (Lagrangian)
density\index{Lagrangian density} of the form $L(\psi_\lambda,\pa_\sigma\psi_\mu)$ in 
Minkowski\index{Minkowski} space,\index{Minkowski space} that provides for the field energy being positive,
if we only consider second-order field equations. I especially recall this here, since from discussions with
certain Academicians wor\-king in theoretical physics, I have seen, that this is unknown even to  them.

Now let us find, as an example, the symmetric tensor of the
electromagnetic field energy-momentum. According to (10.5) the
Lagrangian\index{Lagrangian}
density\index{Lagrangian density}
for this field is
\be
L_f=-\f{1}{16\pi}\sqrt{-\gamma}\,F_{\alpha\beta}F^{\alpha\beta}.
\ee
We write it in terms of the variables $F_{\mu\nu}$ and the metric
coefficients
\be
L_f=-\f{1}{16\pi}\sqrt{-\gamma}\,F_{\alpha\beta}
F_{\mu\nu}\gamma^{\alpha\mu}\gamma^{\beta\nu}.
\ee
According to (11.37) we have
\be
\f{\pa \sqrt{-\gamma}}{\pa \gamma_{\mu\nu}}
=\f{\,1\,}{\,2\,}\sqrt{-\gamma}\,\gamma^{\mu\nu}.
\ee
With the aid of (15.77) we obtain
\be
\f{\pa^\ast L}{\pa \gamma_{\mu\nu}}
=-\f{1}{32\pi}\sqrt{-\gamma}\,\gamma^{\mu\nu}
F_{\alpha\beta}F^{\alpha\beta},
\ee
$\ast$ indicates that differentiation is performed with respect to
$\gamma_{\mu\nu}$, present in expression (15.76).

Similarly
\ba
\ds\f{\pa^\ast L}{\pa \gamma^{\mu\nu}}
=-\f{1}{16\pi}\sqrt{-\gamma}\,F_{\alpha\beta}F_{\sigma\lambda}\times\nonumber\\
\label{15.56}\\
\times\ds\left[\f{\pa \gamma^{\alpha\sigma}}{\pa \gamma^{\mu\nu}}
\cdot\gamma^{\beta\lambda}
+\gamma^{\alpha\sigma}\cdot
\f{\pa \gamma^{\beta\lambda}}
{\pa \gamma^{\mu\nu}}\right].\nonumber
\ea
Since
\[
\f{\pa \gamma^{\alpha\sigma}}{\pa \gamma^{\mu\nu}} =\f{\,1\,}{\,2\,}(\delta_\mu^\alpha \delta_\nu^\sigma
+\delta_\nu^\alpha\delta_\mu^\sigma),\;
\]
then using the antisymmetry properties of the tensor $F_{\alpha\beta}=-F_{\beta\alpha}$, we obtain \be \f{\pa^\ast
L}{\pa \gamma^{\mu\nu}} =-\f{1}{8\pi}\sqrt{-\gamma}\,
F_{\mu\lambda}F_{\nu\sigma}\gamma^{\lambda\sigma}.
\ee In obtaining (15.78) and (15.80) we considered  quantities $\gamma_{\mu\nu},\gamma^{\lambda\sigma}$ as
independent.

Since no derivatives of the metric tensor\index{metric tensor of space} are present in the density of the
electromagnetic field Lagrangian, the density\index{Lagrangian density} of the sym\-met\-ric energy-momentum
tensor will be \be T^{\mu\nu}=-2\f{\pa L}{\pa \gamma_{\mu\nu}}=-2 \left[\f{\pa^\ast L}{\pa \gamma_{\mu\nu}}
+\f{\pa^\ast L}{\pa \gamma^{\alpha\beta}}\cdot
\f{\pa \gamma^{\alpha\beta}}{\pa \gamma_{\mu\nu}}\right].
\ee
From the relation
\be
\gamma^{\alpha\beta}\gamma_{\beta\nu}=\delta_\nu^\alpha,
\ee
we find
\be
\f{\pa \gamma^{\alpha\beta}}{\pa \gamma_{\mu\nu}}
=-\f{1}{2}(\gamma^{\alpha\mu}\gamma^{\beta\nu}
+\gamma^{\alpha\nu}\gamma^{\beta\mu}).
\ee
Substituting this expression into (15.81), we obtain
\be
T^{\mu\nu}=-2\f{\pa L}{\pa \gamma_{\mu\nu}}=-2
\left[\f{\pa^\ast L}{\pa \gamma_{\mu\nu}}
-\f{\pa^\ast L}{\pa \gamma^{\alpha\beta}}
\gamma^{\alpha\mu}\gamma^{\beta\nu}\right].
\ee
Using expressions (15.78) and (15.80) we find the density of
the energy-momentum tensor of the electromagnetic field\label{e-m-tensor}
\be
T^{\mu\nu}=\f{\sqrt{-\gamma}}{4\pi}\left[
-F^{\mu\sigma} F^{\nu\lambda} \gamma_{\sigma\lambda}
+\f{1}{4}\gamma^{\mu\nu} F_{\alpha\beta} F^{\alpha\beta}
\right].
\ee Hence it is readily verified, that the trace of the electromagnetic field energy-momentum tensor turns to
zero, i.\,e.
\[
T=\gamma_{\mu\nu}T^{\mu\nu}=0.
\]

We shall now construct the energy-momentum tensor of substance. The density of the conserved mass or charge is \be
\mu=\sqrt{-\gamma}\,\mu_0 U^0,\;
\pa_\nu (\sqrt{-\gamma}\,\mu_0 U^\nu)=0,
\ee due to Eq.~(11.41), where $\mu_0$ is the density in the rest reference system. The four-dimensional velocity
$U^\nu$ is defined by the expression \be U^\nu=\f{v^\nu}{\sqrt{\gamma_{\alpha\beta} v^\alpha v^\beta}},\;
v^\nu=\f{dx^\nu}{dt},\;v^0 \equiv c.
\ee
Hence, it is clear that
\[
U^\nu U^\lambda \gamma_{\nu\lambda}=1.
\]

Take the variation of expression (15.86) with respect to the
metric tensor.\index{metric tensor of space}
The quantity $\mu$ is independent of the metric
tensor, therefore,
\be
\delta \mu=U^0 \delta (\sqrt{-\gamma}\,\mu_0)
+\sqrt{-\gamma}\,\mu_0 \delta U^0=0,
\ee
here
\be
\delta U^0=-\f{c}{2}\f{v^\alpha v^\beta \delta \gamma_{\alpha\beta}}
{(\gamma_{\alpha\beta} v^\alpha v^\beta)^{3/2}}.
\ee
From expression (15.88) and (15.89) we find
\be
\delta (\sqrt{-\gamma}\,\mu_0)=
\sqrt{-\gamma}\,\mu_0\f{1}{2}
U^\alpha U^\beta \delta \gamma_{\alpha\beta}.
\ee Since the density of the Lagrangian\index{Lagrangian} of substance has the form \be
L=-\sqrt{-\gamma}\,\mu_0 c^2,
\ee the density of the energy-momentum tensor of substance can be determined as \be
t^{\mu\nu}=-2\f{\pa L}{\pa \gamma_{\mu\nu}}.
\ee
On the basis of (15.90) we obtain
\be
t^{\mu\nu}=\,\mu_0 c^2 U^\mu U^\nu.
\ee
Taking into account Eq.~(15.86) we obtain in Cartesian coordinate 
system\index{Galilean (Cartesian) coordinates}:
\be
\pa_\nu t^{\mu\nu}
=\mu_0 c^2\f{\pa U^\mu}{\pa x^\nu}\cdot\f{dx^\nu}{ds}
=\mu_0 c^2\f{dU^\mu}{ds}.
\ee
Let us rewrite Eq.~(10.22) for  mass and charge densities:
\be
\mu_0 c^2\f{dU^\mu}{ds}
=\rho_0 F^{\mu\lambda}U_\lambda =f^\mu.
\ee
After comparing Eqs.~(15.94) and  (15.95) we have
\be
f_\nu=\pa_\alpha t_\nu^\alpha.
\ee
From Eqs.~(8.54) and (15.96) we can see that the law of energy-momentum tensor conservation\index{fundamental conservation laws} for electromagnetic field and sources of charge taken together takes place:
\be
\pa_\alpha (T_\nu^\alpha +t_\nu^\alpha)=0.
\ee

As we noted above, addition to the
Lagrangian density\index{Lagrangian density}
of
a covariant divergence does not alter the field equations.
It is also possible to show~\cite{6}, that it does not alter
the density of the
Hilbert\index{Hilbert} energy-momentum 
tensor,\index{density of  Hilbert energy-momentum tensor}
as well.
On the contrary, the density of the
canonical\index{density of canonical energy-momentum tensor}
tensor (15.49)
does change. But at the same time the divergence of the spin
tensor density\index{density of spin tensor}
changes with it, also. The sum of the canonical
tensor
density and of the divergence of the spin density remains 
intact.

\newpage
\markboth{thesection\hspace{1em}}{}
\section{Lobachevsky velocity space}

Let us remind that the relativistic law of composition of velocities (see Eq.~(9.26)) has the following form:
\be
1-\f{v^{\prime\, 2}}{c^2}=
\f{\left(1-\ds\f{v^2}{c^2}\right)\left(1-\ds\f{u^2}{c^2}\right)}
{\left(1-\ds\f{\vec v\,\vec u}{c^2}\right)^2}.
\ee
Note that this expression is a direct consequence of the existence of the following  invariant
\[
\gamma_u \gamma_v (1-\vec u\vec v)={\rm inv}.
\]
Here \( \gamma_u=(1-u^2)^{-1/2} \),\;\( \gamma_v=(1-v^2)^{-1/2} \).

This invariant has been demonstrated first in the H.\,Poincar\'e\index{Poincar\'e}  article [3] (see \S\,9,
Eq.~(5)),
 where the system of units is taken so that  velocity of light is equal to  1.

It follows just from here that in pseudo-Euclidean space-time the velocity space follows the
Lobachevsky\index{Lobachevsky} geometry\index{Lobachevsky geometry}.

For the further presentation it will be more convenient
to introduce the following notation:
\be
v^\prime=v_a,\;v=v_b,\;u=v_c,
\ee \be \cosh a=\f{1}{\sqrt{1-\ds\f{v_a^2}{c^2}}},\; \sinh a=\f{v_a}{c\,\sqrt{1-\ds\f{v_a^2}{c^2}}},\;
\tanh a=\f{v_a}{c}.
\ee
Substituting (16.2) and (16.3) into (16.1) we obtain
\be
\cosh a=\cosh b\cdot\cosh c-\sinh b\cdot\sinh c\cdot\cos A,
\ee
$A$ is the angle between the velocities $\vec v_b$ and $\vec v_c$.
This is actually nothing, but {\bf the
law of cosines\index{law of cosines}
for a
triangle in
Lobachevsky's\index{Lobachevsky} 
geometry}. It expresses the length of
a side of a triangle in terms of the lengths of the two other sides
and the angle between them. Finding, hence, $\cos A$ and, then,
$\sin A$ etc., one thus establishes the
{\bf law of sines\index{law of sines}
of
the Lo\-ba\-chev\-sky\index{Lobachevsky} 
geometry}\index{Lobachevsky geometry}
\be
\f{\sin A}{\sinh a}=\f{\sin B}{\sinh b}=\f{\sin C}{\sinh c}.
\ee

Below, following
Lobachevsky,\index{Lobachevsky} 
we shall obtain the {\bf law of
co\-si\-nes\index{law of cosines}
for a triangle} in the form
\be
\cos A=-\cos B\cos C+\sin B\sin C\cosh a.
\ee
We write (10.4) in the form
\be
\tanh b\tanh c\cos A=1-\f{\cosh a}{\cosh b\cosh c}.
\ee \markboth{thesection\hspace{1em}}{16. Lobachevsky velocity space} From the law of sines (16.5)
\index{law of sines} we have \be \f{1}{\cosh c}=\f{\sin A}{\sin C}\cdot
\f{\tanh c}{\sinh a}.
\ee Substituting this expression into (16.7) we find \be \tanh b\tanh c\cos A=1-\f{\sin A}{\sin C}\cdot
\f{\tanh c}{\cosh b\tanh a}.
\ee Hence we find $\tanh c$ \be
\tanh c=\f{\tanh a\sin C}{\cos A\sin C\tanh a\tanh b+\ds\f{1}{\cosh b}\sin A}.
\ee With the aid of the law of cosines,\index{law of cosines} 
Lobachevsky\index{Lobachevsky} 
further established the identity \be (1-\tanh b\tanh c\cos A)(1-\tanh a\tanh b\cos C)=
\f{1}{\cosh^2 b}.
\ee Applying (16.10), we find \be 1-\tanh b\,\tanh c\cos A=\f{\ds\f{1}{\cosh b}\sin A}
{\ds\cos A\sin C\tanh a\tanh b+\f{1}{\cosh b}\sin A}.
\ee Substitution of this expression into identity (16.11) yields \be \f{1}{\cosh b}=\f{\sin A-\sin A\cos C\tanh
a\tanh b}
{\ds\cos A\sin C\tanh a\tanh b+\f{1}{\cosh b}\sin A}.
\ee With account for \be
1-\f{1}{\cosh^2 b}=\tanh^2 b,
\ee Eq.~(16.13) assumes the form \be
\f{\tanh b}{\tanh a}-\cos C=\cot A\f{\sin C}{\cosh b}.
\ee
In a similar manner one obtains the relation
\be
\f{\tanh a}{\tanh b}-\cos C=\cot B\f{\sin C}{\cosh a}.
\ee From the {\bf law of sines}\index{law of sines} we have \be \f{1}{\cosh b}=\f{\sin A}{\sin B}\cdot
\f{\tanh b}{\sinh a}.
\ee Substituting this expression into (16.15), we obtain \be 1-\f{\tanh a}{\tanh b}\cos C=
\f{\cos A\sin C}{\cosh a \sin B}.
\ee
Applying expressions (16.16) in (16.18), we find
\be
\cos A=-\cos B\cos C+\sin B\sin C\cosh a.
\ee In a similar manner one obtains the relations: \ba
\cos B=-\cos A\cos C+\sin A\sin C\cosh b\,,\nonumber\\
\label{16.20}\\
\cos C=-\cos A\cos B+\sin A\sin B\cosh c\,.\nonumber \ea

{\bf Thus, the space of velocities in
pseudo-Euclidean\index{pseudo-Euclidean geometry of space-time}
geometry
is the Lobachevsky\index{Lobachevsky} 
space}.

For a rectangular triangle $C=\ds\f{\,\pi\,}{\,2\,}$,
according to (16.4) we have
\be
\cosh c=\cosh a\cosh b.
\ee
From the theorems of sines, (16.5), and of cosines, (16.4)
we obtain
\be
\sin A=\f{\sinh a}{\sinh c},\; \cos A=\f{\tanh b}{\tanh c}.
\ee In line with the obvious equality \be
\sin^2 A+\cos^2 A=1
\ee
one can, making use of expressions (16.22) and (16.21),
obtain the relation
\be
\sin^2 A\cosh^2 b+\cos^2 A\f{1}{\cosh^2 a}=1.
\ee

Consider, as an example~\cite{16},\index{Smorodinsky} the phenomenon of
\textbf{light aber\-ration},\index{aberration of light} i.\,e. the change in direction of a beam of light, when
tran\-sition occurs from one inertial reference system to another. So, in two reference systems, moving with
respect to each other, the directions toward one and the same source $C$ will differ. Let $\theta$ and
$\theta^\prime$ be the angles at which the light from the source at point $C$ is seen from two inertial reference
systems $A$ and $B$, moving with respect to each other with a velocity $v$. In the Lobachevsky\index{Lobachevsky}
 velocity space we shall construct the triangle $ACD$ (see Fig.\,1), with angle $C$
equal to zero, since light has the limit velocity.

Now, we join points $A$ and $B$ by a line, and we drop a per\-pen\-di\-cu\-lar
to this line from point $C$. It will intersect the line at point $D$.
We denote the distance from point $A$ to point $D$ by $x$ and the
distance from point $D$ to $B$ by $y$.

\begin{figure}[th]
\setlength{\unitlength}{1mm}
\begin{center}
\begin{picture}(60,80)
\put(0,0){\line(1,0){60}}
\put(15,0){\line(1,5){15}}
\multiput(30,0)(0,3){25}{\line(0,1){1}}
\put(45,0){\line(-1,5){15}}
\put(13,-5){\bf \textit{A}}
\put(29,-5){\bf \textit{D}}
\put(44,-5){\bf \textit{B}}
\put(29,77){\bf \textit{C}}
\put(11,2){\mathversion{bold}\( \theta^\prime \)}
\put(17,2){\mathversion{bold}\( \alpha \)}
\put(22,2){\mathversion{bold}\( \pi/2 \)}
\put(40,2){\mathversion{bold}\( \theta \)}
\put(22,-5){\bf \textit{x}}
\put(37,-5){\bf \textit{y}}
\put(24,-10){\bf Fig. 1}
\end{picture}
\end{center}
\end{figure}
\vspace*{0.1mm}

Applying for given triangle $ACD$ the
{\bf law of co\-si\-nes} (16.20),\index{law of cosines}
we obtain
\be
\cosh x=\f{1}{\sin \alpha},\;\sinh x=\f{\cos \alpha}{\sin \alpha},
\ee hence \be \tanh x=\cos \alpha=\cos (\pi-\theta^\prime)=
-\cos \theta^\prime,
\ee
similarly
\be
\tanh y=\cos \theta.
\ee In accordance with formula (16.3), $\tanh (x+y)$ is the  velocity of one reference system with respect to the
other in units of the velocity of light \be \f{\,v\,}{\,c\,}=\tanh (x+y)=\f {\tanh x+\tanh y}{1+\tanh x\tanh y}=\f
{\cos \theta -\cos \theta^\prime}{1-\cos \theta\cdot \cos \theta^\prime}.
\ee
Hence follow the known formulae for
{\bf aberration}\index{aberration of light}
\be
\cos \theta^\prime=\f
{\ds\cos \theta-\f{\,v\,}{\,c\,}}{\ds 1-\f{\,v\,}{\,c\,}\cos \theta},
\ee
\be
\sin \theta^\prime=\sqrt{1-\f{v^2}{c^2}}\cdot
\f{\sin \theta}{\left(1-\ds\f{\,v\,}{\,c\,}\cos \theta\right)}.
\ee
Applying formulae (16.29) and (16.30) we obtain
\be
\cos (\theta-\theta^\prime)=\f
{\left(\cos \theta-\ds\f{\,v\,}{\,c\,}\right)\cos \theta+\sqrt{1-\ds\f{v^2}{c^2}}\sin^2 \theta}
{1-\ds\f{\,v\,}{\,c\,}\cos \theta}.
\ee

Let us determine the square distance between infinitesimally close
points in the
Lobachevsky\index{Lobachevsky} 
space.\index{Lobachevsky space}
From (16.1) we find
\be
\vec v^{\,\prime\, 2}=\f{(\vec u-\vec v)^2-\ds\f{1}{c^2}[\vec u, \vec v]^2}
{\left(1-\ds\f{\vec u\vec v}{c^2}\right)^2},
\ee
$v^\prime$ is the relative velocity.

Setting $\vec u=\vec v+d\vec v$ and substituting into (16.32)
we find
\be
(d\ell_v)^2=c^2\f
{(c^2-v^2)(d\vec v)^2+(\vec v d\vec v)^2}
{(c^2-v^2)^2}.
\ee

Passing to spherical coordinates in
velocity space\index{Lobachevsky space}
\be
v_x=v\sin \theta\cos \phi,\;v_y=v\sin \theta\sin \phi,\;
v_z=v\cos \theta,
\ee
we obtain
\be
\!\!\!(d\ell_v)^2= 
\ds c^2\left[\f{c^2(dv)^2}{(c^2-v^2)^2}+
\f{v^2}{(c^2-v^2)}
(d\theta^2+\sin^2 \theta d\phi^2)\right].
\ee

Hence it is evident that the ratio between the length of the circle
and the radius is
\be
\f{\,\ell\,}{\,v\,}=
\f{2\pi}{\sqrt{1-\ds\f{v^2}{c^2}}},
\ee
and is always greater than $2\pi$.

We now introduce the new variable
\be
r=\f{cv}{\sqrt{c^2-v^2}},
\ee
the range of which extends from zero to infinity. In the new
variables we have
\be
d\ell_v^2=\f{dr^2}{1+\ds\f{r^2}{c^2}}+
r^2(d\theta^2+\sin^2 \theta d\phi^2);
\ee
if we introduce the variable
\be
r=c\sinh Z,
\ee
we obtain
\be
d\ell_v^2=c^2dZ^2+c^2\sinh^2 Z(d\theta^2+\sin^2 \theta d\phi^2).
\ee
Usually the space metric in cosmology is written this form,
when dealing with the open
Universe.\index{Universe}

Further we shall dwell, in a descriptive man\-ner, on certain theorems of Lobachevsky's\index{Lobachevsky}
geometry,\index{Lobachevsky geometry} following the book 
by N.\,V.\,Efi\-mov\index{Efimov} ({\sf ``Higher geometry''} M.: Nauka, 1978 (in Russian)) and the
lectures of N.\,A.\,Chernikov\index{Chernikov}  delivered at the
Novosibirsk State University and published as a preprint in 1965.

In the Lobachevsky\index{Lobachevsky} 
geometry,\index{Lobachevsky geometry}
through point
{\bf \textit{A}},
not lying on
the straight line
{\bf \textit{a}},
there pass an infinite number of
straight lines, that do not intersect line
{\bf \textit{a}},
but not all
these straight lines are considered to be parallel to line
{\bf \textit{a}}.
Let $\mathbf{a}$ be a straight line in the plane, and let
{\bf \textit{A}}
be a point outside it (see Fig.\,2), {\bf \textit{b}}
and
{\bf \textit{c}}
are
boundary straight lines that do not intersect straight line
{\bf \textit{a}}.
Any straight line passing through point
{\bf \textit{A}}
inside angle
{\mathversion{bold}\( \beta \)}
will also not intersect straight line
{\bf \textit{a}},
while any straight line passing through point
{\bf \textit{A}}
inside the
angle containing point
{\bf \textit{B}}
will necessarily
intersect straight line
{\bf \textit{a}}.
The straight line
{\bf \textit{b}}
is called the right boundary straight line, and
{\bf \textit{c}}
the left
boundary straight line. It turns out to be that this property is conserved
for any point lying on straight line
{\bf \textit{b}}.
Precisely such a
boundary straight line
{\bf \textit{b}}
is parallel to
{\bf \textit{a}}
in the right-hand direction, and
{\bf \textit{c}}
in the left-hand direction.
\begin{figure}[th]
\setlength{\unitlength}{1mm}
\begin{center}
\begin{picture}(63,30)
\put(0,0){\thicklines\line(1,0){65}}
\put(2,15){\thicklines\line(6,1){60}}
\put(32,-5){\line(0,1){40}}
\put(2,25){\thicklines\line(6,-1){60}}
\put(33,22){\bf \textit{A}}
\put(27,-4){\bf \textit{B}}
\put(32,0){\circle*{1.5}}
\put(32,20){\circle*{1.5}}
\put(48,19){\mathversion{bold}\( \beta \)}
\put(33,16){\mathversion{bold}\( \alpha \)}
\put(63,1){\bf \textit{a}}
\put(2,12){\bf \textit{c}}
\put(2,26){\bf \textit{b}}
\put(28,9){\bf \textit{x}}
\put(24,-10){\bf Fig. 2}
\end{picture}
\end{center}
\end{figure}
Thus, two straight lines parallel to
{\bf \textit{a}}
can be drawn through any one point: one going to the right and the other to the left. In the
Lobachevsky\index{Lobachevsky} geometry,\index{Lobachevsky geometry} the
reciprocity theorem is proven: if one of two straight lines is parallel to the other in a certain direction, then
the second straight line is parallel to the first in the same direction. In a similar manner, it is established,
that two straight lines parallel to a third in a certain direction are parallel to each other, also, in the same
direction. Two straight lines, perpendicular to a third straight line, diverge. Two divergent straight lines
always have one common per\-pen\-di\-cu\-lar, to  both sides of which they diverge indefinitely from each another.

Parallel straight lines, indefinitely receding from each other in one
direction, asymptotically approach each other in the other. The angle
{\mathversion{bold}\( \alpha \)}
is called the parallelism angle at point
{\bf \textit{A}}
with respect to straight line {\bf \textit{a}}.

From the law of cosines\index{law of cosines}
(16.6) we find
\[
1=\sin \alpha\cosh x.
\]
In obtaining this expression we took into account that straight
line $\mathbf{b}$ asymptotically approaches straight line $\mathbf{a}$,
so, therefore, the angle between straight lines $\mathbf{a}$ and
$\mathbf{b}$ is zero. Hence we obtain
Lo\-ba\-chev\-sky's\index{Lobachevsky} 
formula\index{Lobachevsky formula}
\[
\alpha (x)=2\arctan e^{-x},
\]
here $a$ is the distance from point $\mathbf{A}$ to straight line
$\mathbf{a}$. This function plays a fundamental part in the
Lobachevsky\index{Lobachevsky} 
geometry.\index{Lobachevsky geometry}
This is not seen from our exposition, because we obtained the
Lo\-ba\-chev\-sky\index{Lobachevsky}
geometry\index{Lobachevsky geometry}
as the geometry
of velocity space. proceeding from
the pseudo-Euclidean\index{pseudo-Euclidean geometry of space-time} geometry of space-time. Function $\alpha (x)$
de\-cre\-a\-ses monotonously. The area of the triangle is
\be
S=d^2\cdot (\pi-A-B-C),
\ee here $d$ is a constant value. Below we shall derive this formula. From the formula it is evident that in the
Lobachevsky\index{Lobachevsky}  geometry\index{Lobachevsky geometry} similar
triangles do not exist.

Following
Lobachevsky,\index{Lobachevsky} 
we express the function
\be
\cos\bigtriangleup,\quad {\rm where}\quad
2\bigtriangleup=A+B+C,
\ee
via the sides of the triangle. Applying the
law of cosines (16.6)\index{law of cosines}
and, also, the formulae
\be
\sin^2\f{\,A\,}{\,2\,}=\f{1-\cos A}{2},\quad
\cos^2\f{\,A\,}{\,2\,}=\f{1+\cos A}{2},
\ee
we find
\be
\sin^2\f{\,A\,}{\,2\,}=
\f{\sinh (p-b)\cdot\sinh(p-c)}{\sinh b \sinh c},
\ee
\be
\cos^2\f{\,A\,}{\,2\,}=
\f{\sinh p\cdot\sinh (p-a)}{\sinh b\sinh c},
\ee
here $p$ is the half-perimeter of the triangle
\[
2p=a+b+c.
\]
With the aid of formulae (10.44) and (10.45) we obtain
\be
\sin\f{\,A\,}{\,2\,}\cos\f{\,B\,}{\,2\,}=
\f{\sinh (p-b)}{\sinh c}\cos\f{\,C\,}{\,2\,},
\ee
\be
\sin\f{\,B\,}{\,2\,}\cos\f{\,A\,}{\,2\,}=
\f{\sinh (p-a)}{\sinh c}\cos\f{\,C\,}{\,2\,}.
\ee Hence we have \be \sin\f{A+B}{2}= \f{\cosh\left(\ds\f{a-b}{2}\right)}
{\ds\cosh\f{\,c\,}{\,2\,}}\cos\f{\,C\,}{\,2\,}.
\ee
Applying the formulae
\be
\cos\f{\,A\,}{\,2\,}\cos\f{\,B\,}{\,2\,}=
\f{\sinh p}{\sinh c}\sin\f{\,C\,}{\,2\,},
\ee
\be
\sin\f{\,A\,}{\,2\,}\sin\f{\,B\,}{\,2\,}=
\f{\sinh (p-c)}{\sinh c}\sin\f{C}{2},
\ee we find \be \cos\f{A+B}{2}= \f{\cosh\left(\ds\f{a+b}{2}\right)}{\cosh\ds\f{\,c\,}{\,2\,}}
\sin\f{\,C\,}{\,2\,}.
\ee From (16.48) and (16.51) we have \be \cos\bigtriangleup= 2\f{\sinh\ds\f{a}{2}\sinh\f{b}{2}}{\cosh\ds\f{c}{2}}
\sin\f{\,C\,}{\,2\,}\cos\f{\,C\,}{\,2\,}.
\ee Replacing $\ds\sin\f{\,C\,}{\,2\,}\cos\f{\,C\,}{\,2\,}$ in (16.52) by the expressions from\break  Eqs.~(16.44) and (16.45) we find
\be \cos\bigtriangleup= \f{\sqrt{\sinh p\cdot\sinh (p-a)\sinh (p-b)\sinh (p-c)}}
{2\cosh\ds\f{\,a\,}{\,2\,}\cosh\f{\,b\,}{\,2\,}\cosh\f{\,c\,}{\,2\,}}.
\ee
From (16.41) we have the equality
\be
\sin\f{S}{2d^2}=\cos\bigtriangleup.
\ee Comparing (16.53) and (16.54) we obtain \be \sin\f{S}{2d^2}= \f{\sqrt{\sinh p\cdot\sinh (p-a)\sinh (p-b)\sinh
(p-c)}}
{2\cosh\ds\f{a}{2}\cosh\f{b}{2}\cosh\f{c}{2}}.
\ee In our formulae the sides $a,b, c$ are dimensionless quantities, in accordance with definition (16.3).
Eq.~(16.55) is the analog of the Heron\index{Heron} 
formula\index{Heron's formula} in
Euclidean\index{Euclid} geometry.\index{Euclidean geometry} From (16.52) the expression for the
area of the triangle can be written, also, in the form \be \sin\f{S}{2d^2}= \f{\sinh\ds\f{\,a\,}{\,2\,}\sinh\f{\,b\,}{\,2\,}}
{\cosh\ds\f{c}{2}}\sin C.
\ee
The area $S$ is expressed in dimensionless units, since the sides of
the triangle are dimensionless. In our exposition, the constant $d$ is
unity, on the basis of the
law of cosines (16.4).\index{law of cosines}

From formula (16.41) it follows that in the
Lobachevsky\index{Lobachevsky} 
geometry\index{Lobachevsky geometry}
the
area of a triangle cannot be indefinitely large, it is restricted to
the quantity $d^{\,2}\pi$. Thus, admitting the existence of a triangle
of indefinitely large area is equivalent to
Euclid's\index{Euclid}
parallelism axiom.
The areas of polygons can be indefinitely large in the
Lobachevsky\index{Lobachevsky} 
geometry.\index{Lobachevsky geometry}

The area of a spherical triangle in Euclidean\index{Euclid} geometry is \be
S_\bigtriangleup =R^2(A+B+C-\pi),
\ee
here $R$ is the radius of the sphere. Comparing this expression with
formula (16.41), we see that formula (16.41) can be derived from
formula (16.57), if the radius of the sphere is chosen to be imaginary
and equal to the value $R=id$. This circumstance was already noted by
Lambert.\index{Lambert} 

If one introduces the variables
\be
x=\f{v_x}{c},\quad y=\f{v_y}{c},\quad z=\f{v_z}{c},
\ee
then formula (16.33), for the
Lobachevsky\index{Lobachevsky} 
geometry,\index{Lobachevsky geometry}
in the $x, y$ plane
assumes the form
\be
\!\!\!\!\!\!\!\!(d\ell_v)^2=
c^2\f
{\ds (1-y^2)\cdot(dx)^2+
2xydxdy+
(1-x^2)\cdot(dy)^2}{\displaystyle(1-x^2-y^2)^2},
\ee
the quantities $x, y$ are called
Beltrami\index{Beltrami} 
coordinates\index{Beltrami coordinates}
 in the
Lo\-ba\-chev\-sky\index{Lobachevsky} 
geometry.\index{Lobachevsky geometry}

Passing to new variables $\xi, \eta$ with the aid of formulae
\be
x=\tanh \xi,\quad y=\f{\tanh \eta}{\cosh \xi},
\ee
and calculating the differentials
\[
dx=\f{\xi}{\cosh^2\xi},\quad dy= \f{1}{\cosh^2\eta\cosh\xi}d\eta - \f{\tanh\eta}{\cosh^2\xi}\sinh\xi d\xi,
\]
upon performing the required computations, we find
\be
(d\ell_v)^2=c^2(cosh^2\eta d\xi^2+d\eta^2).
\ee
The net of coordinate lines
\be
\xi={\rm const},\quad \eta={\rm const},
\ee
is orthogonal. The area of the triangle in these variables is
\be
S=\int\limits_{\;(\Delta)}\!\!\!\!\!\int\cosh\eta d\xi d\eta.
\ee For calculating the area of a triangle by formula (16.63) it is ne\-ce\-ssary to find the geodesic (extremal)
line\index{geodesic line} in the Lo\-ba\-chev\-sky\index{Lobachevsky} 
geometry\index{Lobachevsky geometry} in coordinates $\xi, \eta$. To this end we shall take advantage of the
principle of stationary action.

Length is
\ba
&&L=\ds\int ds=\int\sqrt{\cosh^2\eta\cdot d\xi^2+d\eta^2}=\nonumber\\
\label{16.64}\\
&&=\ds\int\limits_{\eta_1}^{\eta_2}d\eta\sqrt{\cosh^2\eta\cdot\xi^{\prime\,2}+1}.\nonumber \ea Hence the extremal
curve is found in accordance with the condition \be \delta L=\ds\int\limits_{\eta_1}^{\eta_2}
\f{\xi^\prime\cdot\cosh^2\eta\cdot\delta(\xi^\prime)} {\sqrt{\cosh^2\eta\cdot\xi^{\prime\,2}+1}}\cdot
d\eta=0,\quad
\xi^{\prime}=\f{d\xi}{d\eta}.
\ee
The variation $\delta$ commutes with differentiation, i.\,e.
\be
\delta(\xi^\prime)=(\delta\xi)^\prime;
\ee taking this into account and integrating by parts in the integral (16.65) we obtain \be \delta
L=-\ds\int\limits_{\eta_1}^{\eta_2}d\eta \delta\xi\f{d}{d\eta}\left(\f{\cosh^2\eta\cdot\xi^\prime}
{\sqrt{\cosh^2\eta\cdot\xi^{\prime\,2}+1}}\right)=0.
\ee
Here, it is taken into account that the variations $\delta\xi$
at the limit points of integration are zero.

From equality (16.67), owing to the variation $\delta\xi$ being arbitrary, it follows \be
\f{d}{d\eta}\left(\f{\cosh^2\eta\cdot\xi^\prime}
{\sqrt{\cosh^2\eta\cdot\xi^{\prime\,2}+1}}\right)=0.
\ee Hence we find the equation for the geodesic line\index{geodesic line} \be \f{\cosh^2\eta\cdot\xi^\prime}
{\sqrt{\cosh^2\eta\cdot\xi^{\prime\,2}+1}}=c;
\ee geodesic lines, as the shortest in the Lobachevsky\index{Lobachevsky} geometry,\index{Lobachevsky geometry} are straight lines in it.

Resolving this equation, we obtain
\be
\xi-\xi_0=\pm c\ds\int\f
{d\eta}{\cosh\eta\sqrt{\cosh^2\eta-c^2}}.
\ee
Changing the variable of integration
\be
u=\tanh\eta,
\ee
we find
\ba
&&\xi-\xi_0=\pm \ds\int\f{cdu}{\sqrt{(1-c^2)+c^2u^2}}=\nonumber\\
\label{16.72}\\
&&=\pm\ds\int\f{dv}
{\sqrt{1+v^2}}=\pm\arcsh v.\nonumber
\ea
Here
\be
v=\f{cu}{\sqrt{1-c^2}}.
\ee
It is suitable to take for variable $c$ the following notation:
 \be
c=\sin\delta.
\ee
Thus, the equation of a
geodesic line\index{geodesic line}
has the form
\be
\sinh(\xi-\xi_0)=\pm\tan\delta\cdot\tanh\eta.
\ee

Let us, now, construct a triangle in the $\xi,\,\eta$ plane (Fig. 3). The lines $AB$ and $AC$ are geodesic
lines,\index{geodesic line} that pass through point ($\xi_0,\,0$). The angles $A_1$ and $A_2$ are inferior to the
parallelism angle $\alpha$:
\[
A=A_1+A_2.
\]
From expression (16.75) we find the derivative of the
geodesic line\index{geodesic line}
$AC$ at point $\xi_0$
\be
\xi_0^\prime=-\tan\delta_2.
\ee
\vspace*{11mm}
\begin{figure}[th]
\setlength{\unitlength}{0.9mm}
\begin{flushleft}
\hspace*{1.5cm}
\begin{picture}(25,50)
\put(-4,0){\epsfxsize=8cm \epsfbox{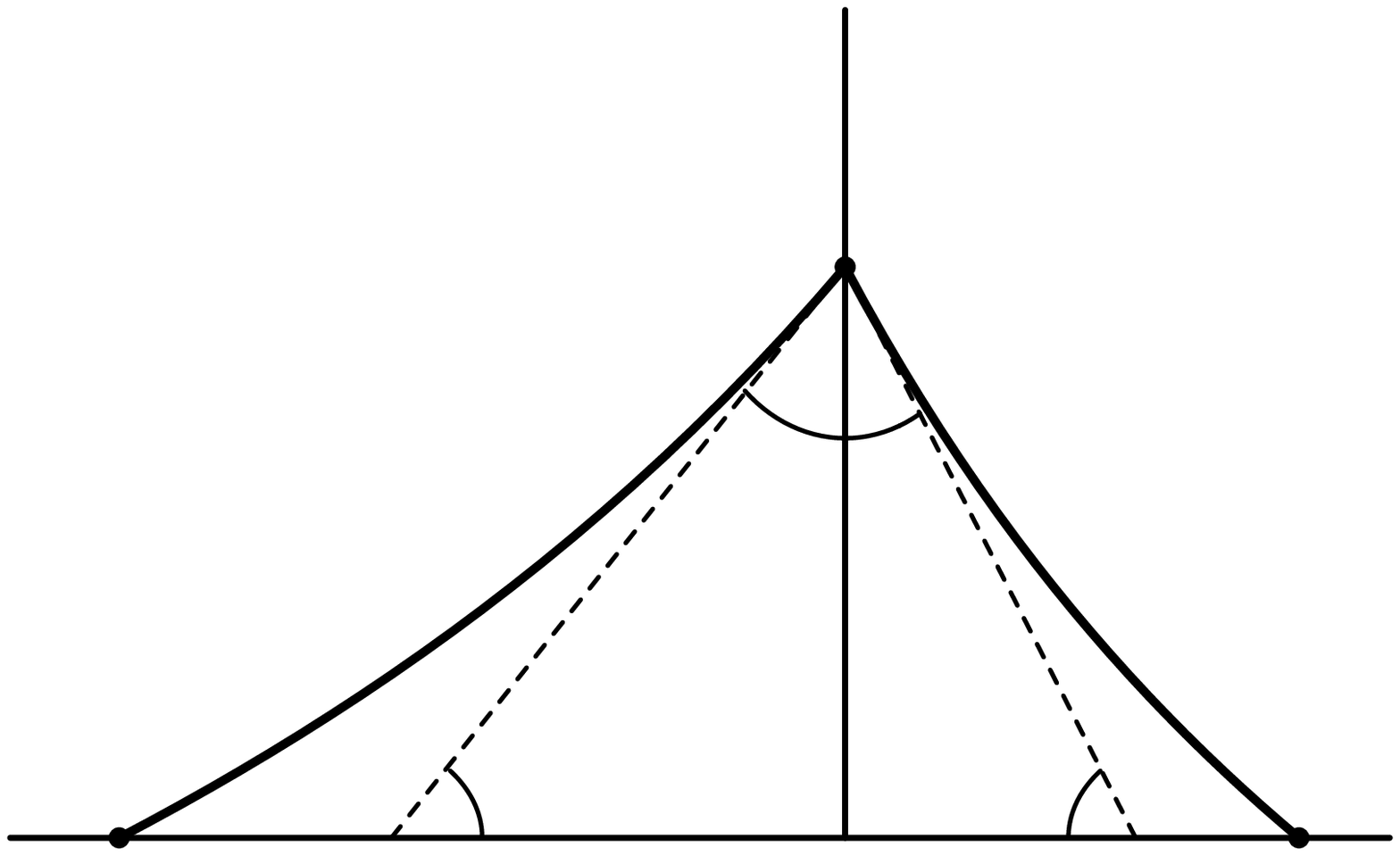}}
\put(52,50){\mathversion{bold}\( \xi \)}
\put(85,0){\mathversion{bold}\( \eta \)}
\put(44,39){\bf \textit{A}}
\put(51,39){\mathversion{bold}\( \xi_0 \)}
\put(41.5,22){\mathversion{bold}\( A_1 \)}
\put(50,22){\mathversion{bold}\( A_2 \)}
\put(2,-4){\bf \textit{B}}
\put(18,-4){\bf \textit{K}}
\put(28,4){\mathversion{bold}\( \delta_1 \)}
\put(20,20){\mathversion{bold}\( \eta_1 (\xi) \)}
\put(47,-4){\bf \textit{P}}
\put(65,-4){\bf \textit{L}}
\put(57,4){\mathversion{bold}\( \delta_2 \)}
\put(61,20){\mathversion{bold}\( \eta_2 (\xi) \)}
\put(75,-4){\bf \textit{C}}
\end{picture}
\end{flushleft}
\vspace*{7mm}
\centerline{\bf Fig. 3}
\end{figure}
Hence and from $\Delta ALP$  we have \be
\delta_2=\f{\pi}{2}-A_2,
\ee
similarly, from $\Delta AKP$, we also find for the
geodesic line $AB$\index{geodesic line}
\be
\delta_1=\f{\pi}{2}-A_1.
\ee
Thus, the constant {\bf \textit{c}} for each geodesic is expressed via the
angles $A_1,\,A_2$. The
geodesic lines\index{geodesic line}
$AB$ and $AC$ intersect the $\eta$
axis at points $\eta_1^0,\,\eta_2^0$.

In accordance with (16.63) the area of the triangle $ABC$ is \be S_\Delta=\ds\int\limits_0^{\xi_0}d\xi
\ds\int\limits_{\eta_1(\xi)}^{\eta_2(\xi)}\cosh\eta\cdot d\eta= \ds\int\limits_0^{\xi_0}\{\sinh\eta_2(\xi)-
\sinh\eta_1(\xi)\}d\xi.
\ee Taking advantage of expression (16.75), we find \be \sinh\eta=\pm\f{\sinh(\xi-\xi_0)}
{\sqrt{\cos^{-2}\delta-\cosh^2(\xi-\xi_0)}}.
\ee Hence we find \be \sinh\eta_2(\xi)=-\f{\sinh(\xi-\xi_0)}
{\sqrt{\sin^{-2}A_2-\cosh^2(\xi-\xi_0)}},
\ee \be \sinh\eta_1(\xi)=\f{\sinh(\xi-\xi_0)}
{\sqrt{\sin^{-2}A_1-\cosh^2(\xi-\xi_0)}}.
\ee Then the intersection points of the geodesic lines\index{geodesic line} with the straight line $\eta(\xi=0)$
are \be \sinh\eta_2^0=\f{\sinh\xi_0}
{\sqrt{\sin^{-2}A_2-\cosh^2\xi_0}},
\ee \be \sinh\eta_1^0=-\f{\sinh\xi_0}
{\sqrt{\sin^{-2}A_1-\cosh^2\xi_0}}.
\ee

From the
law of sines (16.5)\index{law of sines}
we have
\be
\sin B=\sinh\xi_0\cdot\f{\sin A_1}{\sinh|\eta_1^0|}.
\ee
Substituting into this expression the value of $\eta_1^0$ (16.84)
we obtain
\be
\sin B=\sqrt{1-\sin^2A_1\cosh^2\xi_0}\,,\quad\cos B=\sin A_1\cosh\xi_0.
\ee
Similarly
\be
\cos C=\sin A_2\cosh\xi_0.
\ee
Introducing the variable
\be
u=\cosh(\xi-\xi_0)
\ee in the integral (16.79), we obtain \be S_\Delta=\ds\int\limits_1^{\cosh\xi_0}
\left\{\f{1}{\sqrt{\sin^{-2}A_1-u^2}}+
\f{1}{\sqrt{\sin^{-2}A_2-u^2}}\right\}du.
\ee
Hence follows
\ba
&&S_\Delta=\arcsin (\sin A_1\cosh\xi_0)+\nonumber\\
\label{16.90}\\
&&+\arcsin (\sin A_2\cosh\xi_0)-(A_1+A_2).\nonumber \ea Taking into account (16.86) and (16.87), we obtain \be
S_\Delta=\arcsin (\cos B)+\arcsin (\cos C)-A.
\ee
Ultimately, we have
\be
S_\Delta=\pi-A-B-C.
\ee
We have obtained the expression for the area of a triangle $S_\Delta$
in the
Lobachevsky\index{Lobachevsky} 
geometry,\index{Lobachevsky geometry}
that we earlier (16.41) made use of in
finding formula (16.55).

From the above we saw that the
Lobachevsky\index{Lobachevsky} 
geometry,\index{Lobachevsky geometry}
created by him
as an ``imaginary geometry'', has become a composite part of the physics
of relativistic motions, as the geometry of velocity space.

The discovery of Lobachevsky\index{Lobachevsky}  had a great impact on the
de\-ve\-lop\-ment of various parts of mathematics. Thus, for example, the French mathematician
G.\,Hadamard,\index{Hadamard} in the book {\sf ``Non-Euclidean geometry''} in Section
devoted to the theory of automorphic functions noted:
\begin{quote}
{\it\hspace*{5mm}``We hope we have succeeded in showing, how
Lo\-ba\-chev\-sky's\index{Lobachevsky} 
discovery permeates throughout
Poincar\'{e}'s\index{Poincar\'e} 
entire remarkable creation,
for which it served, by the idea of
Poincar\'{e}\index{Poincar\'e} 
himself, as the foundation.
We are sure that
Lobachevsky's\index{Lobachevsky} 
discovery will play a great part, also,
at the further stages of development of the theory we have considered''.}
\end{quote}

Beltrami\index{Beltrami} 
raised the question: {\it ``Is it possible to realize
Lo\-ba\-chev\-sky\index{Lobachevsky} 
planimetry in the form of an internal geometry of a certain surface in
Euclidean\index{Euclid}
space?''}
Hilbert\index{Hilbert} 
has shown, that in
Euclidean\index{Euclid}
space no surface
exists, that is isometric to the {\bf entire}
Lobachevsky\index{Lobachevsky} 
plane. However,
part of the plane of the
Lo\-ba\-chev\-sky\index{Lobachevsky} 
geometry\index{Lobachevsky geometry}
can be realized in
Euclidean\index{Euclid}
space.

\newpage
\markboth{thesection\hspace{1em}}{}
\section*{Problems and exercises}\addcontentsline{toc}{section}{Problems and exercises}
\begin{center}
{\bf Section 2}\\[-0.1mm]
\end{center}
2.1. An electric charge is in a falling elevator. Will it emit
elec\-tro\-mag\-ne\-tic waves?\\
\index{electromagnetic waves}
\noindent
2.2. A charge is in a state of weightlessness in a space ship.
Will it radiate?\\[-1mm]
\begin{center}
{\bf Section 3}\\[-0.1mm]
\end{center}
3.1. Let the metric tensor\index{metric tensor of space} 
of Minkowski\index{Minkowski} 
space\index{Minkowski space} in a 
non-inertial\index{non-inertial reference systems} coordinate system have the
form $\gamma_{\mu\nu}(x)$. Show that there exists a coordinate system $x^\prime$, in which the metric tensor has
the same form $\gamma_{\mu\nu}(x^\prime)$, and that  nonlinear trans\-for\-mations relating these systems
constitute a
group.\index{group}\\[-1mm]
\begin{center}
{\bf Section 4}\\[-0.1mm]
\end{center}
4.1. Is the following statement correct: ``In a moving reference
system (with a constant velocity $v$) time flows slower, than in
a reference system at rest''?\\
\noindent
4.2. Is the
Lorentz\index{Lorentz} 
contraction\index{Lorentz contraction}
of a rod (4.13) real or
apparent?\\
\noindent 4.3. Is it possible, by making use of the Lorentz\index{Lorentz} effect
of contraction, to achieve a high density of substance by accelerating
a rod?\\[-1mm]
\begin{center}
{\bf Section 8}\\[-0.1mm]
\end{center}
8.1. The electric charge of a body is independent on the choice of reference system. On the basis of this
assertion find the trans\-for\-mation law of charge density, when transition occurs from one inertial
reference system to another.\\
\noindent
8.2. With the aid of
Lorentz\index{Lorentz} 
transformations find the field
of a charge undergoing uniformly accelerated motion.\\[-1mm]
\markboth{thesection\hspace{1em}}{Problems and exercises}
\begin{center}
{\bf Section 9}\\[-0.1mm]
\end{center}
9.1. Three small space rockets $A, B$ and $C$ are drifting freely in a region of space distant from other matter,
without rotation and without relative motion, and $B$ and $C$ are 
equidistant 
from $A$. When a signal is
received from $A$, the engines of $B$ and $C$ are switched on, and they start to smoothly accelerate. Let the
rockets $B$ and $C$ be identical and have identical programs of acceleration. Suppose $B$ and $C$  have been
connected from the very beginning by a thin thread. What will happen to the
thread? Will it break or not?\\
(Problem by J.\,Bell)\index{Bell} \\
\noindent
9.2. Let some device emits electromagnetic energy with power 6000 Watt in a definite direction. What force is
required due to the recoil to hold the device at rest?
\begin{center}
{\bf Section 10}\\[-0.1mm]
\end{center}
10.1. Applying the principle of stationary action obtain
the following formula for the
Lorentz\index{Lorentz}
force:
\[
\vec f=\rho\vec E+\f{\rho}{c}[\vec v, \vec H\,],
\]
where $\rho$ is the electric charge density.\\[-2mm]
\begin{center}
{\bf Section 11}\\[-0.1mm]
\end{center}
\noindent 11.1. Does a charge, moving along a geodesic 
line\index{geodesic line} in a uniformly accelerating
reference
system, radiate?\\
\noindent 11.2. Does a charge, moving along a geodesic 
line\index{geodesic line} in an arbitrary non-inertial
reference system,\index{non-inertial reference systems}
radiate?\\
\noindent 11.3. Does a charge, that is at rest in a non-inertial 
reference\index{non-inertial reference systems}
system, radiate?\\
\noindent
11.4. Does an elevator, the rope of which has been torn, represent
an inertial reference system?\\[-1mm]
\begin{center}
{\bf Section 12}\\[-0.1mm]
\end{center}
\noindent
12.1. Find the space geometry on a disk, rotating with a
constant angular velocity $\omega$.\\
\noindent 12.2. Consider an astronaut in a space ship moving with constant acceleration $a$ away from the Earth.
Will he be able to receive
in\-for\-mation from the Control Center during his trip?\\[-1mm]
\begin{center}
{\bf Section 16}\\[-0.1mm]
\end{center}
\noindent 16.1. Find a surface in the Lobachevsky\index{Lobachevsky} 
geometry,\index{Lobachevsky geometry} on which the Eucli\-de\-an\index{Euclid} planimetry is realized.\\
\noindent 16.2. Explain the Thomas\index{Thomas} precession with the aid of the
Lo\-ba\-chev\-sky\index{Lobachevsky} 
geometry.\index{Lobachevsky geometry}\\
\noindent 16.3. Does a triangle exist in the Lobachevsky\index{Lobachevsky}  geometry,\index{Lobachevsky geometry} all angles
of which equal zero?\\
\noindent 16.4. Find the area of a triangle on a sphere of radius $R$ in   Eucli\-de\-an\index{Euclid}
geometry.\\[-1mm]

\newpage
\markboth{thesection\hspace{1em}}{}

\newpage
\def\thispagestyle#1{}
\markboth{thesection\hspace{1em}}{}
\renewcommand{\indexname}{\Large\bf\textrm{Author Index}}

\begin{theindex}
  \markboth{thesection\hspace{1em}}{}
   \item {\it \textbf{A}do} Yurii Mikhailovich, born\break 1927, {\sl 4}
   \item {\it Aharonov} Yakir, born 1932,  {\sl 150}
   \item \textit{Ambartzumyan} Victor Ama\-zas\-po\-vich, 1908--1996,  
   {\sl 112}
   \indexspace
   \item {\it \textbf{B}ell} John Stewart, 1928--1990,  {\sl 240}
  \item {\it Beltrami} Eugenio, 1835--1900,  {\sl 232, 238}
  \item {\it Blokhintsev} Dmitrii Ivanovich,\break 1908--1979, {\sl 146}
  \item {\it Bohm} David, 1917--1992, {\sl 150}
  \item {\it Bolyai} J\'anos, 1802--1860, {\sl 6}
  \item {\it Born} Max, 1882--1970, {\sl 3, 121}
  \item {\it Broglie} Louis de, 1892--1987,\break {\sl 127, 133--135, 137}
  \indexspace
   \item {\it\textbf{C}ahn} William, 1912--1976, {\sl 121}
  \item {\it Chernikov} Nikolai Alek\-san\-dro\-vich, born 1928, 
  {\sl 227}
  \item {\it Christoffel} Elwin Bruno, 1829--1900, {\sl 159, 172}
  \item {\it Chugreev} Yurii Viktorovich,\break born 1960, {\sl 177}
  \newpage
  \item {\it Copernicus} Nicholas, 1473--1543, {\sl 50}
  \indexspace
  \item {\it \textbf{D}irac} Paul Adrien Maurice,\linebreak 1902--1984, 
  {\sl 112, 135, 136, 154}
  \item {\it Doppler} Christian Adreas, 1803--1853, 
                {\sl 110, 111}  
  \indexspace
  \item {\it \textbf{E}fimov} Nikolai Vladimirovich, 1910--1982, 
  {\sl 227}    
  \item {\it Einstein} Albert, 1879--1955, {\sl 3, 23, 25--28, 31, 33, 37,                   39, 40, 46, 
                48, 50, 52--64, 66, 74, 88, 89, 94, 101, 102, 
                107--109, 121--124, 128--130, 132--137, 141--143, 145, 146,
                168, 187}
   \item {\it Euclid}, born 325 BC, {\sl 5--7, 22, 57, 62, 103, 124, 231,                 238, 241}
   \item {\it Euler} Leonhard, 1707--1783, {\sl 16, 203}   
  \indexspace    
  \item {\it \textbf{F}araday} Michael, 1791--1867, {\sl 22}
  \item \textit{Fermi} Enrico, 1901--1954, {\sl 112}
  \item {\it Feynman} Richard Phillips, 1918--1988, {\sl 136}
  \item {\it FitzGerald} George Francis,\break 1851--191, {\sl 68}
  \item {\it Fock} Vladimir Alexandrovich,\break 1898--1974, {\sl 74, 164}
  \item {\it Frederix} Vsevolod Kon\-stan\-ti\-no\-vich, 1885--1944, 
  {\sl 146}
      \markboth{thesection\hspace{1em}}{Author Index}
  \indexspace    
   \item {\it \textbf{G}alilei} Galileo, 1564--1642, {\sl 7, 10, 23, 25,                  26, 41,  
                44, 45, 47, 48, 70, 104, 108, 111, 126, 129, 139, 164, 
                166, 176, 184, 189, 193, 195, 211}
  \item {\it Gauss} Carl Friedrich, 1777--1855, {\sl 90, 203}
  \item {\it Gershtein} Semion Solomonovich, born 1929, {\sl 4}
  \item {\it Ginzburg} Vitaly Lazarevich, born 1916, {\sl 51, 94, 127, 
                135, 137--139, 141, 143}
  \item {\it Goldberg} Stanley, 1934--1996,\break {\sl 130--132}    
   \indexspace     
   \item {\it \textbf{H}adamard} Jacques, 1865--1963, {\sl 237}
   \item {\it Hamilton} William Rowan, 1805--1861, {\sl 17--21}
  \item {\it Heron}, I AD, {\sl 231}
  \item {\it Hertz} Heinrich Rudolf, 1857--\break 1894, {\sl 120, 121}
    \item {\it Hilbert} David, 1862--1943, {\sl 206, 208, 210, 219, 238}
  \item {\it Hoffmann} Banesh, born 1906,\break {\sl 101, 144}
  \item {\it Holton} Gerald, born 1922, 
               {\sl 145}
  \item {\it Hooke} Robert, 1635--1703, {\sl 9}     
  \indexspace
  \item {\it \textbf{I}vanenko} Dmitri Dmitrievich\break 1904--1994, 
  {\sl 112, 146}
  \item {\it Ives} Herbet E.), 1882--1953, 
                {\sl 122, 124}
 \indexspace
  \item {\it \textbf{J}acobi} Carl Gustav Jacob, 1804--1851, 
  {\sl 20}
  \item {\it Jammer} Max, born. 1915, 
               {\sl 121}
    \indexspace
   \item {\it \textbf{K}illing} Wilhelm
    Karl Joseph,\break 1847--1923,        {\sl 213--215}
  
  \item {\it Kronecker} Leopold, 1823--1891, {\sl 78, 192}
  \indexspace
  \item {\it \textbf{L}agrange} Joseph Louis, 1736--1813, {\sl 12--15, 17--19, 21, 
                148--150, 203}
  \item {\it Lambert} Johann Heinrich, 1728--1777, {\sl 231}
    \item {\it Langevin} Paul, 1872--1946, {\sl 124, 142, 145, 146, 177}
  \item {\it Larmor} Joseph, 1857--1942, {\sl 39, 66, 142, 152}
  \item {\it Lebedev} Pjotr Nikolaevich, 1866--1912, {\sl 109}
  \item {\it Levi-Civita} Tullio, 1873--1941, {\sl 90}
  \item {\it Li\'{e}\-nard}   Alfred-Marie, 1869--1958,
{\sl 154}
  \item {\it Lie}, Marius Sophus, 1842--1899, {\sl 81, 205--207}
  \item {\it Lifshitz} Yevgeni Mikhailovich,\break 1915--1985, {\sl 51}
  \item {\it Lobachevsky} Nikolai Ivanovich, 1793--1856, {\sl 6, 75, 
                 220, 221, 223, 225, 227--229, 231--233, 237, 238, 241}
  \item {\it Lorentz} Hendrik Anton, 1853--1928, {\sl 3, 4, 23--28,                 30--34, 36--41, 43--55, 59, 61--63, 66--71, 73, 75, 80--85, 88, 89, 93--95, 97,      99--104, 113, 128--131, 133, 134, 136--140, 142, 143, 145, 146, 148--150, 153,       154, 163, 186, 197, 201, 212, 239, 240}
   \item {\it Lorenz} Ludwig Valentin, 1829--1891, {\sl 84}
  \indexspace
   \item {\it \textbf{M}ach} Ernst, 1838--1916, {\sl 11, 12}
  \item {\it Mandel'stam} Leonid Isaacovich, 1879--1944, {\sl 50, 51, 60}
  \item {\it Matveev} Viktor Anatolievich,\break born 1941, {\sl 147}
  \item {\it Maxwell} James Clerk, 1831--1879, {\sl 22, 27, 32, 33, 39, 44, 46, 
                48, 55, 83, 87, 88, 91--93, 99, 101, 116, 121, 128, 129,
                                148--150, 163, 186, 201}
  \item {\it Michelson} Albert Abraham, 1852--1931, {\sl 24, 38, 68, 177, 178}
  \item {\it Minkowski} Hermann, 1864--1909, {\sl 3, 4,  28, 45, 55, 56,                 58, 59, 62, 63, 76, 78, 87, 103, 108, 133, 146, 159,                  162, 164, 186, 188, 189, 191--193, 212, 214, 215, 239}
    \item {\it Morley} Edward Williams, 1838--1923, {\sl 68}   
      \indexspace     
    \item {\it \textbf{N}ewton} Isaac, 1643--1727, 
    {\sl 6--13, 105}
                \indexspace       
    \item {\it \textbf{P}ais} Abraham, 1918--2000, {\sl 128, 129, 132, 141,
    142}
  \item {\it Pauli} Wolfgang, 1900--1958, {\sl 3, 66, 132, 133, 135--137,
   141}
  \item {\it Petrov} Vladimir Alexeyevitch,\break born 1947, {\sl 4, 34, 113}
  \item {\it Planck} Max Karl Ernst Ludwig, 1858--1947, {\sl 110, 118,\break 124,
  143}
   \item {\it Poincar\'e} Henri, 1854--1912, {\sl 3, 4, 8, 22--34, 
                36, 40, 41, 44--47, 50--52, 55, 56, 61--64, 69,  
                80, 81, 85, 86, 89, 93--95, 97--99, 101, 102, 104, 106, 107,                 112,  118--121, 124, 127--147, 186, 187, 193, 212,
                220, 237}            
  \item {\it Poisson} Simeon Denis, 1781--\break 1840, {\sl 20, 21}
   \item {\it Ponsot} M., {\sl 142}     
  \item {\it Pontryagin} Lev Semionovich,\break 1908--1988, {\sl 147}
   \item {\it Poynting} John Henry), 1852--1914, {\sl 90--93, 117}   
  \item {\it Pythag\'oras}, born 570 BC, {\sl 5}          
  \indexspace            
  \item {\it \textbf{R}eichenbach} Hans, 1891--1953, {\sl 190}
    \item {\it Riemann} Georg Friedrich Bern\-hard, 1826--1866, {\sl 6, 58,                60, 62,  79, 157} 
  \item {\it Rybakov} Yurii Petrovich, born\break 1939, {\sl 156}
  \indexspace      
  \item {\it \textbf{S}agnac} Georges Marc Marie,\break 1869--1928, {\sl 4, 56,               177, 178, 181, 183}
  \item {\it Samokhin} Anatoli Petrovich, born 1950, {\sl 4}
  \item {\it Smorodinsky} Yakov Abramovich, 1917--1992, {\sl 223}
  \item {\it Sommerfeld} Arnold Johannes\break Wilhelm, 1868--1951,\break {\sl 63, 178}
  \item {\it Stark} Johannes, 1874--1957, {\sl 109}
  \newpage
  \item {\it\textbf{T}amm} Igor' evgen'evich, 1895--1971, 
               {\sl 91} 
  \item {\it Taylor} Brook, 1685--1731, {\sl 205, 207}
  \item {\it Terletsky} Yakov Petrovich, 1912--1993, {\sl 156}
  \item {\it Thomas} Llewellyn Hilleth, 1903--1992, {\sl 4, 55, 73, 200,                         241}
  \item {\it Tyapkin} Alexei Alexeevich, 1926--2003, {\sl 146, 191}
    \item {\it Tyurin} Nikolai Yevgenievich, born 1945, {\sl 4}
  \indexspace
  \item {\it \textbf{U}mov} Nikolaj Alekseevich, 1846--1915, {\sl 91}
  \indexspace
   \item {\it \textbf{V}oigt} Woldemar, 1850--1919,
    {\sl 28}
   \indexspace
    \item {\it \textbf{W}eyl} Hermann, 1885--1955, {\sl 50, 51, 101, 102}
  \item {\it Whittaker} Edmund Taylor, 1873--1956, {\sl 63}
  \indexspace
  \item {\it \textbf{Z}angger} Heinrich, 1874--1957, {\sl 101}
  \item {\it Zel'dovich} Yakov Borisocich, 1914--1987, {\sl 143}
  
\end{theindex}

\newpage
\markboth{thesection\hspace{1em}}{}
\renewcommand{\indexname}{\Large\bf\textrm{Subject Index}}
\begin{theindex}
\markboth{thesection\hspace{1em}}{}
  \textbf{A}
  \item aberration of light, 111, 223, 225
  \item absolute motion, 11, 25, 41
  \item absolute space, 7, 11, 12, 131
  \item absolute time, 7, 22, 131
  \item antiparticles, 112

  \indexspace
  \textbf{B}
  \item Beltrami coordinates, 232
  \item binding energy, 124

  \indexspace
   \textbf{C}
  \item ``clock paradox'', 4, 55, 167, 168
  \item Christoffel symbol, 159, 172
  \item conservation law of current, 83, 150
  \item contravariant tensor, 77
  \item contravariant vector, 77, 84, 85, 161
  \item coordinate quantities, 189, 195
  \item coordinate time, 182, 190, 192, 195
  \item coordinate velocity of light, 35, 194
  \item covariant derivative, 160--164,\break 206, 207
  \item covariant tensor, 77, 161
  \item covariant vector, 77, 86, 111, 196

  \indexspace
  \textbf{D}
  \item density of  Hilbert energy-mo\-men\-tum tensor, 208, 219
  \item density of canonical energy-mo\-men\-tum tensor, 209, 219
  \item density of spin tensor, 209, 219
  \item differential conservation laws, 204
  \item Dirac-Lorentz equation, 154
  \item displacement current, 22, 83
  \item Doppler effect, 110, 111

  \indexspace
  \textbf{E}
  \item electrodynamic constant, 42, 186, 193
  \item electromagnetic field invariants, 89
  \item electromagnetic waves, 23, 239
  \item equations of charge conservation, 22, 84
  \item equations of relativistic mechanics, 52, 106, 112, 121, 134, 143,
  162, 200
  \item ether, 25, 26, 28, 32, 33, 137, 139, 140, 144
  \item Euclidean geometry, 5, 6, 22,\break 124, 231
\markboth{thesection\hspace{1em}}{Subject Index}
  \indexspace
  \textbf{F}
  \item forces of gravity, 133, 159
  \item forces of inertia, 159
  \item forces of reaction, 154, 159
  \item four-dimensional space-time, 6, 29, 42, 55--58, 60, 76, 99, 
                103, 193
  \item four-force, 44, 97, 105, 124, 130
  \item four-momentum, 109, 110, 130
  \item four-vector, 85, 96, 97, 104, 105, 111, 125, 148, 153,
                   159, 196--198, 204 
  \item four-velocity, 44, 96, 105, 111, 125
  \item fundamental conservation laws, 56, 57, 101, 108, 219

  \indexspace
   \textbf{G}
  
  \item Galilean (Cartesian) coordinates, 4, 41, 44, 45, 47, 48, 104, 
                111, 129, 164, 166, 176, 184, 189,
                193, 195, 211, 214, 218
  \item Galilean transformations, 23, 26, 48, 104, 108
  \item Gauss theorem, 90, 203
  \item generalized coordinates, 13, 15
  \item generalized momentum, 17
{\baselineskip 13pt
  \item geodesic line, 126, 159, 171, 176, 232--236, 240
  \item gravitational waves, 46, 133, 134
  \item group, 30, 44, 46, 47, 51, 54, 62, 73, 77, 79--82, 88, 89, 99--101, 130, 
                134, 139, 145, 146, 153, 201, 212, 239
      \item group, Lorentz, 44, 46, 47, 51, 54, 62, 73, 81, 82, 99--101, 
                130, 134, 145, 146, 153, 201, 212
   \item group, Poincar\'{e}, 44, 62
  \indexspace
  \textbf{H}
  \item Hamiltonian, 17, 18
  \item Hamiltonian equations, 19
  \item Heron's formula, 231
  \item Hilbert energy-momentum tensor density, 212, 213
  \item Hooke's law, 9

  \indexspace
  \textbf{I}
  \item inert mass, 112, 121
  \item inertial reference systems, 4, 10--13, 24, 25, 27, 28, 30, 
                39, 41, 43--47, 49, 51, 55, 57, 59, 65--69, 73, 85,  
                111, 134, 138--140, 144, 179, 187, 190--193
 }
  \item integral conservation laws, 212
  \item interval, 45, 59, 66--68, 100, 107, 146, 157, 
                 166--169, 171, 173, 176, 179--182, 184, 185, 190, 191, 193
  \indexspace
  \textbf{J}
  \item Jacobian, 76, 205
  \indexspace
  \textbf{K}
  \item Killing equation, 213
  \item Killing vector, 214, 215
  \item Kronecker symbol, 78, 192
  \indexspace
  \textbf{L}
   \item Lagrangian, 202, 203, 205, 215, 218
  \item Lagrangian density, 15, 17, 148, 150, 201, 202, 205, 215, 
           216,  219
  \item Lagrangian equations, 15, 17--19, 148
  \item Lagrangian method, 21, 204
  \item Larmor formula, 152
  \item law of cosines, 221, 224, 228, 229, 231
  \item law of inertia, 126
  \item law of sines, 221, 222, 236
  \item Levi-Civita tensor, 90
  \item L.\,Lorenz condition, 84
  \item Lobachevsky formula, 228
  \item Lobachevsky geometry, 6, 220, 221, 227, 228, 231--233, 237, 238,
  241
  \item Lobachevsky space, 75, 225, 226
  \item local time, 27, 28, 31, 32, 34, 36, 37, 40, 41, 50--52, 134, 187
  \item Lorentz contraction, 4, 55, 69, 129, 239
{\baselineskip 13.5pt
  \item Lorentz force, 54, 93, 95, 150, 154
  \item Lorentz invariance, 52, 134
   \item Lorentz transformations, 4, 28, 30, 36--39, 41--44,
           46--50, 53, 55, 59, 61, 66, 67, 70, 71, 75, 80--82, 84, 85, 
                88, 89, 94, 95, 97, 99, 103, 104, 129--131, 133, 134, 136, 138--140,
                142, 197 
  \indexspace
  \textbf{M}
  \item mass defect, 124
  \item Maxwell-Lorentz equations, 33, 44, 46, 48, 55, 83, 99,  
                101, 128, 129, 134, 148--150, 163, 186, 201
  \item metric tensor of space, 58, 59, 62, 78, 162, 
                 184, 185, 188, 202, 207, 212, 216, 218, 239
  \item Minkowski space, 45, 56, 58, 62, 76, 78, 103, 108,  
                 159, 162, 164, 186, 188, 189, 191--193, 212, 214, 215, 239
  \indexspace
  \textbf{N}
  \item Newton's force of gravity, 9
  \item Newton's laws, 7--9
  \item non-inertial reference systems, 4, 55, 58, 59, 126, 167, 
                171, 173, 176, 178, 185, 192, 193, 239--241
}
  \indexspace
  \textbf{P}
  \item physical time, 27, 65, 66, 185, 191, 192
   \item physical vacuum, 26, 112
  \item Poincar\'{e}'s equations of mechanics, 46, 52, 54, 104, 106, 121,                      133, 134, 162
  \item Poisson bracket, 21
  \item potential, 12, 43, 44, 83--85, 89, 149, 150
  \item Poynting equation, 90--93, 117
  \item Poynting vector, 90, 93, 116
  \item principle of constancy of velocity of light, 3, 4, 39, 40, 50,  
                129, 131, 193
   \item proper time, 166, 169, 173, 174, 180, 182, 188, 196
  \item pseudo-Euclidean geometry of space-time, 4, 23, 46, 49, 54--57, 60, 62, 68, 69, 133, 134, 168, 
                176, 184, 187, 193, 200, 206, 212, 214, 215, 223, 228
  \item Pythagorean theorem, 5
  \indexspace
  \textbf{R}
  \item relating velocities, 75, 86, 177
  \newpage
  \item relativity principle, 3, 4, 10, 23, 25--30, 32, 33, 38, 40, 41, 
                44, 46, 47, 49--51, 54--56, 59, 65, 66, 69,
                                 85, 93, 94, 
                101, 104, 122, 128--131, 133, 134, 137--140, 143, 144, 146, 
                164, 187, 193
  \item Riemannian curvature tensor, 62, 157
  \item Riemannian geometry, 6, 60, 79
  \indexspace
  \textbf{S}
  \item simultaneity, 3, 32, 33, 37, 46, 50, 51, 68, 102, 134, 186, 
                 188--191,  194
  \item synchronization of clocks, 32, 36, 46, 50, 51, 189, 190, 194, 195
  \indexspace
  \textbf{U}
  \item Universe, 8, 45, 127, 227
  \indexspace
  \textbf{V}
  \item variation, Lie, 205--207
  \item variational derivative, 16, 203
  \item vector, spin, 196, 198, 200
  \item vector-potential, 150
  \indexspace
  \textbf{W}
  \item wave equation, 27, 41--43, 46

\end{theindex}

\newpage
\thispagestyle{empty}
\section*{Contents}\addcontentsline{toc}{section}{Contents}
 \markboth{thesection\hspace{1em}}{}
\thispagestyle{empty}
\vbox{ Preface  \dotfill 3}
\hspace*{-3mm} {1.\;Euclidean geometry \dotfill 5}\\
\hspace*{3mm} {2.\;Classical Newtonian mechanics \dotfill 7}\\
\hspace*{3mm} {3.\;Electrodynamics. Space-time geometry \dotfill 22}\\
\hspace*{3mm} {4.\; The relativity of time and the contraction of length \dotfill 65}\\
\hspace*{3mm} {5.\;Adding velocities \dotfill 75}\\
\hspace*{3mm} {6.\; Elements of vector and tensor analysis in Minkow-\\
\hspace*{11mm} {ski space \dotfill 76}\\
\hspace*{3mm} {7.\;Lorentz group \dotfill 80}\\
\hspace*{3mm} {8.\;Invariance of Maxwell-Lorentz equations \dotfill 83}\\
\hspace*{3mm} {9.\; Poincar\'{e}'s relativistic mechanics \dotfill 104}\\
\hspace*{3pt} {10.\;The principle of stationary action in electrodynamics
\dotfill 148}\\
\hspace*{3pt} {11.\;Inertial motion of а test body. Covariant
 {differentiation \dotfill 157}\\
\hspace*{3pt} {12.\;Relativistic motion with constant acceleration. The\\
\hspace*{11mm} {clock paradox. Sagnac effect \dotfill 165}\\
\hspace*{3pt} {13.\; Concerning the limiting velocity \dotfill 184}\\
\hspace*{3pt} {14.\;Thomas precession \dotfill 196}\\
\hspace*{3pt} {15.\;The equations of motion and conservation laws in\\
\hspace*{11mm} {classical field theory \dotfill 201}\\
\hspace*{3pt} {16.\;Lobachevsky velocity space \dotfill 220}\\
{Problems and exercises \dotfill 239}\\ 
{Bibliography \dotfill 242}\\ 
{Author Index \dotfill 244}\\ 
{Subject Index \dotfill 248} 

\end{document}